\documentclass[11pt]{article}

\usepackage{graphicx}
\usepackage{afterpage}
\usepackage{epsfig,cite}
\usepackage{amssymb}
\usepackage{amsmath}
\usepackage{dsfont}
\usepackage{multirow}

\usepackage{url}
\usepackage{hyperref}

\newcommand{\be}{\begin{equation}}
\newcommand{\ee}{\end{equation}}
\newcommand{\bea}{\begin{eqnarray}}
\newcommand{\eea}{\end{eqnarray}}
\newcommand{\bi}{\begin{itemize}}
\newcommand{\ei}{\end{itemize}}
\newcommand{\ben}{\begin{enumerate}}
\newcommand{\een}{\end{enumerate}}
\newcommand{\la}{\left\langle}
\newcommand{\ra}{\right\rangle}
\newcommand{\lc}{\left[}
\newcommand{\rc}{\right]}
\newcommand{\lp}{\left(}
\newcommand{\rp}{\right)}
\newcommand{\as}{\alpha_s}

\def\frac#1#2{{{#1}\over {#2}}}
\def\gsim{\mathrel{\rlap{\lower4pt\hbox{\hskip1pt$\sim$}}
    \raise1pt\hbox{$>$}}}         %greater than or approx. symbol
\def\lsim{\mathrel{\rlap{\lower4pt\hbox{\hskip1pt$\sim$}}
    \raise1pt\hbox{$<$}}}         %less than or approx. symbol

\newcommand{\dat}{\mathrm{dat}}

\newcommand{\rep}{\mathrm{rep}}
\newcommand{\net}{\mathrm{net}}

\newcommand{\tot}{\mathrm{tot}}

\newcommand{\gen}{\mathrm{gen}}

\newcommand{\val}{\mathrm{val}}
\newcommand{\tr}{\mathrm{tr}}

\newcommand{\draft}[1]{}

\def\nn{\nonumber}
\def \so{\sigma_I^{DIS}(x_I,Q^2_I)}

\def\sdy{\frac{d\sigma^{\mathrm{DY}}}{dQ_I^2dY_I}}
\def \npdf{N_{\mathrm{pdf}}}

\def \n0{N_j^{(0)}}
\def \a{\alpha}
\def \b{\beta}
\def \g{\gamma}
\def \c{\xi}
\def \z{\zeta}
\def\lapprox{\lower .7ex\hbox{$\;\stackrel{\textstyle <}{\sim}\;$}}
\def\gapprox{\lower .7ex\hbox{$\;\stackrel{\textstyle >}{\sim}\;$}}

\def\d{{\rm d}}

\def\xaa{x_{1}^{0}}
\def\xbb{x_{2}^{0}}

\begin{document}
\begin{flushright}
Edinburgh 2010/05\\
IFUM-952-FT\\
FR-PHENO-2010-014\\
CP3-10-08\\
\end{flushright}
\begin{center}
{\Large \bf A first unbiased global NLO determination\\ of parton distributions   and their uncertainties}
\vspace{0.8cm}

{\bf  The NNPDF Collaboration:}\\
Richard~D.~Ball$^{1}$,
 Luigi~Del~Debbio$^1$, Stefano~Forte$^2$,\\ Alberto~Guffanti$^3$, 
Jos\'e~I.~Latorre$^4$, 
Juan~Rojo$^2$ and Maria~Ubiali$^{1,5}$.

\vspace{1.cm}
{\it ~$^1$ School of Physics and Astronomy, University of Edinburgh,\\
JCMB, KB, Mayfield Rd, Edinburgh EH9 3JZ, Scotland\\
~$^2$ Dipartimento di Fisica, Universit\`a di Milano and
INFN, Sezione di Milano,\\ Via Celoria 16, I-20133 Milano, Italy\\
~$^3$  Physikalisches Institut, Albert-Ludwigs-Universit\"at Freiburg
\\ Hermann-Herder-Stra\ss e 3, D-79104 Freiburg i. B., Germany  \\
~$^4$ Departament d'Estructura i Constituents de la Mat\`eria, 
Universitat de Barcelona,\\ Diagonal 647, E-08028 Barcelona, Spain\\
~$^5$ Center for Particle Physics Phenomenology CP3,
Universit\'{e} Catholique de Louvain,\\ Chemin du Cyclotron, 1348 Louvain-la-Neuve, Belgium\\}
\end{center}

\vspace{0.8cm}

\begin{center}
{\bf \large Abstract:}
\end{center}

We present a determination of the parton distributions of
the nucleon from a global set of 
hard scattering data using the NNPDF 
methodology: 
NNPDF2.0. Experimental data include
deep--inelastic scattering with the combined HERA-I dataset, 
fixed target Drell-Yan production,
collider weak boson production and inclusive jet
production. Next--to--leading order QCD is used throughout without
resorting to $K$--factors. 
We present and utilize an improved fast algorithm for the solution
of  evolution equations and the computation of general hadronic
processes. We introduce improved techniques for the training of the neural
networks which are used as  parton parametrization, and we use a
novel approach for the proper treatment of normalization
uncertainties. We assess quantitatively the impact of individual datasets 
on PDFs. We find very good consistency of all datasets with 
each other and
with NLO QCD, with no evidence of tension between datasets. 
Some PDF combinations relevant for LHC
observables turn out to be determined rather more accurately
than in any other parton fit.

\clearpage

\tableofcontents

\clearpage

%----------------------------
%
% \section{Introduction}
%
%---------------------------------
\section{Introduction}

\label{sec-intro}

Over the last several years, we have developed a 
novel approach~\cite{Forte:2002fg} 
to the determination of parton
distribution functions (PDFs), which combines a 
Monte Carlo representation of the
probability measure in the space of PDFs with the use
of neural networks as a set of unbiased basis functions (the NNPDF
methodology, henceforth). The method 
was developed, refined,
and applied to problems of increasing complexity:  the parametrization
of a single structure function~\cite{Forte:2002fg}, of 
several structure functions~\cite{DelDebbio:2004qj} and
 the determination
of the nonsinglet 
parton distribution~\cite{DelDebbio:2007ee}. Eventually, in 
Ref.~\cite{Ball:2008by} a first complete set
of
parton distributions was constructed,
using essentially all the then--available deep--inelastic
scattering (DIS) data. This parton set, NNPDF1.0, included
five independent parton distributions (the two lightest flavours
and antiflavours and the gluon). It was then extended in
Refs.~\cite{Rojo:2008ke,Ball:2009mk} 
to also include an independent parametrization
of
the strange and antistrange quarks, with heavier flavours
determined dynamically (NNPDF1.2 parton set). All NNPDF  parton sets
are  available
through the LHAPDF interface~\cite{LHAPDFurl,Bourilkov:2006cj}. 
In these works, as well as in studies for the HERA--LHC
workshop~\cite{Dittmar:2009ii}, it was shown that PDFs determined using
the NNPDF methodology enjoy
several desirable features: the Monte Carlo behaves in a statistically
consistent way (e.g., uncertainties scale as expected with the size of
the sample)~\cite{Ball:2008by,Ball:2009mk}; results are demonstrably independent of the parton
parametrization~\cite{Ball:2008by,Ball:2009mk}; PDFs behave as 
expected upon the addition  of new
data (e.g. uncertainties expand when data are removed and 
shrink when they are added unless the new data is incompatible with
the old)~\cite{Ball:2008by,Dittmar:2009ii} and results are even stable
upon the addition of new independent PDF
parametrizations~\cite{Rojo:2008ke,Ball:2008by}. 

With PDF uncertainties under control, detailed 
precision physics studies become
possible,   
 such as for instance the determination of CKM
matrix elements~\cite{Ball:2009mk}. However, the requirements of
precision physics are such that it is mandatory to exploit all the
available information in PDF determination. Specifically, it has been
known for a long time (see Ref.~\cite{Ball:2008by} for references to
the earlier literature) that DIS data are insufficient to determine
accurately many aspects of PDFs, such as the flavour decomposition
of the quark and antiquark sea  or the gluon distribution, especially at
large $x$: indeed, the current state--of--the--art PDF determinations,
such as CTEQ6.6~\cite{Nadolsky:2008zw} and
MSTW2008~\cite{Martin:2009iq} are based on global fits, in which
hadronic data are included along with DIS data. 

In this paper we present a PDF determination using NNPDF methodology
based on a global fit. The data used for fitting include, on top of
all the data used in Ref.~\cite{Ball:2009mk} (DIS data and ``dimuon''
charm neutrino production data) also hadronic data, specifically
Drell--Yan (DY), W and Z production and Tevatron inclusive jets.
We also replace
the separate ZEUS and H1 datasets with the recently published
HERA-I combined dataset~\cite{H1:2009wt}. The dataset used
in this parton determination  is thus
comparable in variety and size (and is in fact slightly larger) to that used
by the CTEQ~\cite{Nadolsky:2008zw} and MSTW
groups~\cite{Martin:2009iq}. 

The PDF determination presented here is based on a consistent use of
NLO QCD. This is novel in the context of a global parton
determination: indeed, in other parton fits such as 
Refs.~\cite{Nadolsky:2008zw,Martin:2009iq} only DIS data are 
treated using fully NLO QCD, while several sets of hadronic data are treated
using LO theory improved through $K$--factors. The main bottleneck in the
use of NLO theory for hadronic processes is the speed in the
computation of hadron--level observables, which requires a convolution of the PDF of
both incoming hadrons with parton--level cross sections. The use of
Mellin--space techniques (as e.g. in Ref.~\cite{Alekhin:2006zm})
solves this problem, but at the cost of limiting the flexibility of
the acceptable PDF parametrization: specifically, the very flexible
neural network method of Refs.~\cite{Ball:2008by,Ball:2009mk}
parametrizes PDFs in $x$ space. Efficient fast methods to overcome
this hurdle have been suggested~(see~\cite{Carli:2005ji},
in~\cite{Dittmar:2005ed}), based on the idea of
precomputing and storing the convolution with a set of basis functions
over which any PDF can be expanded. These methods have been
implemented in fast public codes for specific processes, such
as
FastNLO~\cite{Kluge:2006xs}  for jet production, and very recently in
a general--purpose interface APPLGRID~\cite{Carli:2009rw}. 

In this paper, we use similar ideas to fully exploit the 
powerful parton evolution method introduced in
Ref.~\cite{DelDebbio:2007ee}, based on the convolution of PDFs 
with a pre--computed kernel, determined using Mellin--space
techniques. This gives us a new 
approach, which we call the
FastKernel method,  which we use
both for parton evolution, and for the computation of DIS and DY
physical observables.
The FastKernel method leads to
a considerable increase in speed in comparison to
Refs.~\cite{Ball:2008by,Ball:2009mk} for DIS data, and it makes possible
for the first time to
use exact NLO theory for DY in a global parton fit.

Thanks to the FastKernel method, 
we are able to produce a first fully NLO global
parton set using NNPDF methodology: the NNPDF2.0 parton set. This
parton determination
enjoys the same desirable features of the previous NNPDF1.0 and
NNPDF1.2 PDF sets, with which in particular it is fully compatible,
though uncertainties are now significantly smaller, and in fact 
sometimes also rather smaller than those of other existing global 
fits. Thanks to the use of a
Monte Carlo methodology, it is possible to perform a  
detailed comparison of NNPDF2.0 PDFs with those of previous NNPDF fits, and in
particular to assess the impact of the various new 
aspects of this parton determination, both due to  improved
methodology and the use of more precise data and a wider dataset.
Perhaps the most striking feature of the NNPDF2.0 parton determination is
 the fact that it is free of tension between different datasets and
NLO QCD: in fact, whereas the addition of new data leads to sizable
error reduction, we do not find any evidence of any individual dataset
being incompatible with the others, nor for the
distribution of fit results  to contradict statistical
expectations. Specifically, any combination or subset of the data
included in  the global analysis can be fitted using the same
methodology, and results obtained fitting to various subsets of data
are all compatible with each other.

Whereas we refer to the previous NNPDF
papers~\cite{Ball:2008by,Ball:2009mk}  for a general introduction to
the NNPDF methodology, all the new aspects of the 
NNPDF2.0 parton determination are fully documented
in this paper. In particular, in Sect.~\ref{sec:expdata} we
discuss the features of the new data used here, and specifically the
kinematics of DY and jet data.
In Sect.~\ref{sec:evolution} we discuss in detail the FastKernel
method, and its application to parton evolution and the computation of
DIS and DY observables.
In Sect.~\ref{sec-minim} we discuss several improvements in the
techniques that ensure that the quality of the fit to different data
is balanced, which are made necessary by the greater complexity of the
NNPDF2.0 dataset.

Readers who are not interested in the details of parton determination
and the NNPDF methodology, and mostly interested in PDF  use 
should skip directly to Sect.~\ref{sec:results}, where
our results are presented. In this section,
after  comparing the NNPDF2.0 PDF set both with previous NNPDF sets and
with current MSTW and CTEQ PDFs, we turn to a series of studies of its
features. Specifically, we study possible non--gaussian
behaviour of our results by comparing standard deviations with
confidence level intervals; we assess one by one the impact on the new fit of
the aforementioned improved  fitting method, of an improved treatment
of normalization uncertainties discussed elsewhere and used
here~\cite{t0}, of the new combined HERA data, and of the addition of
either jet or DY data; we discuss the impact of positivity
constraints; and we discuss the dependence of our results on the value
of $\alpha_s$. 

Finally, in Sect.~\ref{sec:pheno} we perform
some preliminary studies of the phenomenological implications of this
PDF determination: after briefly summarizing the quality of the
agreement between data and theory for the processes used in the fit, we
reassess the implication of our improved strangeness determination for
the so--called NuTeV anomaly~\cite{Mason:2007zz,Davidson:2001ji}, and
we discuss some LHC standard candles. Some statistical tools and a
brief summary of factorization and kinematics for the DY
process are collected in appendices.

%------------------------------
%
%\section{Experimental data}
%
%------------------------------
%----------------------------------------------------------

\section{Experimental data}
\label{sec:expdata}

The NNPDF2.0 parton determination includes both deep-inelastic (DIS) data
and  hadronic Tevatron data for fixed--target Drell-Yan and collider 
weak vector boson and inclusive jet
production. The DIS dataset only differs from that used in the previous
NNPDF1.2~\cite{Ball:2009mk} 
PDF determination in the replacement of separate H1 and ZEUS datasets 
with the combined HERA-I dataset of Ref.~\cite{H1:2009wt}.

The treatment of experimental data in the present fit follows
Ref.~\cite{Ball:2008by}, with the  exception of 
normalization uncertainties, which are treated using the improved
method presented in Ref.~\cite{t0}, the so-called
$t_0$ method. 
All information on correlated systematic errors,when available, is included in
our fit.

In this section first  we introduce the dataset and the way
we construct the experimental covariance matrix. Then we discuss
the details of the new datasets used in the NNPDF2.0 analysis as
compared to previous work, and finally we show how the 
Monte Carlo generation of replicas of experimental data is used
to construct the sampling of the available experimental information.

\subsection{Dataset, uncertainties and correlations}
 \label{sec:dataset}

The dataset used for the present fit is summarized in
Table~\ref{tab:exp-sets}, where experimental data is separated into
DIS data, fixed target Drell-Yan production, collider
weak boson production and inclusive jet production.
For each dataset we provide the number of
points both before and after kinematic cuts, and their kinematic
ranges.  
The same kinematical cuts
as in~\cite{Ball:2008by} are applied to DIS data, while no cuts
are applied to the hadronic data: we impose
$Q^2 \ge Q_0^2=m_c^2=2$ GeV$^2$ and $W^2 \ge 12.5$ GeV$^2$.

\begin{figure}[ht]
\begin{center}
\epsfig{width=\textwidth,figure=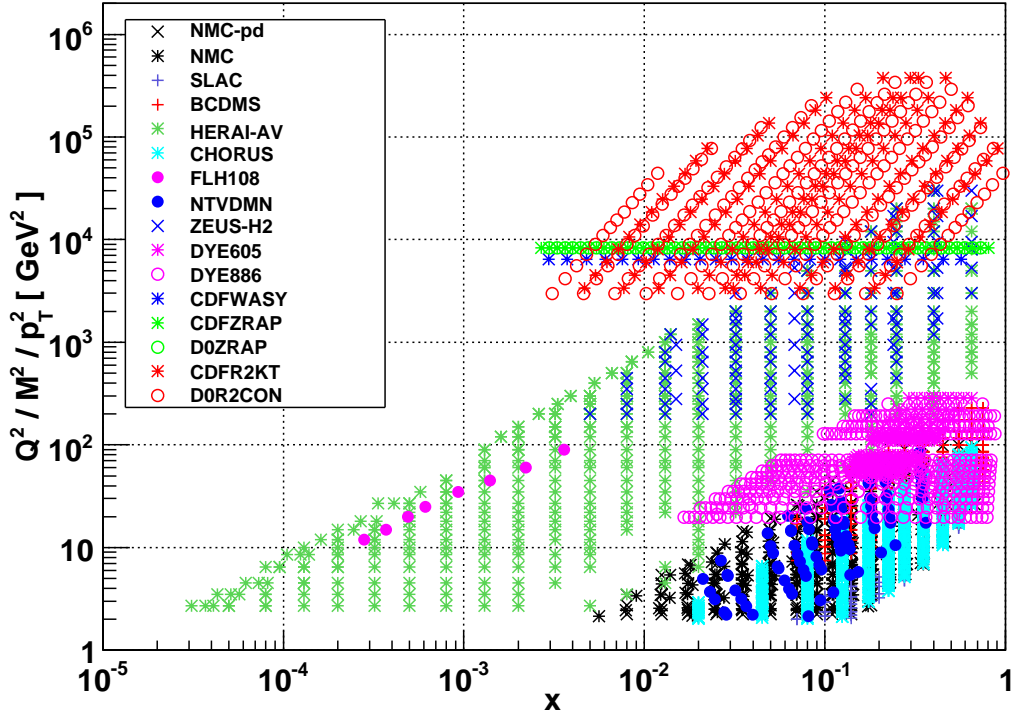}
\caption{ \small Experimental data
which enter the NNPDF2.0 analysis 
(Table~\ref{tab:exp-sets}). For hadronic
data, the values of $x_1$ and $x_2$ determined by
leading order partonic kinematics  
(Eqs.~(\ref{eq:ydef}),~(\ref{eq:xfdef2}) and~(\ref{eq:jetLO}))
are plotted (two values per data point).
\label{fig:dataplot}} 
\end{center}
\end{figure}

For hadronic data we use the LO partonic
kinematics to estimate the effective range of Bjorken-$x$ which
eaech dataset span (see Sect.~\ref{sec:expnewdata} below for a 
definition of the pertinent kinematic variables).
eeIn Fig.~\ref{fig:dataplot} we show a scatter plot of the data, which
demonstrates that the kinematic coverage is now
much more extended than in  the DIS--only NNPDF1.2 fit.

The DIS data of Table~\ref{tab:exp-sets} and Fig.~\ref{fig:dataplot}
differ from the NNPDF1.2 set because of the replacement of all ZEUS
and H1 data from the HERA-I run
with the combined set of Ref.~\cite{H1:2009wt}. 
The combined HERA-I dataset has a better accuracy than
that expected on purely statistical grounds from the combination of
previous H1 and ZEUS data because of the reduction of systematic errors from
the cross-calibration of
the two experiments.  These data are given with 110 correlated
systematic uncertainties 
and three correlated procedural uncertainties, which we
fully include in the covariance matrix.
The remaining  DIS data are the same as in
Ref.~\cite{Ball:2009mk}, to which we refer for further
details. Hadronic data are discussed in greater detail in
Sect.~\ref{sec:expnewdata} below.

In Table~\ref{tab:exp-sets-errors} we show the percentage
average experimental uncertainties for each dataset, where
uncertainties are separated into statistical (which includes
uncorrelated systematic), correlated systematic and
normalization uncertainties. As in the case of Table~\ref{tab:exp-sets},
for the DIS datasets we provide the values with and without kinematical
cuts, if different.

The covariance matrix is computed for all the data included in the fit, as
discussed in Ref.~\cite{Ball:2008by}. An important difference
in comparison to~\cite{Ball:2008by} is the improved treatment of
normalization uncertainties. Following~\cite{t0},
the covariance matrix for each experiment is computed from
the knowledge of statistical, systematic and normalization
uncertainties as follows:
\bea
\label{eq:covmat_t0}
({\rm cov_{t_0}})_{IJ}=
\lp\sum_{l=1}^{N_c}\sigma_{I,l}\sigma_{J,l}
+\delta_{IJ}\sigma_{I,s}^2\rp F_{I} F_{J}+
\lp
\sum_{n=1}^{N_a} \sigma_{I,n}\sigma_{J,n}+
\sum_{n=1}^{N_r} \sigma_{I,n}\sigma_{J,n}
\rp F_{I}^{(0)} F_{J}^{(0)}\ ,
\eea
where $I$ and $J$ run over the experimental points,
$F_{I}$ and $F_{J}$ are the measured
central values for the observables $I$ and
 $J$, and $F_{I}^{(0)}$, $F_{J}^{(0)}$ are the corresponding observables
as determined from some previous fit.  

The uncertainties, given as relative values, are:
$\sigma_{I,l}$, the $N_c$ correlated systematic uncertainties;
$\sigma_{I,n}$, the $N_a$ ($N_r$) absolute (relative) normalization
uncertainties; $\sigma_{I,s}$ the statistical uncertainties
(which includes uncorrelated systematic uncertainties). The values
of $F^{(0)}_I$ have been determined iteratively, by repeating the fit
and using for $F^{(0)}_I$ at each iteration the results of the previous
fit. In practice, convergence of the procedure is very fast and
the final values of $F^{(0)}_I$
used in Eq.~(\ref{eq:covmat_t0}) do not differ significantly from the 
final NNPDF2.0 fit results. Note that thanks to this iterative
procedure,  
normalization
uncertainties can be included in the covariance matrix as all other
systematics and therefore they do not require the fitting of shift parameters.

The use of this treatment of normalization uncertainties is necessary
because of the presence in the fit of data affected by disparate
normalization uncertainties: indeed, the simpler method used in
Refs.~\cite{Forte:2002fg}-\cite{Ball:2009mk} is only accurate~\cite{t0} when all
normalization uncertainties have a similar size.

% ----------------------------------------------------------------
 \begin{table}
 \tiny
 \centering
 \begin{tabular}{|c|c|c|c|c|c|c|c|}
  \hline
\multicolumn{8}{|c|}{\bf Deep-Inelastic scattering}\\
\hline
  Experiment & Set & Ref. &$N_{\rm dat}$ & $x_{\rm min}$ & $x_{\rm max}$
 & $Q^2_{\rm min}$ [GeV$^2$]  & $Q^2_{\rm max}$ [GeV$^2$]   \\ \hline
\hline
 NMC-pd          & & &          260 (153)  & & & & \\ \hline
 &  NMC-pd          & \cite{Arneodo:1996kd}& 260 (153) &      0.0015 (0.008) &      0.68 &  0.2 (2.2) &  99.0 \\
 \hline
 NMC             & &  &        288 (245)  & & & &  \\ \hline
 &  NMC             &  \cite{Arneodo:1996qe} & 288 (245) &      0.0035 (0.0056) &      0.47 &  0.8 (2.1) &  61.2  \\
 \hline
 SLAC            & &  &        422 (93)  & & & & \\ \hline
 &  SLACp           & \cite{Whitlow:1991uw} & 211 (47) &      0.07 &      0.85 (0.55) &        0.58 (2.0) &   29.2  \\
 \hline
 &  SLACd           & \cite{Whitlow:1991uw} & 211 (46) &      0.07 &      0.85 (0.55) &        0.58  (2.0)&   29.1  \\
 \hline
 BCDMS           & &   &       605 (581)  & & & &\\ \hline
 &  BCDMSp          &\cite{bcdms1}  &351 (333) &      0.07 &      0.75 &        7.5 &      230.0     \\
 \hline
 &  BCDMSd          &  \cite{bcdms1}  &254 (248) &      0.07 &      0.75 &        8.8 &      230.0     \\
 \hline
 HERAI-AV        & & &          741 (608) & & & & \\ \hline
 &  HERA1-NCep      & \cite{H1:2009wt} & 528 (395) &      $6.2\,10^{-7}$ ($3.1\,10^{-5}$) &      0.65 &        0.045 (2.7) &    30000  \\
 \hline
 &  HERA1-NCem      & \cite{H1:2009wt} & 145 &      $1.3\,10^{-3}$ &      0.65 &       90.000 &    30000  \\
 \hline
 &  HERA1-CCep      & \cite{H1:2009wt} &  34 &      0.008 &      0.4 &      300.0 &    15000 \\
 \hline
 &  HERA1-CCem      & \cite{H1:2009wt} &  34 &      0.013 &      0.4 &      300.0 &    30000  \\
 \hline
 CHORUS          & &  &       1214 (942) & & & & \\ \hline
 &  CHORUSnu        & \cite{Onengut:2005kv} & 607 (471) &      0.02 &      0.65 &        0.3 (2.0) &       95.2  \\
 \hline
 &  CHORUSnb        & \cite{Onengut:2005kv} & 607 (471)&      0.02 &      0.65 &        0.3 (2.0) &       95.2 \\
 \hline
 FLH108          & &  &          8  & & & & \\ \hline
 &  FLH108          & \cite{h1fl} &  8 &      0.00028 &      0.0036 &       12.0 &       90.000 \\
 \hline
 NTVDMN          & &   &        90 (84) & & & &\\ \hline
 &  NTVnuDMN        & \cite{Goncharov:2001qe,MasonPhD} & 45 (43) &      0.027 &      0.36 &        1.1 (2.2)&      116.5 \\
 \hline
 &  NTVnbDMN        & \cite{Goncharov:2001qe,MasonPhD} & 45 (41)&      0.021 &      0.25 &        0.8 (2.1) &       68.3 \\
 \hline
 ZEUS-H2         & &   &       127  & & & & \\ \hline
 &  Z06NC           &  \cite{Chekanov:2009gm}& 90 &      $5\,10^{-3}$ &      0.65 &      200 &    $3\,10^5$  \\
 \hline
 &  Z06CC           & \cite{Chekanov:2008aa} & 37 &      0.015 &      0.65 &      280 &    $3\,10^5$ \\
 \hline
 \hline
\multicolumn{8}{|c|}{\bf Fixed Target Drell-Yan production}\\
\hline
 Experiment & Set & Ref. & $N_{\rm dat}$ & $\lc y/x^F_{\rm min},y/x^F_{\rm max}\rc $ &  $\lc x_{\rm min}, x_{\rm max}\rc$
 & $M^2_{\rm min}$ [GeV$^2$]  & $M^2_{\rm max}$ [GeV$^2$]
   \\ \hline
\hline
 DYE605          & &           &  119 & & &\\ \hline
 &  DYE605          & \cite{Moreno:1990sf} & 119 &     $\lc -0.20,0.40\rc$  & $\lc 0.14,0.65\rc$  &    50.5 &      286 \\
 \hline
 DYE866          & &   &       390  & & & &\\ \hline
 &  DYE866p         & \cite{Webb:2003ps,Webb:2003bj} &184 &     $\lc  0.0, 0.78\rc$ &  $\lc 0.017,0.87\rc$ &     19.8 &      251.2\\
 \hline
 &  DYE866r         & \cite{Towell:2001nh} & 15 &     $\lc 0.05 ,0.53\rc$  & $\lc 0.025,0.56\rc$ &       21.2 &      166.4 \\
 \hline
 \hline
\multicolumn{8}{|c|}{\bf Collider vector boson production}\\
\hline
 Experiment & Set & Ref.& $N_{\rm dat}$ & $\lc y_{\rm min},y_{\rm max}\rc $ &  $\lc x_{\rm min}, x_{\rm max}\rc$
 & $M^2_{\rm min}$ [GeV$^2$] & $M^2_{\rm max}$ [GeV$^2$]
  \\ \hline
 CDFWASY         & &   &        13  & & & & \\ \hline
 &  CDFWASY         & \cite{Aaltonen:2009ta}  & 13 &      $\lc 0.10, 2.63\rc$ &  $\lc 2.9\,10^{-3},0.56\rc$ &    6463 &     6463   \\
 \hline
 CDFZRAP         & &    &       29  & & & & \\ \hline
 &  CDFZRAP         & \cite{Abazov:2007jy} &  29 &     $\lc  0.05 ,2.85\rc$ & $\lc 2.9\,10^{-3},0.80\rc$ &     8315 &     8315 \\
 \hline
 D0ZRAP          & &   &        28  & & & &\\ \hline
 &  D0ZRAP          & \cite{Aaltonen:2009pc} & 28 &     $\lc  0.05, 2.75\rc $ & $\lc 2.9\,10^{-3},0.72\rc$ &    8315 &     8315 \\ 
 \hline
 \hline
\multicolumn{8}{|c|}{\bf Collider inclusive jet production}\\
\hline
 Experiment & Set & Ref. & $N_{\rm dat}$ &  $\lc y_{\rm min},y_{\rm max}\rc $ &  $\lc x_{\rm min}, x_{\rm max}\rc$
 & $p^{2}_{T,\rm min}$ [GeV$^2$] & $p^2_{T,\rm max}$   [GeV$^2$] \\ \hline
 CDFR2KT         & &    &       76  & & & &\\ \hline
 &  CDFR2KT         & ~\cite{Abulencia:2007ez}  & 76 &      $\lc 0.05 ,  1.85\rc$ & 
$\lc 4.6\,10^{-3},0.90\rc$  &     3364 &   $3.7\,10^5$  \\
 \hline
 D0R2CON         & &    &      110  & & & &\\ \hline
 &  D0R2CON         & \cite{D0:2008hua} & 110 &     $\lc  0.20, 2.20\rc $ & $\lc 3.1\,10^{-3},0.97\rc$  & 3000 & $3.4\,10^5$ \\
 \hline
 \hline
\multicolumn{8}{|c|}{\bf Total}\\
\hline
 Experiment  & &&  $N_{\rm dat}$ & $x_{\rm min}$ & $x_{\rm max}$
 & $Q^2_{\rm min}$ [GeV$^2$]  & $Q^2_{\rm max}$ [GeV$^2$]    \\ \hline
 TOTAL           &&& 4520 (3415)   &  $3.1\,10^{-5}$  & 0.97 & 2.0 & $3.7\,10^5$\\ \hline

 \end{tabular}
\caption{\small \label{tab:exp-sets} Experimental datasets included in the NNPDF2.0 global analysis. For DIS experiments we provide in each case the number
of data points and the ranges of the kinematical variables
before and after (in parenthesis) kinematical cuts. 
For hadronic
data we  show the ranges of parton $x$ covered for each
set (denoted by $\lc x_{\rm min}, x_{\rm max} \rc$), determined using 
leading order parton kinematics
(Eqs.~(\ref{eq:ydef}),~(\ref{eq:xfdef2}) and~(\ref{eq:jetLO})). 
Note that  hadronic data are unaffected by
kinematic cuts. The values of
$x_{\rm min}$ and $Q^2_{\rm min}$ for the total dataset hold after
imposing
kinematic cuts.}
 \end{table}
% -----------------------------

% ----------------------------------------------------------------
 \begin{table}
 \small
 \centering
 \begin{tabular}{|c|c|c|c|c|}
  \hline
\multicolumn{5}{|c|}{\bf Deep-Inelastic scattering}\\
\hline
 Set 
& $\la \sigma_{\rm stat}\ra$ (\%) & 
   $\la \sigma_{\rm sys}\ra$  (\%) & $\la \sigma_{\rm norm}\ra$  (\%)
& $\la \sigma_{\rm tot}\ra$ (\%)  \\ \hline
\hline
   NMC-pd          &   2.0 (1.7) &   0.4 (0.2)  &   0.0 &   2.1 (1.8) \\
   NMC             &  3.7 (3.7) &   2.3 (2.1)&   2.0 &   5.0 (4.9)\\
   SLACp           &   2.7 (3.8)&   0.0 &   2.2 &   3.6 (4.5) \\
   SLACd           &   2.5 (3.4)&   0.0 &   1.8 &   3.1 (3.9) \\
  BCDMSp          &  3.2 (3.1) &   2.0 (1.7) &   3.2 &   5.5 (5.2) \\
   BCDMSd         &   4.5 (4.4)&   2.3  (2.1)&   3.2 &   6.6 (6.4) \\
   HERA1-NCep      & 4.0 &   1.9 (1.5) &   0.5 &   4.7 (4.5) \\
   HERA1-NCem      &   10.9  &   1.9 &   0.5 &  11.2 \\
   HERA1-CCep      &  11.2 &   2.1 &   0.5 &  11.4 \\
   HERA1-CCem      &  22.3 &   3.5 &   0.5 &  22.7 \\
   CHORUSnu        &   4.2 (4.1) &   6.4 (5.8) &   7.9 (7.6) &  11.2 (10.6) \\
   CHORUSnb        &  13.8 (14.9)&   7.8 (7.5) &   8.7 (8.2)&  18.7 (19.1) \\
   FLH108          &  47.2 &  53.3 &   5.0 &  71.9 \\
   NTVnuDMN        &  16.2 (16.0) &   0.0 &   2.1 &  16.3 (16.2) \\
   NTVnbDMN        &  26.6 (26.4) &   0.0 &   2.1 &  26.7 (26.5) \\
   Z06NC           &   3.8 &   3.7 &   2.6 &   6.4 \\
   Z06CC           &  25.5 &  14.3 &   2.6 &  31.9 \\
 \hline
 \hline
\multicolumn{5}{|c|}{\bf Fixed Target Drell-Yan production}\\
\hline
Set & $\la \sigma_{\rm stat}\ra$ (\%)  & 
   $\la \sigma_{\rm sys}\ra$ (\%)  & $\la \sigma_{\rm norm}\ra$ (\%) 
& $\la \sigma_{\rm tot}\ra$ (\%)  \\ \hline
 DYE605          &  16.6 &   0.0 &  15.0 &  22.6 \\
  DYE866p        &  20.4 &   0.0 &   6.5 &  22.1 \\
  DYE866r        &  3.6 &   1.0 &   0.0 &   3.8 \\
 \hline
 \hline
\multicolumn{5}{|c|}{\bf Collider vector boson production}\\
\hline
  Set &$\la \sigma_{\rm stat}\ra$  (\%) & 
   $\la \sigma_{\rm sys}\ra$  (\%) & $\la \sigma_{\rm norm}\ra$  (\%)
& $\la \sigma_{\rm tot}\ra$ (\%)   \\ \hline
  CDFWASY       &   4.2 &   4.2 &   0.0 &   6.0 \\
   CDFZRAP         &   5.1 &   6.0 &   6.0 &  11.5 \\
   D0ZRAP          &   7.6 &   0.0 &   6.1 &  10.2 \\
 \hline
 \hline
\multicolumn{5}{|c|}{\bf Collider inclusive jet production}\\
\hline
 Set 
& $\la \sigma_{\rm stat}\ra$ (\%) & 
   $\la \sigma_{\rm sys}\ra$ (\%)  & $\la \sigma_{\rm norm}\ra$  (\%)
& $\la \sigma_{\rm tot}\ra$  (\%) \\ \hline
   CDFR2KT         &    4.5 &  21.1 &   5.8 &  23.0 \\
 D0R2CON         &    4.4 &  14.3 &   6.1 &  16.8 \\
 \hline
 \end{tabular}
\caption{\small \label{tab:exp-sets-errors} Average
statistical, systematic and normalization uncertainties for
each of the experimental datasets included in NNPDF2.0. Uncorrelated systematic uncertainties are
considered as part of the statistical uncertainty.
All uncertainties are given in percentage. Details on the number of
points and the kinematics of each dataset are provided in
Table~\ref{tab:exp-sets}. For DIS experiments average uncertainties
are given both before and (in parenthesis) after cuts.}
 \end{table}
% -----------------------------

\subsection{New experimental observables}
\label{sec:expnewdata}

The hadronic observables used in the NNPDF2.0 PDF determination
correspond to three classes of 
processes: Drell--Yan production in fixed target
experiments, collider weak vector boson
production, and collider inclusive jet production. For
each type of process we briefly introduce the leading order structure
of the observables and 
kinematics  used in 
Tab.~\ref{tab:exp-sets} and Fig.~\ref{fig:dataplot}, 
then discuss the features of the data. Full NLO
expressions for Drell-Yan observables are summarized in
Appendix~\ref{sec:dyobservables}, and their fast implementation is
presented in detail in Sect.~\ref{sec:evolution}. For jet observables,
we interfaced our code with FastNLO~\cite{Kluge:2006xs}, by direct
inclusion of the precomputed tables from this reference, to which we
refer for explicit expressions for the cross--sections.

\subsubsection{Drell-Yan production on a fixed target} 

We consider data for the double--differential distribution 
 in $M$, the invariant mass of the Drell-Yan lepton pair,
and  either the rapidity of the pair $y$ or Feynman $x_F$,
respectively defined in terms of the hadronic kinematics as
\begin{equation}\label{xfydef}
 y
\equiv \frac{1}{2} \ln\frac{q_0+q_z}{q_0-q_z};\quad  x_F \equiv
\frac{2q_z}{\sqrt{s}} , 
\end{equation} 
where $\sqrt{s}$ is the hadron--hadron center-of-mass energy,
$q$ is the four-vector of the Drell-Yan pair and $q_z$ is its
projection on the longitudinal axis. 

At leading order, the parton kinematics is entirely fixed in terms of
hadronic variables by
\begin{equation}
\xaa = \sqrt{\tau} e^{y}=\frac{M}{\sqrt{s}}e^y \ ,
\qquad \xbb = \sqrt{\tau} e^{-y}=\frac{M}{\sqrt{s}}e^{-y} \ , 
\label{eq:ydef}
\end{equation}
or equivalently
\be
\xaa = \frac{1}{2} \left( x_F + \sqrt{x_F^2+4\tau} \right), \qquad
\xbb = \frac{1}{2} \left( -x_F + \sqrt{x_F^2+4\tau} \right) \ .
\label{eq:xfdef2}
\ee
The corresponding inverse relations are
\begin{equation}
 \tau =\xaa\xbb; \quad M^2=s\xaa\xbb 
\label{eq:invkin}
\end{equation}
and
\begin{equation}
y = \frac{1}{2} \ln \frac{\xaa}{\xbb};\qquad
x_F \equiv x_1^0 - x_2^0
\label{eq:yxfdefinv}
\end{equation}

At leading order, the $y$ or $x_F$ Drell-Yan differential 
distribution is given by
\begin{eqnarray}
\label{eq:dylo-data}
\frac{\d   \sigma}{\d M^2 \d y}(M^2,y) &=& \frac{4\pi \alpha^2}{9 M^2 s} 
\sum_i e_i^2 \lc   q_i(x_1,M^2) \bar{q}_i(x_2,M^2) 
+ \bar{q}_i(x_1,M^2)  q_i(x_2,M^2)\rc\ ,\\
\frac{\d   \sigma}{\d M^2 \d x_F}(M^2,y) &=&
\frac{1}{x_1^0+x_2^0}\frac{\d   \sigma}{\d M^2 \d y}(M^2,y) ,
\end{eqnarray}
where $\alpha$ is the fine--structure  constant and
$e_i$ the quark electric charges.

The fixed--target Drell-Yan data used for our parton determination are:

\begin{itemize}

\item E605 

This experiment
provides the absolute cross section
for DY production from a  proton beam on a copper
target~\cite{Moreno:1990sf}. The double
differential distribution in $y$ and $M^2$ is given.
 No correlation matrix is provided, 
and only a total systematic
uncertainty $\sigma_{\rm sys}= 10\%$ 
is given. Therefore, we will add statistical
and total systematic errors in quadrature. The only source
of correlation between the data points comes from the
absolute normalization uncertainty of 15\%. We do not apply any
nuclear corrections, which we expect~\cite{Ball:2009mk} to be small.

\item E866

This experiment, also known as NuSea, is based
on the experimental set-up of the previous DY experiments
E605~\cite{Moreno:1990sf} and E772~\cite{McGaughey:1994dx}. 
The absolute cross section measurements
on a proton target is
described in~\cite{Webb:2003ps,Webb:2003bj}, while the cross section ratio
between deuteron and proton targets can be found in~\cite{Towell:2001nh}.
Double
differential distributions in  $x_F$ and $M$ are provided. 
No correlation matrix is provided, and only a total systematic
uncertainty is given, so we add statistical
and total systematic errors in quadrature. The only source of correlation
comes from the $6.5\%$ absolute normalization uncertainty, which
cancels in the cross--section 
ratio~\cite{Towell:2001nh}.

\end{itemize}

Note that we do  not include fixed target Drell-Yan data from the E772
experiment~\cite{McGaughey:1994dx} nor from the deuteron
data of E866~\cite{Webb:2003ps,Webb:2003bj}. These datasets
have been shown to have poor compatibility with other
Drell-Yan measurements~\cite{Alekhin:2006zm}
and thus do not add additional
information to the global PDF analysis. As we have shown
elsewhere~\cite{Forte:2002fg}-\cite{Ball:2009mk}, within NNPDF
methodology the addition of incompatible data only increases uncertainties,
and thus these data are not included. The issue of their
compatibility with other Drell-Yan data will be addressed elsewhere.

\subsubsection{Weak boson production}

We consider the  rapidity distributions for 
 $W$ and $Z$ production. At  leading order, the parton  kinematics is
as
in Eqs.~(\ref{xfydef})-(\ref{eq:yxfdefinv}), and the differential 
distribution is given by
\be
\label{eq:wprod-data}
\frac{\d \sigma}{\d y} =  \frac{\pi G_F M_V^2\sqrt{2}}{3 s} \sum_{i,j} c_{ij} 
 \lc  q_i(x_1,M_V^2) \bar{q}_j(x_2,M_V^2) +  
\bar{q}_i(x_1,M_V^2) q_j(x_2,M_V^2)\rc,
\ee
where $M_V$ denotes either $M_W$ or $M_Z$;  the  electroweak
couplings are
\begin{eqnarray}
& c_{ij} = |V_{ij}| & \mbox{for $W^{\pm}$}\; ,\nonumber \\
& c_{ij} = (v_i^2 + a_i^2) \delta_{ij} & \mbox{for $Z^0$ unpolarized  }\ ,
\label{eq:ewcoup}
\end{eqnarray}
where $|V_{ij}|$ are  CKM matrix elements and $v_i$, $a_i$ the
$Z-$boson  vector and axial couplings.

The weak boson production data included in our parton determination are:
\begin{itemize}

\item D0 $Z$ rapidity distribution

This measurement, performed at Tevatron Run II and
described in Ref.~\cite{Abazov:2007jy}, gives
the $Z/\gamma^*$ rapidity distribution in the
range $71 \le M_{ee} \le 111$ GeV. The contribution from the
$Z^0/\gamma^*$ interference terms is well below the experimental
uncertainties and it is  neglected.
No correlation matrix is provided, so
we add in quadrature systematic and statistical
uncertainties. The only correlated systematic error is the
absolute normalization uncertainty from the
Tevatron luminosity, $6.1\%$.

\item CDF $Z$  rapidity distribution 

This observable is analogous to its D0 counterpart, and it
is described in Ref.~\cite{Aaltonen:2009pc}.
For this experiment, $N_{\rm sys}=11$ 
independent correlated systematic
uncertainties are provided, which have been used in the construction
of the covariance matrix.

\item CDF $W$ boson asymmetry 

This measurement, also performed at Tevatron Run II, is
described in Ref.~\cite{Aaltonen:2009ta}.
For this dataset, $N_{\rm sys}=7$ independent 
correlated systematic uncertainties are quoted, from which
the experimental correlation matrix can be constructed.
The physical observable is the rapidity asymmetry 
\be
\label{eq:asyw1}
A\lp y_W\rp\equiv \frac{ d\sigma^{W^+}/dy_W - d\sigma^{W^-}/dy_W}{
d\sigma^{W^+}/dy_W + d\sigma^{W^-}/dy_W} \ .
\ee
Since the $A(y_W)$ distribution is symmetric at the Tevatron, the experimental
data is folded onto positive rapidities to improve
statistics.

\end{itemize}

Because of the lack of a fast analytic implementation,
we do not include lepton--level data, such as the Tevatron
$W$ asymmetries
Refs.~\cite{Acosta:2005ud,Abazov:2008qv}, which have been included in
recent parton fits~\cite{Martin:2009iq,Nadolsky:2009xu} using
$K$--factors. The recent development of the
APPLGRID~\cite{Carli:2009rw}  interface is likely to facilitate the
future inclusion of these data in our fits.

\subsubsection{Inclusive jet production}

We include data for the  inclusive jet production cross section as a
function  of the transverse momentum $p_T$ of the jet  for fixed
rapidity bins $\Delta \eta$. The leading--order parton kinematics is fixed by
\be
\label{eq:jetLO}
x_1^0=\frac{p_T}{\sqrt{s}}e^{\eta} \ , 
\quad x_2^0=\frac{p_T}{\sqrt{s}}e^{-\eta},
\ee
while a simple leading--order expression for the cross--section is
not available because of the need to provide a jet algorithm.

We include the following data:

\begin{itemize}
\item CDF Run II --- $k_T$ algorithm 

This data is obtained using the $k_T$ algorithm with 
$R=0.7$. The dataset and the various sources
of systematic uncertainties have been described in 
Ref.~\cite{Abulencia:2007ez}. We choose to use the
$k_T$ algorithm measurements rather than the cone
algorithm measurements~\cite{Aaltonen:2008eq}, since the 
latter are not infrared safe. Data at $R=0.7$ are
preferable to available measurements at $R=0.5$
or $R=1$ since at Tevatron energies $R=0.7$ optimizes
the interplay between sensitivity to perturbative
radiation and impact of non-perturbative effects like 
Underlying Event~\cite{Dasgupta:2007wa,Cacciari:2008gd}.

The data is provided in bins of rapidity $\Delta\eta$ and
transverse momentum $p_T$. The kinematical coverage can be
seen in Table~\ref{tab:exp-sets}. On top of the absolute
normalization uncertainty of $5.8\%$, which is fully correlated among all
bins, there are $N_{\rm sys}=28$ 
sources of systematic uncertainty,
fully correlated among all bins of $p_T$ and $\eta$, used
to construct the covariance matrix.

\item D0 Run II --- midpoint algorithm

This dataset is obtained using the MidPoint algorithm with 
$R=0.7$. The dataset and the various sources
of systematic uncertainties have been described in Ref.~\cite{D0:2008hua}. 
While the MidPoint algorithm is IRC unsafe, the effects
of such unsafety in inclusive distributions are smaller than
typical uncertainties~\cite{Salam:2007xv} and thus it is safe to
include this dataset into the analysis.

The data is provided in bins of rapidity $\Delta\eta$ and
transverse momentum $p_T$. The kinematical coverage can be
seen in Table~\ref{tab:exp-sets}. On top of the absolute
normalization uncertainty of $6.1\%$, which is fully correlated among all
bins, there are $N_{\rm sys}=23$ 
sources of systematic uncertainty.

\end{itemize}

No inclusive jet measurements from Run
I~\cite{Abbott:2000kp,Affolder:2001fa} 
are included. Although their
consistency with Run I data has been debated in the
literature~\cite{Pumplin:2009nk,Martin:2009iq}, Run II data have
increased statistics, are obtained with a better understanding of the detector,
and are provided with the different sources of systematic
uncertainties. The issue of the Tevatron jet data compatibility will
be discussed elsewhere; for the time being, we have checked
that the NNPDF2.0 fit yields a description of Run I
jet data which is  reasonably close to that of 
CTEQ6.6~\cite{Nadolsky:2009xu}, which included
such datasets. This suggests that no tension between data should arise
when these older data are included in  the fit.

\subsection{Generation of the pseudo--data sample}

Following Ref.~\cite{Ball:2008by},
error propagation from experimental data to the fit is handled
by a Monte Carlo sampling of the probability distribution defined by
data. The statistical sample is obtained by generating $N_{\rm rep}$
artificial replicas of data points following a multi-gaussian
distribution centered on each data point with the variance given by
the experimental uncertainty as discussed in Sect.~2.4 of Ref.~\cite{Ball:2008by}.

Appropriate statistical estimators have been devised in
Ref.~\cite{DelDebbio:2007ee} in order to quantify the accuracy of the
statistical sampling obtained from a given ensemble of replicas (see
Appendix~B of Ref.~\cite{DelDebbio:2007ee}). Using these estimators, 
we have verified that a Monte Carlo sample of
pseudo-data with $N_{\rep}=1000$ is sufficient to reproduce the mean
values, the 
variances, and the correlations of experimental data with a 1\%
accuracy for all the experiments.
The statistical estimators for the Monte Carlo generation of
artificial replicas of the experimental data are shown for
each of the datasets included in the fit in 
Tables~\ref{tab:mc1} and~\ref{tab:mc2}.

\begin{table}
\centering
\begin{tabular}{|c|c|}
\hline
$r\lc F\rc$ & 1.00000  \\
\hline
 $\la \sigma^{(\exp)}
\ra_{\dat}$(\%) &  11.3\\
 $\la \sigma^{(\gen)}
\ra_{\dat}$(\%)&  11.4\\
 $r\lc \sigma^{(\gen)}
 \rc$ &  0.99996  \\
\hline
 $\la \rho^{(\exp)}
\ra_{\dat}$ &  0.176\\
 $\la \rho^{(\gen)}
\ra_{\dat}$&  0.179\\
 $r\lc \rho^{(\gen)}
 \rc$ &  0.99676  \\
\hline
\end{tabular}
\caption{\small \label{tab:mc1} Table of statistical estimators for the
Monte Carlo sample of $N_{\rm rep}=
1000$ replicas. All estimators are defined 
in Appendix B of Ref.~\cite{DelDebbio:2007ee}. Note that
uncertainties are given as percentages.}
\end{table}

{
\begin{table}
\centering
\small
\begin{tabular}{|c|c|c|c|c|c|c|c|}
\hline 
Experiment   & $ r[F] $   & $\la \sigma^{(\exp)}\ra_{\dat}$ (\%) & $\la \sigma^{(\gen)}\ra_{\dat}$ (\%) & $ r[\sigma] $   & $\la \rho^{(\exp)}\ra_{\dat}$ & $\la \rho^{(\gen)}\ra_{\dat}$ & $ r[\rho] $  \\
\hline 
NMC-pd          &  1.000&   1.78&   1.72& 0.999& 0.03& 0.03& 0.963\\
\hline
NMC             &  1.000&   4.91&   4.89& 0.998& 0.16& 0.16& 0.987\\
\hline
SLAC            &  1.000&   4.20&   4.16& 0.999& 0.31& 0.29& 0.986\\
\hline
BCDMS           &  1.000&   5.73&   5.70& 0.999& 0.47& 0.46& 0.994\\
\hline
HERAI-AV        &  1.000&   7.52&   7.53& 1.000& 0.07 & 0.07  & 0.951\\
\hline
CHORUS          &  1.000&  14.83&  14.92& 0.999& 0.09 & 0.09 & 0.998\\
\hline
FLH108          &  1.000&  71.90&  70.78& 1.000& 0.64& 0.63& 0.997\\
\hline
NTVDMN          &  1.000&  21.22&  21.10& 0.998& 0.03& 0.03 &  0.978 \\
\hline
ZEUS-H2         &  1.000&  13.79&  13.56& 1.000& 0.28& 0.28& 0.994\\
\hline
DYE605          &  1.000&  22.60&  23.11& 1.000& 0.47& 0.48& 0.983\\
\hline
DYE866          &  1.000&  20.76&  20.73& 1.000& 0.20& 0.19& 0.989\\
\hline
CDFWASY         &  1.000&   5.99&   6.06& 0.999& 0.55& 0.53& 0.995\\
\hline
CDFZRAP         &  1.000&  11.51&  11.52& 1.000& 0.82& 0.82 & 0.999\\
\hline
D0ZRAP          &  1.000&  10.23&  10.50& 0.999& 0.53& 0.54& 0.995 \\
\hline
CDFR2KT         &  1.000&  22.97&  22.92& 1.000& 0.77& 0.77& 0.998 \\
\hline
D0R2CON         &  1.000&  16.82&  17.18& 1.000& 0.78& 0.78& 0.997 \\
\hline
\end{tabular}
\caption{\small \label{tab:mc2} Same as Table \ref{tab:mc1} for 
individual experiments.  Note that
uncertainties are given as percentages.}
\end{table}
}

%--------------------------------------------------------

% ------------------------------------------------------
%
%\section{Theoretical treatment}
%
% ----------------------------------------------------
%-----------------------------------------------------------------------------------------------------------

\section{The FastKernel method}
\label{sec:evolution}

One of the main upgrades in the NNPDF analysis framework  used for
this paper has been a new fast implementation of the method for the solution of
DGLAP evolution equations and the computation of factorized
observables developed in
Refs.~\cite{DelDebbio:2007ee,Ball:2008by}, which we call the
FastKernel method. The method of Refs.~\cite{DelDebbio:2007ee,Ball:2008by} 
is based on the
idea of pre--computing a Green function which takes PDFs from their
initial scale to the scale of physical observable.  The Green function
can be determined in $N$ space, thereby requiring a single
(complex--space) integration for the solution of the evolution
equation. Furthermore, the Green function can be pre--combined
with the hard cross sections (coefficient functions) 
into a suitable kernel, in such a way that 
the computation of any observable is reduced to the
determination of the convolution of this kernel with the pertinent
parton distributions, which are parametrized in $x$ space using neural
networks as discussed in Refs.~\cite{DelDebbio:2007ee,Ball:2008by}.
For hadronic observables, which depend on two PDFs, a double
convolution must be performed.

The main bottleneck of this method is the computation of these 
convolutions. In the FastKernel method, the convolution is
sped up  by means of the use of 
interpolating polynomials, thereby leading to both fast evolution and
fast computation of all observables for which the kernels have been
determined. This allows us to use in the fit an exact computation of
the Drell--Yan (DY) process, which in other
current global 
PDF fits~\cite{Nadolsky:2008zw,Martin:2009iq} is instead treated  using 
a $K$--factor approximation to the NLO (and even
NNLO) result, due to lack of a
fast--enough implementation.

Several tools for fast evaluation of hadronic observables
have been developed recently, based on an idea of
Ref.~\cite{Carli:2005ji}. These have been implemented 
for the case of jet production and
related observables in 
the FastNLO framework~\cite{Kluge:2006xs}. More recently, the 
general--purpose interface  APPLGRID based on the same idea has been
constructed~\cite{Carli:2009rw}. Also, the method has been used in the
fast $x$--space DGLAP evolution code HOPPET~\cite{Salam:2008qg}. A related
approach in the case of polarized observables is presented in
Ref.~\cite{deFlorian:2009vb}.
The method which is presented in this paper is based on similar ideas,
and  it allows for the first time the fast and accurate
computation of fixed target Drell--Yan cross--sections and of collider
weak boson production.

In this section we start with a description of the new strategy
used to solve the PDF evolution equations in the
present analysis, as well as the associated technique to compute DIS
structure functions. Then we turn to discuss how analogous
techniques can be used for  the fast and accurate
computation of hadronic observables. 
Although the method is completely general, for simplicity
we restrict the discussion to the Drell--Yan process, since 
for inclusive jets FastNLO will be used instead~\cite{Kluge:2006xs}.

\subsection{Fast PDF evolution}
\label{sec:fastpdfevol}

The notation we adopt here
is similar to that of Ref.~\cite{Ball:2008by}; however here we use
the index $I$ to denote both the kinematical variables which define
an experimental point $(x,Q^2)$ and the type of observable, while
in
Ref.~\cite{Ball:2008by}
$I$ was only labelling observables.

Before sketching the construction of the observables, we look at PDF
evolution. PDFs can be written in terms of the basis defined in
Ref.~\cite{Ball:2008by}:
\begin{equation}
\label{eq:basis}
f_j=\{\Sigma,g,V,V_3,V_8,V_{15},V_{24},V_{35},T_3,T_8,T_{15},T_{24},T_{35}\}.
\end{equation}
As in Ref.~\cite{Ball:2008by}, we do not consider the possibility of
intrinsic heavy flavours, so that only seven of these basis functions
(the six lightest flavours and the gluon) need to be independently
parametrized. 
If $\Gamma_{jk}$ is the matrix of  DGLAP evolution kernels and
$(x_I,Q_I^2 )$ defines the kinematics of a given experimental point,
we can write the PDF evolved from a fixed initial scale $Q_0^2$ to the
scale of the experimental point as 
\begin{equation} 
f_j(x_I,Q_I^2)\, =
\sum_{k=1}^{\npdf}\,\int_{x_I}^1\,\frac{dy}{y}\,
\Gamma_{jk}\left(\frac{x_I}{y},Q_0^2,Q_I^2\right)\,f_k(y,Q_0^2).
\label{initq}
\end{equation}
In  Ref.~\cite{Ball:2008by}, the
integral in Eq.~(\ref{init}) was performed numerically by means of a
gaussian sum on a grid of points distributed between $x_I$ and 1,
chosen according to the value of $x_I$.  Here instead  we use
 a single grid in $x$, independent of
the $x_I$ value. We label the set of
points in the grid as $x_{\a}$ by $\a=1,...,N_x$, with
\be
x_{\mathrm{min}}\equiv x_1<x_2<...<x_{N_x-1}<x_{N_x}\equiv 1.
\nonumber
\ee 
Having chosen a grid of points, we define a set of interpolating functions $\mathcal{I}^{(\a)}$ such that:
\bea
\mathcal{I}^{(\a)}(x_{\a})&=& 1 \nonumber \\
\mathcal{I}^{(\a)}(x_{\b})&=& 0\, ,\b\neq \a \nonumber \\
\sum_{\a=1}^{N_x}\mathcal{I}^{(\a)}(y)&=& 1\, ,\forall y .
\label{propE}
\eea
\begin{figure}[t!]
\begin{center}
\includegraphics[width=9cm]{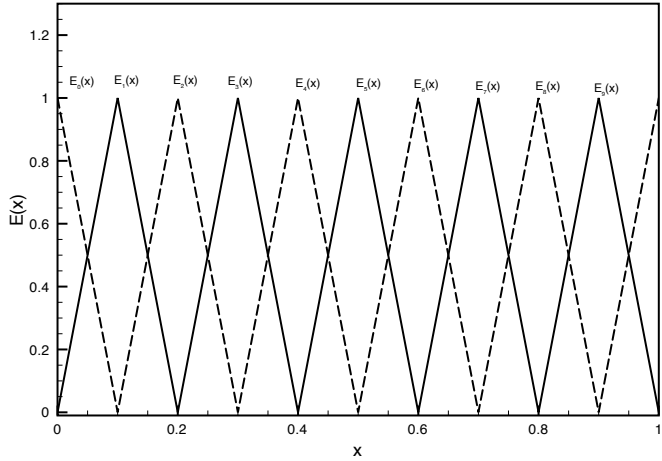}
\end{center}
\caption{Set of interpolating triangular basis functions.} 
\label{triang}
\end{figure}
An illustrative example is given by the  basis of functions
drawn in Fig.~\ref{triang}. Each function $E^{(\a)}$ has a triangular
shape centered in $x_{\a}$ and it vanishes outside the interval
$(x_{\a-1},x_{\a+1})$. For any $y$, only two triangular
functions are non zero and their sum is always equal to one.

With a general  interpolation basis, PDFs
at the initial scale can be approximated as
\begin{equation}
f_k(y,Q_0^2)\equiv f^0_k(y)=\sum_{\a=1}^{N_x}f^0_k(x_{\a})\,\mathcal{I}^{(\a)}(y)+\mathcal{O}[(x_{\alpha+1}-x_{\alpha})^p]\,,
\label{approx}
\end{equation}
where $p$ is the lowest order neglected in the interpolation. With the
(linear) triangular basis Fig.~\ref{triang} $p=2$.
Dropping for simplicity the dependence on $Q_0^2$ and $Q^2_I$, 
Eq.~(\ref{initq}) becomes
\bea
f_j(x_I,Q^2_I)\,\equiv f_j(x_I) &=& \sum_{k=1}^{\npdf}\sum_{\a=1}^{N_x}\,f^0_k(x_{\a})\,\int_{x_I}^1\,\frac{dy}{y}\,
\Gamma_{jk}\left(\frac{x_I}{y}\right)\,\mathcal{I}^{(\a)}(y) +\mathcal{O}[(x_{\alpha+1}-x_{\alpha})^p]\nn\\
f_j(x_I) &=& \sum_{k=1}^{\npdf}\sum_{\a=1}^{N_x}\hat\sigma^{Ij}_{\a k}\,f_k^0(x_{\a})+\mathcal{O}[(x_{\alpha+1}-x_{\alpha})^p],
\label{fastlikeq}
\eea
where
\be
\hat\sigma^j_{\a k}(x_I,Q_0^2,Q^2_I)\equiv\hat\sigma^{Ij}_{\a k}=\,\int_{x_I}^1\frac{dy}{y}\,
\Gamma_{jk}\left(\frac{x_I}{y}\right)\,\mathcal{I}^{(\a)}(y).
\label{coefq}
\ee
In our notation $I$ specifies the data point, $\a$ runs over the points in the $x$--grid and $(j,k)$ run over the PDFs which evolve coupled to each other. 
Having precomputed the $\hat\sigma^{Ij}_{\a k}$ coefficients for each
point $I$, the evaluation of the PDFs only requires $N_x$
evaluations of the PDFs at the initial scale, independent of the
point at which the evolved PDFs are needed, thereby
reducing the computational cost of evolution.

\begin{figure}[t!]
\begin{center}
\includegraphics[width=9cm]{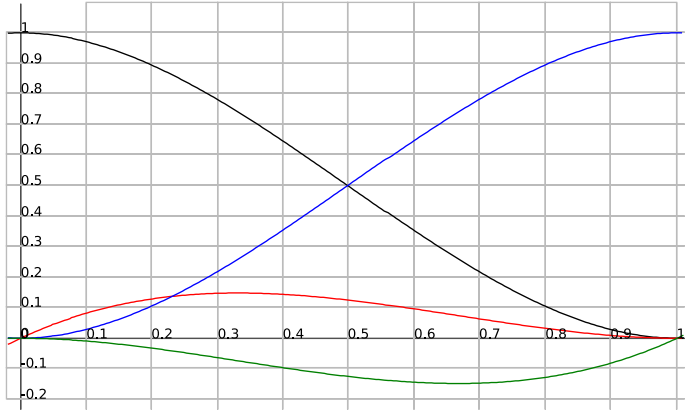}
\end{center}
\caption{Set of interpolating Hermite cubic functions in the [0,1] interval.} 
\label{cub}
\end{figure}

If the interpolation is performed  on a  more complicated
set of functions than the triangular basis Fig.~\ref{triang},
better accuracy can be obtained with a smaller
number of points and thus a reduced
computational cost. 
For PDF evolution we will use the cubic
Hermite interpolation drawn in Fig.~\ref{cub}.
With this choice, 
for each interval $y\in [x_{\a},x_{\a+1})$ the function to be
  approximated 
can be written as
\begin{eqnarray}
 f^0_k(y) &=& h_{00}(t)f^0_k(x_{\a}) + h_{10}(t)h_{\a}m_{\a} + h_{01}(t) f^0_k(x_{\a+1}) + h_{11}(t)h_{\a}m_{\a+1}\nn\\
 &&+\mathcal{O}[(x_{\alpha+1}-x_{\alpha})^4],\nn
\label{cubic}
\end{eqnarray}
where
\begin{equation}
 h_{\a} = g(x_{{\a}+1})-g(x_{\a}),\qquad 
 t = \frac{g(y)-g(x_{\a})}{h_{\a}},
\end{equation}
and $g(y)$ is a monotonic function in  [0,1] which
determines the distribution of points in the interval 
(linear, logarithmic, etc.);  $m_{\a}$ and $m_{\a+1}$ are 
derivatives of the interpolated function at the right and left--hand
side of the interval, which can be defined as  finite differences:
\begin{equation}
 m_{\a} =
 \begin{cases} 
\frac{f^0_k(x_{\a})-f^0_k(x_{{\a}-1})}{2h_{\a-1}} + \frac{f^0_k(x_{\a+1})-f^0_k(x_{\a})}{2h_{\a}}, 
&{\small \mbox{for } 2\le \a \le N_x-1 }\\ 
\frac{f^0_k(x_{\a+1})-f^0_k(x_{\a})}{h_{\a}} , 
&{\small \mbox{for } \a = 1 }\\
\frac{f^0_k(x_{\a})-f^0_k(x_{\a-1})}{h_{\a-1}}, 
&{\small \mbox{for } \a = N_x.}
\end{cases}
\end{equation}
Finally the functions $h$ are 3$^{\mathrm{rd}}$--order polynomials
drawn in Fig.~\ref{cub} and defined as
\begin{eqnarray}
h_{00}(t) &=& 2t^3-3t^2+1=(1+2t)(1-t)^2 \\
h_{10}(t) &=& t^3-2t^2+t =t(t-1)^2 \nn\\
h_{01}(t) &=& -2t^3+3t^2 =t^2(3-2t) \nn\\
h_{11}(t) &=& t^3-t^2=t^2(t-1) \nn
\label{h} 
\end{eqnarray}

Collecting all  terms, Eq.~(\ref{cubic}) becomes
\begin{eqnarray}
f^0_k(y) &=& f^0_k(x_{\a-1})\,A^{(\a)}(y) + f^0_k(x_{\a})\,B^{(\a)}(y) +f^0_k(x_{\a+1})\,C^{(\a)}(y) 
\\ && +f^0_k(x_{\a+2})\,D^{(\a)}(y) +\mathcal{O}[(x_{\alpha+1}-x_{\alpha})^4]\nn. 
 \label{cubicexpansion}
\end{eqnarray}
Hence the function, at any given point $y$ is obtained as a linear 
combination of $f^0$ at the four nearest points in the grid. 
The coefficients of such combination are given by:
\begin{alignat}{5}
 A^{(\a)}(y) & = 
 \begin{cases} 0, & \mbox{for } \a = 1 \\ -h_{10}(t)\frac{h_{\a}}{h_{\a-1}}, & \mbox{for } \a \ne 1 \end{cases}\\
B^{(\a)}(y) &= 
 \begin{cases} h_{00}(t) - h_{10}(t) -\frac{h_{11}(t)}{2}, & \mbox{for } \a = 1 \\ 
 h_{00}(t) -\frac{h_{10}(t)}{2}\left(1-\frac{h_{\a}}{h_{\a+1}}\right) - h_{11}(t) , & \mbox{for } \a = N_x-1\\
 h_{00}(t) -\frac{h_{10}(t)}{2}\left(1-\frac{h_{\a}}{h_{\a+1}}\right) -\frac{h_{11}(t)}{2}, & \mbox{for } \a \ne 1,N_x-1
 \end{cases}\notag\\
 C^{(\a)}(y)  &= 
 \begin{cases} h_{01}(t) + \frac{h_{11}(t)}{2}\left(1-\frac{h_{\a}}{h_{\a+1}}\right) + h_{10}(t), & \mbox{for }  \a = 1 \\ 
  h_{01}(t) + h_{11}(t) + \frac{h_{10}(t)}{2}, & \mbox{for } \a = N_x-1\\
h_{01}(t) + \frac{h_{11}(t)}{2}\left(1-\frac{h_{\a}}{h_{\a+1}}\right) +\frac{h_{10}(t)}{2}, & \mbox{for } \a \ne 1,N_x-1
 \end{cases}\notag\\
 D^{(\a)}(y) &= 
 \begin{cases}
  0, & \mbox{for } \a = N_x-1\\
 h_{11}(t)\frac{h_{\a}}{2h_{\a+1}}, & \mbox{for } \a \ne N_x-1
 \end{cases}\notag
\end{alignat}

If we substitute Eq.~(\ref{cubicexpansion}) into the integral for the
evolution of the PDFs, with $\c$ the index such that 
\be
x_{\c}\le x_I < x_{\c+1},\nn
\ee
we end up with the following expressions for the 
$\hat\sigma$ coefficients 
\begin{eqnarray}\label{eq:tab5}
\hat\sigma^{Ij}_{\a k}=
\begin{cases}
\int_{x_I}^{x_{\c+1}}\,\frac{dy}{y}\,\Gamma_{jk}\left(\frac{x_I}{y}\right)\,A^{(\c)}(y),& \mbox{for~} \a=\c,\\
\int_{x_I}^{x_{\c+1}}\,\frac{dy}{y}\,\Gamma_{jk}\left(\frac{x_I}{y}\right)\,B^{(\c)}(y)&\\
\quad +\theta(N_x-(\c+2))\int_{x_{\c+1}}^{x_{\c+2}}\,\frac{dy}{y}\,\Gamma_{jk}\left(\frac{x_I}{y}\right)\,A^{(\c+1)}(y),&\mbox{for~} \a = \c+1,\\
\int_{x_I}^{x_{\c+1}}\,\frac{dy}{y}\,\Gamma_{jk}\left(\frac{x_I}{y}\right)\,C^{(\c)}(y)&\\
\quad  +\theta(N_x-(\c+2))\int_{x_{\c+1}}^{x_{\c+2}}\,\frac{dy}{y}\,\Gamma_{jk}\left(\frac{x_I}{y}\right)\,B^{(\c+1)}(y)&\\
\quad +\theta(N_x-(\c+3))\int_{x_{\c+2}}^{x_{\c+3}}\,\frac{dy}{y}\,\Gamma_{jk}\left(\frac{x_I}{y}\right)\,A^{(\c+2)}(y),&\mbox{for~} \a = \c+2,\\
\theta(N_x-(I-1))\int_{x_{\a-2}}^{x_{\a-1}}\,\frac{dy}{y}\,\Gamma_{jk}\left(\frac{x_I}{y}\right)\,D^{(\a-1)}(y) &\\
\quad  +\theta(N_x-\a)\int_{x_{\a-1}}^{x_{\a}}\,\frac{dy}{y}\,\Gamma_{jk}\left(\frac{x_I}{y}\right)\,C^{(\a-1)}(y) & \\
\quad +\theta(N_x-(\a+1))\int_{x_\a}^{x_{\a+1}}\,\frac{dy}{y}\,\Gamma_{jk}\left(\frac{x_I}{y}\right)\,B^{(\a)}(y) & \\
\quad +\theta(N_x-(\a+2))\int_{x_{\a+1}}^{x_{\a+2}}\,\frac{dy}{y}\,\Gamma_{jk}\left(\frac{x_I}{y}\right)\,A^{(\a+1)}(y), &\mbox{for~} \c+3 \le \a \le N_x+1,\\
0&\mbox{for~} \a <\c.\\
\end{cases}
\end{eqnarray}
Despite the complicated bookkeeping, these expressions can be easily pre--computed and input into the fit.

A final remark: 
because of the divergent behaviour of the $x$--space evolution kernel
at $x=1$, the integrals including $x_I$ in the integration interval
need to be regularized in $y\sim x_I$. If we consider for instance the
first integral of $A^{(\a)}$ in Eqn.(\ref{eq:tab5}), we can perform
the same subtraction as in Ref.~\cite{Ball:2008by} in order to have a
consistent definition of all precomputed coefficients: 
\bea
&&\mbox{$\int_{x_I}^{x_{\c+1}}\frac{dy}{y}\,\Gamma_{jk}\left(\frac{x_I}{y}\right)\,A^{(\c)}(y)$}\nn\\
&&\qquad =
\mbox{$\int_{x_I}^{x_{\c+1}}\frac{dy}{y}\,\Gamma_{jk}\left(\frac{x_I}{y}\right)\,\left(A^{(\c)}(y)\,-\,\frac{x_I}{y}A^{(\c)}(x_I)\right)+\,A^{(\c)}(x_I)\,\int_{x_I}^{x_{\c+1}}\frac{dy}{y^2}\,\Gamma_{jk}\left(\frac{x_I}{y}\right)$}\nn\\
&&\qquad =\mbox{$\int_{x_I}^{x_{\c+1}}\frac{dy}{y}\,\Gamma_{jk}\left(\frac{x_I}{y}\right)\,
\left(A^{(\c)}(y)-\,\frac{x_I}{y}A^{(\c)}(x_I)\right)+\,A^{(\c)}(x_I)\,\int_{x_I/x_{\c+1}}^{1}dz\,\Gamma_{jk}(z)$}\nonumber\\
&&\qquad =\mbox{$\int_{x_I}^{x_{\c+1}}\frac{dy}{y}\,\Gamma_{jk}\left(\frac{x_I}{y}\right)\,
\left(A^{(\c)}(y)-\,\frac{x_I}{y}A^{(\c)}(x_I)\right)$}\nn\\
&&\qquad\qquad\qquad\qquad +\mbox{$\,A^{(\c)}(x_I)\,\big[ \Gamma_{jk}(N)\big|_{N=2}-\int_0^{x_I/x_{\c+1}}dz\,\Gamma_{jk}(z)\big]$}.
\label{int_cub}
\eea
As a result all $\hat\sigma$ are regularized; they can be stored once and for all for each experimental point, given that they do not depend on the PDF at the initial scale.

%%%%%%%%%%%%%%%%%%%%%
\begin{figure}[ht!]
\begin{center}
\includegraphics[width=0.83\textwidth]{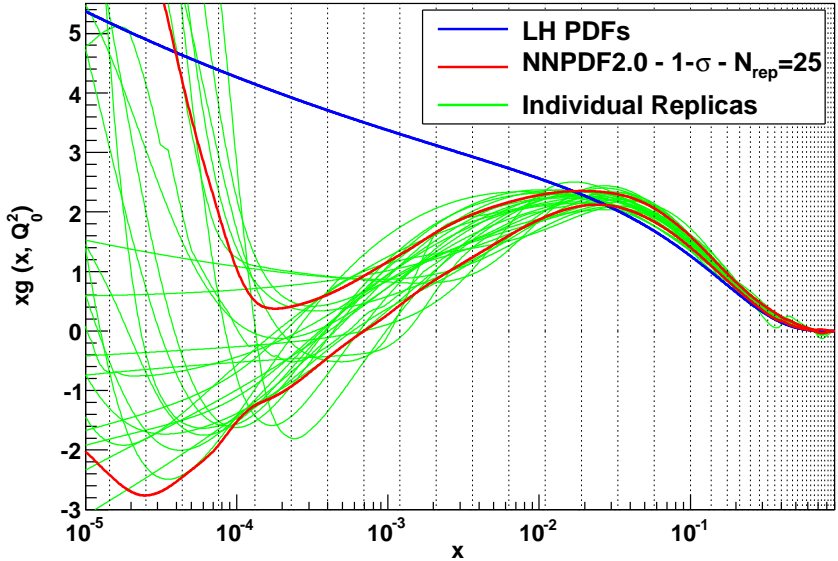}
\end{center}
\vspace*{-1cm}
\caption{\small The sampling grid of $x$ values used for evolution
  superposed to the 
Les Houches gluon and 25 replicas of
  the gluon distribution from the NNPDF2.0 set at the starting scale.}
\label{gridcomp}
\end{figure}
%%%%%%%%%%%%%%%%%%
The accuracy of our PDF evolution code, described above, has been
cross--checked against the Les Houches PDF evolution benchmark
tables, originally produced from the
comparison of the HOPPET~\cite{Salam:2008qg} and
PEGASUS~\cite{pegasus} codes. In order to perform a meaningful comparison, 
we use the same
settings described in detail in Ref.~\cite{Dittmar:2005ed}.  We show
in Table~\ref{tab:lh}  the relative difference for various combinations of PDFs
between our PDF evolution and the benchmark tables of
Ref.~\cite{Dittmar:2005ed} at NLO in the ZM--VFNS, for three different
grids. In each grid, the interval $[x_{\mathrm{min}},1]$ is
divided into a log region at small $x$ and a linear region
medium--high $x$. As we can see, the choice of a relatively small grid of
50 points leads to reproducing the Les Houches tables with an accuracy
of $\mathcal{O}(10^{-5})$, more than enough for the precision
phenomenology we aim to. Note that even though of course each
individual replica has more structure than the average PDF, and much
more structure than the simple Les Houches toy PDFs, they are still
quite smooth on the scale of this grid, as it can be seen in
Fig.~\ref{gridcomp}. This ensures that benchmarking with the Les
Houches table is adequate to guarantee the accuracy of our evolution code.

\begin{table}
                            \begin{center}
\begin{tabular}{|c||c|c|c|c|}
\hline
x (30 pts) & $e_{\mathrm{rel}}(u_v)$ & $e_{\mathrm{rel}}(d_v)$& $e_{\mathrm{rel}}(\Sigma)$& $e_{\mathrm{rel}}(g)$\\
\hline
\hline
$1\cdot 10^{-7} $ &$2.5 \cdot 10^{-4}$  & $3.5\cdot 10^{-4}$& $2.1\cdot 10^{-4}$ & $2.1\cdot 10^{-4}$ \\
$1\cdot 10^{-6} $ &$1.6 \cdot 10^{-3}$  & $1.5\cdot 10^{-3}$& $2.3\cdot 10^{-4}$ & $2.5\cdot 10^{-4}$ \\
$1\cdot 10^{-5} $ &$1.5 \cdot 10^{-3}$  & $1.4\cdot 10^{-3}$& $2.5\cdot 10^{-4}$ & $2.8\cdot 10^{-4}$ \\
$1\cdot 10^{-4} $ &$6.5 \cdot 10^{-4}$  & $5.1\cdot 10^{-4}$& $3.0\cdot 10^{-4}$ & $3.4\cdot 10^{-4}$ \\
$1\cdot 10^{-3} $ &$6.5 \cdot 10^{-4}$  & $4.7\cdot 10^{-4}$& $3.4\cdot 10^{-4}$ & $3.9\cdot 10^{-4}$ \\
$1\cdot 10^{-2} $ &$1.4 \cdot 10^{-3}$  & $1.9\cdot 10^{-3}$& $3.4\cdot 10^{-4}$ & $5.3\cdot 10^{-4}$ \\
$1\cdot 10^{-1} $ &$7.0 \cdot 10^{-4}$  & $1.0\cdot 10^{-3}$& $1.1\cdot 10^{-4}$ & $4.1\cdot 10^{-4}$ \\
$3\cdot 10^{-1} $ &$1.9 \cdot 10^{-5}$  & $8.6\cdot 10^{-5}$& $1.3\cdot 10^{-5}$ & $5.8\cdot 10^{-5}$ \\
$5\cdot 10^{-1} $ &$1.5 \cdot 10^{-4}$  & $1.8\cdot 10^{-4}$& $1.0\cdot 10^{-4}$ & $1.1\cdot 10^{-4}$ \\
$7\cdot 10^{-1} $ &$3.8 \cdot 10^{-4}$  & $3.9\cdot 10^{-5}$& $3.1\cdot 10^{-4}$ & $2.8\cdot 10^{-4}$ \\
$9\cdot 10^{-1} $ &$8.5 \cdot 10^{-3}$  & $9.5\cdot 10^{-2}$& $3.4\cdot 10^{-3}$ & $2.0\cdot 10^{-2}$ \\
\hline
\end{tabular}

\vspace*{0.5cm}
\begin{tabular}{|c||c|c|c|c|}
\hline
x (50 pts) & $e_{\mathrm{rel}}(u_v)$ & $e_{\mathrm{rel}}(d_v)$& $e_{\mathrm{rel}}(\Sigma)$& $e_{\mathrm{rel}}(g)$\\
\hline
\hline
$1\cdot 10^{-7} $ &$2.1 \cdot 10^{-4}$  & $2.3\cdot 10^{-4}$& $2.7\cdot 10^{-5}$ & $4.7\cdot 10^{-6}$ \\
$1\cdot 10^{-6} $ &$8.9 \cdot 10^{-5}$  & $8.4\cdot 10^{-5}$& $3.0\cdot 10^{-5}$ & $2.1\cdot 10^{-5}$ \\
$1\cdot 10^{-5} $ &$9.3 \cdot 10^{-5}$  & $6.0\cdot 10^{-5}$& $2.3\cdot 10^{-5}$ & $2.0\cdot 10^{-5}$ \\
$1\cdot 10^{-4} $ &$4.5 \cdot 10^{-5}$  & $2.8\cdot 10^{-5}$& $4.4\cdot 10^{-5}$ & $4.2\cdot 10^{-5}$ \\
$1\cdot 10^{-3} $ &$3.0 \cdot 10^{-5}$  & $1.7\cdot 10^{-5}$& $4.0\cdot 10^{-5}$ & $3.5\cdot 10^{-5}$ \\
$1\cdot 10^{-2} $ &$7.9 \cdot 10^{-5}$  & $6.8\cdot 10^{-5}$& $4.5\cdot 10^{-5}$ & $5.8\cdot 10^{-5}$ \\
$1\cdot 10^{-1} $ &$1.7 \cdot 10^{-4}$  & $2.1\cdot 10^{-4}$& $1.6\cdot 10^{-5}$ & $3.9\cdot 10^{-5}$ \\
$3\cdot 10^{-1} $ &$9.1 \cdot 10^{-6}$  & $3.9\cdot 10^{-5}$& $1.1\cdot 10^{-5}$ & $1.9\cdot 10^{-7}$ \\
$5\cdot 10^{-1} $ &$2.4 \cdot 10^{-5}$  & $2.2\cdot 10^{-5}$& $2.2\cdot 10^{-5}$ & $2.2\cdot 10^{-5}$ \\
$7\cdot 10^{-1} $ &$9.1 \cdot 10^{-5}$  & $1.5\cdot 10^{-5}$& $7.8\cdot 10^{-5}$ & $1.2\cdot 10^{-4}$ \\
$9\cdot 10^{-1} $ &$1.0 \cdot 10^{-3}$  & $3.3\cdot 10^{-3}$& $8.0\cdot 10^{-4}$ & $2.8\cdot 10^{-3}$ \\
\hline
\end{tabular}

\vspace*{0.5cm}
\begin{tabular}{|c||c|c|c|c|}
\hline
x (100 pts) & $e_{\mathrm{rel}}(u_v)$ & $e_{\mathrm{rel}}(d_v)$& $e_{\mathrm{rel}}(\Sigma)$& $e_{\mathrm{rel}}(g)$\\
\hline
\hline
$1\cdot 10^{-7} $ &$3.2 \cdot 10^{-5}$  & $5.0\cdot 10^{-5}$& $5.4\cdot 10^{-6}$ & $2.0\cdot 10^{-5}$ \\
$1\cdot 10^{-6} $ &$2.6 \cdot 10^{-6}$  & $1.3\cdot 10^{-6}$& $5.7\cdot 10^{-6}$ & $5.9\cdot 10^{-6}$ \\
$1\cdot 10^{-5} $ &$1.1 \cdot 10^{-5}$  & $2.2\cdot 10^{-5}$& $3.7\cdot 10^{-6}$ & $1.0\cdot 10^{-5}$ \\
$1\cdot 10^{-4} $ &$1.8 \cdot 10^{-5}$  & $3.3\cdot 10^{-6}$& $1.3\cdot 10^{-5}$ & $6.9\cdot 10^{-6}$ \\
$1\cdot 10^{-3} $ &$1.3 \cdot 10^{-6}$  & $4.9\cdot 10^{-6}$& $4.7\cdot 10^{-6}$ & $7.7\cdot 10^{-6}$ \\
$1\cdot 10^{-2} $ &$1.6 \cdot 10^{-5}$  & $1.7\cdot 10^{-5}$& $4.8\cdot 10^{-6}$ & $1.1\cdot 10^{-6}$ \\
$1\cdot 10^{-1} $ &$3.4 \cdot 10^{-5}$  & $2.9\cdot 10^{-5}$& $8.7\cdot 10^{-6}$ & $2.1\cdot 10^{-6}$ \\
$3\cdot 10^{-1} $ &$2.0 \cdot 10^{-6}$  & $2.5\cdot 10^{-5}$& $7.9\cdot 10^{-6}$ & $3.9\cdot 10^{-6}$ \\
$5\cdot 10^{-1} $ &$1.7 \cdot 10^{-5}$  & $1.3\cdot 10^{-5}$& $1.7\cdot 10^{-5}$ & $3.1\cdot 10^{-5}$ \\
$7\cdot 10^{-1} $ &$7.1 \cdot 10^{-5}$  & $8.3\cdot 10^{-6}$& $6.3\cdot 10^{-5}$ & $1.3\cdot 10^{-4}$ \\
$9\cdot 10^{-1} $ &$3.9 \cdot 10^{-5}$  & $3.8\cdot 10^{-4}$& $2.5\cdot 10^{-5}$ & $1.7\cdot 10^{-3}$ \\
\hline
\end{tabular}
\end{center}
\label{tab:lh}
\caption{Relative accuracy of FastKernel evolution compared to the Les Houches benchmark tables for 
  PDFs evolved to the scale $Q^2=10^4$ GeV$^2$. 
  The interpolation is performed on cubic Hermite polynomials and the grid is composed of 30 points (top),
  50 points (middle), or 100 points (bottom), distributed logarithmically in the small--$x$ region and 
  linearly in the medium-- and large--$x$ region. }
\end{table}

\subsection{Fast computation of DIS observables}
\label{sec:fastdis}

Using the strategy described in the previous section,
we can easily write down the expression for the DIS observables
included in our fit and show explicitly how their computation works on
the interpolation basis. The basic
idea~\cite{DelDebbio:2007ee,Ball:2008by}
 is that, starting with the
standard factorized expression
\be
\so = \sum_{k=1}^{\npdf} C_{Ik}\otimes f_k(x_I,Q^2_I) \equiv\sum_{k=1}^{\npdf}\,\int_{x_I}^1\,\frac{dy}{y}\,
C_{Ik}\left(\frac{x_I}{y},\as(Q^2_I)\right)\,f_k(y,Q^2_I).
\label{fact}
\ee
(where $I$ denotes  both the observable and the
kinematic point), we can  absorb the coefficient function $C_{Ik}$
into a modified evolution kernel $K_{Ij}$ which can be precomputed before
starting the fit (see Appendix A of Ref.~\cite{Ball:2008by}): 
\be
K_{Ij}(x_I,\as(Q_I^2),\as(Q_0^2))=\sum_{k=1}^{\npdf}C_{Ik}\otimes\Gamma_{kj}(x_I,\as(Q_I^2),\as(Q_0^2)).
\label{kerdef}
\ee 
The kernel acts on the $j$--th PDFs at the initial scale, and it is an
observable--dependent linear combination of products of coefficient
functions and evolution kernels: 
\be
\so = \sum_{j=1}^{\npdf}\,K_{Ij}\otimes f_j^0\equiv\sum_{j=1}^{\npdf}\,\int_{x_I}^1\,\frac{dy}{y}\,
c_k\,K_{Ij}\left(\frac{x_I}{y},\as(Q_I^2),\as(Q_0^2)\right) f^0_j(y,Q_0^2).
\label{init}
\ee
If we substitute Eq.~(\ref{fastlikeq}) into the expression for the observable, we can write it as:
\bea
\so &=&
\sum_{j=1}^{\npdf}\sum_{\a=1}^{N_x}\,f_j^0(x_{\a})\,\int_{x_I}^1\,\frac{dy}{y}\,
K_{Ij}\left(\frac{x_I}{y}\right)\,\mathcal{I}^{(\a)}(y)\\
&=&\sum_{j=1}^{\npdf}\sum_{\a=1}^{N_x}f^0_j(x_{\a})\,\hat\sigma^I_{\a j}+\mathcal{O}[(x_{\alpha+1}-x_{\alpha})^p],\nn
\label{fastlike}
\eea
where
\be
\hat\sigma^I_{\a j}(x_I,Q_0^2,Q_I^2)
\equiv\hat\sigma^I_{\a j}=
\,\int_{x_I}^1\frac{dy}{y}\,K_{Ij}\left(\frac{x_I}{y},\as(Q_I^2),\as(Q_0^2)\right)\,\mathcal{I}^{(\a)}(y).
\label{coef}
\ee
Now the only index running over the PDF basis is $j$ because the other
index $k$ is contracted in the definition of $K$.

Consider for example the expression for the deuteron structure function. We can write down explicitly the terms 
of Eq.~(\ref{fastlike}) as:
\be
\label{f2d}
F_2^d(x_{I},Q_I^2) = \sum_{\a=1}^{N_x}\,\sigma_{\a 10}^I\,f_{10}(x_{\a})+\sigma_{\a 1}^I\,f_{1}(x_{\a})+\sigma_{\a 2}^I\,f_{2}(x_{\a}) +\mathcal{O}[(x_{\alpha+1}-x_{\alpha})^p],
\ee
with
\bea
\sigma_{\a 10}&=&\mbox{$\int_{x_I}^1\,\frac{dy}{y}\,\frac{1}{18}
\left(C_{2,q}\otimes\Gamma^-\right)\left(\frac{x_I}{y}\right)\,\mathcal{I}^{(\a)}(y)$} \nn\\
\sigma_{\a 1}&=&\mbox{$ \int_{x_I}^1\,\frac{dy}{y}\,\big[-\frac{1}{18}\left(C_{2,q}\otimes\Gamma^{15,q}\right)\left(\frac{x_I}{y}\right) +\frac{1}{30}\left(C_{2,q}\otimes\Gamma^{24,q}\right)\left(\frac{x_I}{y}\right)$} \nn\\
&&\hspace{0.3cm} \mbox{$-\frac{1}{30}\left(C_{2,q}\otimes\Gamma^{35,q}\right)\left(\frac{x_I}{y}\right)+\frac{5}{18}\left(C_{2,q}\otimes\Gamma^{S,qq}\right)\left(\frac{x_I}{y}\right)$}\nn\\
&&\hspace{0.3cm} \mbox{$ -c_g(n_f)\left(C_{2,g}\otimes\Gamma^{S,gq}\right)\left(\frac{x_I}{y}\right)\big]\mathcal{I}^{(\a)}(y)$}\nn\\
\sigma_{\a 2}&=& \mbox{$\int_{x_I}^1\,\frac{dy}{y}\,\big[-\frac{1}{18}\left(C_{2,q}\otimes\Gamma^{15,q}\right)\left(\frac{x_I}{y}\right)+\frac{1}{30}\left(C_{2,q}\otimes\Gamma^{24,g}\right)\left(\frac{x_I}{y}\right)$}\nn\\
&&\hspace{0.3cm} -\mbox{$\frac{1}{30}\left(C_{2,q}\otimes\Gamma^{35,g}\right)\left(\frac{x_I}{y}\right)+\frac{5}{18}\left(C_{2,q}\otimes\Gamma^{S,qg}\right)\left(\frac{x_I}{y}\right)$}\nn\\
&&\hspace{0.3cm} -\mbox{$ c_g(n_f)\left(C_{2,g}\otimes\Gamma^{S,gg}\right)\left(\frac{x_I}{y}\right)\big]\mathcal{I}^{(\a)}(y)$}
\eea
where all kernels and coefficient functions are defined in Ref.~\cite{Ball:2008by} and
\be
f^0_{10} =T_{8,0}\qquad f^0_1=\Sigma_0 \qquad f^0_2=g_0 \nn
\ee
in the evolution basis of Eq.~(\ref{eq:basis}).

\subsection{Fast computation of hadronic observables}
\label{sec:fastdy}

The FastKernel  implementation 
of hadronic observables requires a double convolution
of the coefficient function with two parton distributions. We could
follow the same strategy used for DIS: construct a kernel for each
observable and each
pair of initial PDFs, and then compute the double convolution with a
suitable generalization of the method introduced in
Sect.~\ref{sec:fastdis}. However, for hadronic observables, we adopt
a somewhat different strategy, which allows us to treat in a more symmetric
way processes for which a fast interface already exists (such as
jets) and those (such as DY) for which we have to develop our own
interface. Namely, instead of including the coefficient function into
the kernel according to Eq.~(\ref{kerdef}), we compute the convolution
Eq.~(\ref{fact}) using the fast interpolation method.
 
To see how this works, consider first the case of a process with only 
one parton in the initial state. 
Starting from Eq.~(\ref{fact}), we can project the evolved
PDF $f_k$ onto an interpolation basis as follows: 
\be
\so =\sum_{k=1}^{\npdf}\sum_{\a=1}^{N_y}\,f_k(y_{\a},Q^2_I)\int_{x_I}^1\,\frac{dy}{y}\,
C_{Ik}\left(\frac{x_I}{y},\as(Q^2_I)\right)\,\mathcal{I}^{\a}(y)+\mathcal{O}[(y_{\alpha+1}-y_{\alpha})^q],
\label{rb}
\ee
where $q$ indicates the first order neglected in the interpolation of the evolved PDFs.
This defines 
another grid  of points, $  \lbrace y_{\a}\rbrace $, upon which the coefficients can be pre--computed before starting the fit:
\be
\int_{x_I}^1\,\frac{dy}{y}\,
C_{Ik}\left(\frac{x_I}{y},\as(Q^2_I)\right)\,\mathcal{I}^{\a}(y) \equiv C_{Ik}^{\a}.
\label{def}
\ee
If, on top of this interpolation, we interpolate the parton distributions at the initial scale on the $ \lbrace x_{\a}\rbrace$ grid as we did in the previous subsection, we get
\bea
\so &=& \sum_{k=1}^{\npdf}\,\sum_{\a=1}^{N_y}f_k(y_{\a},Q^2_I)\,C_{Ik}^{\a}+\mathcal{O}[(y_{\alpha+1}-y_{\alpha})^q]\\
 &=& \sum_{k,n=1}^{\npdf}\sum_{\a=1}^{N_y}\sum_{\b=1}^{N_x} \,C_{Ik}^{\a} \,\hat\sigma^{\a,I}_{\b kn}\,f^{0}_n(x_{\b})+\mathcal{O}[(y_{\alpha+1}-y_{\alpha})^q(x_{\b+1}-x_{\b})^{p}].\nn
\label{difdis}
\eea
Notice that the two interpolations are independent of each other. 
The number of points  $N_{x}$ and $N_{y}$ in each grid, the
interpolating functions, and the interpolation orders
$p$ and $q$  are 
not necessarily the same.

We now apply this  to the rapidity--differential Drell--Yan cross section,
introduced in Sect.~\ref{sec:expnewdata}, 
to exemplify the procedure. The NLO cross section is given by
\bea
\sdy &=& n(Q_I^2) \sum_{j=1}^{N_q} e_j^2 \,\int_{x^0_1}^1\,dx_1 \,\int_{x^0_2}^1\,dx_2 \\
&&\,\,\bigg\lbrace\left[q_j(x_1,Q_I^2)\bar{q}_j(x_2,Q_I^2)+
q_j(x_2,Q_I^2)\bar{q}_j(x_1,Q_I^2)\right]\,\left(D^{\mathrm{qq}}(x_1,x_2,Y_I)\right)\nonumber\\
&&\hspace{2cm}+g(x_1,Q_I^2)\left[q_j(x_2,Q_I^2)+\bar{q}_j(x_2,Q_I^2)\right]\,\left(D^{\mathrm{gq}}(x_1,x_2,Y_I)\right)
\nonumber\\
&&\hspace{2cm}+g(x_2,Q_I^2)\left[q_j(x_1,Q_I^2)+\bar{q}_j(x_1,Q_I^2)\right]\,\left(D^{\mathrm{qg}}(x_1,x_2,Y_I)\right)\bigg\rbrace,\nn
\label{dy}
\eea
where the normalization factor is explicitly written in 
Sect.~\ref{sec:expnewdata} and the coefficient functions can be 
found in Refs.~\cite{Gehrmann:1997ez,Gehrmann:1997pi}
(see also Appendix~\ref{sec:dyobservables}).

For each point of the interpolation grid, we define a set of
two--dimensional interpolating functions as the product of
one--dimensional functions defined in Eq.~(\ref{propE}): 
\be
\mathcal{I}^{(\a,\b)}(x_1,x_2)\,\equiv\,\mathcal{I}^{(\a)}(x_1)\mathcal{I}^{(\b)}(x_2).
\label{eigen2d}
\ee	
The product of two functions can be approximated by means of these interpolating functions as
\be
f(y_1)h(y_2)=\sum_{\a,\b=1}^{N_y}f(y_{1,\a})h(y_{2,\b})\,\mathcal{I}^{(\a,\b)}(y_1,y_2)+\mathcal{O}[(y_{1,\a+1}-y_{1,\a})^{q}(y_{2,\b+1}-y_{2,\b})^{q}].
\label{approx2d}
\ee
Applying Eq.~(\ref{approx2d}) to the PDFs in Eq.~(\ref{dy}), we get
\bea
\sdy &=& n(Q_I^2) \sum_{j=1}^{N_q} e_j^2 \sum_{\a,\b=1}^{N_x}\,
\big[q_j(y_{1,\a})\bar{q}_j(y_{2,\b})+\bar{q}_j(y_{1,\a})q_j(y_{2,\b})\big]\\
&&\hspace{1.3cm}\int_{x^0_1}^1\,dx_1 \,\int_{x^0_2}^1\,dx_2 \mathcal{I}^{(\a,\b)}(x_1,x_2)\,D^{\mathrm{qq}}(x_1,x_2,Y_I)\nn\\
&&\hspace{-2.3cm}
+\big[g(y_{1,\a})(q_j(y_{2,\b})+\bar{q}_j(y_{2,\b}))\big]
\int_{x^0_1}^1\,dx_1 \,\int_{x^0_2}^1\,dx_2 \mathcal{I}^{(\a,\b)}(x_1,x_2)\,D^{\mathrm{gq}}(x_2,x_1,Y_I)\nn\\
&&\hspace{-2.3cm}
+\big[g(y_{1,\a})(q_j(y_{2,\b})+\bar{q}_j(y_{2,\b}))\big]\int_{x^0_1}^1\,dx_1 \,\int_{x^0_2}^1\,dx_2 \mathcal{I}^{(\a,\b)}(x_1,x_2)\,D^{\mathrm{gq}}(x_1,x_2,Y_I)\nn\\
&&\hspace{3.0cm}+\mathcal{O}[(y_{1,\a+1}-y_{1,\a})^{q}(y_{2,\b+1}-y_{2,\b})^{q}],\nn
\label{dy_eigen}
\eea
where at next--to--leading order 
$D^{\mathrm{qg}}(x_1,x_2,Y_I)=D^{\mathrm{gq}}(x_2,x_1,Y_I)$.
Therefore, we can define
\be
C_{I,ij}^{(\a,\b)} \equiv  \int_{x^0_1}^1dx_1 \,\int_{x^0_2}^1dx_2 \mathcal{I}^{(\a,\b)}(x_1,x_2)\,D^{\mathrm{ij}}(x_1,x_2,Y_I),
\label{defC}
\ee
where $i,j$ run over the non--zero combinations of $q,\bar{q}$ and $g$. By substituting them into Eq.~(\ref{dy_eigen}), we end up with the expression
\bea
\sdy &=& n(Q_I^2) \sum_{j=1}^{N_q} e_j^2 \,\sum_{\a,\b=1}^{N_y}C_{I,qq}^{(\a,\b)} 
\left[q_j(y_{1,\a})\bar{q}_j(y_{2,\b})+\bar{q}_j(y_{1,\a})q_j(y_{2,\b})\right]\nn\\
&&\hspace{1cm}+C_{I,gq}^{(\a,\b)}\left[g(y_{1,\a})(q_j(y_{2,\b})+\bar{q}_j(y_{2,\b}))\right]\nn\\
&&\hspace{1cm}+C_{I,qg}^{(\a,\b)}\left[(q_j(y_{1,\a})+\bar{q}_j(y_{1,\a}))g(y_{2,\b})\right]\nn\\
&&\hspace{1cm}+\mathcal{O}[(y_{1,\a+1}-y_{1,\a})^{q}(y_{2,\b+1}-y_{2,\b})^{q}]\, ,
\label{difdy}
\eea 
which is the analogue of Eq.~(\ref{rb}) for a hadronic observable.
The physical basis $\lbrace q \rbrace_j$ and the evolution basis $\lbrace f \rbrace_j$ are related by a matrix $A$:
\be
q_j=A_{jr}f_r \qquad \bar{q}_j=\bar{A}_{js}f_s\, .
\nonumber
\ee
Each PDF $f$ is evolved at the physical scale of the process, and the evolution matrix $\Gamma$ which relates the initial scale PDFs to the evolved ones is
\be
f_r(x,Q^2)=\Gamma_{rn}(x,Q_0^2,Q^2)\otimes f_n(x,Q_0^2).
\nonumber
\ee
Therefore Eq.~(\ref{difdy}) becomes
\bea
\sdy &=& n(Q_I^2) \sum_{j=1}^{N_q} e_j^2 \,\sum_{\a,\b=1}^{N_y}\sum_{r,s=1}^{\npdf}C_{I,qq}^{(\a,\b)} 
(A_{jr}\bar{A}_{js}+\bar{A}_{jr}A_{js})\,f_r(y_{1,\a})f_s(y_{2,\b})\\ 
&&\hspace{1cm}+\left[C_{I,gq}^{(\a,\b)} 
\delta_{r2}(A_{js}+\bar{A}_{js})+C_{I,qg}^{(\a,\b)}(A_{jr}+\bar{A}_{js})\delta_{q2}\right]\,f_r(y_{1,\a})f_s(y_{2,\b)})\nn\\
&&\hspace{1cm}+\mathcal{O}[(y_{1,\a+1}-y_{1,\a})^{q}(y_{2,\b+1}-y_{2,\b})^{q}]\nn.
\label{fastnlody}
\eea

Defining
\bea
c_{rs}&\equiv& \sum_{j=1}^{N_q} e_j^2 (A_{jr}\bar{A}_{js}+\bar{A}_{jr}A_{js})\\
d_{rs} &\equiv& \sum_{j=1}^{N_q} e_j^2 \left[\delta_{r2}(A_{js}+\bar{A}_{js})+(A_{jr}+\bar{A}_{jr})\delta_{s2}\right]\nn
\label{defcpq}
\eea
and applying Eq.~(\ref{fastlikeq}) to the evolved PDFs, we end up
with a result which is similar to Eq.~(\ref{difdis}):

\bea
\sdy &=& n(Q_I^2)\sum_{\g,\delta=1}^{N_x}\sum_{l,m=1}^{\npdf}\bigg[\sum_{\a,\b=1}^{N_y}\sum_{r,s=1}^{\npdf}c_{rs}C_{I,qq}^{(\a,\b)} 
\hat\sigma^{\a,I}_{\g rl} \hat\sigma^{\b,I}_{\delta sm}\\
&&+[d_{rs}C_{I,gq}^{(\a,\b)}+d_{sr}C_{I,qg}^{(\a,\b)}] 
\hat\sigma^{\a,I}_{\g rl} \hat\sigma^{\b,I}_{\delta sm}\bigg]f^{(0)}_l(x_{1,\g})f^{(0)}_m(x_{2,\delta})\nn\\
&&+\mathcal{O}[(y_{1,\a+1}-y_{1,\a})^{q}(y_{2,\b+1}-y_{2,\b})^{q}(x_{1,\g+1}-x_{1,\g})^{p}(x_{2,\delta+1}-x_{2,\delta})^{p}].\nn
\label{eq:difdy}
\eea

In order to define the coefficients in Eq.~(\ref{dy_eigen}), we have
 to make an explicit choice of an interpolating basis. For the interpolation 
of the evolved PDFs we use the triangular interpolating basis drawn
 in Fig.~\ref{triang}, defined as
\begin{equation}
 E^{(\a)}(y)=\frac{y-y_{\a-1}}{y_{\a}-y_{\a-1}}\theta[(y_{\a}-y)(y-y_{\a-1})]+\frac{y_{\a+1}-y}{y_{\a+1}-y_{\a}}\theta[(y_{\a}-y)(y-y_{\a+1})].
\end{equation}
With this definition, we can project the PDFs on the triangular basis
\begin{equation*}
q(y)=\sum_{\a=1}^{N_x}q(y_{\a})\,E^{(\a)}(y)	+ \mathcal{O}[(y_{\a+1}-y_{\a})^{2}]
\end{equation*}
and define
\be
C_{K,ij}^{(\a,\b)}=\int_{x^0_1}^{1}dx_1\int_{x^0_2}^{1}dx_2\,E^{(\a)}(x_1)\,E^{(\b)}(x_2)D^{(K)}_{ij}(x_1,x_2),
\label{c1lin}
\ee
where $K$ indicates the perturbative order and $i,j$ run over the non--zero combinations of $q,\bar{q}$ and $g$.
To be more explicit, defining the index $\c$ and $\z$ in such a way that
\be
x_{\c}<x_1^0<x_{\c+1} \qquad x_{\z}<x_2^0<x_{\z+1},
\ee 
we can give the precise definition of the NLO coefficients: 
\begin{eqnarray}\label{eq:tab6}
C_{K,ij}^{(\a,\b)}=
\begin{cases}
\int_{x^0_1}^{x_{\a+1}}dx_1\int_{x^0_2}^{x_{\b+1}}dx_2\,E^{(\a)}(x_1)\,E^{(\b)}(x_2)D^{(K)}_{ij}(x_1,x_2),&
%\mbox{for~} 
\,\a=\c,\c+1,\,\b=\z,\z+1\\
\int_{x^0_1}^{x_{\a+1}}dx_1\int_{x^{\b-1}}^{x_{\b+1}}dx_2\,E^{(\a)}(x_1)\,E^{(\b)}(x_2)D^{(K)}_{ij}(x_1,x_2),&
%\mbox{for~} 
\,\a\le\c+1,\,\b\ge\z+2,\\
\int_{x_{\a-1}}^{x_{\a+1}}dx_1\int_{x^0_2}^{x_{\b+1}}dx_2\,E^{(\a)}(x_1)\,E^{(\b)}(x_2)D^{(K)}_{ij}(x_1,x_2), &
%\mbox{for~} 
\,\a\ge\c+2,\,\b\le\z+1,\\
\int_{x_{\a-1}}^{x_{\a+1}}dx_1\int_{x_{\b-1}}^{x_{\b+1}}dx_2\,E^{(\a)}(x_1)\,E^{(\b)}(x_2)D^{(K)}_{ij}(x_1,x_2), &
%\mbox{for~} 
\,\a\ge\c+2,\,\b\ge\z+2,\\
0&
%\mbox{for~}
\, \a \le\c-1,\,\b\le\z-1,\\
\end{cases}
\end{eqnarray}
while the expression for the LO is trivial, given that $D^{(0)}_{q\bar{q}}(x_1,x_2)=\delta(x_1-x^0_1)\delta(x_2-x^0_2)$.

The FastKernel method for hadronic observables is easily
interfaced to other existing fast codes, such as FastNLO for
inclusive jets~\cite{Kluge:2006xs}, by simply using FastKernel for the
interpolation at the initial scale and parton evolution, and exploiting
the existing interface for the convolution of the evolved PDF with the
appropriate coefficient functions. In the particular case of 
the inclusive jet measurements used in the present analysis,
the analogs of the coefficients $C_{I,ij}^{(\a,\b)}$ 
in Eq.~(\ref{difdy}) can be
directly extracted from the FastNLO precomputed tables
through its interface, although in such case
the relevant PDF combinations are different than those of the DY
process  Eq.~(\ref{difdy}).  

\subsection{FastKernel benchmarking}

It is straightforward to extend the FastKernel method described in the 
previous section to all fixed--target DY and collider vector boson production 
datasets described in Sect.~\ref{sec:expnewdata}, using the appropriate 
couplings and PDF combinations. More details on the computation of these
observables can be found in Appendix~\ref{sec:dyobservables}.

In order to assess the accuracy of the method, 
we have benchmarked the results obtained with our code to those 
produced by an independent code~\cite{vicinipriv} 
which computes the exact NLO cross sections for all relevant 
 Drell--Yan distributions. The comparison is performed by using 
a given set of input PDFs and evaluating
the various cross--sections  for all observables included in the fit
in the kinematical points which correspond to the
included data.

\begin{figure}[ht!]
\begin{center}
\includegraphics[width=0.83\textwidth]{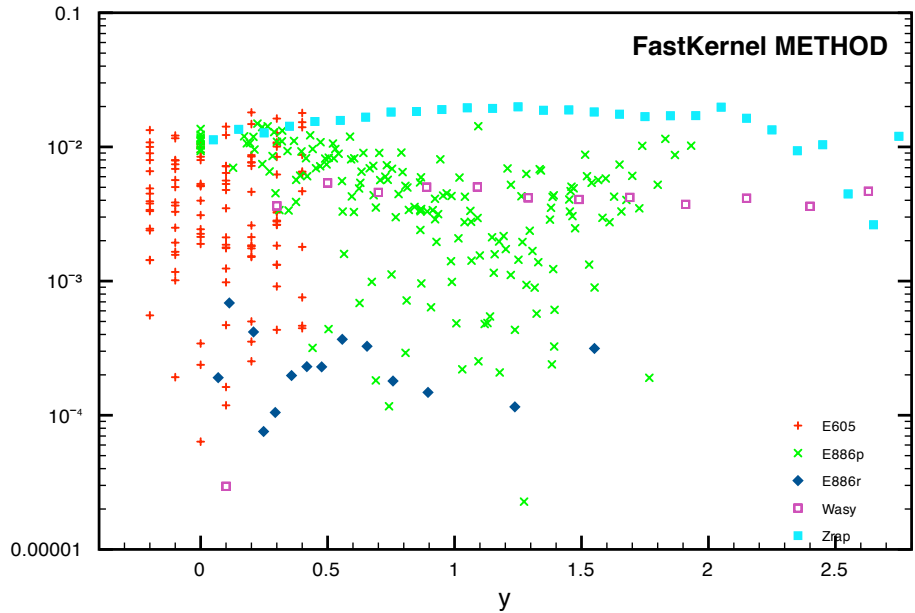}
\end{center}
\vspace*{-1cm}
\caption{\small Relative accuracy for NLO Drell--Yan rapidity distributions
  using the FastKernel method,
compared to the code of~\cite{vicinipriv}, 
as a
  function of rapidity $y$. Each point corresponds to 
the kinematics of a data point included in
the NNPDF2.0 fit.  The accuracy refers to a grid of 100
  points distributed as the root square of the log from
  $x_{\mathrm{min}}$ to 1.}
\label{nlody}
\end{figure}

The benchmarking of the
FastKernel code for the Drell--Yan process 
has been performed for the following observables,
introduced in Sect.~\ref{sec:expnewdata}:
\begin{itemize}
\item Rapidity and $x_F$ distributions and asymmetries for  fixed target 
Drell--Yan in pp and pCu collisions (E605 and E866 kinematics)
\item The $W$ rapidity distribution and asymmetries at hadron colliders
(Tevatron kinematics)
\item The $Z$ rapidity distribution at hadron colliders (Tevatron kinematics)
\end{itemize}
The results of this benchmark comparison are displayed in Fig.~\ref{nlody},
where the relative accuracy between the FastKernel implementation
and the exact code is shown
for all data points included in the NNPDF2.0. 
This accuracy  has been obtained with a grid of 100
  points distributed as the root square of the log from
  $x_{\mathrm{min}}$ to 1.

\begin{figure}[ht!]
\begin{center}
\includegraphics[width=0.83\textwidth]{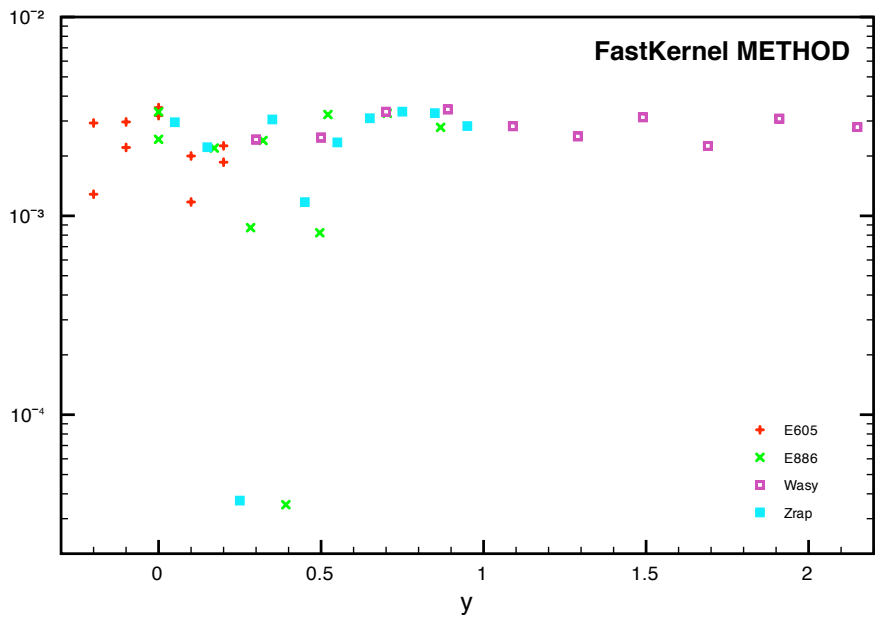}
\end{center}
\vspace*{-1cm}
\caption{\small Same as Fig.~\ref{nlody},  for
40 points in the kinematical range covered by the data points included
in the NNPDF2.0 fit, using a grid of 500 points distributed as
the root square of the log from 
  $x_{\mathrm{min}}$ to 1.}
\label{nlody-big}
\end{figure}

It is clear from Fig.~\ref{nlody} that 
with a linear interpolation performed on a 100--points grid, 
we get a reasonable accuracy for all points, 1\% in the worse
case, which is suitable because the experimental uncertainties
of the available datasets are rather larger 
(see Table~\ref{tab:exp-sets-errors}). 
This accuracy can be improved arbitrarily by 
increasing the number of data points in the grid, with a very
small cost in terms of speed: this is demonstrated in
Fig.~\ref{nlody-big}, where we show the improvement in accuracy
obtained by using a grid of 500 points.

%%%%%%%%%%%%%%%%%%%%%
%%%%%%%%%%%%%%%%%%

%%%%%%%%%%%%%%%%%%%%%
%%%%%%%%%%%%%%%%%%

% -----------------------------------------------
%
% Section on minimization and dynamical stopping
%
% -----------------------------------------------

\section{Minimization and stopping}
\label{sec-minim}

As discussed at length in Ref.~\cite{Ball:2008by}, our parametrization
of PDFs differs from other approaches in that we use an unbiased basis
of functions characterized by a very large and redundant set of
parameters: neural networks.  This requires a detailed analysis of the
fitting strategy.  There are two difficulties that have to be
overcome.  First, it is necessary to devise an algorithm to fit neural
network PDFs: observables depend nonlocally and sometimes nonlinearly
on PDFs through convolutions, and the fitting strategy must deal with
this dependence. We have solved this difficulty by means of genetic
algorithms. Second, any redundant parameterization may accommodate not
only the smooth shape of the ``true'' underlying PDFs, but also
fluctuations of the experimental data.  The best fit form of the set
of PDFs is not just given by the absolute minimum of some figure of
merit: it is the possibility of further decreasing the figure of merit
which guarantees that the best fit is not driven by the functional
form of the parameterization. The best fit is instead given by a
suitable optimal training, beyond which the figure of merit improves
only because one is fitting the statistical noise in the data, which
raises the question of how this optimal fit is determined.  We solve
this through the so--called cross--validation
method~\cite{Bishop:1995}, based on the random separation of data into
training and validation sets.  Namely, PDFs are trained on a fraction
of the data and validated on the rest of the data.  Training is
stopped when the quality of the fit to validation data deteriorates
while the quality of the fit to training data keeps improving. This
corresponds to the onset of a regime where neural networks start to
fit random fluctuations rather than the underlying physics
(overlearning).

\subsection{Genetic algorithm strategy}

The fitting of a set of neural networks (which parameterize the PDFs)
to the data is performed by minimization of a suitably defined figure
of merit~\cite{Ball:2008by}. This is a complex task for two
reasons: we need to find a reasonable minimum in a very large
parameter space, and the figure of merit to be minimized is a nonlocal
functional of the set of functions which are being determined in the
minimization.  Genetic algorithms turn out to provide an efficient
solution to this minimization problem.

The basic idea underlying genetic algorithm minimization is to create
a pool of possible solutions to minimize the figure of merit,
each one characterized by a set of parameters. Genetic algorithms 
work on the parameter space, creating new possible solutions
and discarding those which are far from the minimum. 
As a consequence, the genetic algorithm cycle corresponds to
successive generations where: 
i) we create new possible solutions by mutation and crossing;
ii) we naturally select the best candidates and eliminate the rest.
This strategy has proven to be generally very useful to deal with minimization
of functional forms which are further convoluted to deliver
observables (see Ref.~\cite{tau,GonzalezGarcia:2006ay} for applications unrelated to PDF
fitting).

The fitting of the
neural networks on the individual replicas is performed by minimizing 
the error function~\cite{Ball:2008by}
\begin{equation}
  \label{eq:errfun}
  E^{(k)}=\frac{1}{N_{\mathrm{dat}}}\sum_{I,J=1}^{N_{\rm dat}}
                 \left(F_I^{(\mathrm{art})(k)}-F_I^{(\mathrm{net})(k)}\right)
                 \left(\left({\mathrm{cov_{t_0}}}\right)^{-1}\right)_{IJ}
                 \left(F_J^{(\mathrm{art})(k)}-F_J^{(\mathrm{net})(k)}\right) \ ,
\end{equation}
where $F_I^{\rm (art)(k)}$ is the value of the observable $F_I$ at the
kinematical point $I$ corresponding to the Monte Carlo replica $k$, and
%, Eq.~(\ref{eq:replicas}),
 $F_I^{(\rm net)(k)}$ is
the same observable computed from the neural network PDFs, and where
the $t_0$ covariance matrix ${\rm cov_{t_0}}$ has been defined in
Eq.~(\ref{eq:covmat_t0}).
The details of how genetic algorithm minimization
is applied to the problem of PDFs fitting was
presented in Ref.  \cite{Ball:2008by}. This
strategy has been now improved in order to deal with the
addition of multiple new experimental datasets,
as we shall now discuss.

\subsection{Targeted weighted training}

In order to deal more efficiently with the need of fitting data from
a wide variety of different experiments and different datasets within
an experiment we adopt a dynamical 
weighted fitting technique.  The basic idea is
to construct a  minimization procedure that rapidly converges 
towards a configuration for which the final figure
of merit $E^{(k)}$ is 
as even as possible among all
the experimental sets. Weighted fitting consists of adjusting the
weights of the datasets in the determination of the error function
during the minimization procedure according to their individual figure
of merit: datasets that yield a large contribution to the error
function get a larger weight in the total figure of merit.

In a first epoch of the
genetic algorithms minimization, weighted training is
activated. This means than rather than Eq.~(\ref{eq:errfun}), the
actual function which is minimized is 
\begin{equation}
  \label{eq:weight_errfun}
  E_{\rm wt}^{(k)}=\frac{1}{N_{\mathrm{dat}}}
  \sum_{j=1}^{N_{\mathrm{sets}}}p_j^{(k)} N_{\mathrm{dat},j}E_j^{(k)}\, ,
\end{equation}
where $E_j^{(k)}$ is the error function in Eq.~(\ref{eq:errfun}) restricted to the
dataset $j$, $N_{{\rm dat},j}$ is the number of points of this dataset 
and $p_j^{(k)}$ are weights associated to this dataset which are
adjusted dynamically as described below.

In the present analysis, a different, more refined way of determining
the weights  $p_j^{(k)}$ has been adopted  as  compared 
to Refs.~\cite{DelDebbio:2007ee,Ball:2008by}. 
The idea is the following: in the beginning of the fit,
target values $E_{i}^{\rm targ}$ for the figure of  merit of each
experiment are chosen.
Then, at each generation of the minimization, the weights of individual
sets are updated using the conditions
\begin{enumerate}
\item If $E_{i}^{(k)} \ge E_{i}^{\rm targ}$, then $p_i^{(k)}=\lp E_{i}^{(k)}/E_{i}^{\rm targ}\rp^2$, 
\item If $E_{i}^{(k)} < E_{i}^{\rm targ}$, then $p_i^{(k)}=0$ \ .
\end{enumerate}
Hence, sets which are far above their target value will get a
larger weight in the figure of merit.
On the other hand,  
sets which are below their target are likely to be already learnt properly
and thus are removed from the figure of merit which is being minimized.
The determination of the target values
$E_{i}^{\rm targ}$ for all the sets
which enter into the fit is an iterative procedure that works as follows. 
We start with all $E_{i}^{\rm targ}=1$ and proceed
to a first very long fit. Then, we use the outcome of the fit to produce
a first nontrivial set of $E_{i}^{\rm targ}$ values. This procedure is iterated
until convergence.  In practice, convergence is very fast: 
we have used the values of
$\la E_i\ra$ from a
first batch of 100 replicas, in turn produced using as target values
those of a previous very long fit; these values differ generally by
$2-4$\% (at most 10\% in a couple cases) from the 
values of
$\la E_i\ra$ for the reference fit shown in Table~\ref{tab:estfit2}.
This implementation of  targeted weighted training is such that 
the error function of each dataset
tends smoothly to its ``natural'' value, that is,
$p_i^{(k)}\to 1$ as the minimization progresses. Those 
sets which are harder to fit are given
more weight than the experiments that are learnt faster.

An important feature of weighted training is
that weights are given  to individual datasets (as identified in
Table~\ref{tab:exp-sets}) and not just to experiments. This is
motivated by the fact that typically each dataset covers a distinct,
restricted kinematic region. Hence, the weighting takes care of the
fact that the data in different  kinematic regions carry different
amounts of  information and thus require unequal amounts of training.

\begin{figure}[ht]
\begin{center}
\epsfig{width=0.75\textwidth,figure=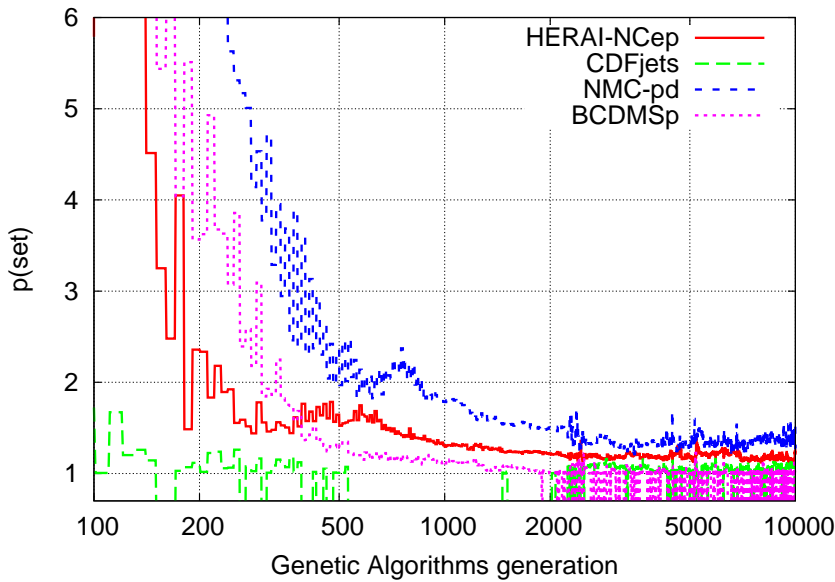}
\caption{\small Illustration of the weighted training
in one particular replica. Individual weights for each dataset converge
to a value of $p_i$ which is close to 1 as the training progresses.
Only the behaviour of representative datasets is shown.
\label{fig:wtexample}} 
\end{center}
\end{figure}

As an illustration of our procedure, we show in Fig.~\ref{fig:wtexample}
the $p_i^{(k)}$ weight
 profiles as a function of the number of genetic algorithm
generations for some sets of a given typical replica. 
Note how, at the early stages of the minimization, 
sets which are harder to learn, such as BCDMSp or NMC-pd
are given more weight than the rest, while at the end
of the weighted training epoch all weights are either $p_i^{(k)}\sim 1$ or
oscillate between 0 and 1, a sign that these sets have been
properly learnt.

The targeted weighted training epoch lasts for $N_{\rm gen}^{\rm wt}$
generations, unless the total error function 
Eq.~(\ref{eq:errfun}) is above
some threshold $E^{(k)} \ge E^{\mathrm{sw}}$. If it is,
weighted training continues until  $E^{(k)}$ falls below the threshold
value. Afterwards, the error function is just the unweighted error
function   Eq.~(\ref{eq:errfun}) computed on experiments.
In this final training epoch, a dynamical stopping of the minimization
is activated, as we shall discuss in the next section.
Going through a final training epoch with the unweighted error function is
in principle
important in order to eliminate any possible residual bias from the choice
of $E_{i}^{\rm targ}$ values in the previous epoch. However, in practice this
safeguard has little effect, as it turns out 
that all weights tend
to unity at the end of the targeted weighted training epoch as they
ought to. 
The whole procedure ensures that  a uniform quality of the fit
for all datasets  is achieved, and that the fit is refined using the correct
figure of merit which includes all the information on correlated
systematics.

%%%%%%%%%%%%%%%%%%%%%%%%%%%%%%%%%%%%%%%%%%%%%%%
%%%%%%%%%%%%%%%%%%%%%%%%%%%%%%%%%%%%%%%%%%%%%%%%%

\subsection{Genetic algorithm parameters}

Genetic algorithms are controlled by some parameters that
can be tuned in order to optimize the efficiency of the whole
minimization procedure. 
The creation of new candidate PDFs that can lower the figure of merit
used in the minimization is implemented using mutations.
That is, each PDF is modified by changing some of the 
parameters that define the neural network. In this work, 
the initial mutation rates $\eta_{i,j}^{(0)}$,
where $i$ labels the PDF and $j$ the specific mutation within this
PDF, for the individual PDFs
are kept the same as in~\cite{Ball:2008by,Ball:2009mk}. 
As training proceeds, 
all mutation rates are adjusted dynamically as a function of the number of iterations $N_{\rm
ite}$
\begin{equation}
  \eta_{i,j} = \eta_{i,j}^{(0)}/N_{\rm ite}^{r_{\eta}} \ . 
\end{equation}
In order to optimally span the range of all possible
beneficial mutations, we introduce an exponent
 $r_{\eta}$ which is randomized between 0 and 1 at
each interation of the genetic algorithm. An analysis of the values of
 $r_{\eta}$ for which mutations are accepted in each generation
 reveals a
flat profile: both large and small mutations are
beneficial at all stages of the minimization.

The number of mutants (new candidate solutions)
in each genetic algorithm generation depends
on the stage of the training.  When the number
of generations is smaller than $N_{\rm gen}^{\rm mut}$,
we use a large population of mutants $N_{\rm mut}^a\gg 1$,
while afterwards we use a much reduced population
 $N_{\rm mut}^b \ll N_{\rm mut}^a $. The rationale for this procedure
is that at early stages of the minimization it is beneficial to
explore as large a parameter space as possible, thus  we need a large
population. Once we are closer to a minimum, a reduced
population helps in propagating the beneficial mutations to
further improve the fitness of the best candidates.
The final choices of parameters of the genetic algorithm which have
been adopted in the NNPDF2.0 parton determination are summarized in
Table~\ref{tab:gapars}.

%%%%%%%%%%%%%%%%%%%%%%%%%%%%%%%%%%%%%%%%
\begin{table}
  \centering
  \begin{tabular}{|c|c|c|c|c|c|}
    \hline 
    $N_{\rm gen}^{\rm wt}$ & $N_{\rm gen}^{\rm mut}$
&   $N_{\rm gen}^{\rm max}$ & $E^{\mathrm{sw}}$ & $N_{\rm mut}^a$ 
&  $N_{\rm mut}^b $\\
    \hline
    $10000$ & 2500 & 30000 & 2.6 & 80 & 10\\
    \hline
  \end{tabular}
  \caption{\small Parameter values for the genetic algorithm.}
  \label{tab:gapars}
\end{table}
%%%%%%%%%%%%%%%%%%%%%%%%%%%%%%%%%%%%%%%%%

\subsection{Preprocessing}

Neural networks can accommodate any functional form, provided they
are made of a large number of layers and sufficient time
is used to train them. Nevertheless, it is customary to use
preprocessing of data to subtract some dominant functional
dependence. Then, smaller neural networks can be trained
in a short time to deal with the deviations with respect
to the dominant function. In our case, we  use preprocessing
to divide out some of the asymptotic small and large $x$
behaviour of PDFs. We avoid possible
 bias related to this by exploring
a large space of preprocessing functions.  

In this work, preprocessing is implemented in the way described in
Sect.~3.1 of Ref.~\cite{Ball:2009mk}, to which we refer for a more
detailed discussion. However, we now adopt in the fit  a 
wider randomized  range of variation
of preprocessing exponents, thus ensuring greater  stability and lack
of bias. 
The range of preprocessing exponents used here is shown in 
Table~\ref{tab:prepexps}.

The explicit independence of results on 
preprocessing exponents within the ranges defined in 
Table~\ref{tab:prepexps}  can be verified by computing the
correlation between the value of a given preprocessing exponent
and the associated value of the
$\chi^{2}$ computed between the $k$--th net and
experimental data, defined by
\begin{equation}
  \label{eq:chi2_errfun}
  \chi^{2(k)}=\frac{1}{N_{\mathrm{dat}}}\sum_{I,J=1}^{N_{\rm dat}}
                 \left(F_I^{(\mathrm{exp})}-F_I^{(\mathrm{net})(k)}\right)
                 \left(\left({\mathrm{cov}}\right)^{-1}\right)_{IJ}
                 \left(F_J^{(\mathrm{exp})}-F_J^{(\mathrm{net})(k)}\right) \ .
\end{equation}
Note that we always include
a factor $\frac{1}{N_{\mathrm{dat}}}$ in the 
definition of the  $\chi^{2}$. Also, 
note that $\left(\left({\mathrm{cov}}\right)^{-1}\right)_{IJ}$ is
the standard covariance matrix, which differs from  the
 $t_0$--covariance matrix Eq.~(\ref{eq:covmat_t0}) because of the
replacement of  $F_{I}^{(0)}$,$F_{J}^{(0)}$ with the measured
values  $F_{I}$,$F_{J}$  in the second term on the right--hand side.

Therefore, we define the
correlation coefficient as follows: considering
for definiteness the large--$x$
preprocessing exponent of the singlet PDF $\Sigma(x,Q^2)$, we have 
\be
\label{eq:preproccorr}
r\lc \chi^2, m_{\Sigma} \rc \equiv \frac{\la \chi^{2}m_{\Sigma}\ra_{\rm rep}
-\la \chi^{2}\ra_{\rm rep}\la m_{\Sigma}\ra_{\rm rep}}{\sigma_{m_{\Sigma}}^2}
 \ .
\ee
This provides the variation $\delta\chi^2$ as
the large--$x$ exponent  $\delta m_{\Sigma}$ is varied around its mean value.
The correlations we find  are very weak as
shown in the last two columns of Table~\ref{tab:prepexps}. It is clear 
that the
$\chi^{2(k)}$ for the
individual replicas is only
marginally affected. This validates quantitatively the  
stability of our results with respect
 to the preprocessing exponents.

%%%%%%%%%%%%%%%%%%
\begin{table}
  \begin{center}
    \begin{tabular}{|c|c|c|c|c|}
      \hline PDF & $\lc m_{\rm min},m_{\rm max}\rc$ & 
$\lc n_{\rm min},n_{\rm max}\rc$ & $r\lc\chi^2,m\rc$ & $r\lc \chi^2, n\rc $\\
      \hline
\hline
      $\Sigma(x,Q_0^2)$  & $\lc 2.55,3.45 \rc$ & 
$\lc 1.05,1.35 \rc$ & -0.018 & 0.131
      \\
      \hline
      $g(x,Q_0^2)$  & $\lc 1.05,1.35 \rc$ & 
      $\lc 1.05,1.35 \rc$  & -0.002  & 0.050\\
      \hline
      $T_3(x,Q_0^2)$  & $\lc 2.55,3.45 \rc$ & 
      $\lc 0,0.5\rc$  &  -0.023 & -0.130 \\
      \hline
      $V_T(x,Q_0^2)$  & $\lc 2.55,3.45 \rc$ & 
      $\lc 0,0.5\rc$  &  0.003& -0.068\\
      \hline
      $\Delta_S(x,Q_0^2)$  & $\lc 12,14\rc$ & 
      $\lc -0.95,-0.65\rc$  & 0.000 & -0.069 \\
      \hline
       $s^+(x,Q_0^2)$  & $\lc 2.55,3.45 \rc$ & 
      $\lc 1.05,1.35 \rc$  & 0.021 & -0.055\\
\hline
      $s^-(x,Q_0^2)$  & $\lc 2.55,3.45 \rc$ & 
      $\lc 0,0.5\rc$  & -0.027 & -0.015 \\
      \hline
    \end{tabular}
    \caption{\small \label{tab:prepexps} The range of
random variation of the large-$x$ and small-$x$ preprocessing exponents
$m$ and $n$ used in the present analysis (the precise form of these 
exponents is given in Sect.~3.1 of Ref.~\cite{Ball:2009mk}). 
The last two columns give the correlation coefficient 
Eq.~(\ref{eq:preproccorr}) between
the $\chi^2$ and respectively the large and small--$x$ 
preprocessing exponents.}
  \end{center}
\end{table}
%%%%%%%%%%%%%%%

\subsection{Positivity constraints}

General theoretical constraints can be imposed during the
minimization procedure, thereby 
guaranteeing that the
fitting procedure only explores the subspace of acceptable 
physical solutions: for example, 
the valence and momentum sum rules are enforced in 
this way~\cite{Ball:2008by}. An important theoretical
constraint is the positivity of physical cross--sections.
As discussed in Ref.~\cite{Altarelli:1998gn}, positivity should 
be imposed on  observable hadronic cross--sections and not on partonic
quantities, which do not necessarily satisfy this constraint.

As in Ref.~\cite{Ball:2008by}, positivity constraints on
relevant physical observables have been  imposed during the genetic
algorithm minimization using a Lagrange multiplier, which strongly
penalizes those PDF configurations which lead to negative observables.
In particular, we impose positivity of $F_L(x,Q^2)$, which
constrains the gluon and the singlet PDFs at small--$x$, as well as
that of the dimuon cross section $d^2\sigma^{\nu,c}/dxdy$~\cite{Ball:2009mk}, 
which
constrains the strange PDFs. Positivity should hold for
any physical cross section which may be measured in principle. In
practice, most PDFs are already well constrained by actual data, so
that positivity is only relevant for PDFs such as the gluon and the
strange distributions which are poorly constrained by the
data. 

Due to the positivity constraints, the minimized
error function Eq.~(\ref{eq:errfun}) (or Eq.~(\ref{eq:weight_errfun})
in the weighted training epoch) is
modified as follows 
\be
\label{eq:poscon}
E^{(k)} \to E^{(k)} - \lambda_{\rm pos} \sum_{I=1}^{N_{\rm dat,pos}}
\Theta\lp -F_{I}^{({\rm net})(k)}\rp F_{I}^{({\rm net})(k)}\, ,
\ee
where $N_{\rm dat,pos}$ is the number of pseudodata points
used to implement the positivity constraints and we choose
$\lambda_{\rm pos}\sim 10^{10}$ as its associate Lagrange multiplier.
Positivity of $F_L(x,Q^2)$ is implemented in
the range $10^{-9}\le x \le 0.005$ and that
of the dimuon cross section in $10^{-9}\le x\le 0.5$,
in both cases at the initial evolution scale $Q^2=2$ GeV$^2$.
This is done because if positivity is enforced at low scales, it will
be preserved by DGLAP evolution.

The impact of the positivity constraints on the NNPDF2.0 PDF determination  will
be quantified in Sect.~\ref{sec:res:positivity}.

\subsection{Determination of the optimal fit}
\label{sec-dynstop}

We now turn to the formulation of the stopping criterion, which is
designed to stop the fit at the point where the fit reproduces the
information contained in the data but not its statistical
fluctuations.  The stopping criterion is applied on the training of
each replica, and it is based on the cross--validation
method, widely used in the context of neural network
training~\cite{Bishop:1995}. Its application to our case has been
described in detail in Refs.~\cite{DelDebbio:2007ee,Ball:2008by}, so
here will mainly focus on the modifications introduced for
NNPDF2.0.

As discussed in the previous section, dynamical stopping is activated after
$N_{\rm gen}^{\rm wt}$ generations of targeted
weighted training. Then,  the weighted
training on sets is switched off and minimization
is done using Eq.~(\ref{eq:errfun}) evaluated with
the error function based on equally weighted experiments.
The dynamical stopping criterion is only activated
if a number of prior conditions are fulfilled. 
We first require that all experiments have an
error function below some reasonable threshold $E_{\mathrm{thres}}$.
Then, it is necessary that a moving average over the error function
for the training and validation sets satisfy
\begin{equation}
\label{eq:trratcond}
r_{\tr} > 1-\delta_{\rm tr};\quad r_{\val} > 1+\delta_{\rm val},
\end{equation}
where
\begin{equation}
  \label{eq:dec-train}
r_{\rm tr}\equiv  \frac{\langle E_{\mathrm{tr}}(i)\rangle}
  {\langle E_{\mathrm{tr}}(i-\Delta_{\mathrm{smear}})\rangle} 
\, ,
\end{equation}
\begin{equation}
  \label{eq:dec-valid}
 r_{\rm val}\equiv \frac{\langle E_{\mathrm{val}}(i)\rangle}
       {\langle E_{\mathrm{val}}(i-\Delta_{\mathrm{smear}})\rangle} \,.
\end{equation}
where the smeared error functions are given by
\begin{equation}
  \label{eq:smearing}
  \langle E_{\mathrm{tr,val}}(i)\rangle\equiv
  \frac{1}{N_{\mathrm{smear}}}
\sum_{l=i-N_{\mathrm{smear}}+1}^iE_{\mathrm{tr,val}}(l)\,,
\end{equation}
with $E_{\mathrm{tr,val}}(l)$ being the figure
of merit Eq.~(\ref{eq:errfun}) restricted to the training
and validation sets for the genetic algorithms generation $l$.

\begin{figure}[ht]
\begin{center}
\epsfig{width=0.76\textwidth,figure=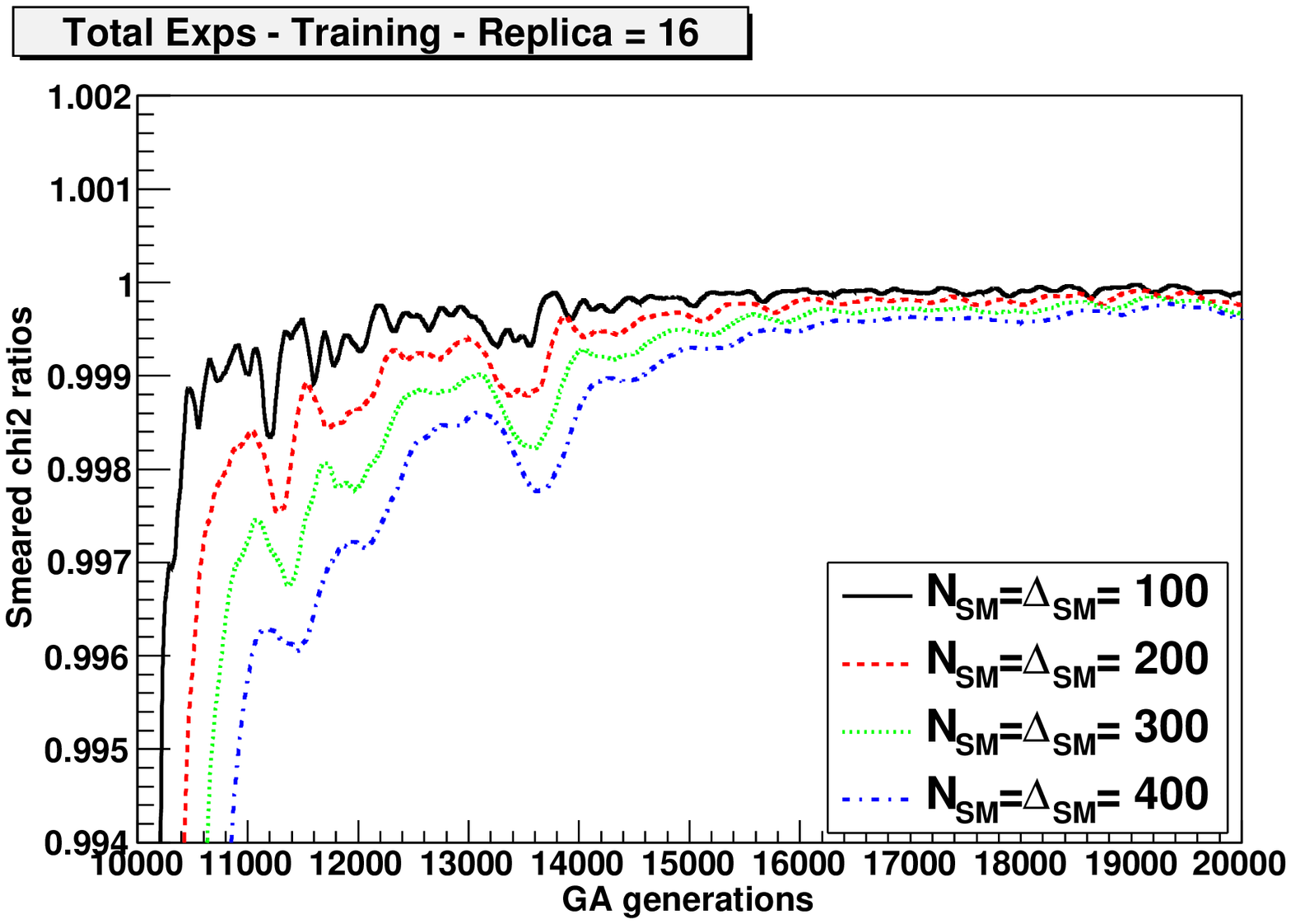}
\epsfig{width=0.76\textwidth,figure=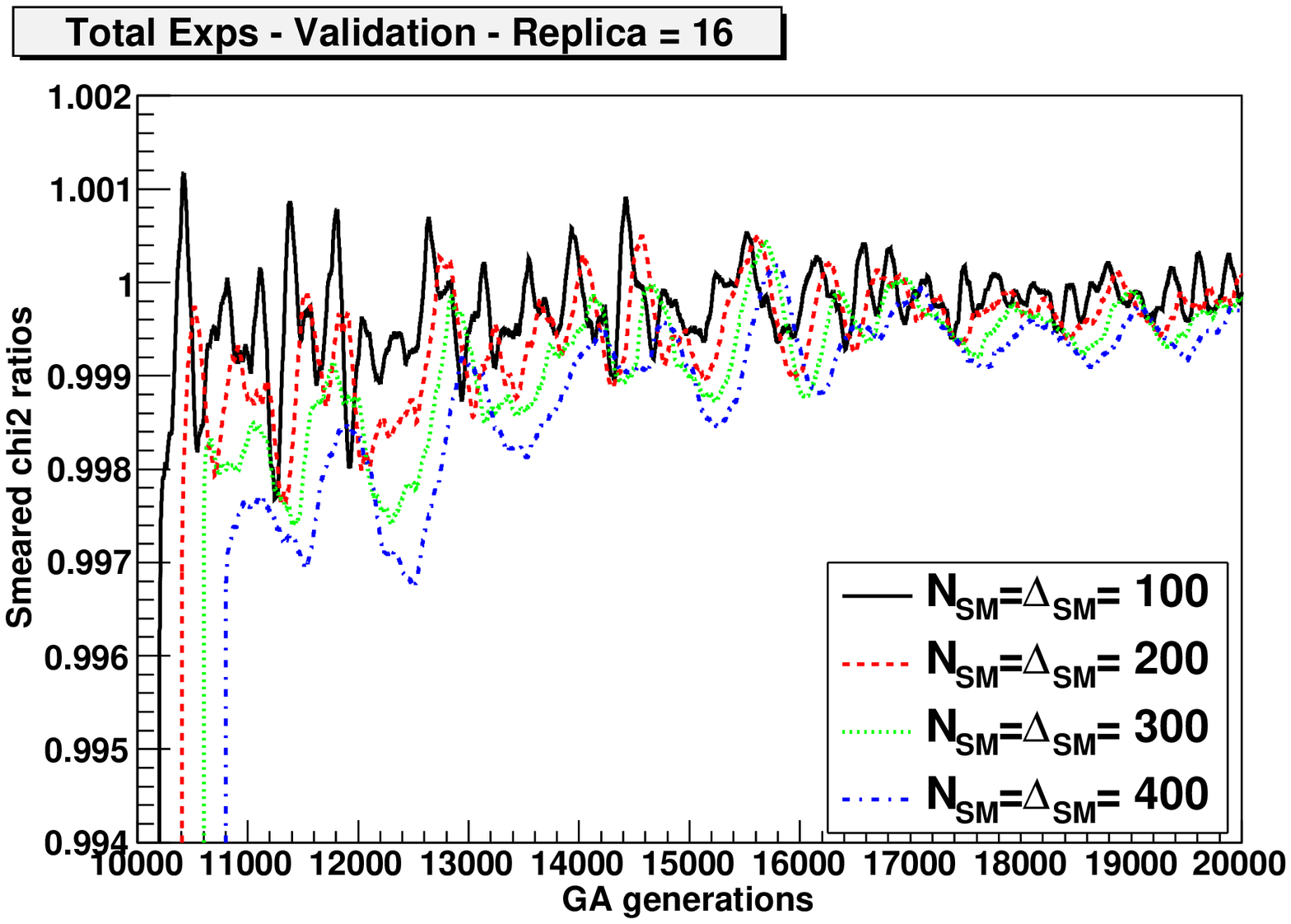}
\caption{\small The training (upper plot) and validation
(lower plot) 
 ratios Eqs.~(\ref{eq:dec-train}- \ref{eq:dec-valid})  
for a particular replica, as a function
of the number of genetic algorithms generations, for various
choices of the smearing parameter
$N_{\rm smear}=\Delta_{\rm smear}$. 
The value $N_{\rm smear}=\Delta_{\rm smear}=200$ is used in the
reference fit (see Table~\ref{tab:dynstop}).
\label{fig:smearrat}} 
\end{center}
\end{figure}

The values of the stopping parameters $\delta_{\rm
  tr}$ and $\delta_{\rm val}$ must be determined by analyzing the
behaviour of the fit for the particular dataset which is being used for 
neural network training.  As an illustration of how this is done in 
practice, we show in
Fig.~\ref{fig:smearrat} the averaged  training and validation $E_{\rm tr/val}$
ratios Eqs.~(\ref{eq:dec-train}-\ref{eq:dec-valid})
for a given replica and different values of the smearing length
$N_{\rm smear}$.  For this particular replica the training has been
artificially prolonged beyond its stopping point.
From Fig.~\ref{fig:smearrat} it is apparent that 
while the training ratio satisfies $r_{\tr} < 1$ always, i.e. that
$E^{(k)}_{\rm tr}$ continues to decrease, after a given number of
generations we have $r_{\val} > 1$, which then oscillates above and
below 1: this is the sign that we have entered an `overlearning'
regime and minimization needs to be stopped. 

The optimal values of the
stopping parameters are chosen to be small enough that overlearning is
avoided, but large enough that the fit does not stop on statistical
fluctuations. The latter condition can be  met only if the value of
$N_{\rm smear}$ is large enough, but if $N_{\rm smear}$ is too large
stopping becomes very difficult and the first condition cannot be
met.  In practice, we have produced a set of 100 replicas with very
long training, and for each
value of $N_{\rm smear}$ we have tried out a range of values of $\delta_{\rm
  tr}$ and $\delta_{\rm val}$, until an optimal set of values which satisfies
all the above criteria has been found. 
The final values of the parameters determined in this way are listed in
Table~\ref{tab:dynstop}. In order to avoid unacceptably long fits,  when
a very large number of 
iterations $N_{\rm gen}^{\rm max}$ is reached
(see Table~\ref{tab:gapars}) training is stopped
anyway. This  leads to a small loss of accuracy of the
corresponding fits which is acceptable provided it only happens for
a small fraction of replicas.

%%%%%%%%%%%%%%%%%%%%%%%%%
\begin{table}
  \centering
  \begin{tabular}{|c|c|c|c|c|c|}
    \hline 
    $N_{\mathrm{smear}} $ & $\Delta_{\mathrm{smear}}$ & $\delta_{\rm tr}$
    & $\delta_{\rm val}$ &  $E_{\mathrm{thres}}$ & $N_{\rm gen}^{\rm max}$\\
    \hline
    $200$ & $200$ & $10^{-4}$&  $3\,10^{-4}$ & 6 & 30000\\
    \hline
  \end{tabular}
  \caption{\small Parameter values for the stopping criterion.}
  \label{tab:dynstop}
\end{table}
%%%%%%%%%%%%%%%%%%%%%%%%%

%%%%%%%%%%%%%%%%%%%%%%%%%%%%%%%%%%%%%%%%%%%%%%%
%%%%%%%%%%%%%%%%%%%%%%%%%%%%%%%%%%%%%%%%%%%%%%%%%

In order to check the consistency of the whole procedure, we have
produced a set of 100 replicas from a fit with the 
same settings as the final reference fit but with
no stopping and a large maximum number of
genetic algorithm  generations $N_{\rm gen}^{\rm max}=50000$. 
This set of 100 replicas allows us thus to verify that the targeted
weighted training and stopping criterion do not bias the fitting
procedure, in that the values of $E_j^{(k)}$ do not drift away from
the target values $E_{j}^{\rm targ}$ when the weighted training is
switched off, and also that the stopping criterion does not introduce
underlearning by stopping the fit at a time when the quality of the
fit is still improving. These conclusions are borne out, and
 in fact, in these
fits for many experiments and replicas 
the value of $E_j^{(k)}$ changes
very little after the target values $E_{j}^{\rm targ}$ are reached 
--- indeed, the target
values were obtained from a very long fit in the first place. Indeed,
the average $\chi^2$ for this fit is only marginally better than that
of the reference fit. However, some experiments do show signs of
overlearning, with an accordingly lower value of the contribution to
the $\chi^2$ . 

\begin{figure}[ht]
\begin{center}
\epsfig{width=0.76\textwidth,figure=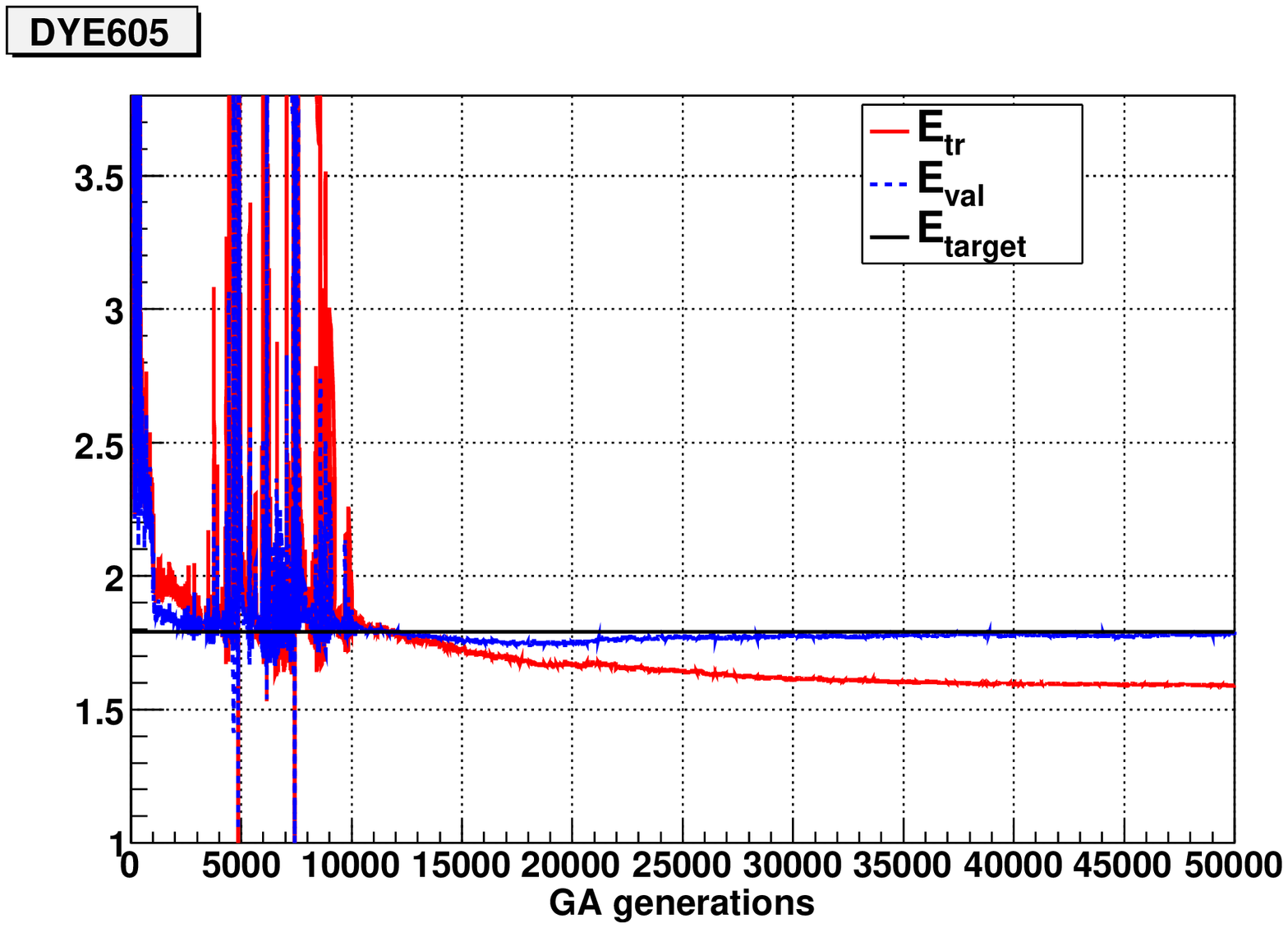}
\epsfig{width=0.76\textwidth,figure=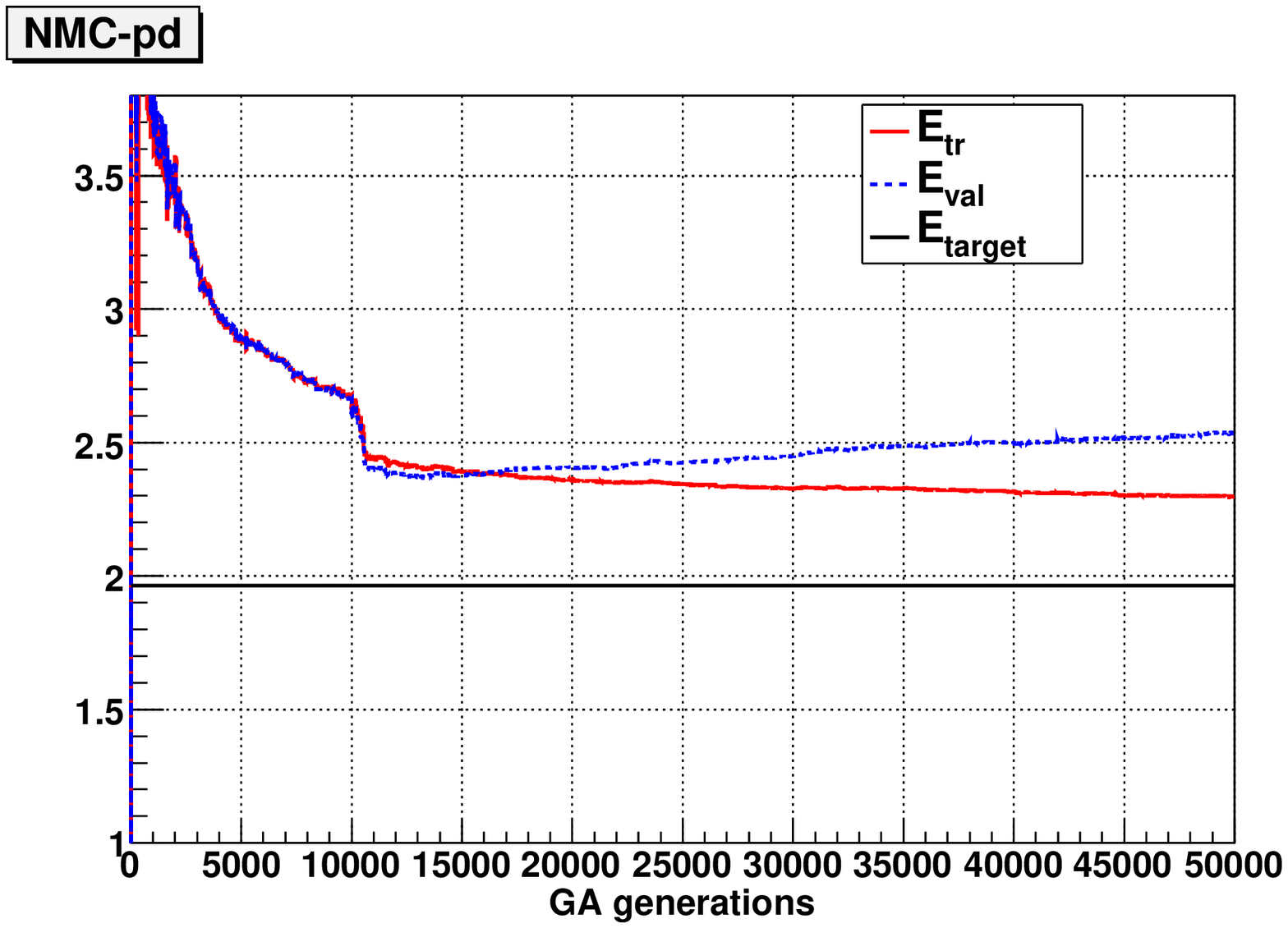}
\caption{\small Two typical examples of  overlearning
behaviour, extracted from a fit with the 
same settings as the final reference fit but with
no stopping and a large maximum number of
genetic algorithm  generations $N_{\rm gen}^{\rm max}=50000$.
The upper plot shows the overlearning of the E605 experiment
observed in one particular replica, and the lower plot corresponds
to the NMC-pd experiment. Note that in these fits weighted training
is switched off at $N_{\rm gen}^{\rm wt}=10000$. 
%In both cases overlearning is seen to set it at about 15000 generations. 
%This is the point where dynamical stopping avoids overlearning.
\label{fig:chi2profiles_verylongfit}} 
\end{center}
\end{figure}

This is illustrated in 
Fig.~\ref{fig:chi2profiles_verylongfit}, where we show 
the $E^{(k)}_i$ profiles for 
two particular experiments (E605 and
NMC-pd) and replicas taken from this fit without stopping.
In the first training epoch, in which the weighted training
Eq.~(\ref{eq:weight_errfun}) is activated, one can see oscillations,
but the downwards trend is clearly visible. Once targeted weighted training
is switched off, minimization proceeds smoothly, and we see
in the two cases that after a given number of genetic algorithms
 generations we
enter in overlearning. For the two
experiments the typical
overlearning behaviour, characterized by the
fact that the validation $E^{(k)}_{\rm tr}$ is rising
while the training $E^{(k)}_{\rm val}$ is still decreasing, sets
in at about 15000 generations. 
This is the point where dynamical stopping avoids overlearning.

%%%%%%%%%%%%%%%%%%%%%%%%%%%%%%%%%%%%%%%%%%%%%%%
%%%%%%%%%%%%%%%%%%%%%%%%%%%%%%%%%%%%%%%%%%%%%%%%%

% --------------------------------------------
%
%\section{Results}
%
%
%-------------------------------------------------
\section{Results}
\label{sec:results}

In this section we present the NNPDF2.0 
parton determination. First we discuss the statistical features
of the fit, then we turn to a comparison of NNPDF2.0 PDFs
and uncertainties with  other
PDF determinations and with
previous NNPDF releases. Next we turn to a study of 
potential deviations from
gaussian behaviour in PDF uncertainty bands.
A detailed comparison between NNPDF2.0 and NNPDF1.2 follows, in which
the impact of each of the differences between these fits is studied in turn:
improved neural network training, treatment of
normalization uncertainties, 
impact of the combined HERA-I dataset,  impact of the inclusion
of jet and Drell-Yan data. Finally we discuss the impact
of the positivity constraints in the PDF determination, and study the
sensitivity of NNPDF2.0 to variations in the value of the strong
coupling $\alpha_s$.

Note that while results for the NNPDF2.0
fit are obtained with  $N_{\rm rep}=1000$ replicas, those for all other
comparisons performed here are done  with $N_{\rm rep}=100$ replicas.

\subsection{NNPDF2.0: statistical features}
\label{sec:res:stat}

%%%%%%%%%%%%%%%%%
\begin{table}
\centering
\begin{tabular}{|c|c|}
\hline 
$\chi^{2}_{\tot}$ &      1.21 \\
$\la E \ra \pm \sigma_{E} $   &       2.32 $\pm$       0.10    \\
$\la E_{\rm tr} \ra \pm \sigma_{E_{\rm tr}}$&       2.29 $\pm$       0.11    \\
$\la E_{\rm val} \ra \pm \sigma_{E_{\rm val}}$&       2.35 $\pm$       0.12    \\
$\la{\rm TL} \ra \pm \sigma_{\rm TL}$   &   16175 $\pm$    6257     \\
\hline
$\la \chi^{2(k)} \ra \pm \sigma_{\chi^{2}} $  &       1.29 $\pm$       0.09    \\
\hline
 $\la \sigma^{(\exp)}
\ra_{\dat}$(\%) &  11.4\\
 $\la \sigma^{(\net)}
\ra_{\dat} $(\%)&  6.0\\
\hline
 $\la \rho^{(\exp)}
\ra_{\dat}$ &  0.18\\
 $\la \rho^{(\net)}
\ra_{\dat}$&  0.54\\
%\hline
% $\la {\rm cov}^{(\exp)}
%\ra_{\dat}$ &  $2.5\,10^2$\\
% $\la  {\rm cov}^{(\net)}
%\ra_{\dat}$&  $4.63$\\
\hline
\end{tabular}
\caption{\small \label{tab:estfit1} Table of statistical estimators
  for NNPDF2.0 with $N_{\rm rep}=
1000$ replicas. The total average uncertainty is given in percentage.}
\end{table}

{
\begin{table}
\centering
\small
\begin{tabular}{|c|c|c|c|c|c|c|}
\hline 
Experiment    & $\chi^2 $& $\la E\ra $   & $\la \sigma^{(\exp)}\ra_{\dat}$(\%) & $\la \sigma^{(\net)}\ra_{\dat}$(\%) & $\la \rho^{(\exp)}\ra_{\dat}$ & $\la \rho^{(\net)}\ra_{\dat}$ \\
\hline 
NMC-pd      & 0.99  &   2.05  & 1.8 & 0.5& 0.03 & 0.36\\
\hline
NMC          & 1.69 &   2.79  & 4.9 & 1.7 & 0.16& 0.77\\
\hline
SLAC         & 1.34 &   2.42  & 4.2 & 1.9 & 0.31& 0.84\\
\hline
BCDMS   &  1.27      &   2.40  & 5.7& 2.6& 0.47& 0.55\\
\hline
HERAI-AV  &  1.14   &   2.25   & 7.5 & 1.3& 0.06& 0.44\\
\hline
CHORUS     &  1.18  &   2.32  & 14.8& 12.8 & 0.09& 0.38\\
\hline
FLH108      &  1.49  &   2.51   & 71.9& 3.3 & 0.65& 0.68\\
\hline
NTVDMN       & 0.67  &   1.90   & 21.1& 14.6 & 0.03 & 0.63\\
\hline
ZEUS-H2     &  1.51 &   2.66  & 13.6 & 1.2& 0.29& 0.58\\
\hline
DYE605      &  0.88  &   1.85 & 22.6& 8.3 & 0.47& 0.75\\
\hline
DYE866      & 1.28  &   2.35  & 20.8& 9.1& 0.20& 0.45\\
\hline
CDFWASY     &  1.85  &   3.09 & 6.0& 4.3& 0.52& 0.72\\
\hline
CDFZRAP     &  2.02  &   2.96  & 11.5& 3.5& 0.83& 0.65\\
\hline
D0ZRAP      & 0.57  &   1.65  & 10.2& 3.0& 0.53& 0.69\\
\hline
CDFR2KT    & 0.80  &   2.22   & 23.0 & 5.2& 0.78& 0.67\\
\hline
D0R2CON   & 0.93   &   1.92  & 16.2 & 6.0& 0.78& 0.64\\
\hline
\end{tabular}
\caption{\small \label{tab:estfit2} Same as Table \ref{tab:estfit1}
  for individual 
individual experiments. Note that
experimental uncertainties are always given in percentage.}
\end{table}
}

%%%%%%%%%%%%%%%%%

The statistical features of the NNPDF2.0 analysis are summarized in
Tables~\ref{tab:estfit1} (for the total dataset) 
and~\ref{tab:estfit2} (for individual experiments).
Note that
$E^{(k)}$ Eq.~(\ref{eq:errfun}) and  ${\chi^2}^{(k)}$
Eq.~(\ref{eq:chi2_errfun})
differ both because in the former
each PDF replica is compared to the data replica it is fitted to,
while in the latter it is compared to the actual data, and also
because of the different treatment of normalization uncertainties as
discussed after Eq.~(\ref{eq:chi2_errfun}). 
The value of
$\chi^{2}_{\tot}$ then refers to the average over replicas (best fit
PDF set), while  the value  $\la \chi^{2(k)}_{\tot} \ra  $ is the
average (and
associate standard deviation) of ${\chi^2}^{(k)}$ computed for each
replica. The average training length $\langle TL\rangle$ (expressed as
a number of generations of the genetic algorithm) is also given
in this table.

%%%%%%%%%%%%%%
\begin{figure}[ht]
\begin{center}
\epsfig{width=0.49\textwidth,figure=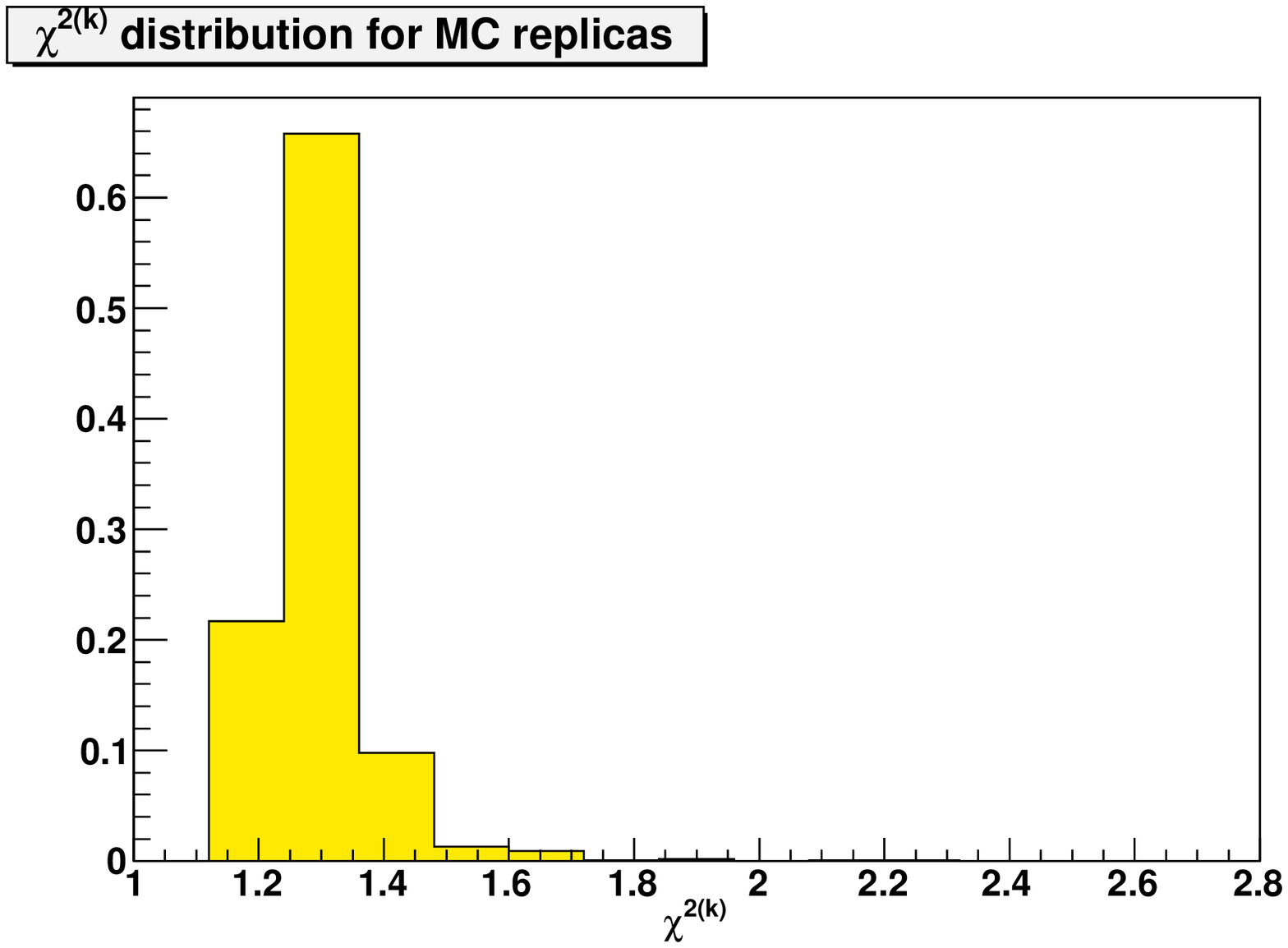}
\epsfig{width=0.49\textwidth,figure=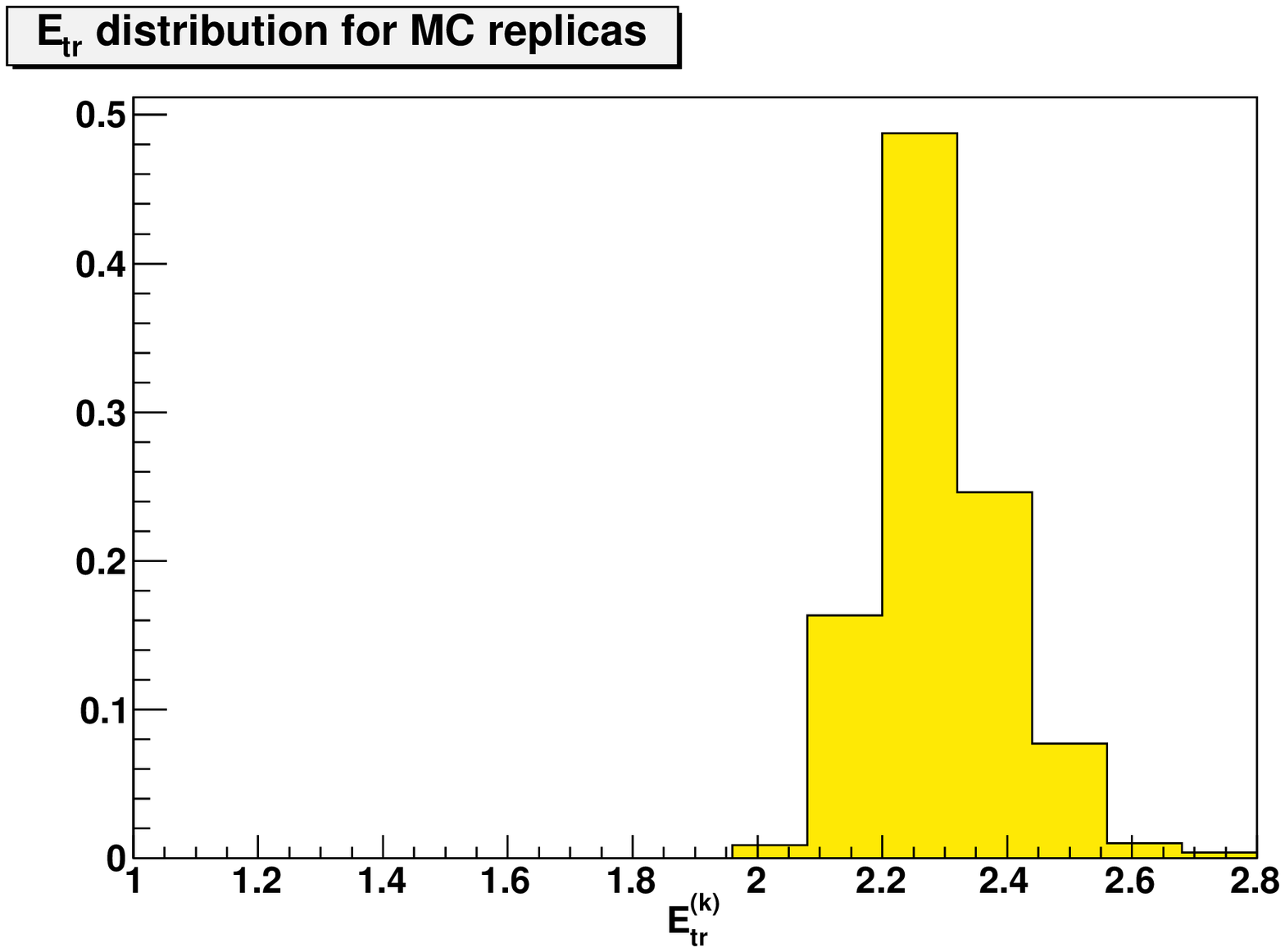}
\caption{\small Distribution of $\chi^{2(k)}$
  Eq.~(\ref{eq:chi2_errfun}) (left) and  $E^{(k)}_{\rm tr}$ 
Eq.~(\ref{eq:errfun}) 
over the sample of $N_{\mathrm rep}=1000$ replicas. \label{chi2histoplots}} 
\end{center}
\end{figure}
\begin{figure}[ht!]
\begin{center}
\epsfig{width=0.49\textwidth,figure=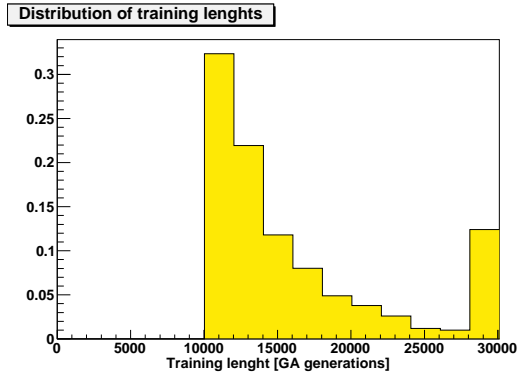}
\caption{\small Distribution of training lengths over the sample of
  $N_{\mathrm rep}=1000$ replicas. 
\label{fig:tl}} 
\end{center}
\end{figure}

%%%%%%%%%%%%%%%%%%
The distribution of 
$\chi^{2(k)}$ Eq.~(\ref{eq:chi2_errfun}), $E^{(k)}_{\rm tr}$
Eq.~(\ref{eq:errfun}) and  training lengths among the $N_{\rm
  rep}=1000$ replicas are shown in Fig.~\ref{chi2histoplots} and
Fig.~\ref{fig:tl} respectively.  While
most of the replicas fulfill the stopping criterion, a small fraction
($\sim12\%$)
of them stop at the maximum training length $N_{\rm gen}^{\rm max}$
which, as discussed in Sect.~\ref{sec-dynstop}, has been introduced in
order to avoid unacceptably long fits. 
This causes some loss of accuracy in outliying fits, but we have
checked that as $N_{\rm gen}^{\rm max}$ is raised more and more of
these replicas would eventually stop, and that the loss of accuracy due to this
choice of value of $N_{\rm gen}^{\rm max}$ is actually very small.

\begin{figure}[ht!]
\begin{center}
\epsfig{width=0.95\textwidth,figure=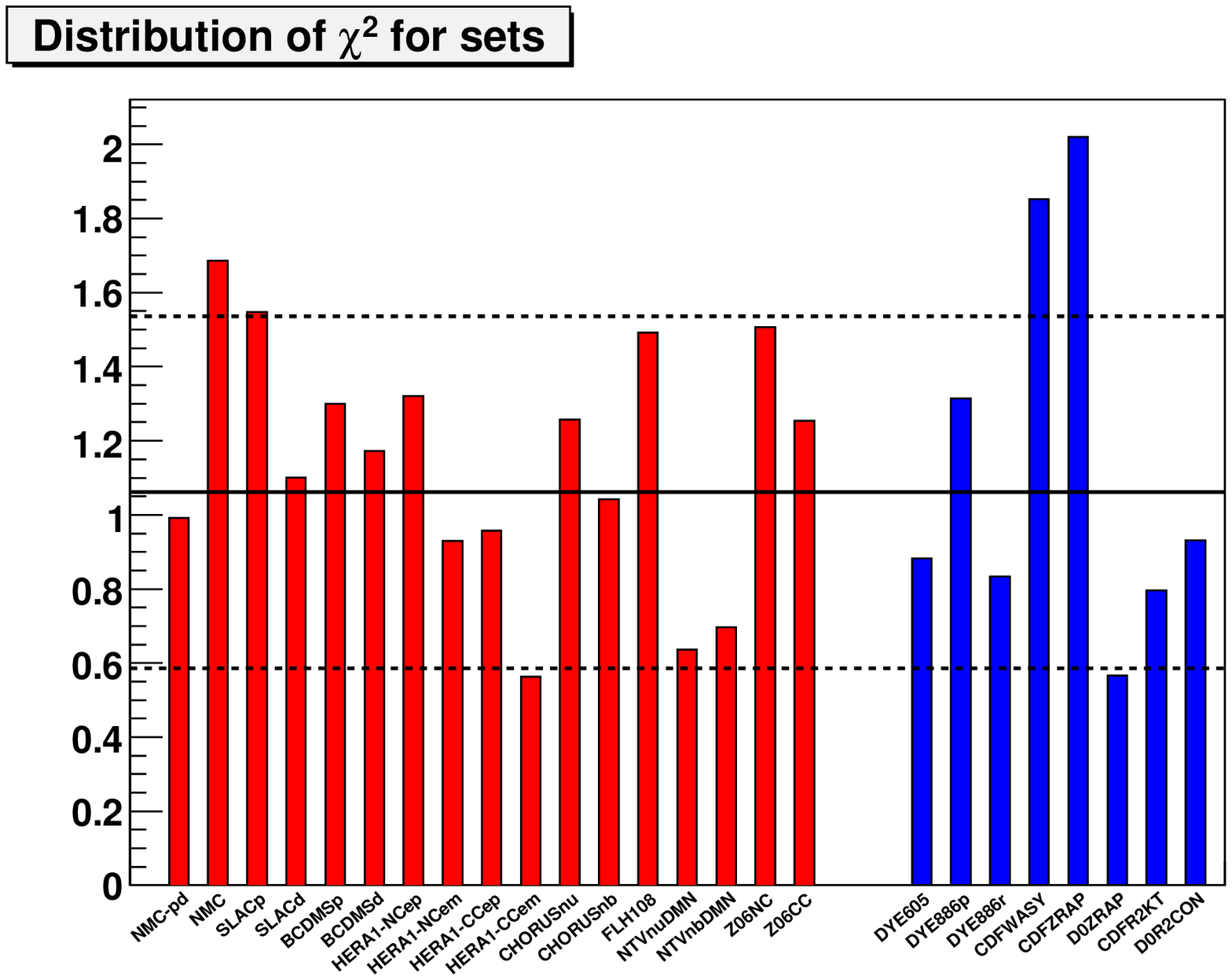}
\caption{\small Values of the
$\chi^2$ (or more properly the $\chi^2$ per data point - see Eq.(52)) 
for the datasets included in the NNPDF2.0 reference fit,
listed in Table~\ref{tab:estfit2}. The
horizontal  line corresponds to the unweighted average
of these $\chi^2$ over the datasets and
and the black dashed line to the one--sigma interval about it:
$\la \chi^2\ra_{\rm sets} = 
1.06$, $\sigma_{\chi^2}=0.40$; DIS and
hadronic datasets are grouped respectively 
to the left and right of the histogram and distinguished  by different colors.
\label{fig:chi2histo}} 
\end{center}
\end{figure}

The features of the fit can be summarized as follows:

\begin{itemize}

\item As in previous fits, the values of $\chi^2_{\tot}$  and 
$\la E \ra $ differ by about one unit, consistent with the expectation that
  the best fit correctly reproduces the underlying true behaviour about which
  data fluctuate, with replicas further fluctuating about
  data. Interestingly, much of the replica fluctuation is already
  removed by neural network training, i.e. when going from $\la E
  \ra$ to 
$\la \chi^{2(k)}\ra  $, with only a further small
  amount of statistical fluctuation being removed when averaging over
  replicas to get the best--fit $\chi_{\tot}$. This reduction 
  was already present in NNPDF1.2 (see the first column of
  Tab.~\ref{tab:estdataset1} below), where however both $\la
  \chi^{2(k)}\ra$ and $\la E \ra $ differed rather more from the best
  fit $\chi^2_{\tot}$ and from each other. The improvement shows that
  the training and stopping algorithm used here and described in
  Sect.~\ref{sec-minim} are more efficient.
\item The quality of the fit as measured by its $\chi^2_{\rm tot}=1.21$ has
  improved in comparison to NNPDF1.2~\cite{Ball:2009mk} despite the
  widening of the dataset to also include hadronic data. As we will
  discuss in greater detail in Sect.~\ref{sec:res:dataset} below (see
  in particular Tab.~\ref{tab:estdataset1}) this improvement is
  largely due to the improvement in training and stopping, 
      and to a lesser extent
    to the improved treatment of normalization uncertainties. The
    inclusion of the very precise combined HERA data then leads to a
    small deterioration in fit quality (possibly because of  the lack
    of inclusion of charm mass effects near charm threshold), while
    the jet and DY data do not lead to any further deterioration. This
    $\chi^2$  value has very low gaussian probability and it is thus  
quite unlikely as a statistical fluctuation: it suggests 
experimental uncertainties might be underestimated at the 10\% level,
or that there might be theoretical uncertainties of the same order.
This appears consistent with the expected accuracy of a NLO treatment
of QCD, and the typical accuracy with which experimental uncertainties
are estimated.
\item The histogram of $\chi^2$ values for each experimental dataset
  is shown in Fig.~\ref{fig:chi2histo}, where the unweighted average
  $\la \chi^2\ra_{\rm sets}\equiv \frac{1}{N_{\rm
      set}}\sum_{j=1}^{N_{\rm set}} \chi^2_{\rm set,j}$ 
  and standard deviation over datasets are also shown.
  We see no evidence of any
  specific dataset being clearly inconsistent with the other, 
and the distribution of values looks broadly
  consistent with statistical expectations, with about five datasets
  with $\chi^2$ at more than one but less than  two sigma from the
  average. Also, we see no obvious difference or tension between
  hadronic and DIS datasets. 
Clearly, the $\chi^2$ values for some experiments if taken
  at face value have low
  gaussian probabilities (though only one, namely NMC, has a
  probability less than 0.01\%). However, they appear to be stable upon
  the inclusion of new data, thus suggesting a lack of tension between
  different datasets. For instance, the $\chi^2$ value of the NMC data
  is very close to that of Refs.~\cite{Forte:2002fg,DelDebbio:2004qj}: 
this value thus appears to reflect
  the internal consistency of these data, not their consistency with
  other data. Some of the issues with specific datasets will be
discussed in somewhat greater detail in this section
  below, while the behaviour of the fit quality as more data are
  included in the fit will be discussed in detail in
  Sect.~\ref{sec:res:dataset}, where strong evidence for the lack of
  tension between datasets will be presented.
\item As in previous NNPDF determinations, the uncertainty of the fit,
  as measured by the average standard deviation $\langle\sigma\rangle$ is
  rather smaller than that of the data: 6.0\% vs. 11.4\%. The uncertainty
  reduction shows that the PDF determination is combining the
  information contained in the data into a determination of an
  underlying physical law. As one would expect the greatest reduction
  is observed in HERA DIS data, but sizable reductions are also seen
  in Drell-Yan and jet data, thus confirming the consistency of these
  data with the global dataset.
\end{itemize}

%%%%%%%%%%%%%%%
%%%%%%%%%%%%%%%
Let us now consider in greater detail the quality of the fit for
some specific experiments whose $\chi^2$ differs by more than one
sigma from the average:

\begin{itemize}
\item The high value of the $\chi^2$ of the NMC $F_2^p$ data has been
  observed in all our previous PDF determinations. It should be
  observed that, as already mentioned, it was first observed in
  Refs.~\cite{Forte:2002fg,DelDebbio:2004qj}, where a parametrization
  of the structure function $F_2^p(x,Q^2)$ was constructed without
  using either PDFs or QCD: hence, this value simply reflects the fact
  that the data within this set are not consistent with each other,
  i.e. they show point-by-point fluctuations which are wider than
  allowed by their declared uncertainty.
\item For dimuon data $\chi^2\sim 0.65$, as
was also the case in NNPDF1.2~\cite{Ball:2009mk}. As discussed
there in detail, this stems from the fact that  
statistical and systematic uncertainties
are added in quadrature for this dataset:
the dominant statistical uncertainty 
is affected by a bin by bin correlation due to the  unfolding procedure
used in  extracting the dimuon cross section from the measured
observable, but the corresponding covariance matrix is not available.

\item The $\chi^2$ of the HERA-I combined data is $\chi^2=1.14$,
  somewhat larger than the value found when fitting
the separate ZEUS and H1 data. The value comes from averaging the
relatively large $\chi^2\sim 1.3$ for the very precise NC positron
dataset, with a low value $\chi^2\sim 0.6$ for CC electron data. 
The reasons for this distribution of values are unclear,
however, we  note that
also in NNPDF1.2~\cite{Ball:2009mk} the $\chi^2$ of the CC datasets was
typically smaller than the average as well. We note also that the
same pattern of $\chi^2$ among the different datasets
has been obtained within the framework of the HERAPDF1.0
analysis of these combined HERA-I dataset~\cite{H1:2009wt,mandyprivate}.
\item The CDF direct $W-$asymmetry measurements have $\chi^2=1.85$.  
The poor compatibility of these data
with the rest of the global fit data was also
  noted in the global analysis of
  Refs.~\cite{Accardi:2009br,accardipriv}.
\item The quality of the fit to Z rapidity distribution data at the
Tevatron differs widely between experiments: 
while an excellent fit is obtained for
D0 data, CDF data are not so well described. 
This suggests that there might be  problem of internal consistency
between the two experiments. A similar pattern was observed in 
the MSTW08 global
fit~\cite{Martin:2009iq}.  Note that
these datasets have a very moderate impact on the global
fit, as proven by the fact that (see Sect.~\ref{sec:res:dataset} below, 
in particular Table~\ref{tab:estdataset1}) the
$\chi^2$ of these data is essentially the same in NNPDF2.0 and  in
NNPDF1.2 (where they are not fitted).
\end{itemize}

Finally, we have checked that if we run a very long fit without
dynamical stopping, the $\chi^2$ of the experiments whose values
exceed the average by more than one sigma does not improve
significantly. This shows that the deviation of these $\chi^2$ values
from the average is not due to underlearning.

\subsection{Parton distributions}
\label{sec:res:pdfs}

%%%%%%%%%%%%%%%%%%%%%%%%%%%%
\begin{figure}[ht]
\begin{center}
\epsfig{width=0.49\textwidth,figure=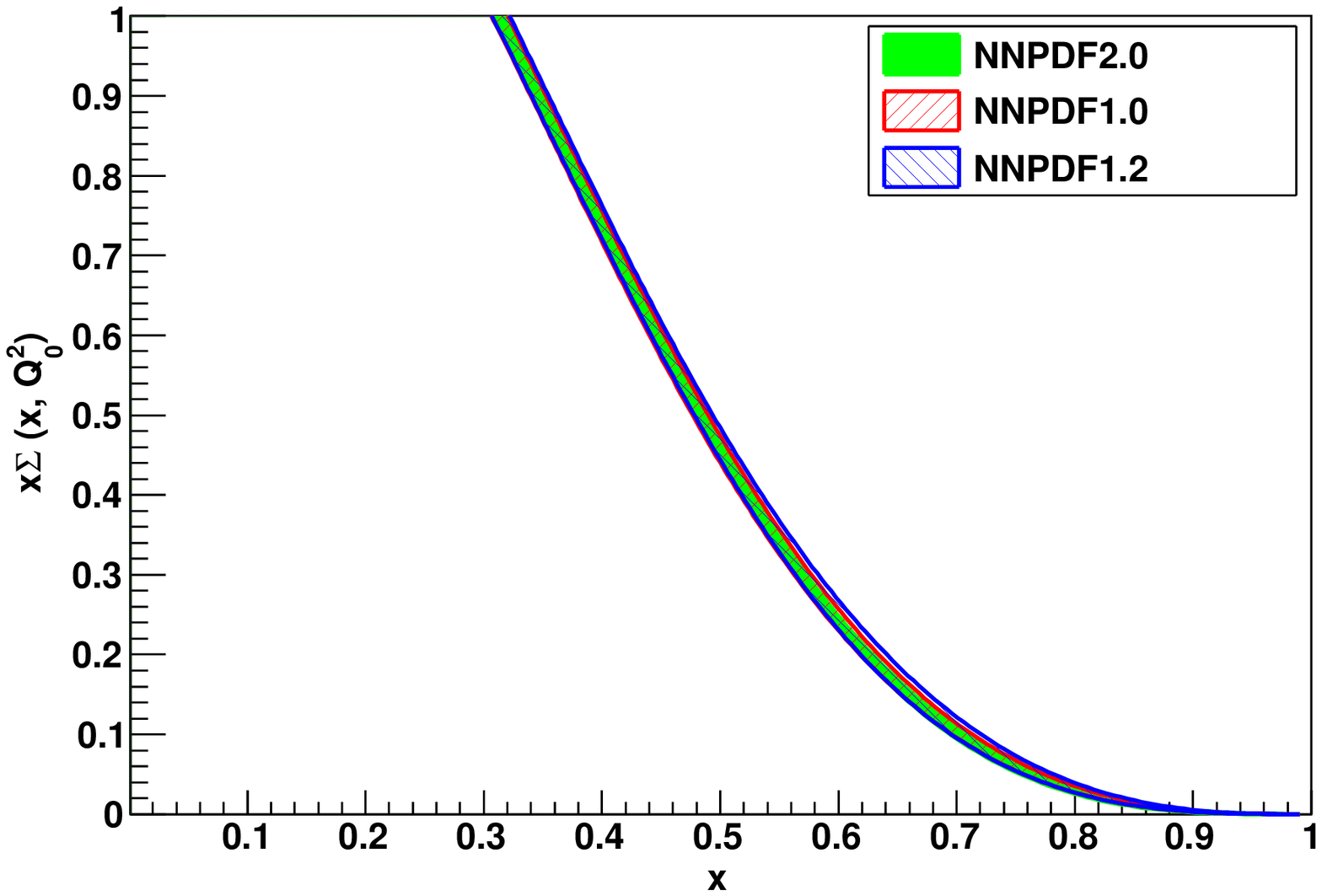}
\epsfig{width=0.49\textwidth,figure=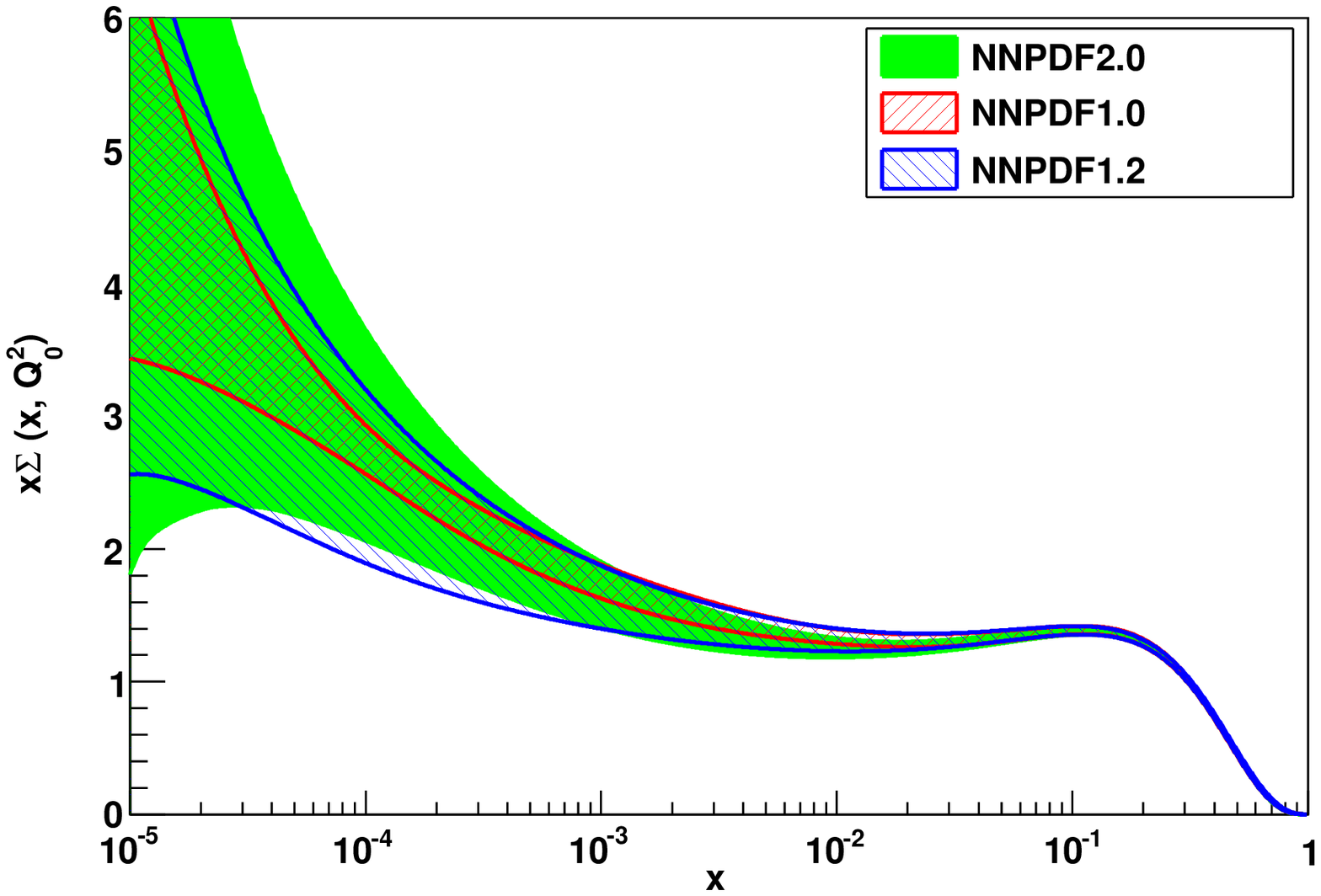}
\epsfig{width=0.49\textwidth,figure=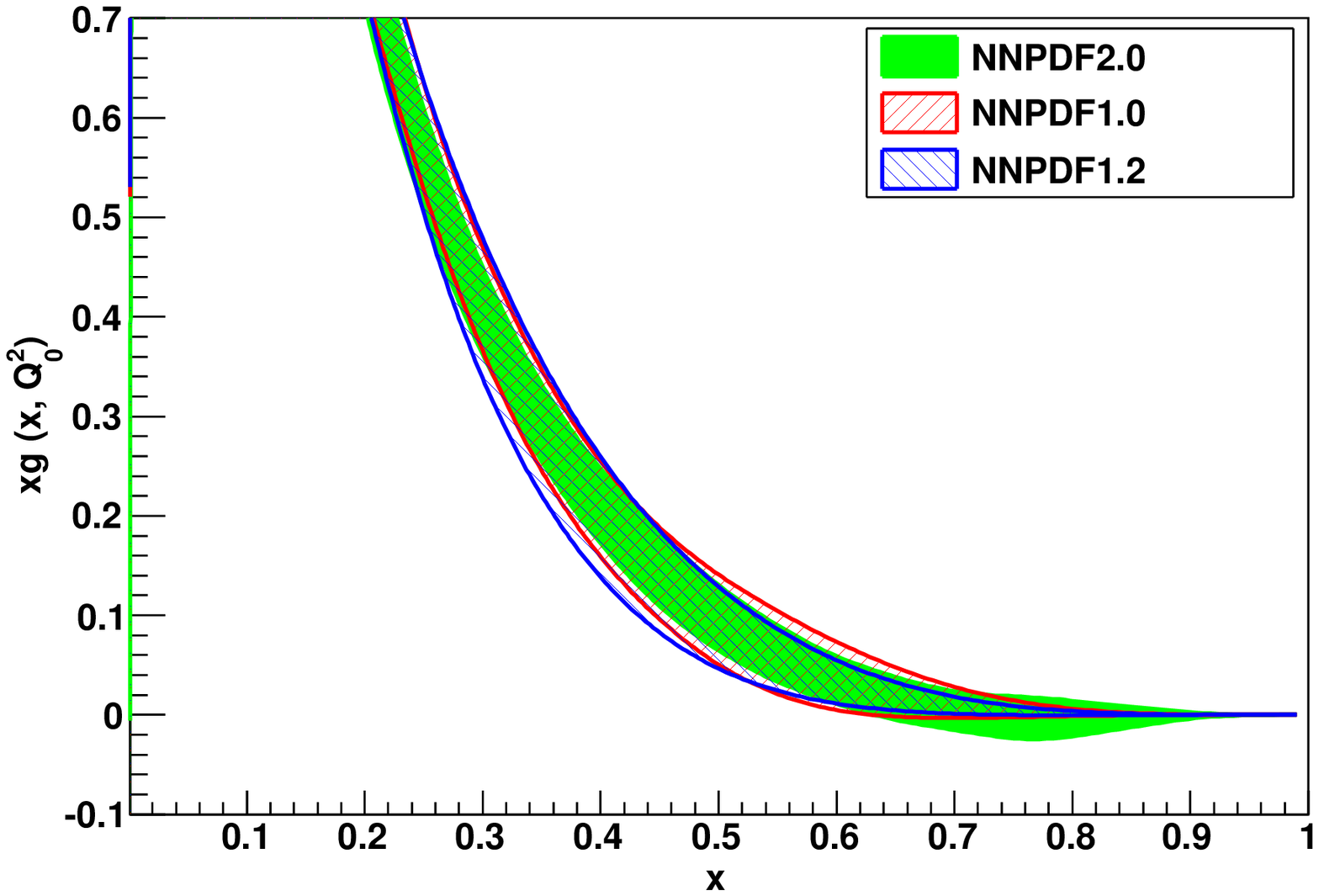}
\epsfig{width=0.49\textwidth,figure=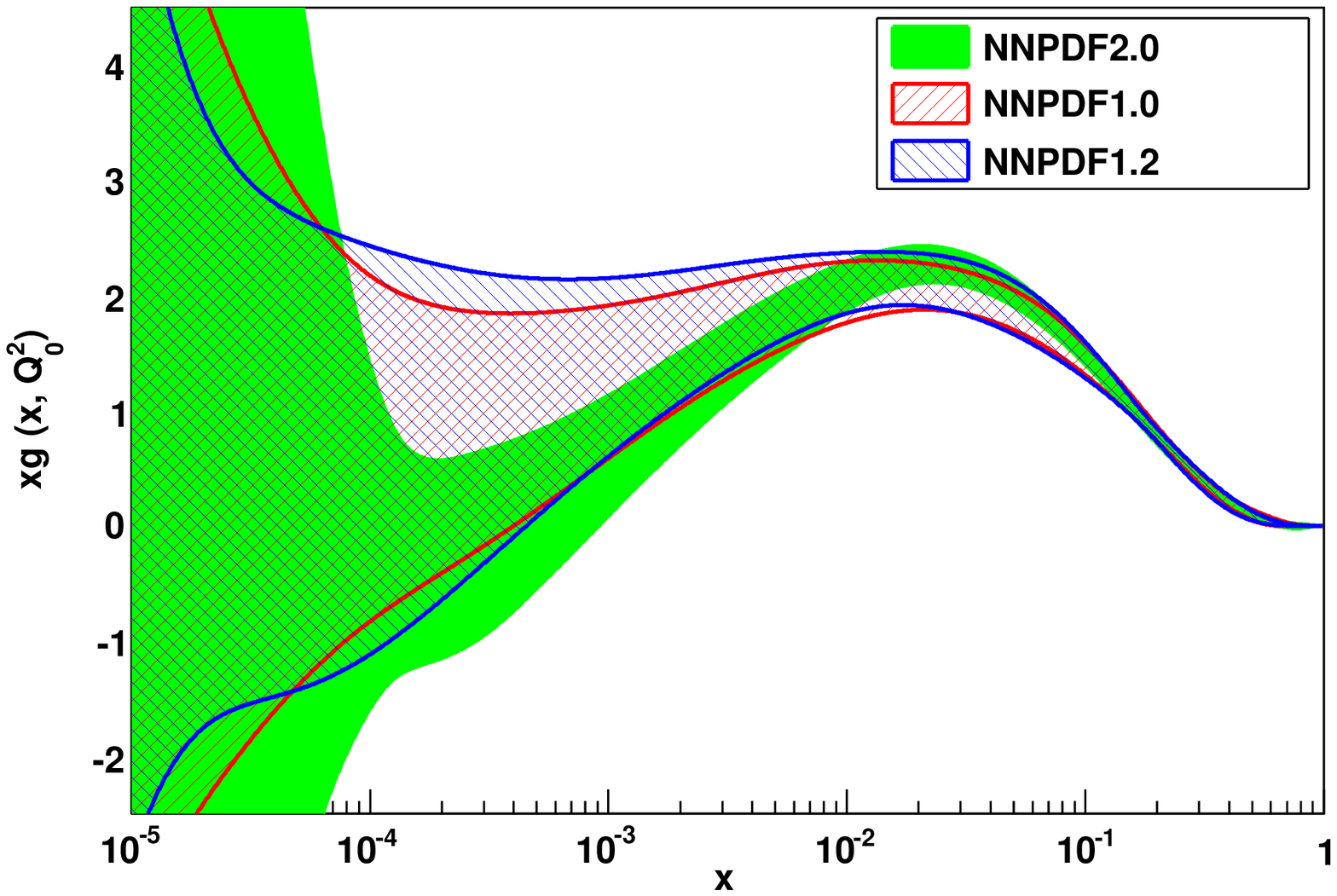}
\epsfig{width=0.49\textwidth,figure=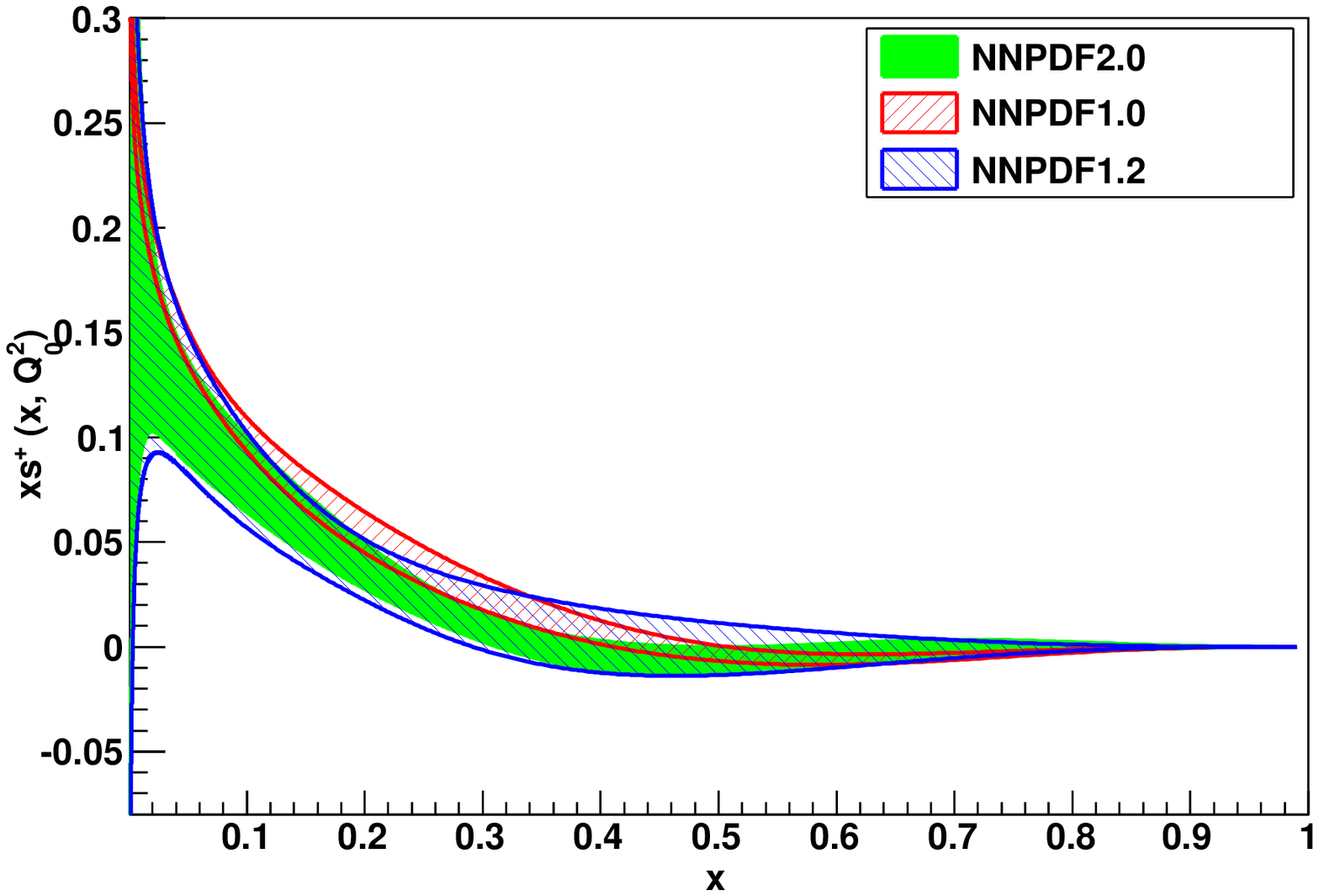}
\epsfig{width=0.49\textwidth,figure=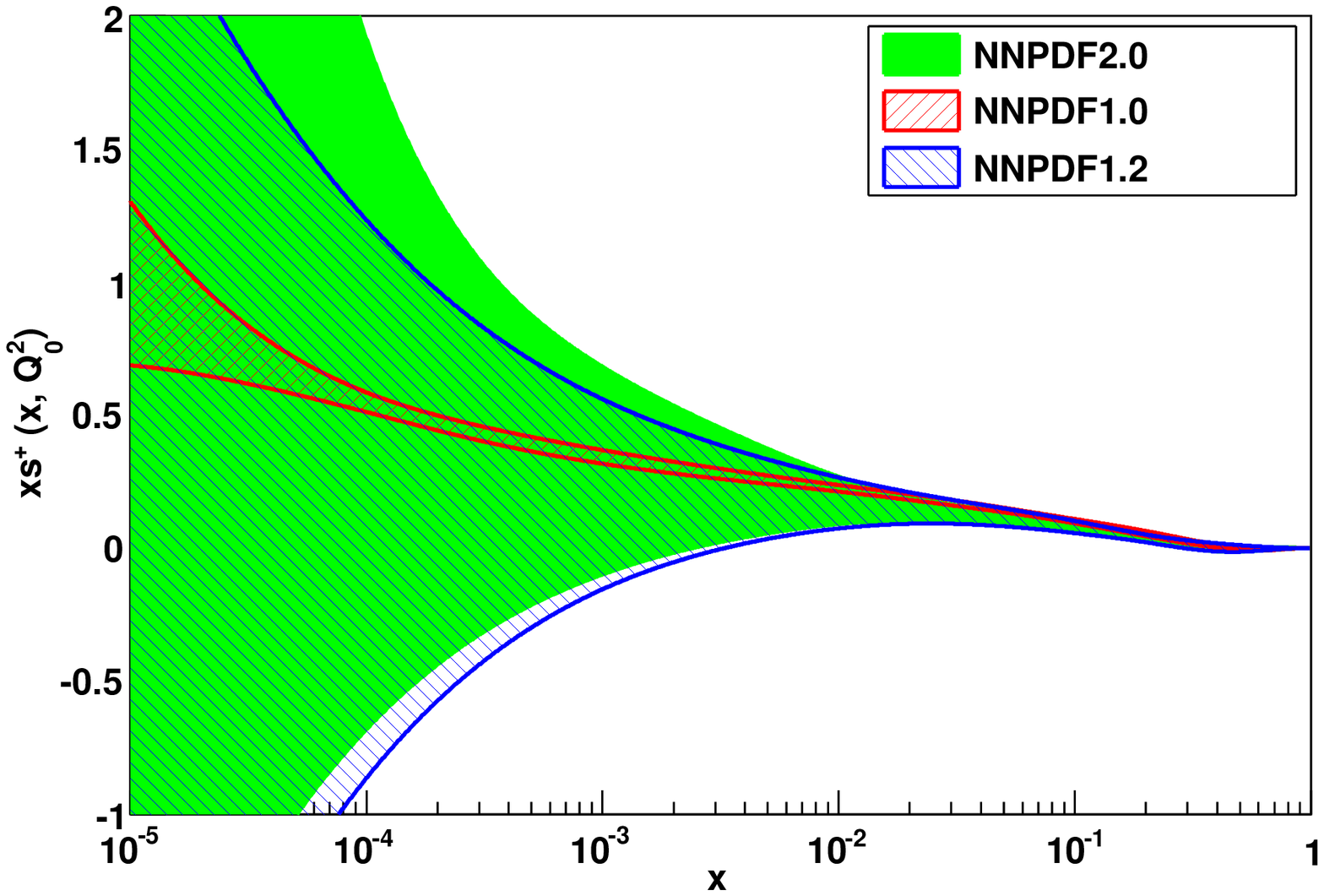}
\caption{\small The singlet $\Sigma=\sum_i (q_i+\bar{q_i})$, gluon $g$ 
and total strangeness $s^+=s+\bar{s}$ at the 
initial scale $Q_0^2=2$ GeV$^2$ from the NNPDF2.0 analysis 
both on linear (left) and
logarithmic (right) scale, compared to the previous NNPDF releases 
NNPDF1.0~\cite{Ball:2008by} and 
NNPDF 1.2~\cite{Ball:2009mk}.
 \label{fig:singletPDFs-nnpdf}} 
\end{center}
\end{figure}

\begin{figure}[ht]
\begin{center}
\epsfig{width=0.49\textwidth,figure=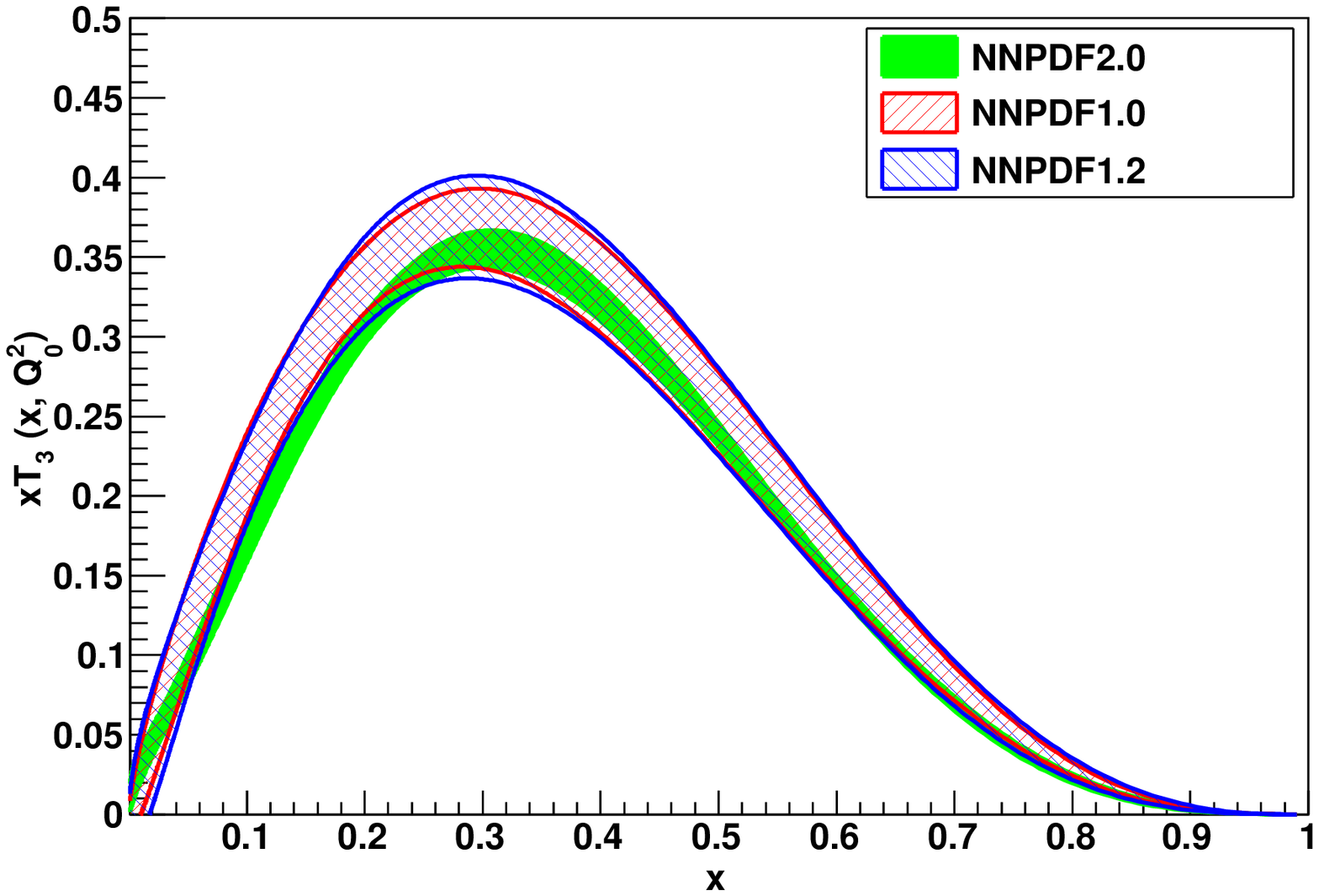}
\epsfig{width=0.49\textwidth,figure=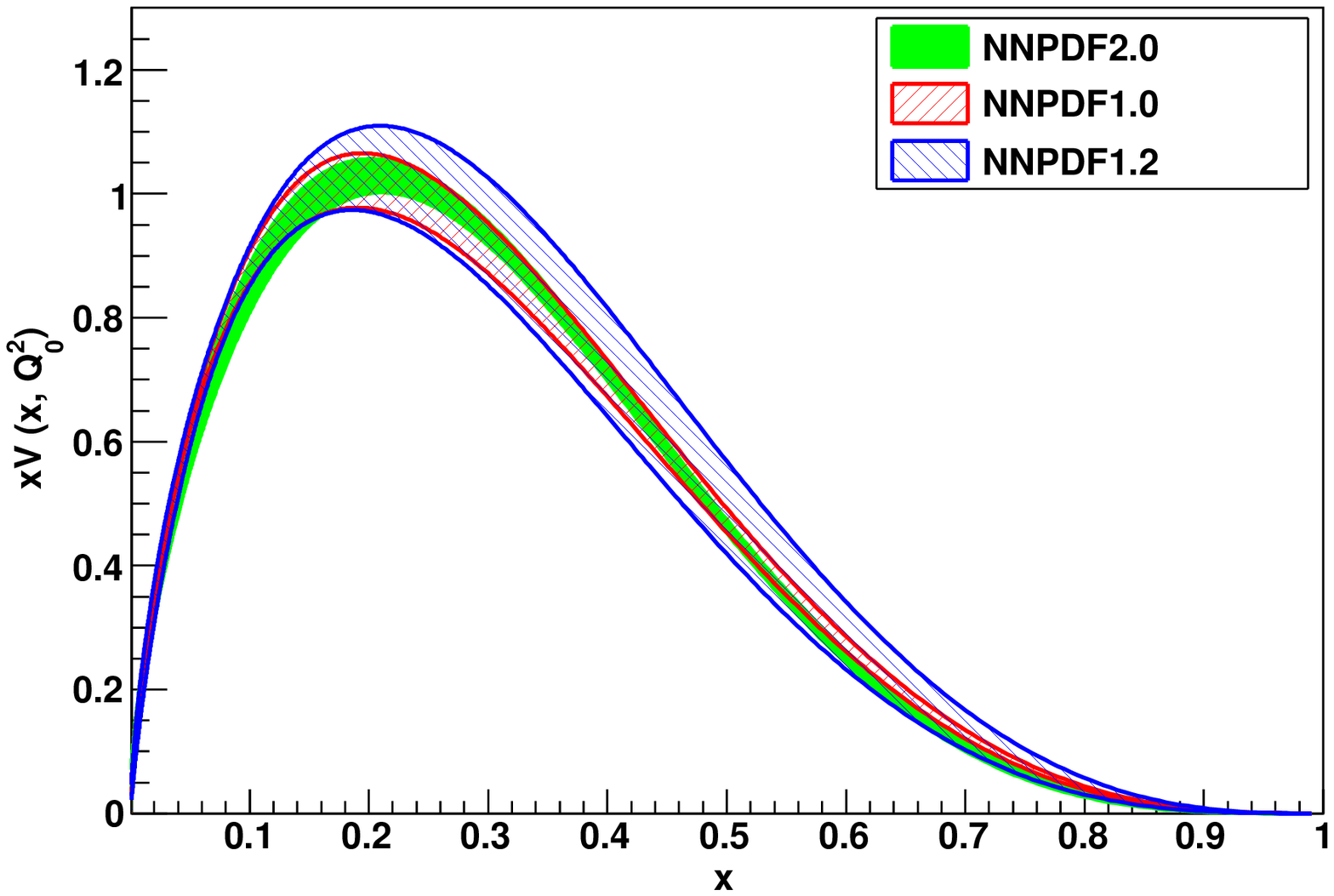}
\epsfig{width=0.49\textwidth,figure=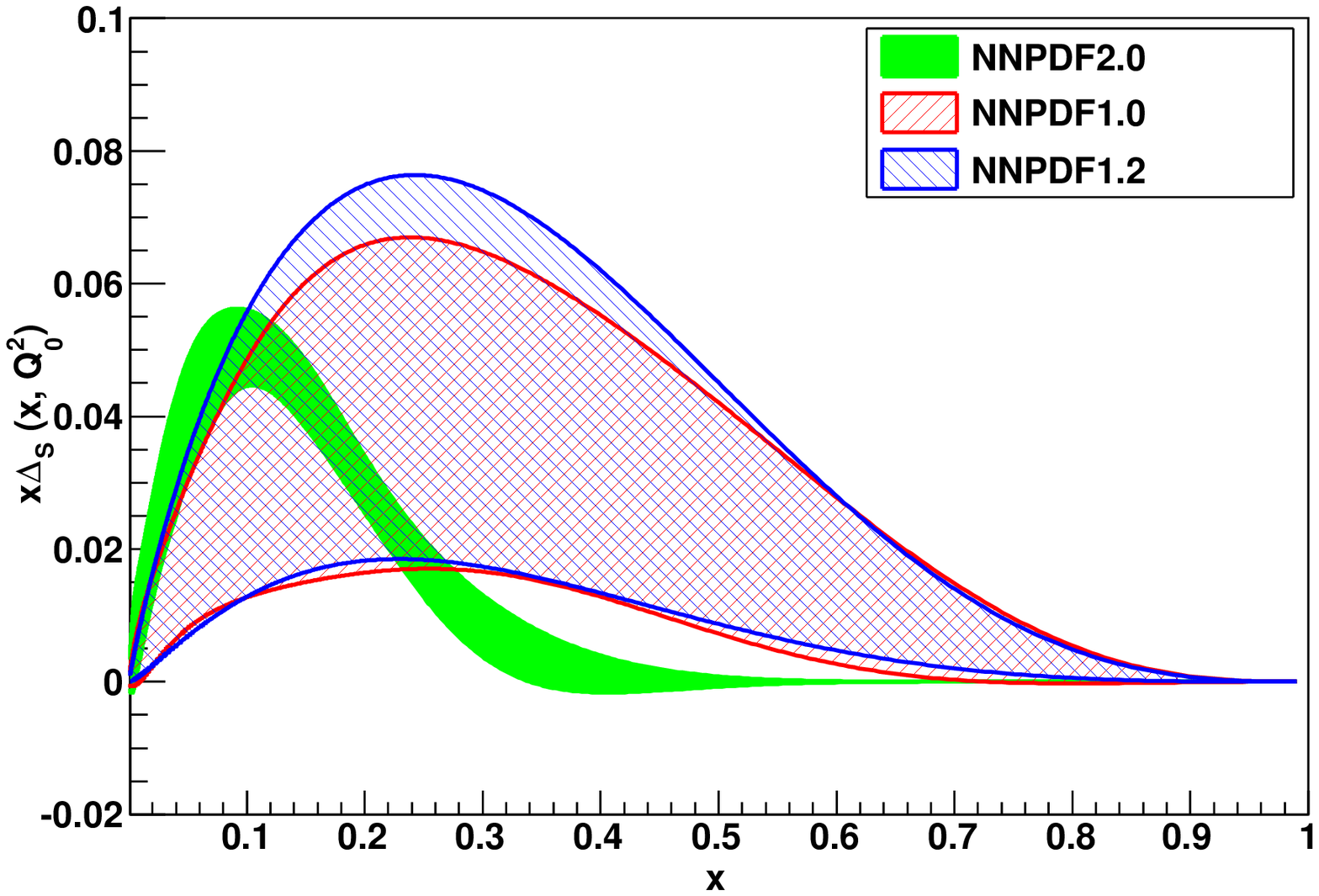}
\epsfig{width=0.49\textwidth,figure=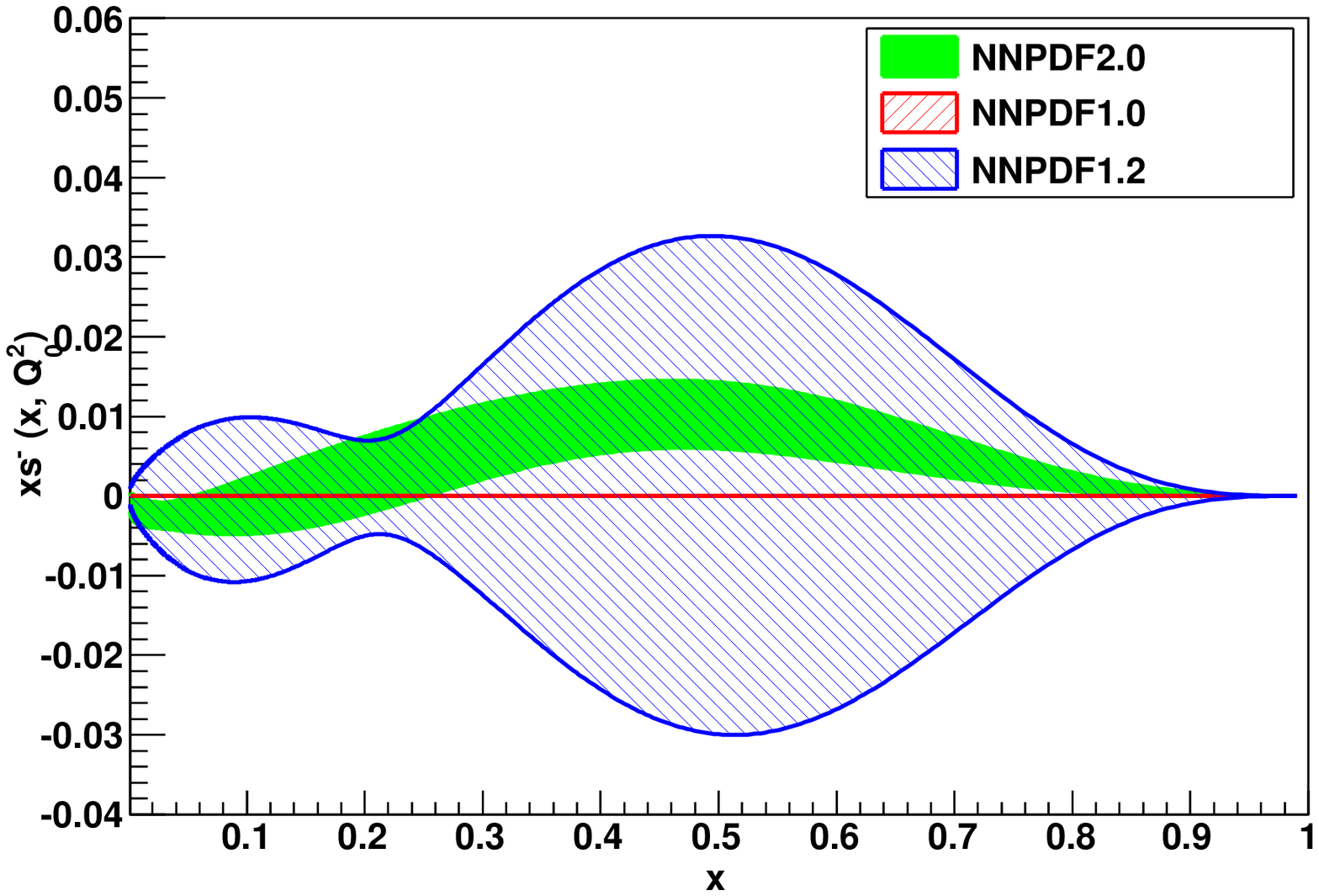}
\caption{\small Same as Fig.~\ref{fig:singletPDFs-nnpdf} for the 
triplet $T_3= u+\bar{u}-d-\bar{d}$,
total valence $V= \sum_i (q_i-\bar{q_i})$, 
sea asymmetry $\Delta_S= \bar{d}-\bar{u}$ 
and strangeness asymmetry $s^- = s-\bar{s}$.
 \label{fig:valencePDFs-nnpdf}} 
\end{center}
\end{figure}
%%%%%%%%%%%%%%%%%%%%%%%%%%

%%%%%%%%%%%%%%%%%%%%%%%%%%%%%%%%%%%%%%%%%%%%%%
\begin{figure}[ht]
\begin{center}
\epsfig{width=0.49\textwidth,figure=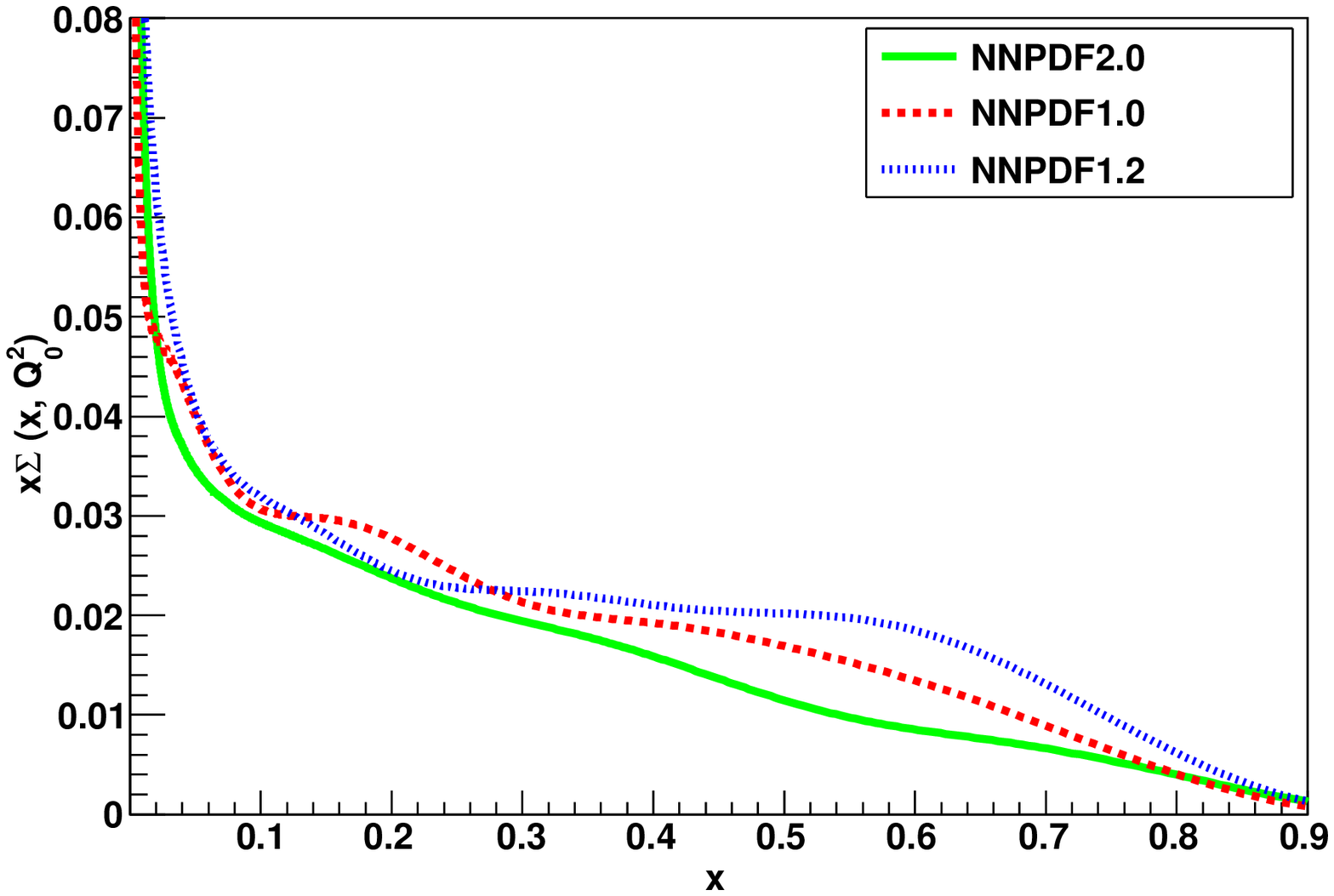}
\epsfig{width=0.49\textwidth,figure=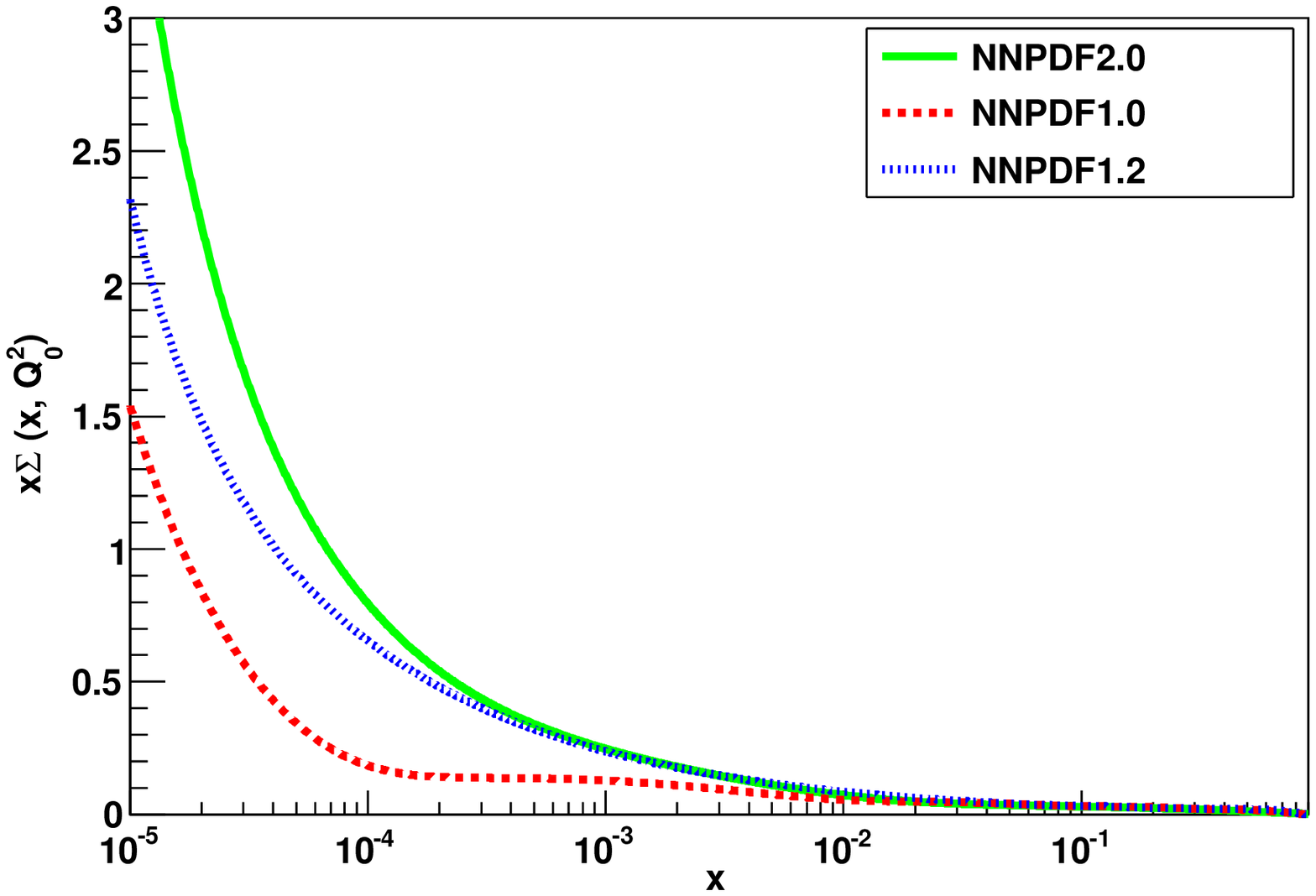}
\epsfig{width=0.49\textwidth,figure=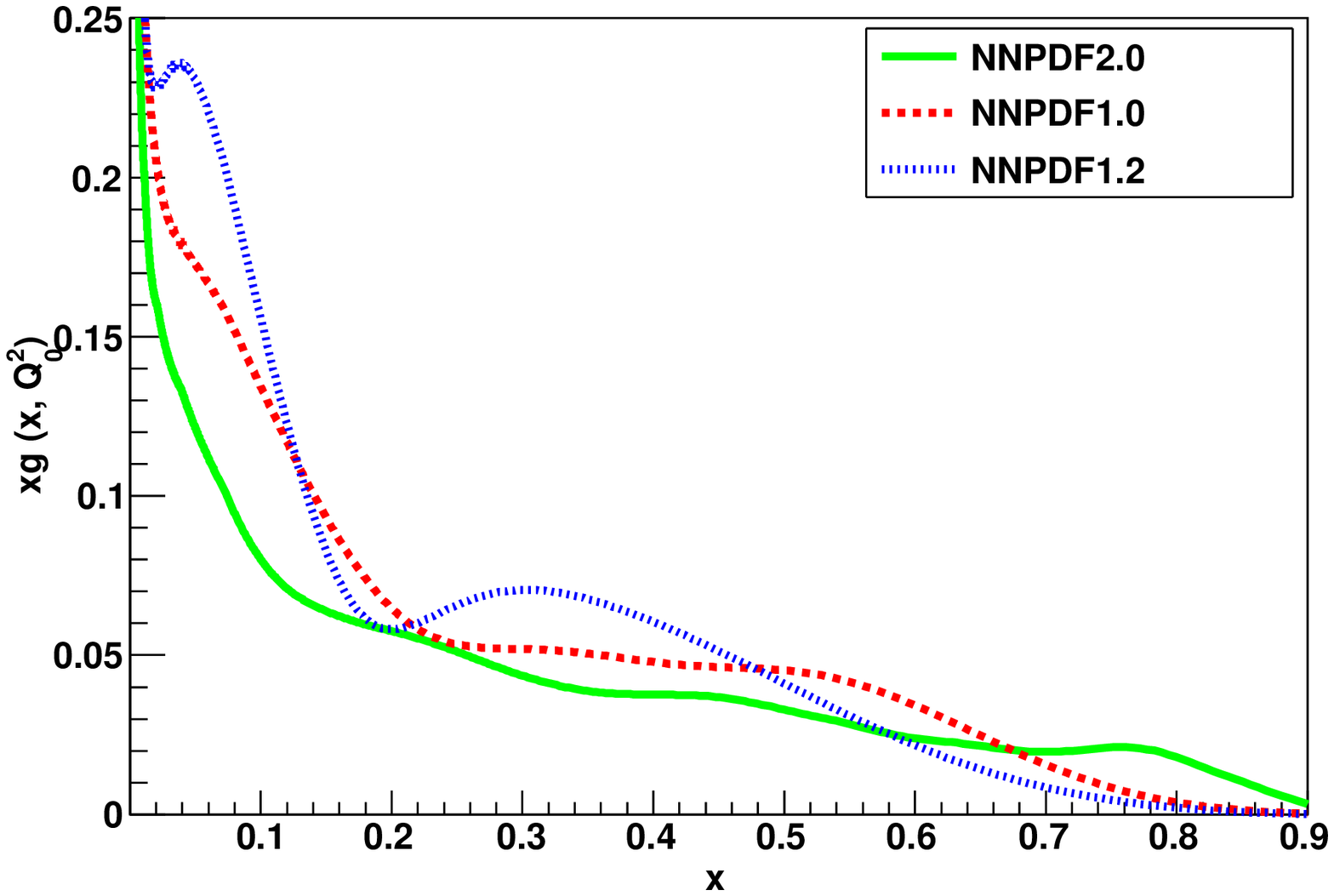}
\epsfig{width=0.49\textwidth,figure=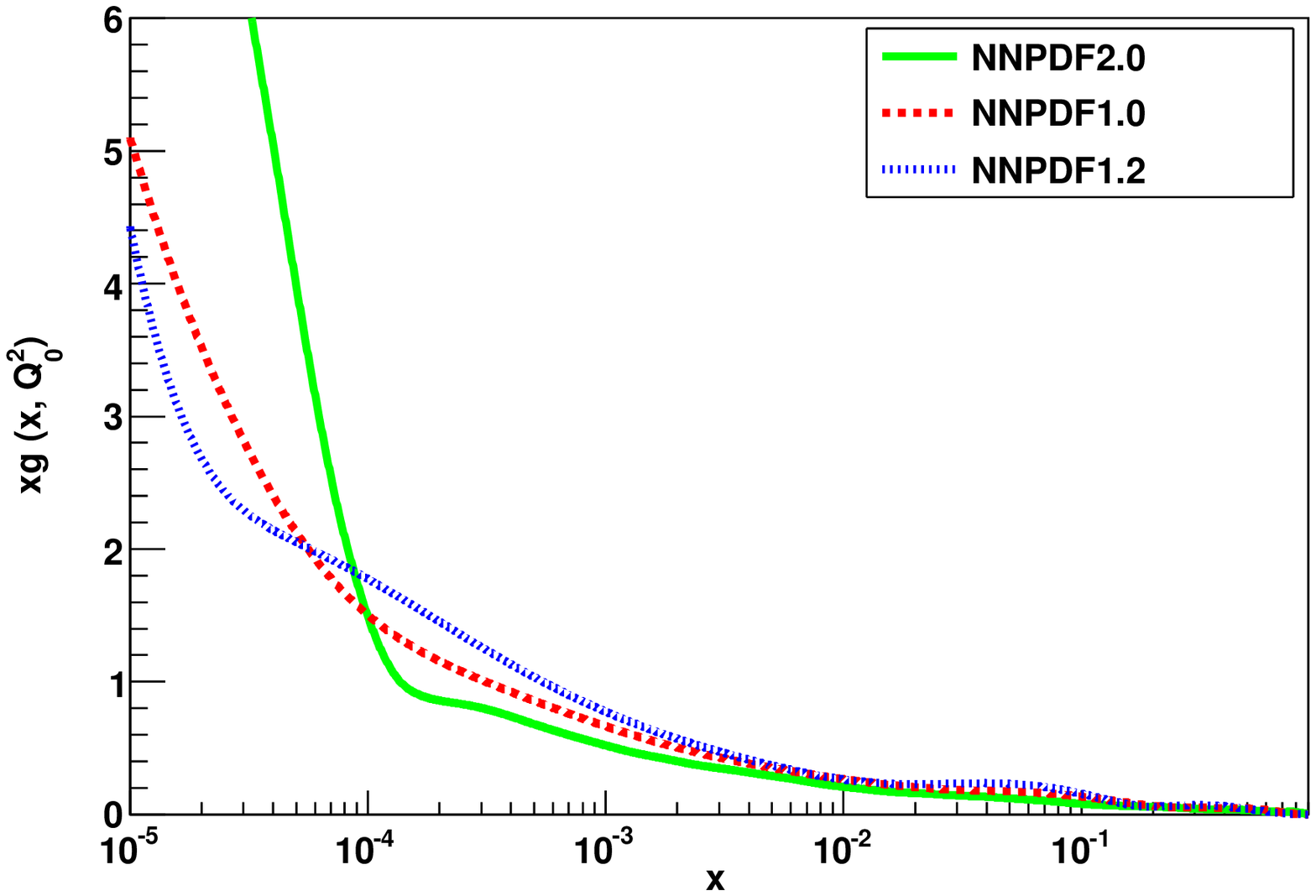}
\epsfig{width=0.49\textwidth,figure=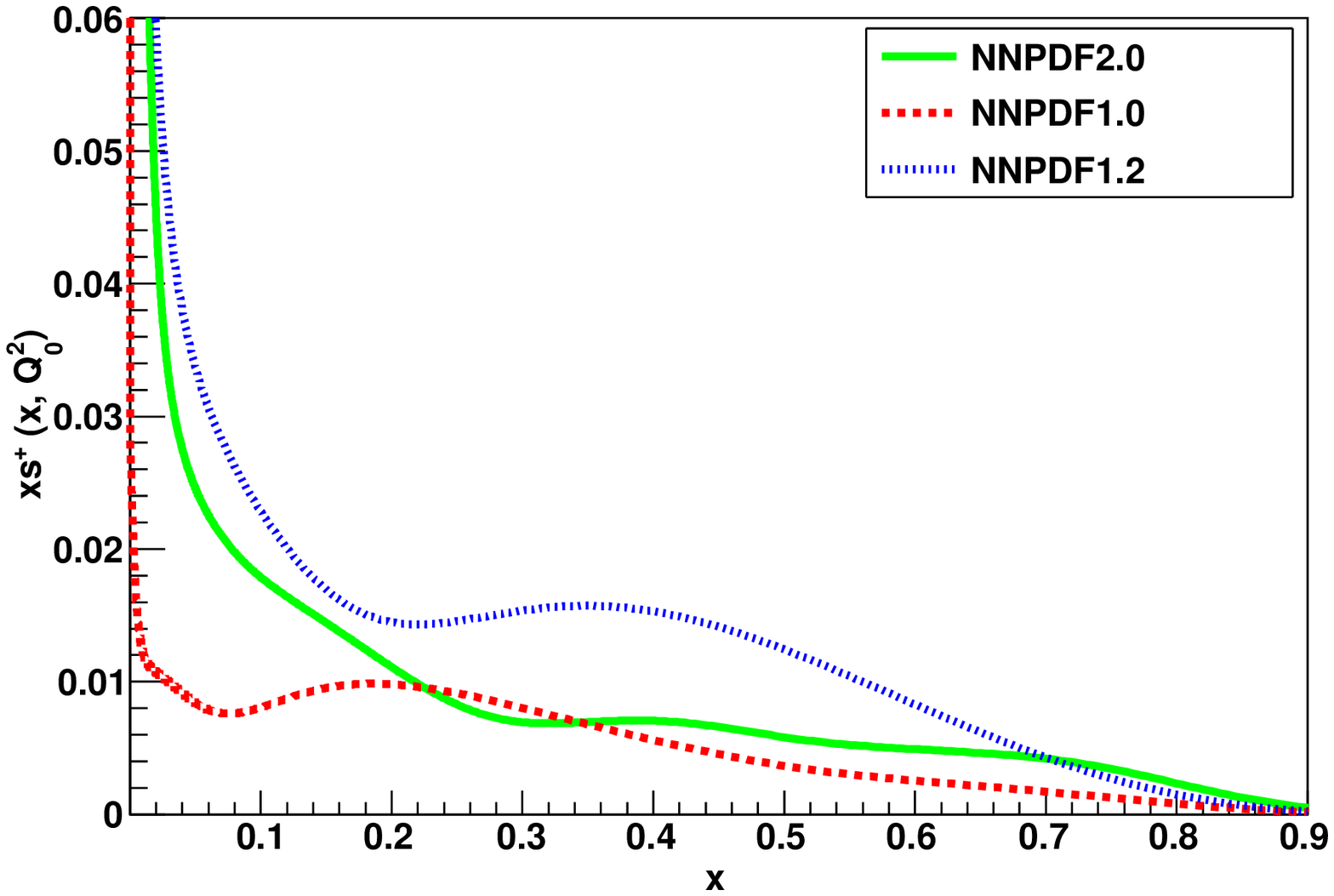}
\epsfig{width=0.49\textwidth,figure=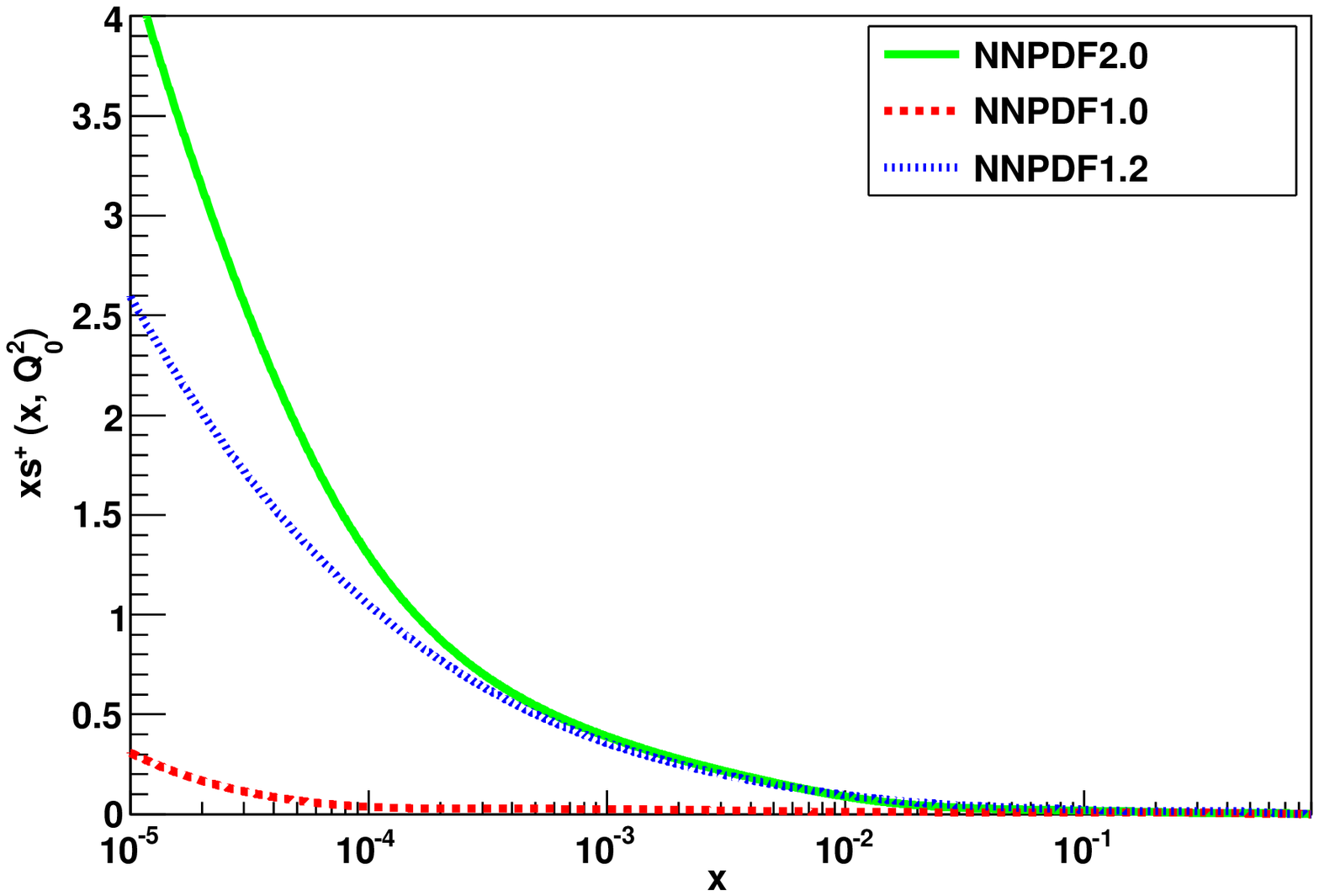}
\caption{\small Absolute uncertainties on the PDFs of Fig.~\ref{fig:singletPDFs-nnpdf}.
\label{fig:pdferrorsabs-nnpdf}} 
\end{center}
\end{figure}
\begin{figure}[ht!]
\begin{center}
\epsfig{width=0.45\textwidth,figure=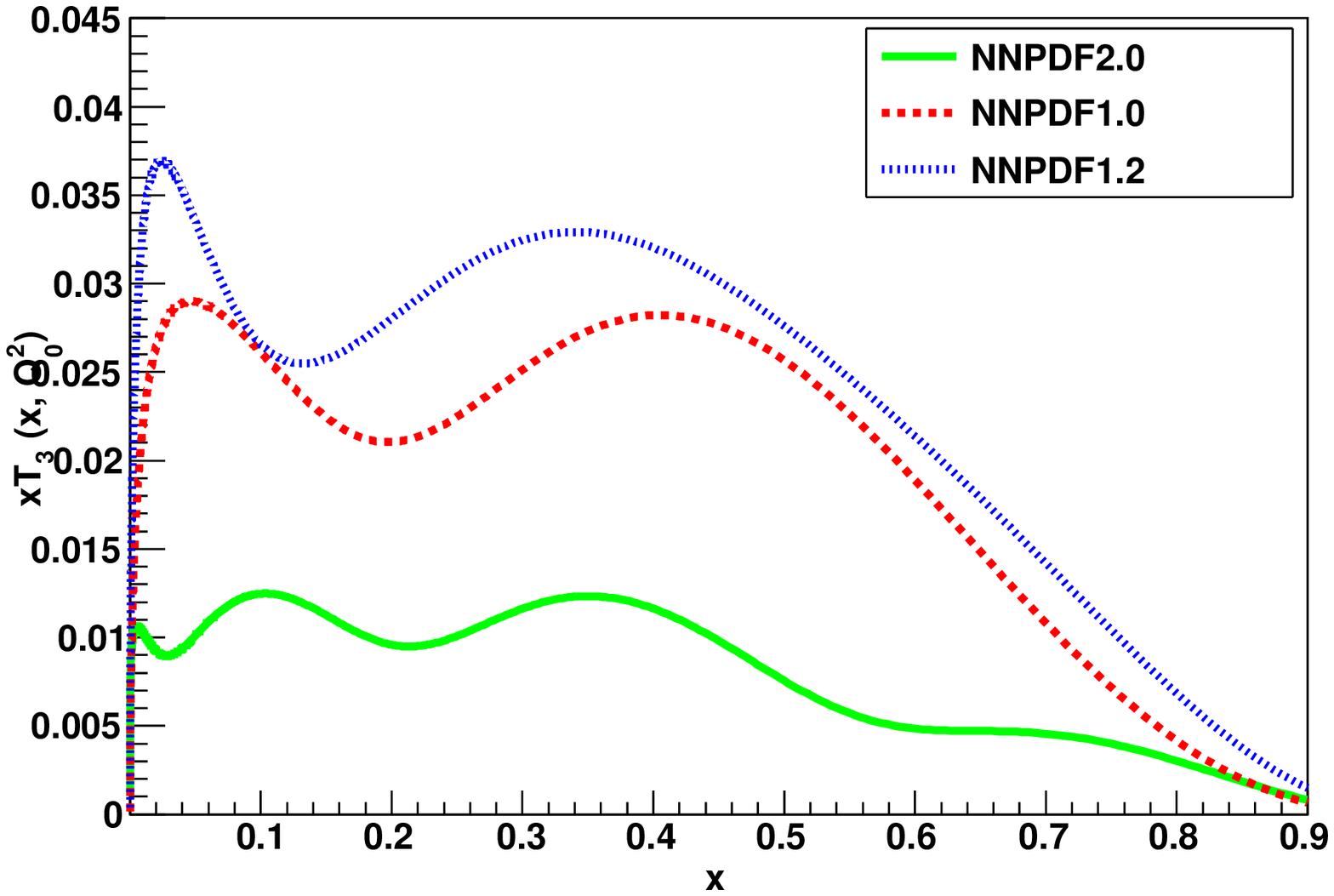}
\epsfig{width=0.45\textwidth,figure=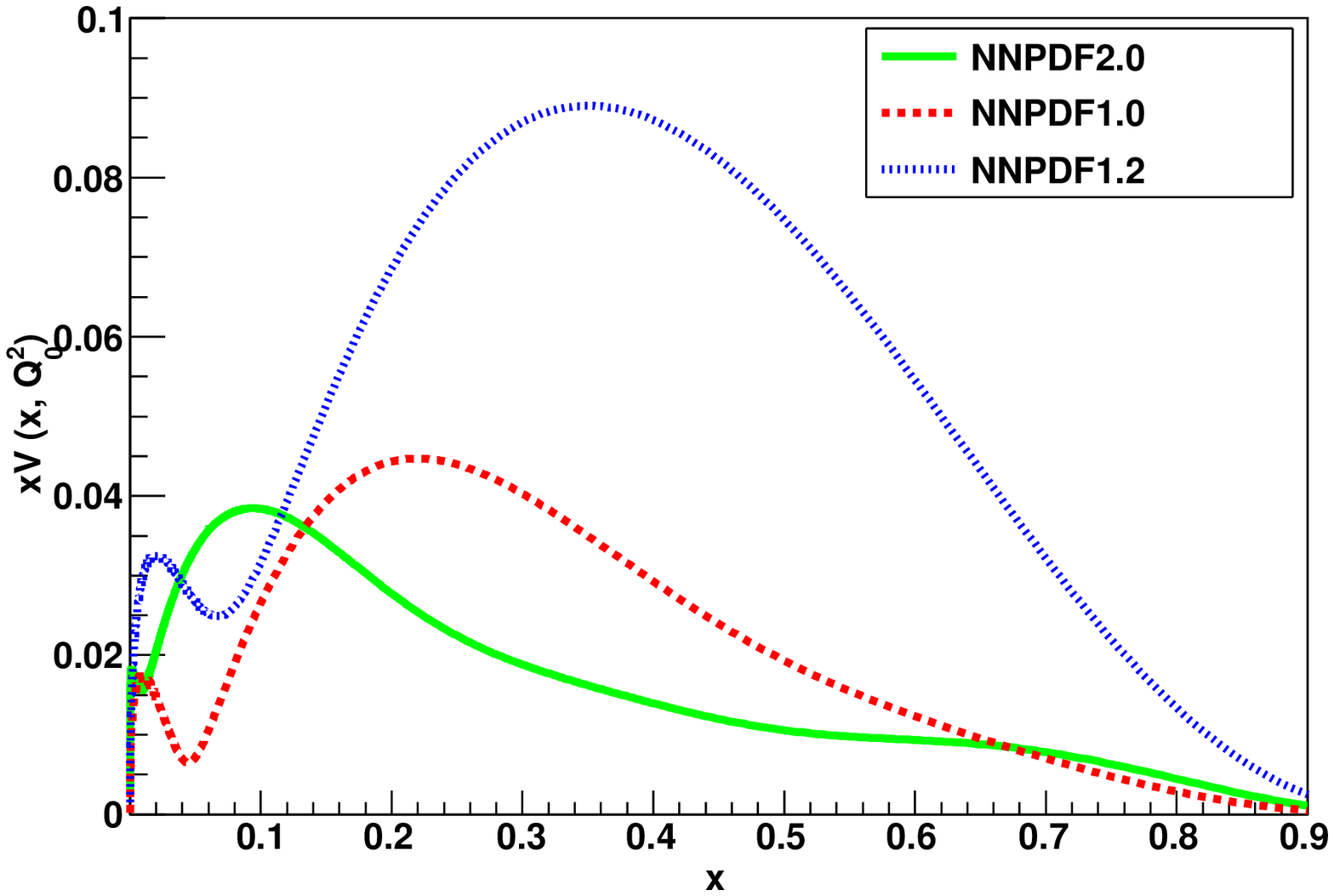}
\epsfig{width=0.45\textwidth,figure=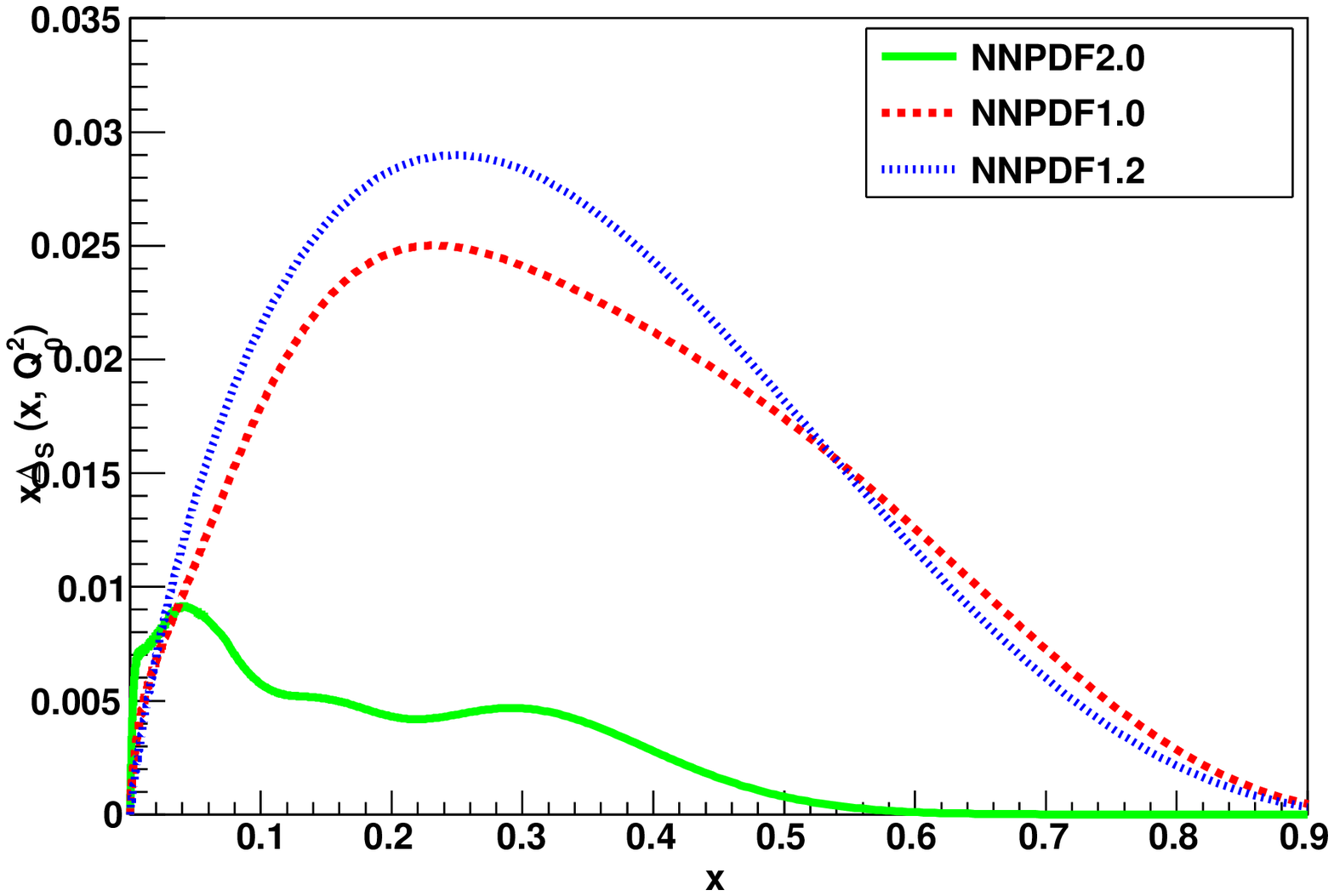}
\epsfig{width=0.45\textwidth,figure=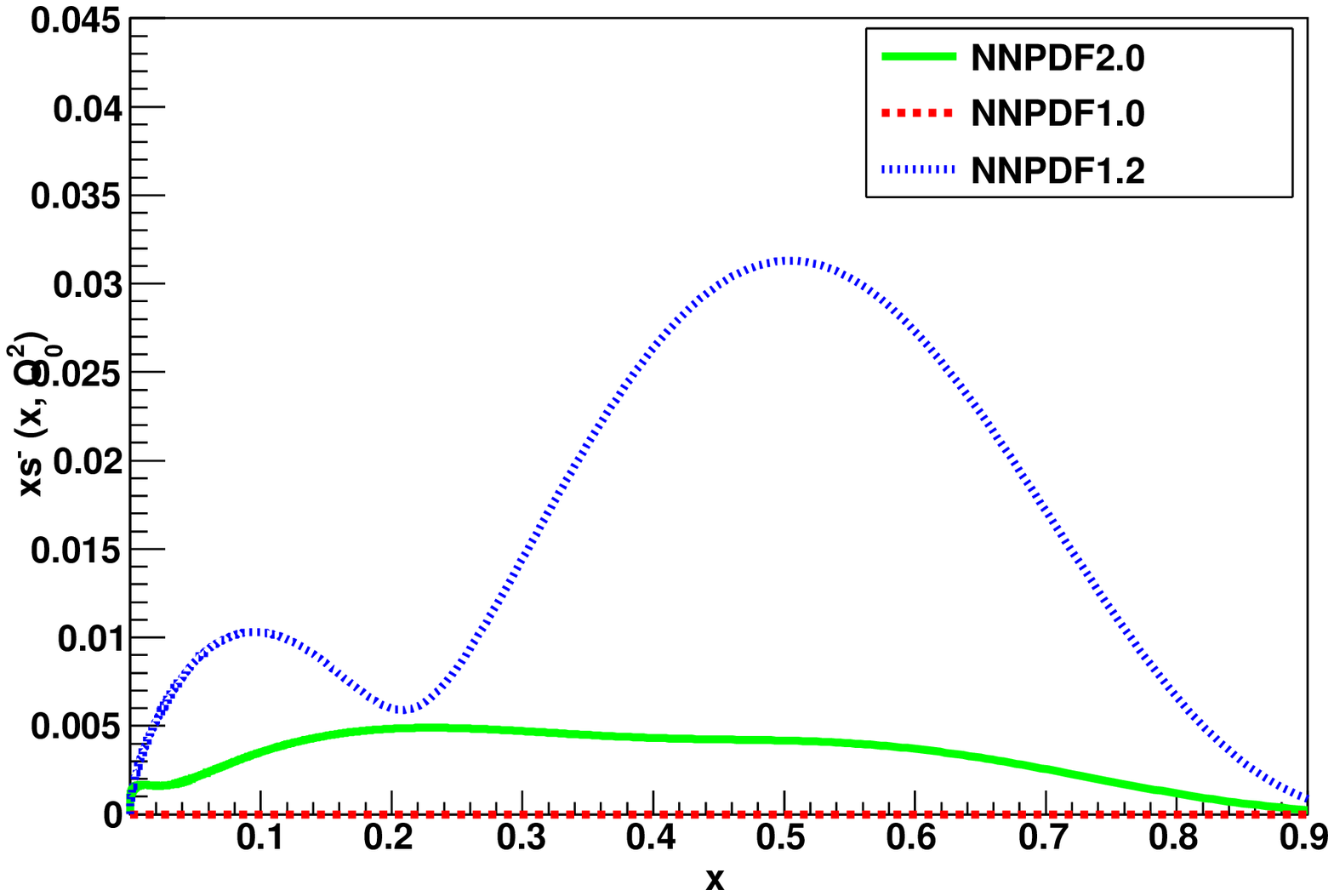}
\caption{\small Absolute uncertainties on the PDFs of Fig.~\ref{fig:valencePDFs-nnpdf}.
\label{fig:pdferrors2abs-nnpdf}} 
\end{center}
\end{figure}
\begin{figure}[hb!]
\begin{center}
\epsfig{width=0.99\textwidth,figure=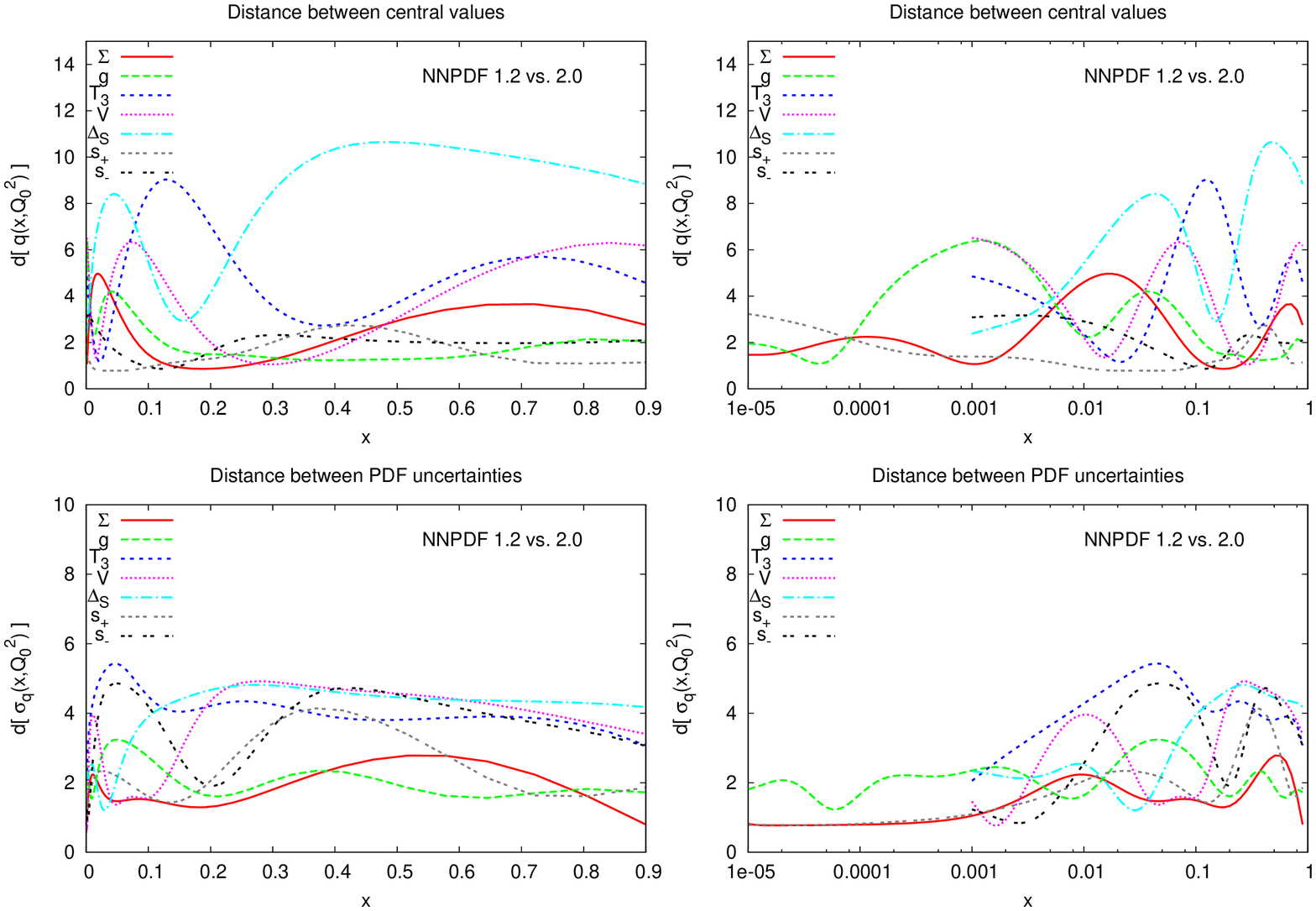}
\end{center}
\caption{\small Distance between
the NNPDF2.0 and the NNPDF1.2  parton sets 
(central values and uncertainties)
for all PDFs as a function of $x$. All distances are
computed from sets of  
$N_{\rm rep}=100$ replicas (see Appendix~\ref{sec:distances}.)
\label{fig:stabtab-20-12}}
%\end{table}
\end{figure}

The NNPDF2.0 PDFs are compared to  the previous NNPDF1.0~\cite{Ball:2008by}  
and NNPDF1.2~\cite{Ball:2009mk} parton sets in
Figs.~\ref{fig:singletPDFs-nnpdf}--\ref{fig:pdferrors2abs-nnpdf}.
All PDF combinations are defined as in
Refs.~\cite{Ball:2008by,Ball:2009mk}. Note that all uncertainty bands
shown are one--sigma; the relation to 68\% confidence levels will be
discussed in Sect.~\ref{sec:cl} below.
 The
consistency between subsequent NNPDF releases, extensively discussed in
previous work~\cite{Ball:2008by,Ball:2009mk} is apparent. Also
apparent is 
the reduction in uncertainty obtained going from NNPDF1.2 to NNPDF2.0;
the causes for
this improvement will be discussed in detail
in Sect.~\ref{sec:res:dataset} below. 
In order to further quantify the differences between
the NNPDF2.0 and NNPDF1.2 parton sets,
the distance (as defined in Appendix~\ref{sec:distances}) between these sets
are shown in Fig~\ref{fig:stabtab-20-12} as a function of $x$: all
PDFs for all $x$ are consistent at the 90\% confidence level, and in
fact almost all are consistent to within one sigma.

%%%%%%%%%%%%%%%%%%%%%%%%%%%%%%%%%%%%%%%%%%%%%%%%%%%%%%%
%%%%%%%%%%%%%%%%%%%%%%%%%%%%%%%%%%%%%%%%%%%%%%%%
%%%%%%%%%%%%%%%%%%%%%%%%%%%%%%%%%%%%%%%%%%%%%%%%%%%%

\begin{figure}[h!]
\begin{center}
\epsfig{width=0.49\textwidth,figure=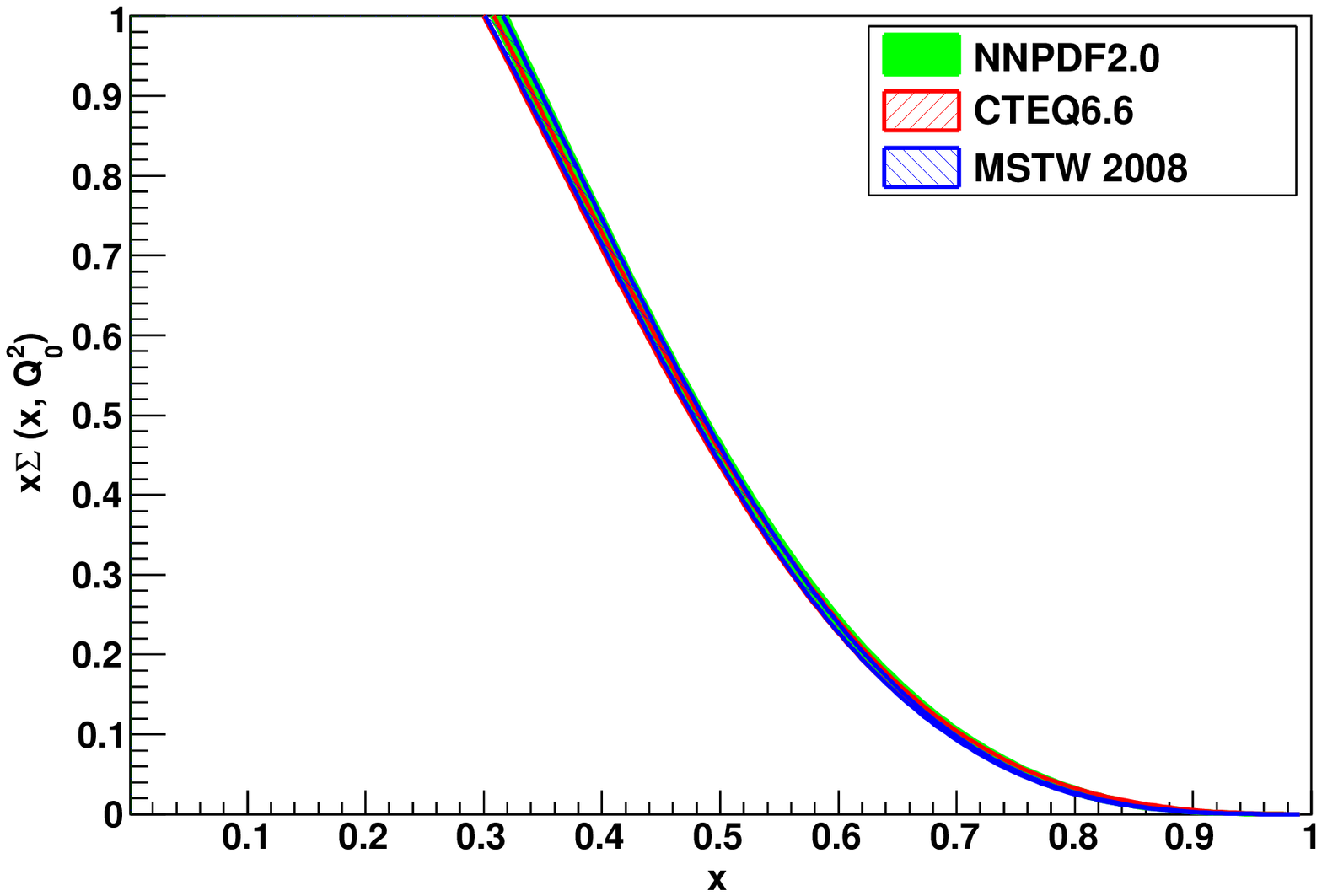}
\epsfig{width=0.49\textwidth,figure=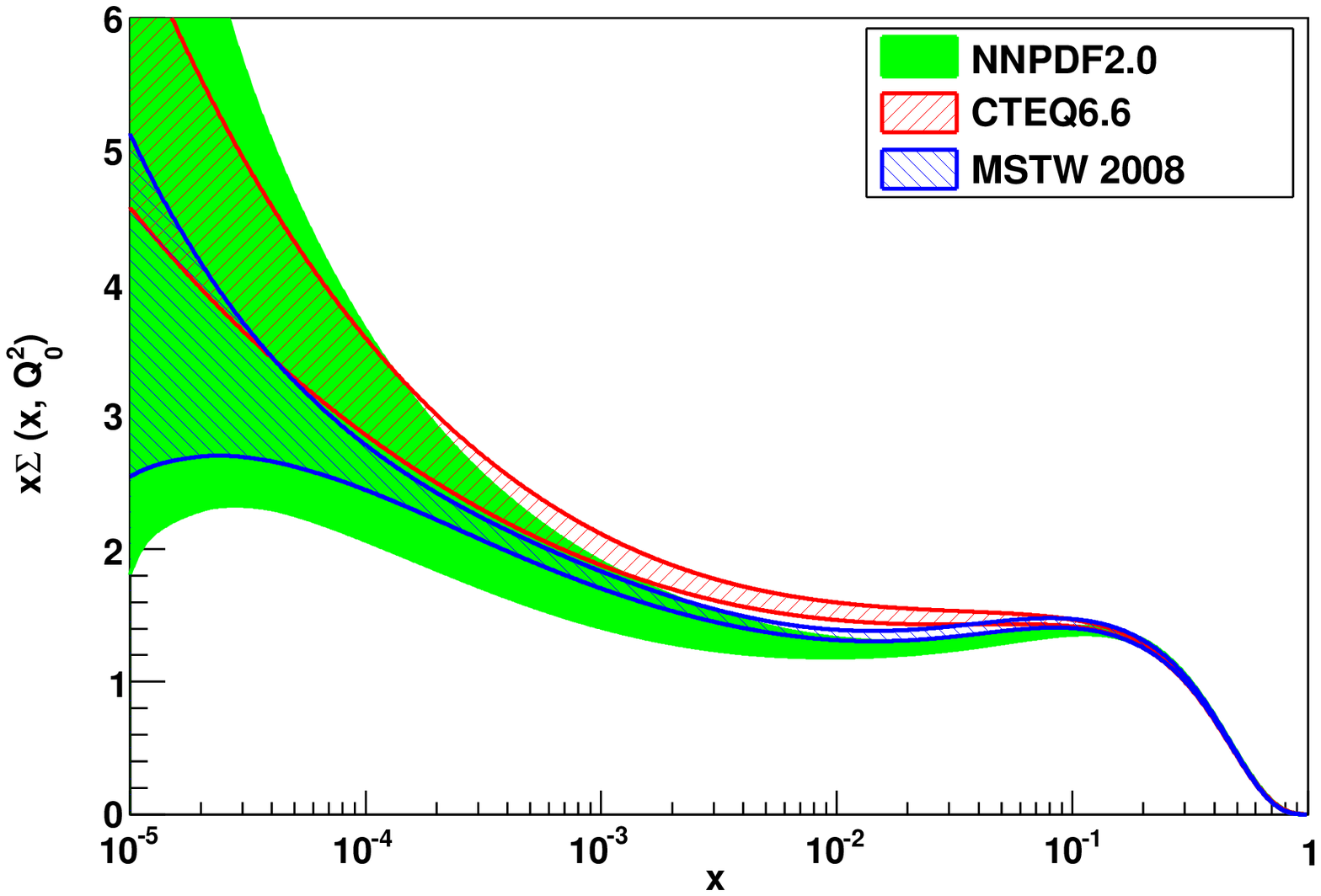}
\epsfig{width=0.49\textwidth,figure=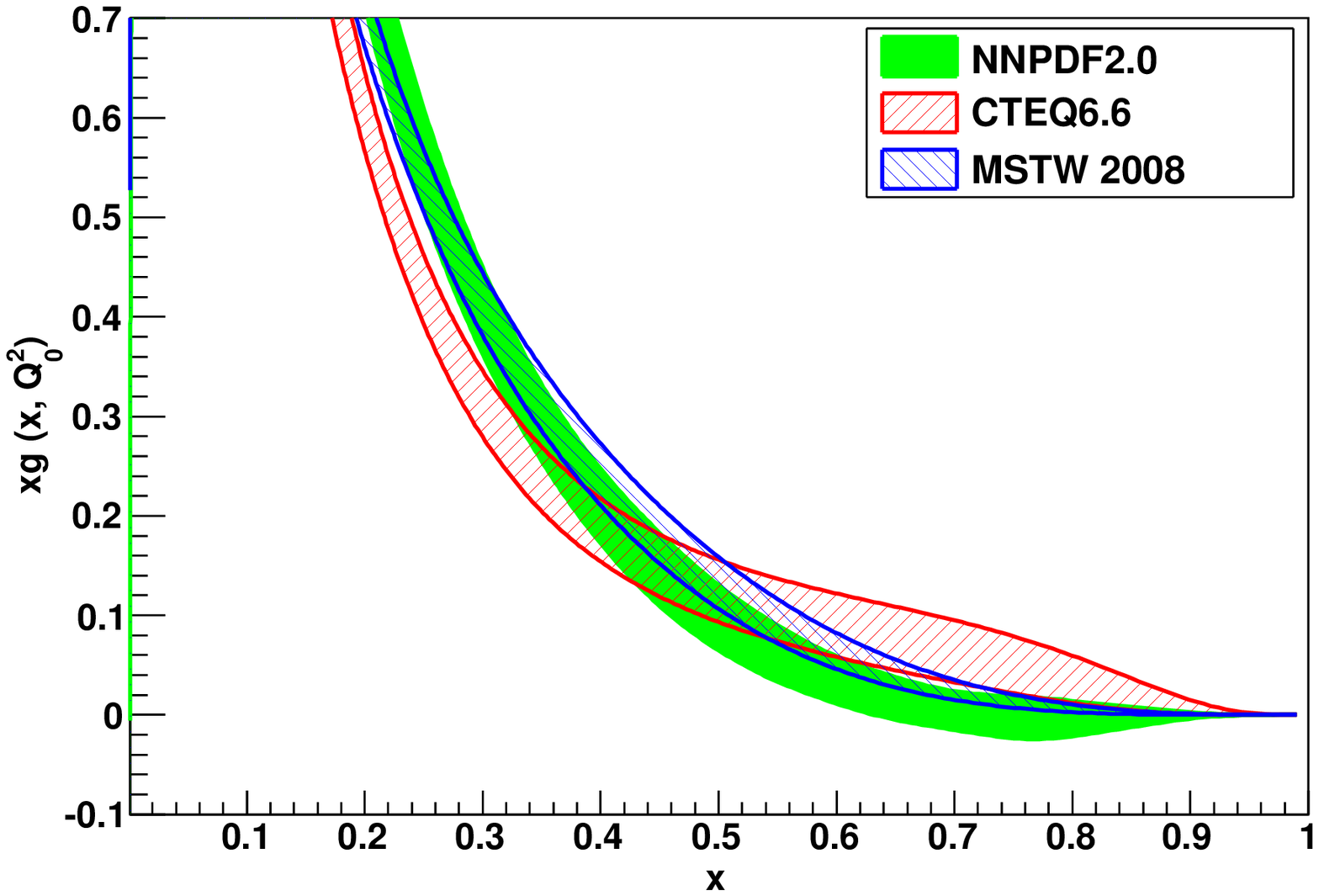}
\epsfig{width=0.49\textwidth,figure=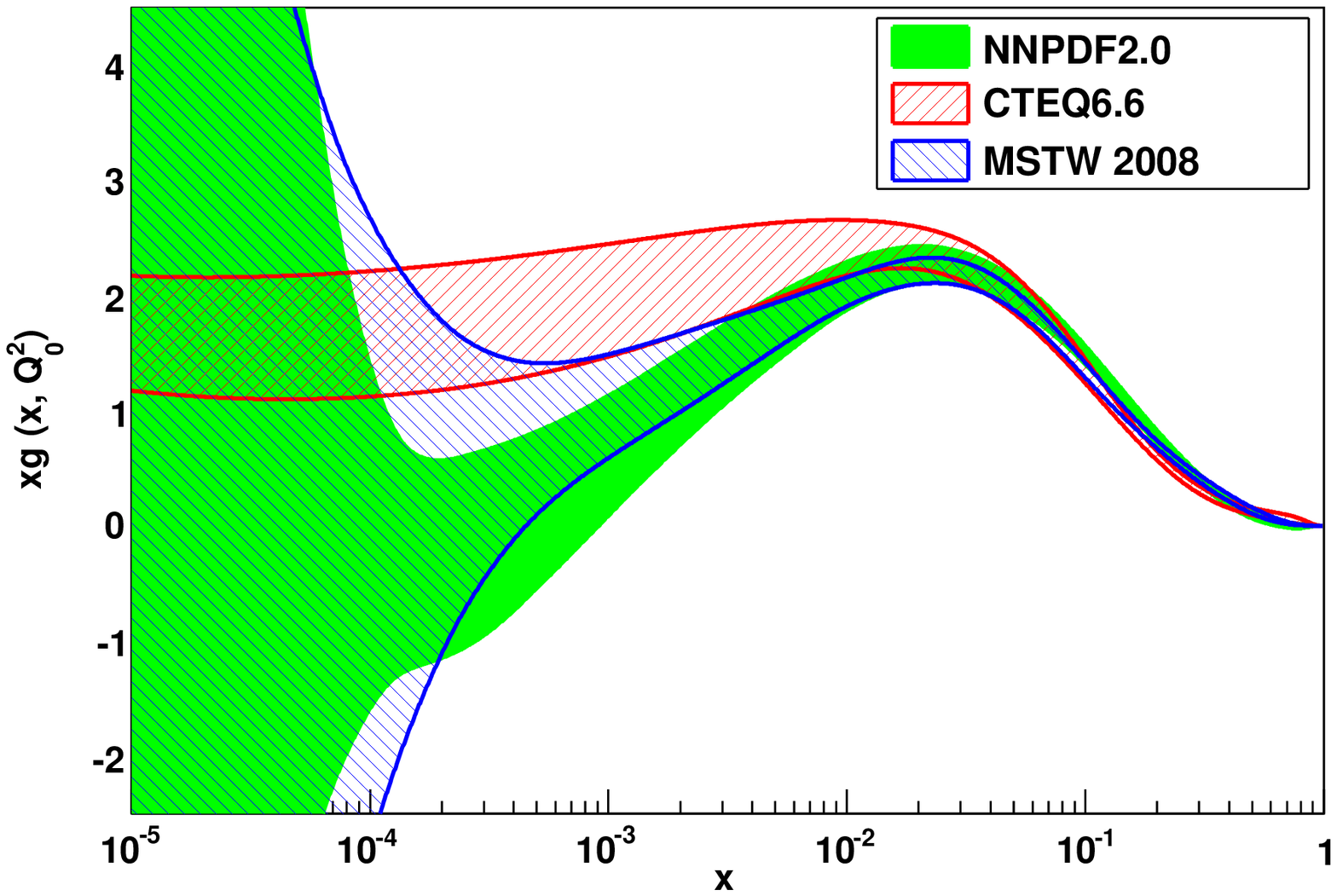}
\epsfig{width=0.49\textwidth,figure=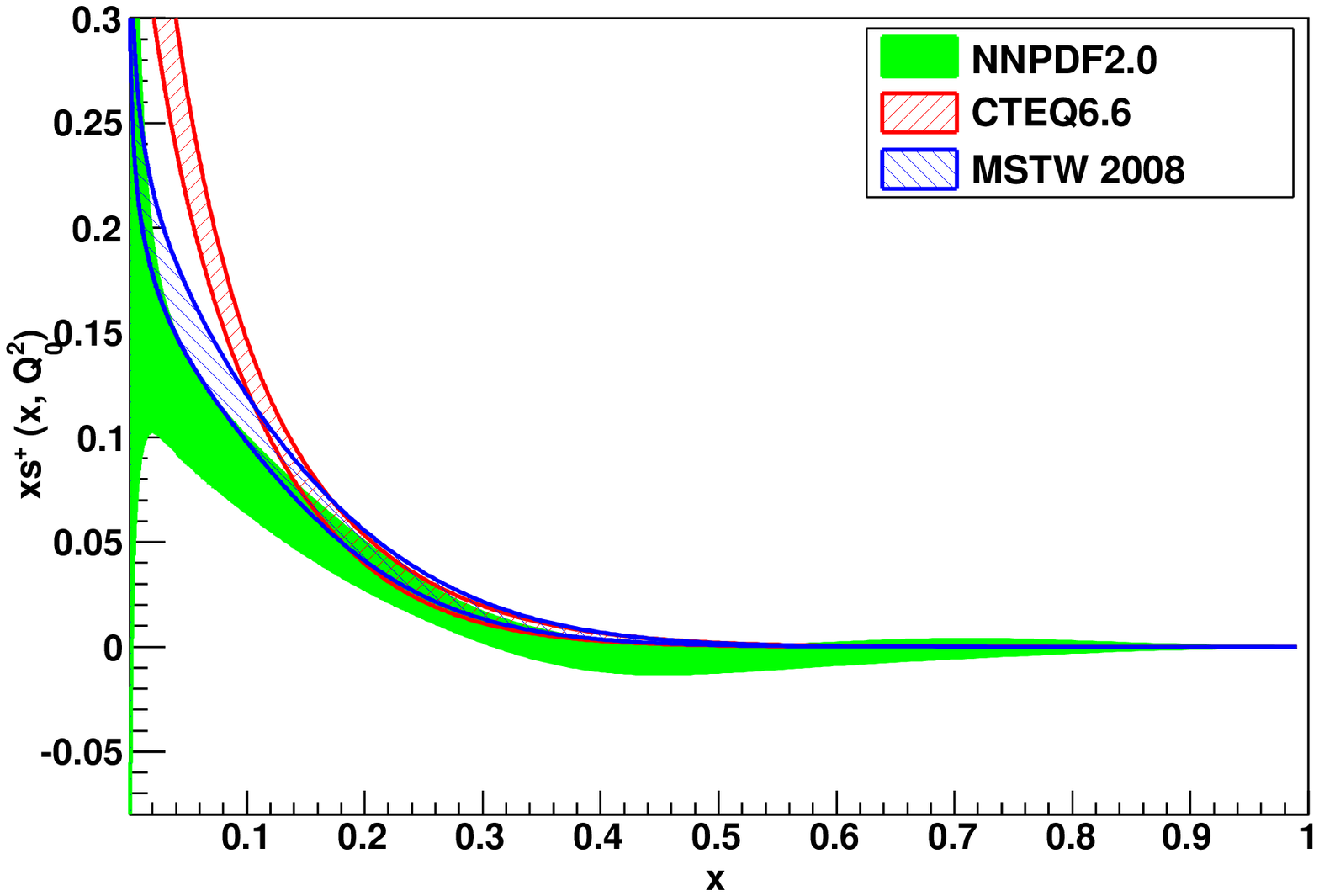}
\epsfig{width=0.49\textwidth,figure=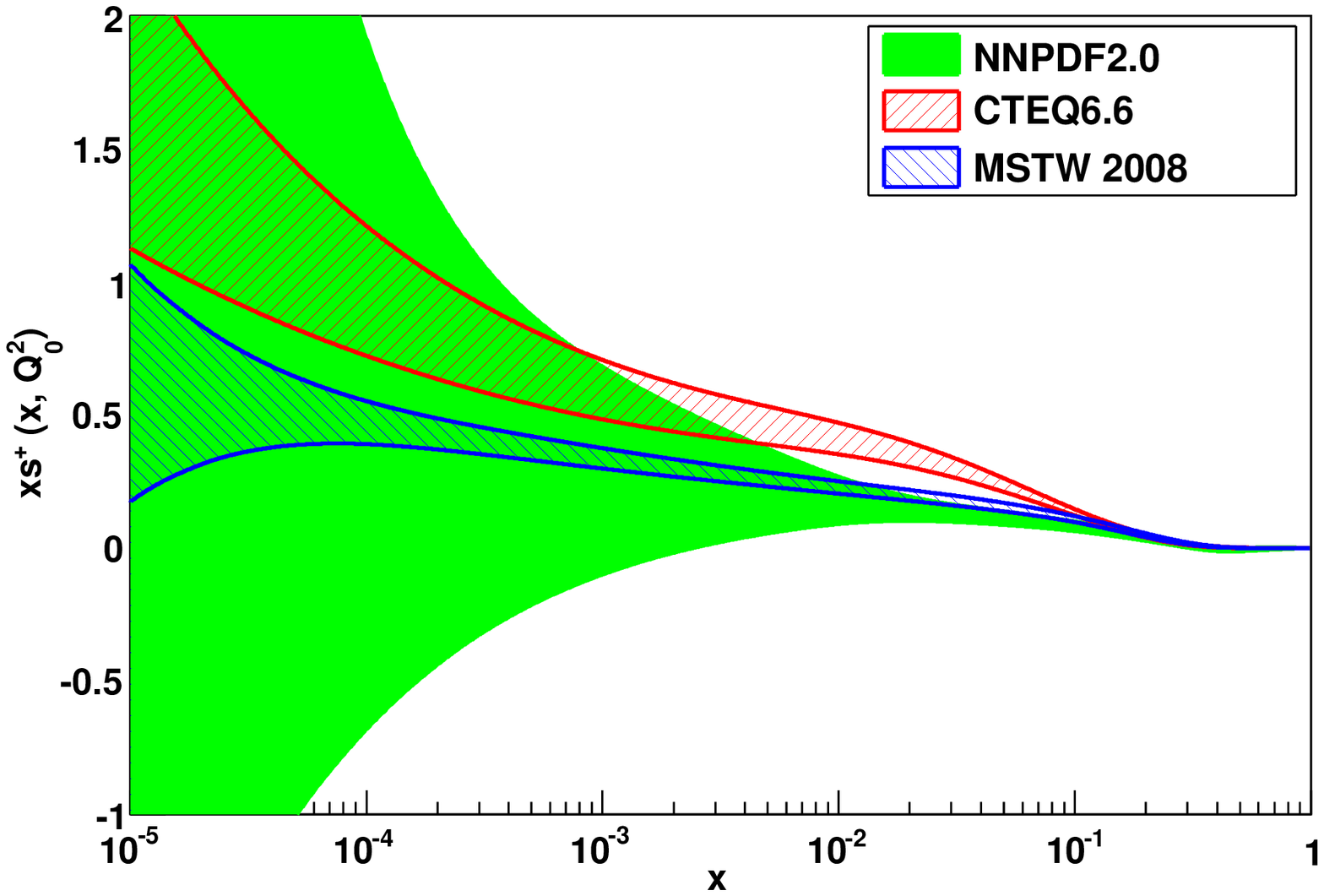}
\caption{\small Same as Fig.~\ref{fig:singletPDFs-nnpdf}, but compared to 
 MSTW08~\cite{Martin:2009iq} and CTEQ6.6~\cite{Nadolsky:2008zw} PDFs.
 \label{fig:singletPDFs}} 
\end{center}
\end{figure}

\begin{figure}[h!]
\begin{center}
\epsfig{width=0.49\textwidth,figure=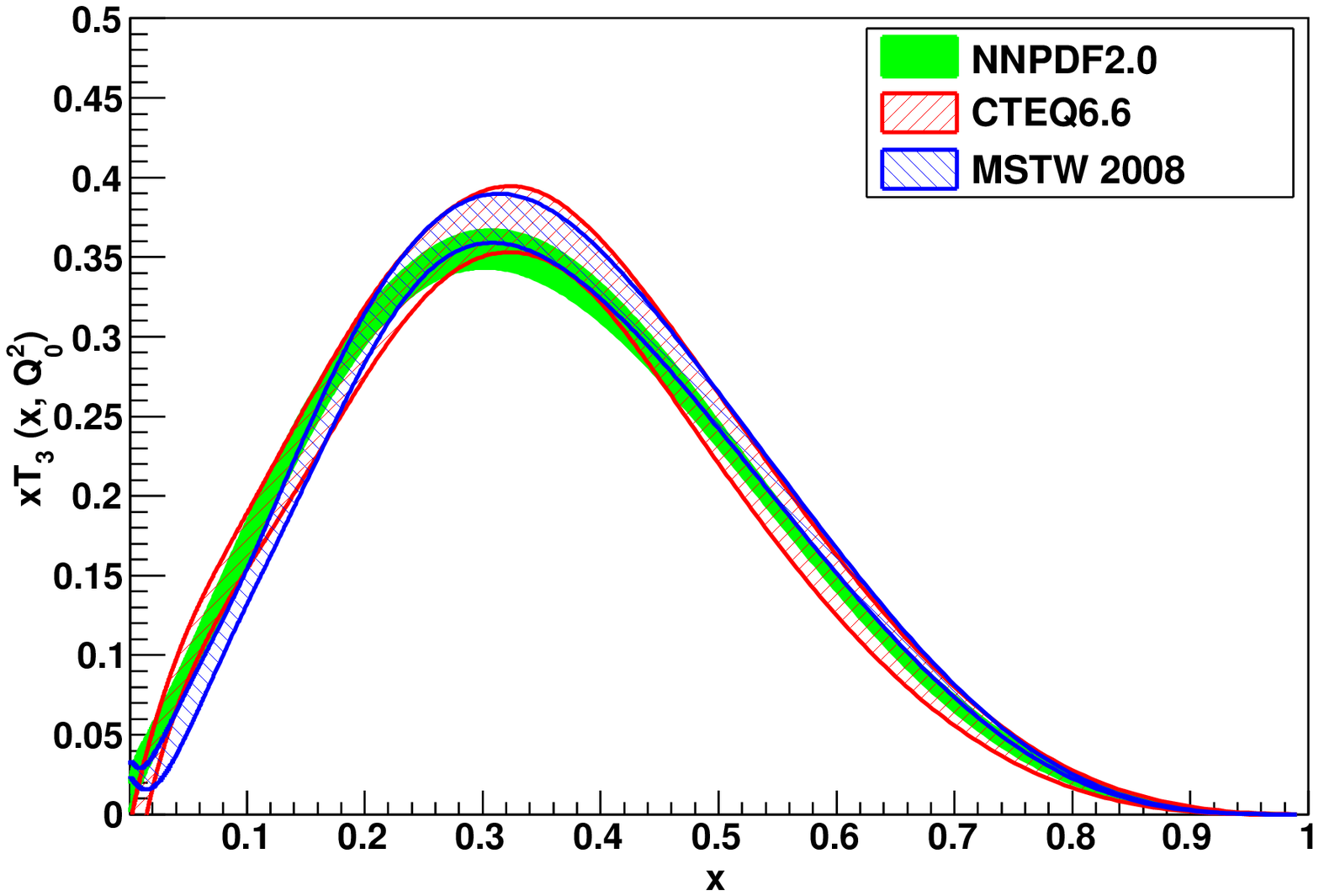}
\epsfig{width=0.49\textwidth,figure=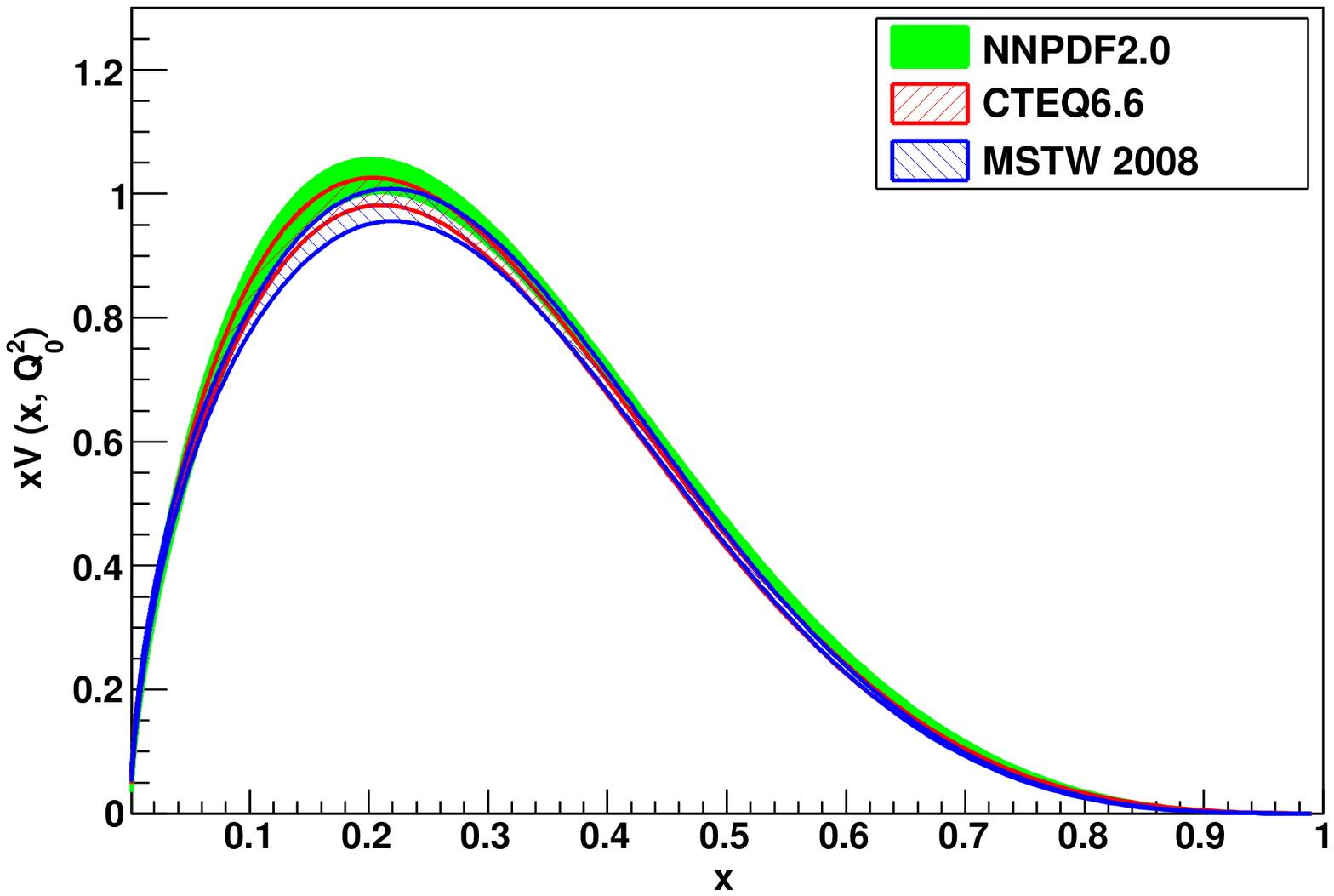}
\epsfig{width=0.49\textwidth,figure=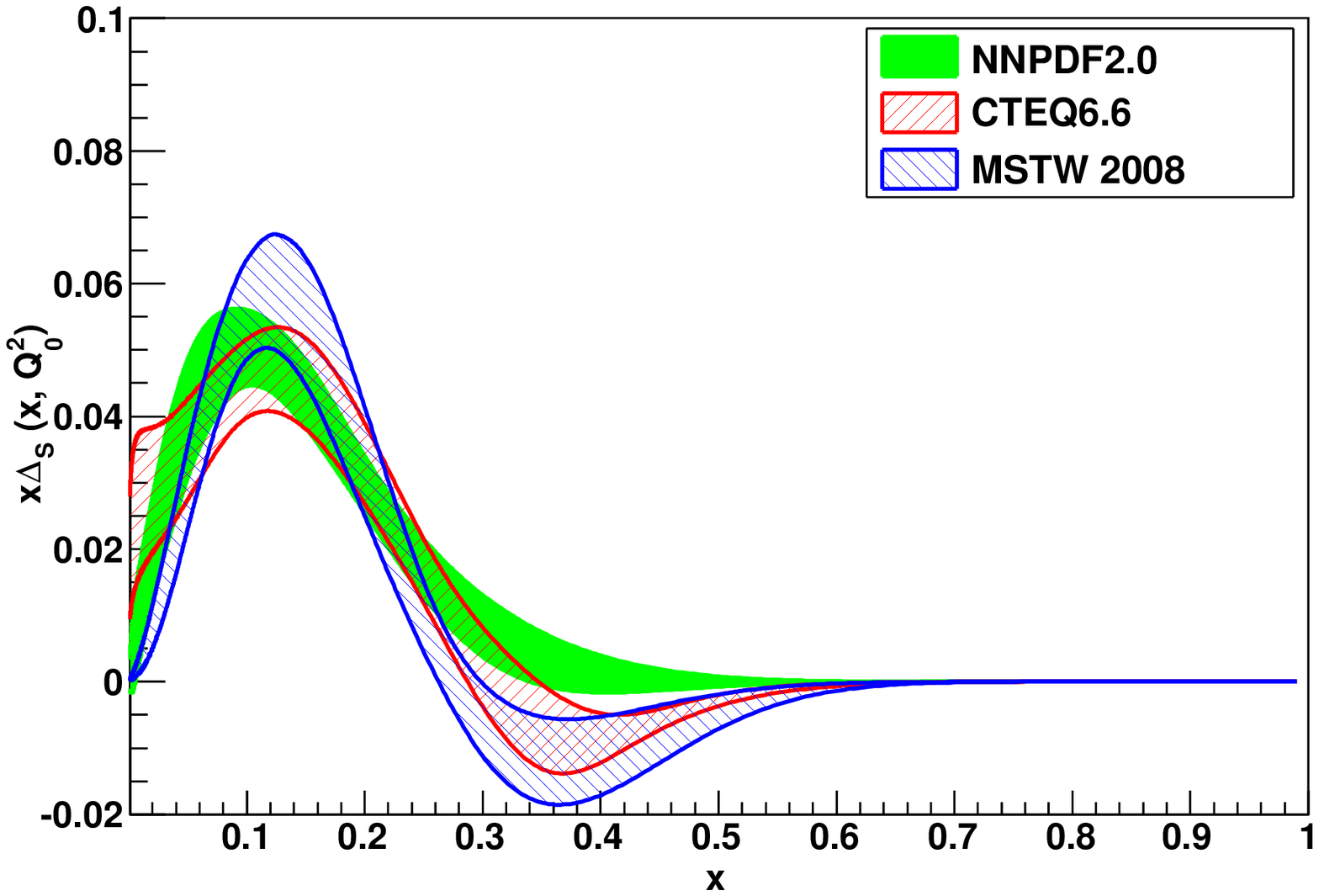}
\epsfig{width=0.49\textwidth,figure=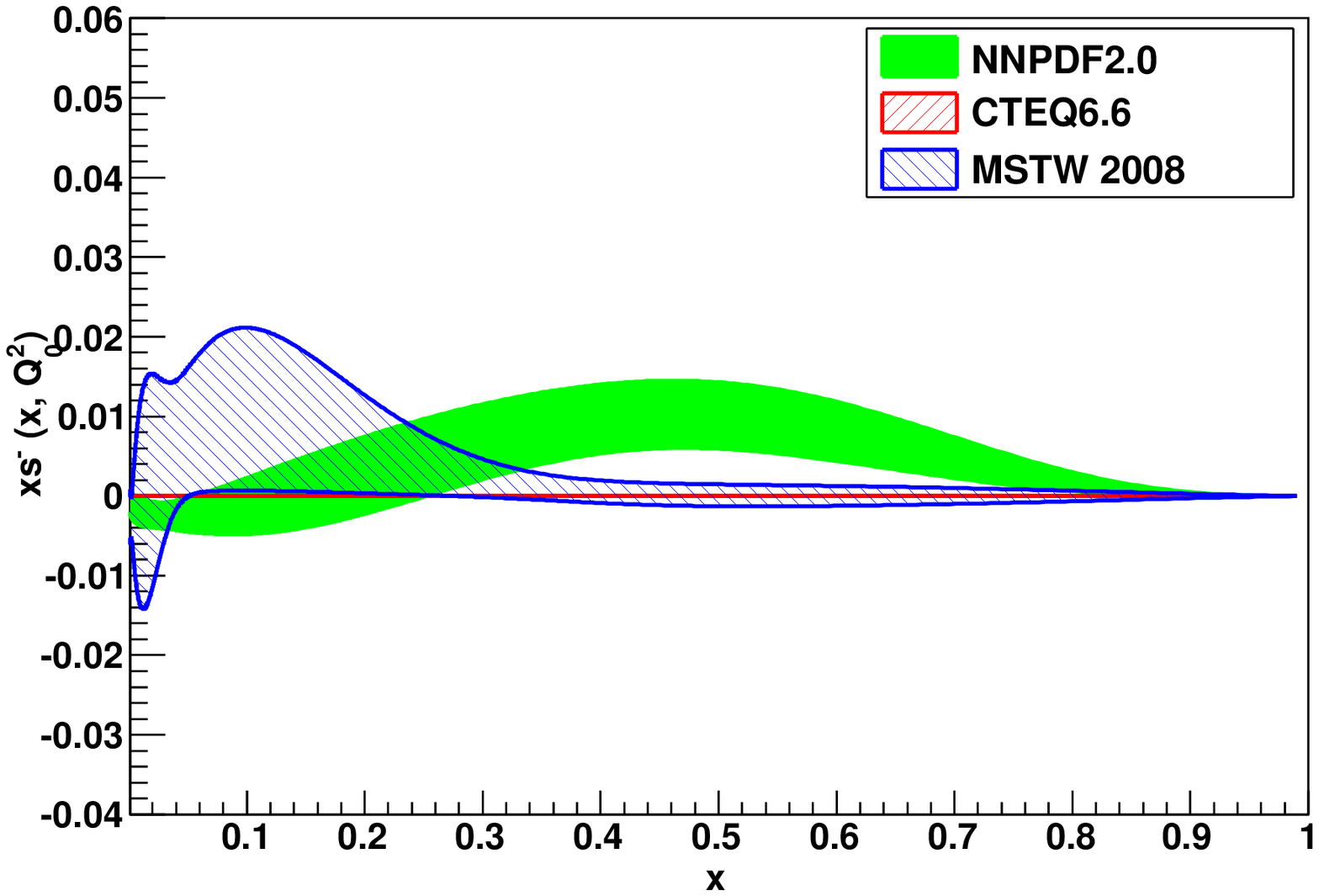}
\caption{\small Same as Fig.~\ref{fig:valencePDFs-nnpdf}, but compared to 
 MSTW08~\cite{Martin:2009iq} and CTEQ6.6~\cite{Nadolsky:2008zw} PDFs.
 \label{fig:valencePDFs}} 
\end{center}
\end{figure}

\begin{figure}[h!]
\begin{center}
\epsfig{width=0.49\textwidth,figure=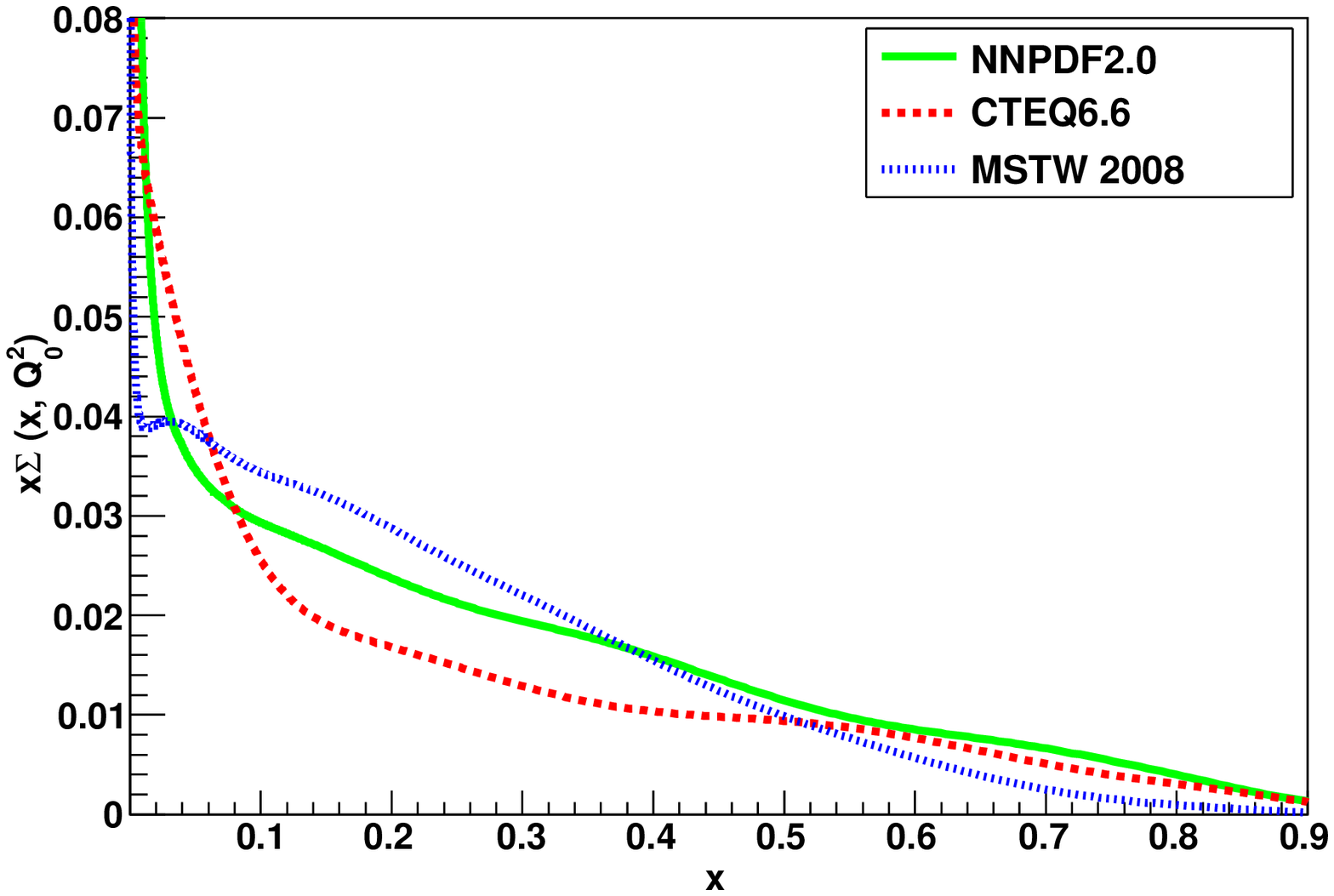}
\epsfig{width=0.49\textwidth,figure=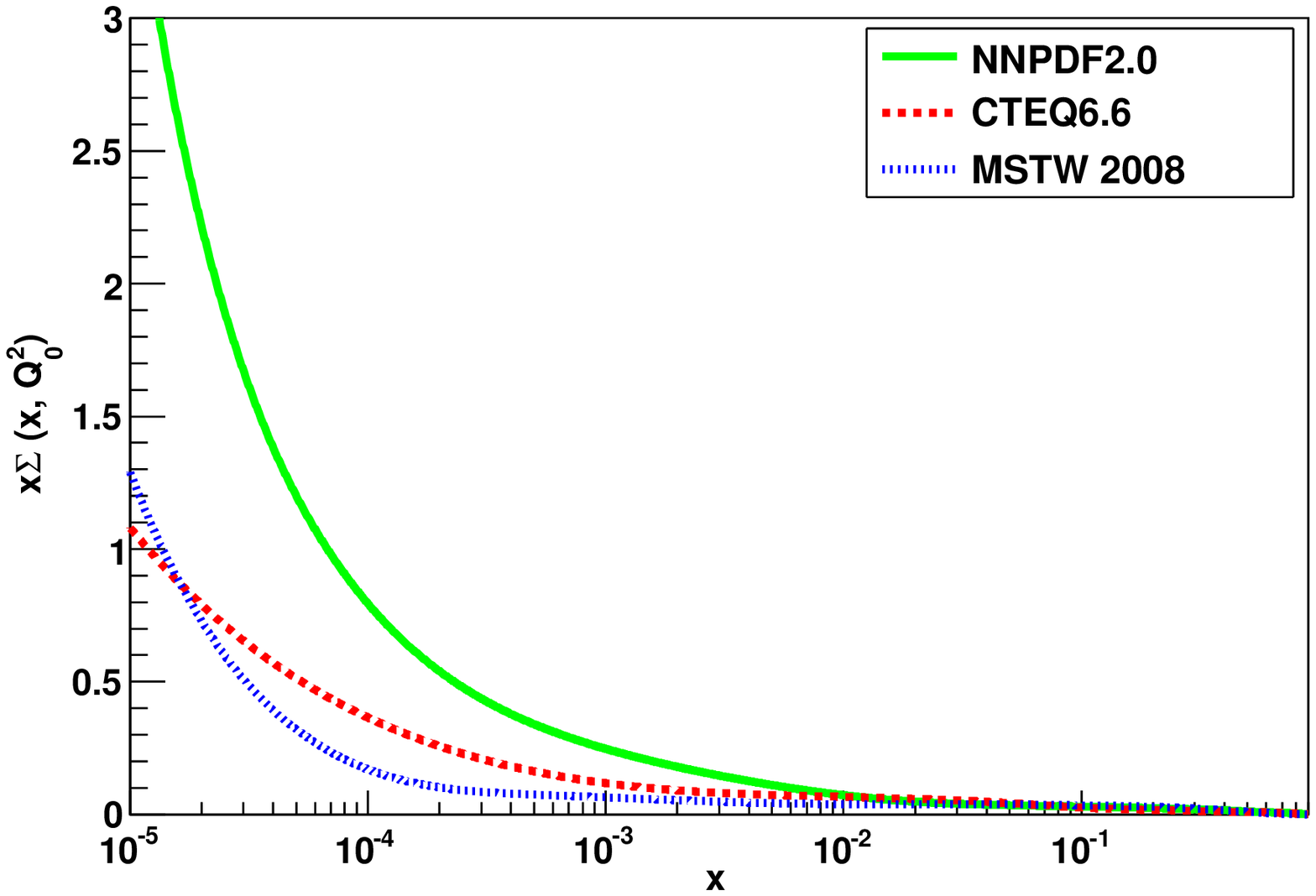}
\epsfig{width=0.49\textwidth,figure=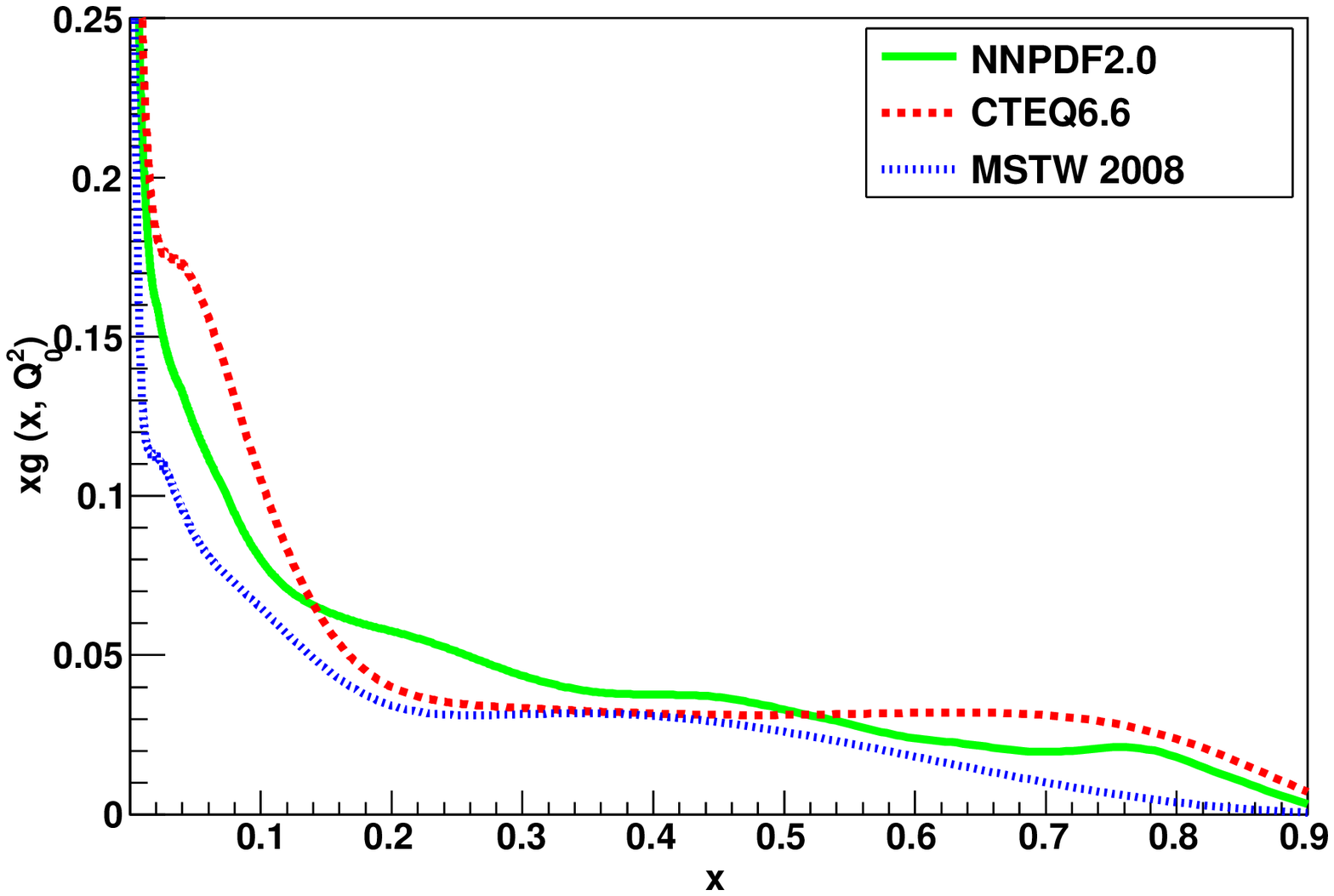}
\epsfig{width=0.49\textwidth,figure=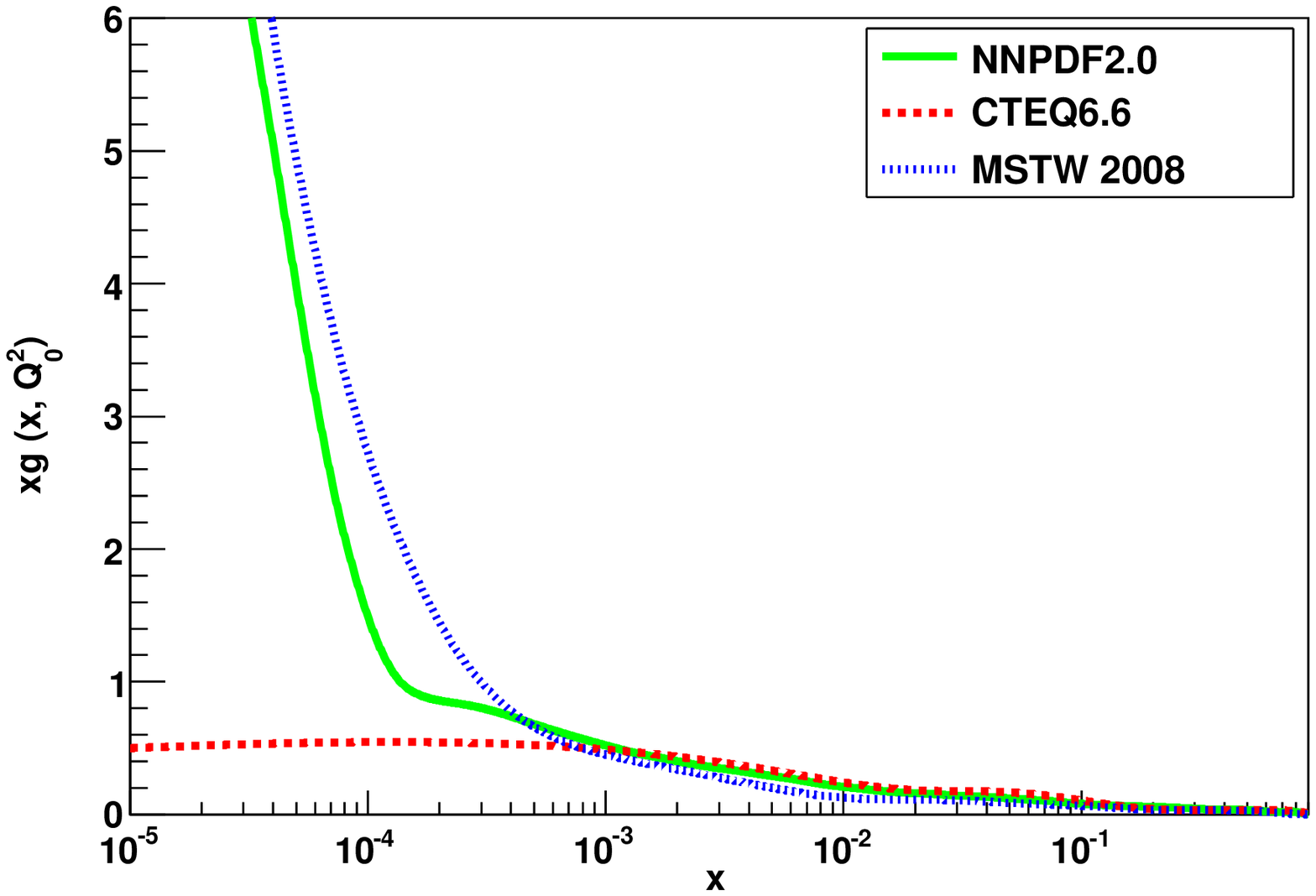}
\epsfig{width=0.49\textwidth,figure=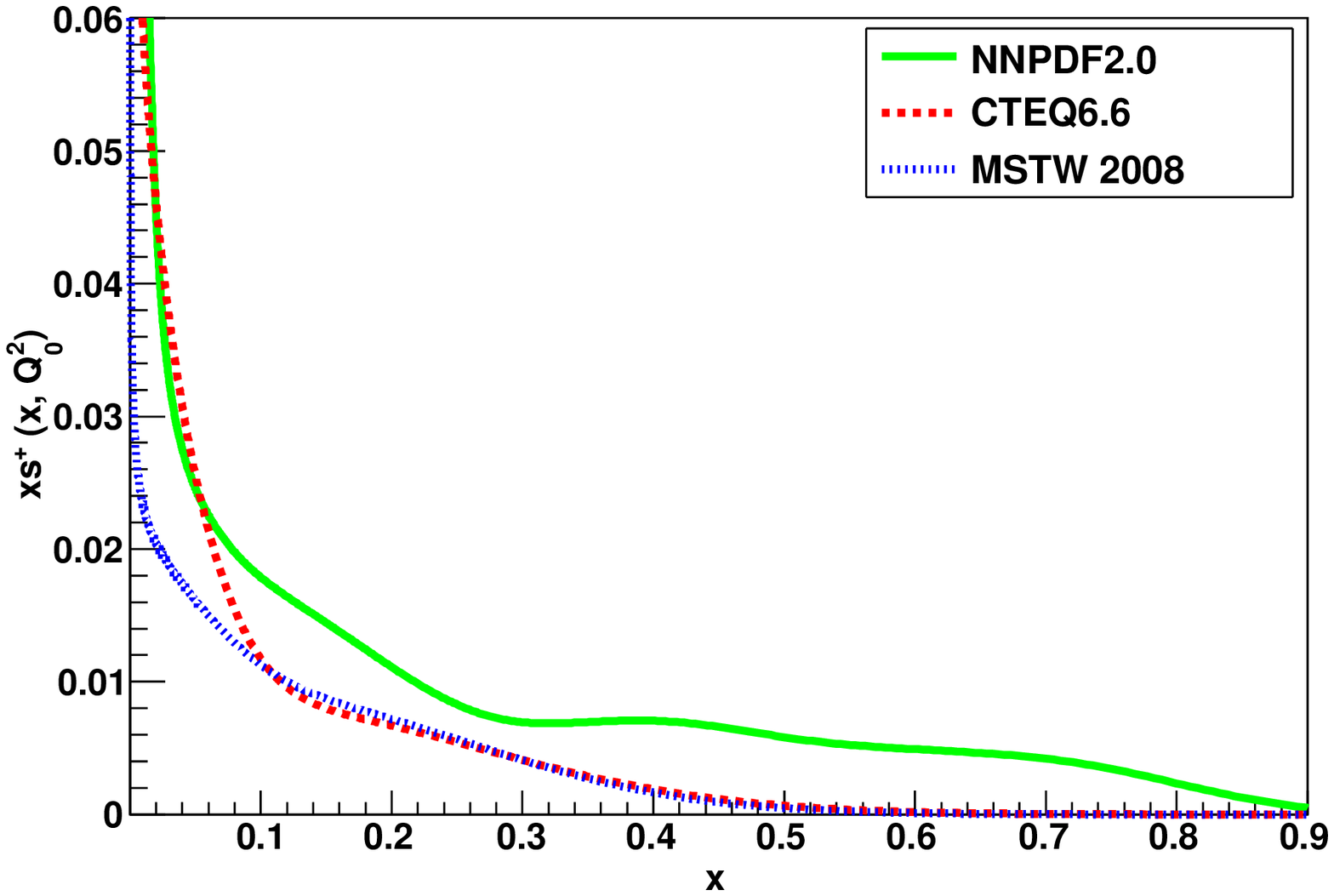}
\epsfig{width=0.49\textwidth,figure=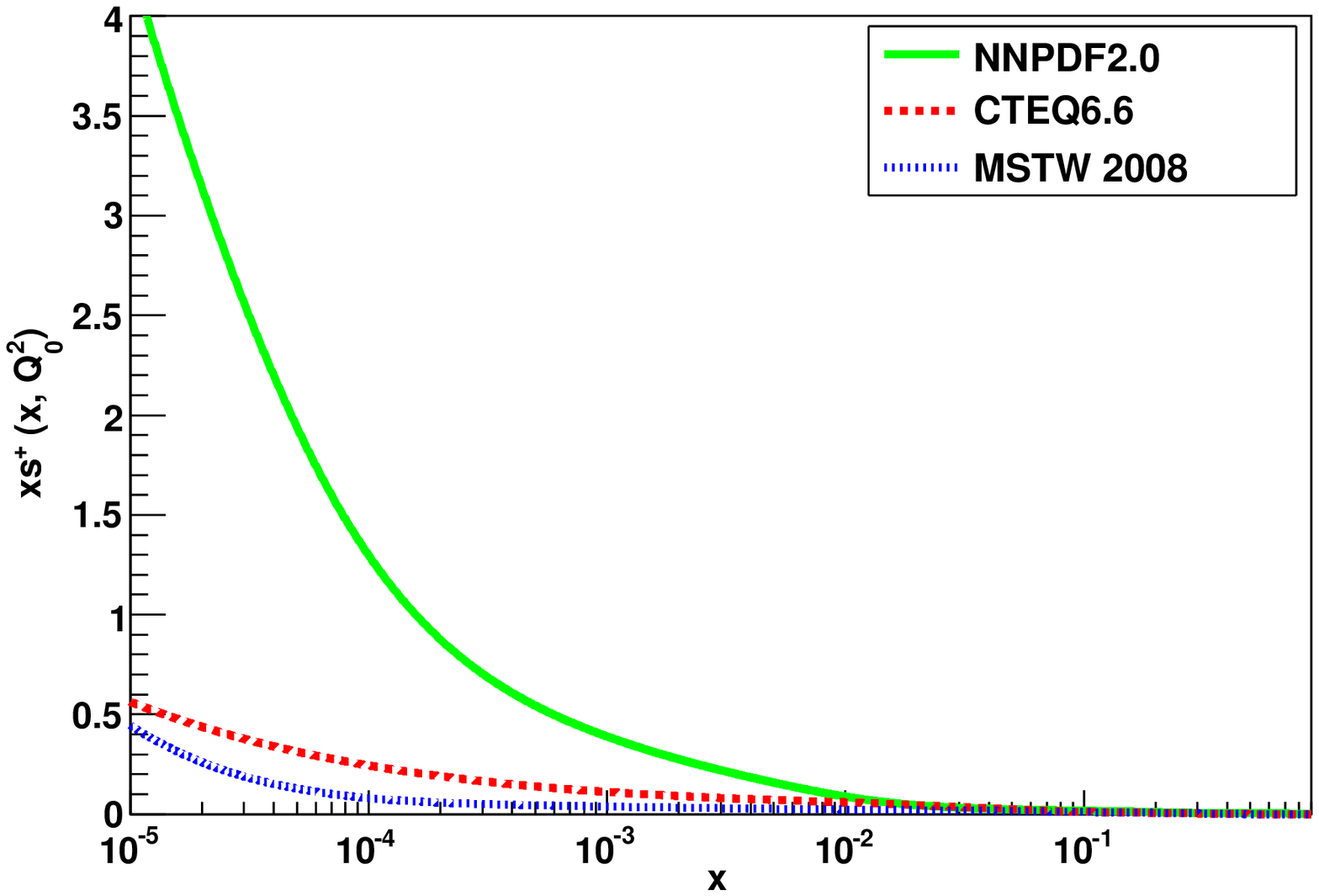}
\caption{\small  Same as Fig.~\ref{fig:pdferrorsabs-nnpdf}, but compared to 
 MSTW08~\cite{Martin:2009iq} and CTEQ6.6~\cite{Nadolsky:2008zw} PDFs.
 \label{fig:pdferrorsabs}}
\end{center}
\end{figure}

\begin{figure}[h!]
\begin{center}
\epsfig{width=0.49\textwidth,figure=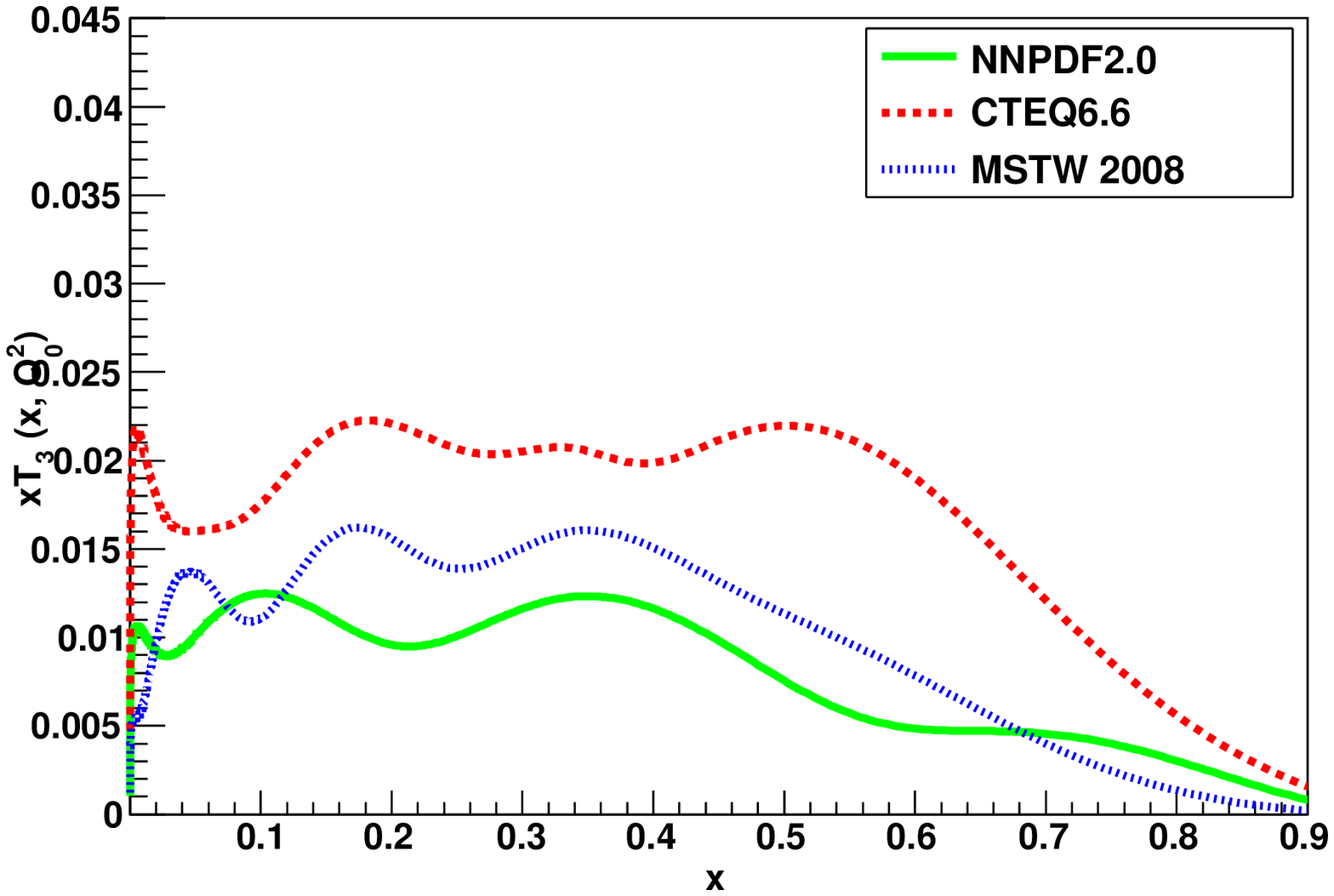}
\epsfig{width=0.49\textwidth,figure=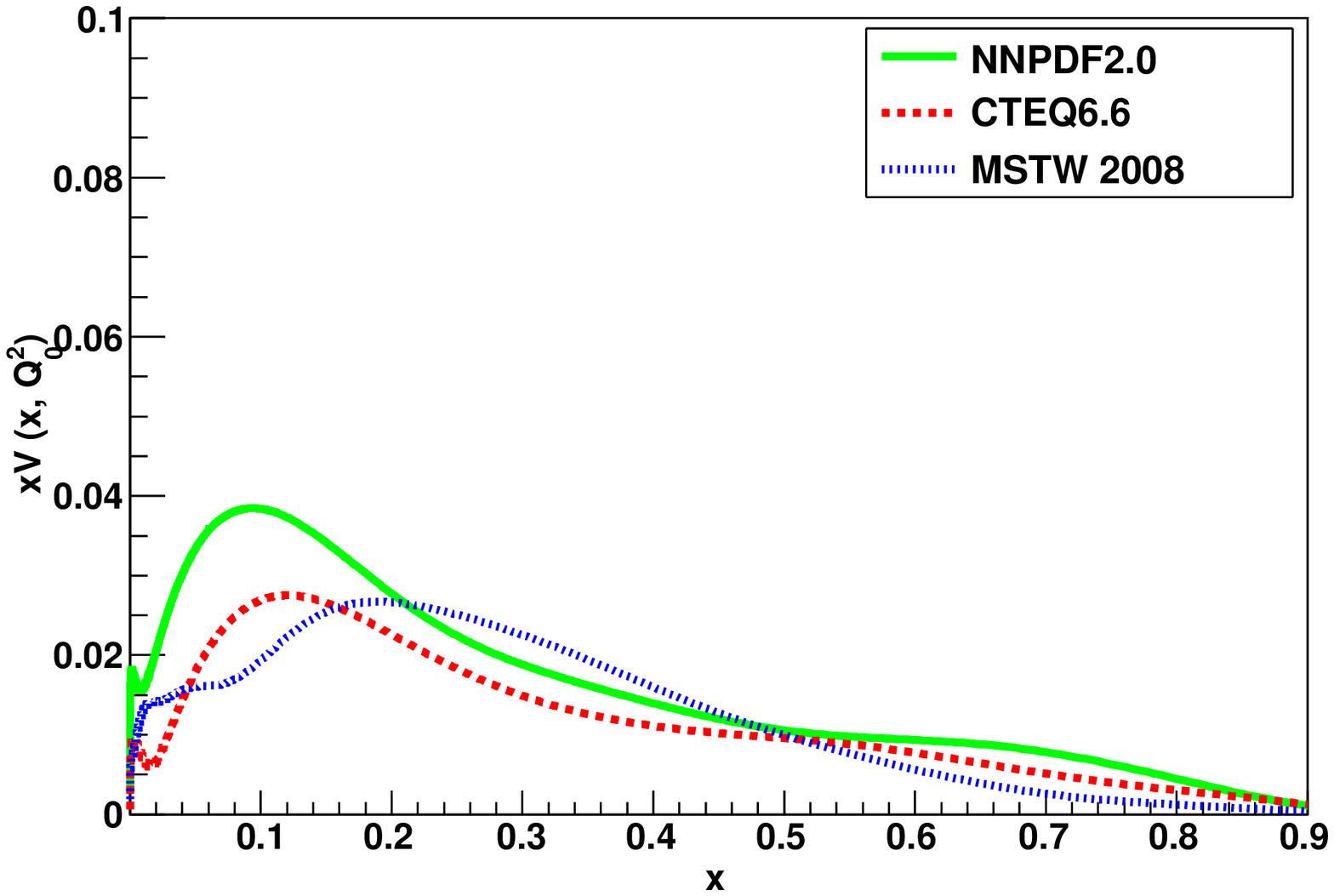}
\epsfig{width=0.49\textwidth,figure=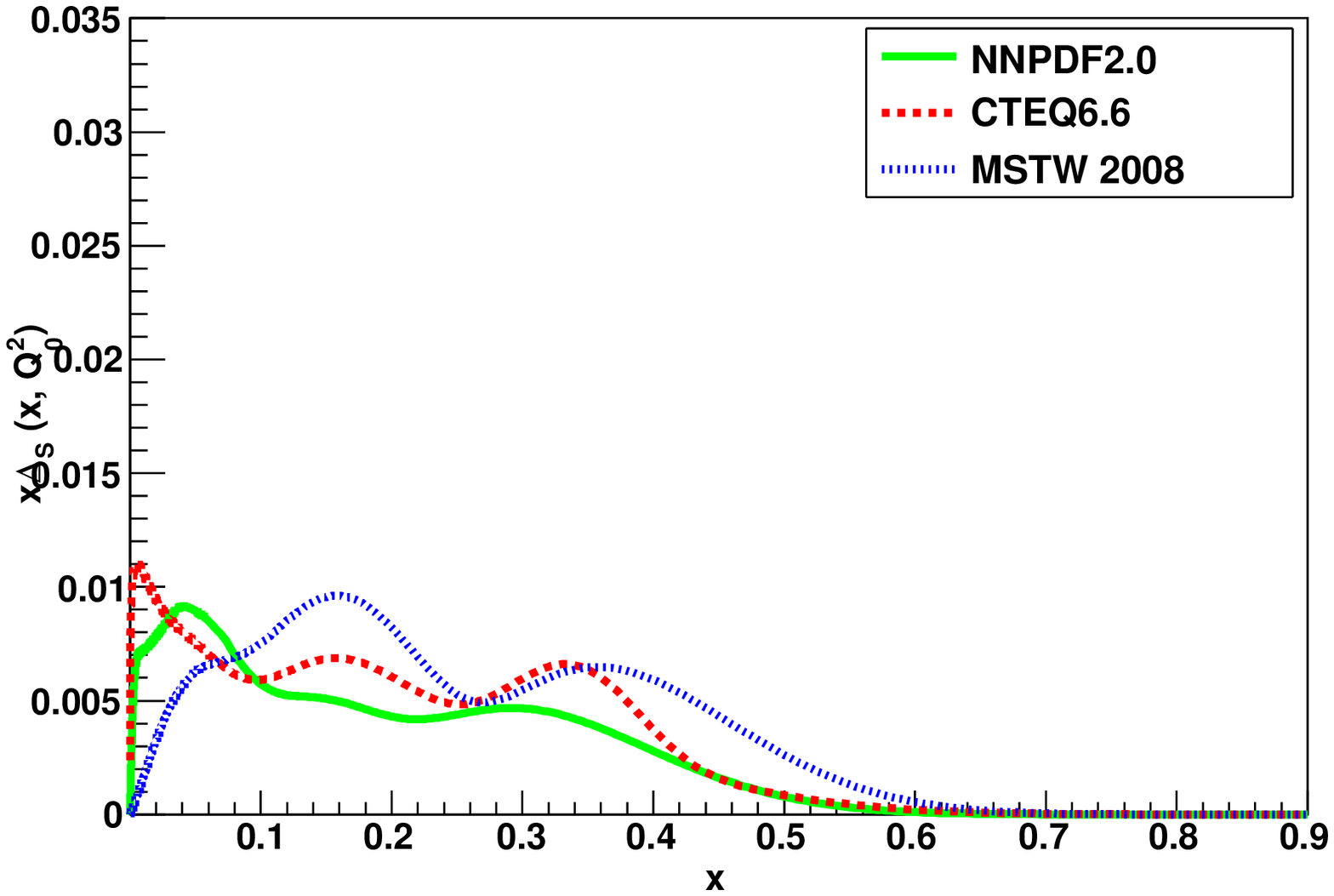}
\epsfig{width=0.49\textwidth,figure=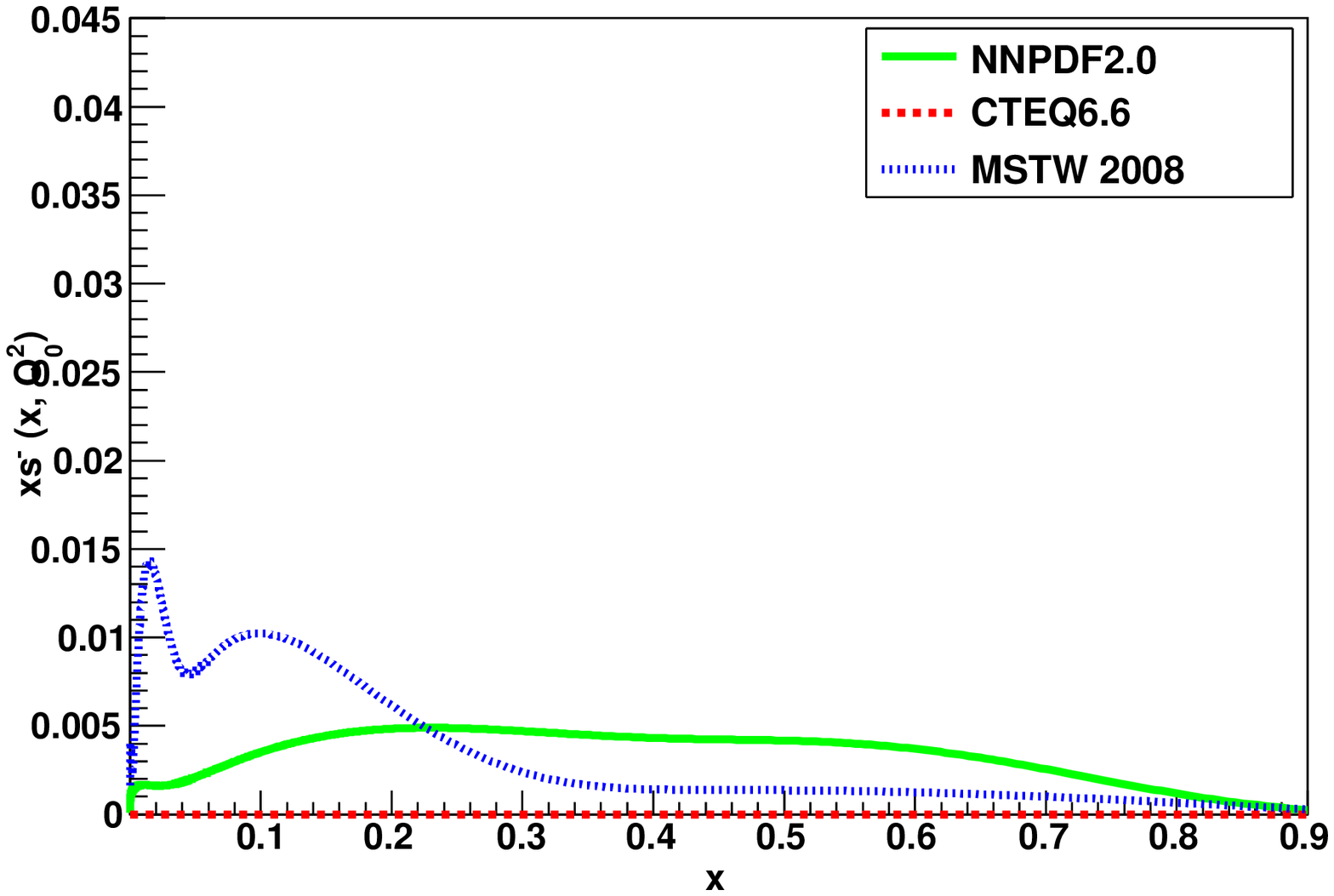}
\caption{\small  Same as Fig.~\ref{fig:pdferrors2abs-nnpdf}, but compared to 
 MSTW08~\cite{Martin:2009iq} and CTEQ6.6~\cite{Nadolsky:2008zw} PDFs.
 \label{fig:pdferrorsabs2}}
\end{center}
\end{figure}

The  NNPDF2.0 PDFs are also compared to CTEQ6.6~\cite{Nadolsky:2008zw} and
MSTW08~\cite{Martin:2009iq} PDFs in 
Figs.~\ref{fig:singletPDFs}--\ref{fig:pdferrorsabs2}.
Most NNPDF2.0 uncertainties are comparable to the CTEQ6.6 and MSTW08 ones;
there are however some interesting exceptions. The uncertainty on
strangeness, which NNPDF2.0 parametrizes with as many parameters as
any other PDF, is rather larger than those of MSTW08 and CTEQ6.6, in
which these PDFs are parametrized with a very small number of
parameters. The NNPDF2.0 uncertainty on total quark singlet (which
contains a sizable strange contribution) is also larger. The
uncertainty on the small $x$ gluon is significantly larger than that
found by CTEQ6.6, but  comparable to that MSTW08, which has an
extra parameter to describe the small $x$ gluon in comparison to
CTEQ6.6. The uncertainty on the triplet combination is rather smaller
in NNPDF2.0 than either MSTW08 or CTEQ6.6.  As we shall see in
Sect.~\ref{sec:res:dataset}, this small uncertainty is largely due to
the impact of Drell-Yan data (which are found to be completely
consistent with DIS data within our NLO treatment): hence, the fact
that we find it to be smaller than MSTW08 or CTEQ6.6 does not appear
to be due to the choice of dataset.
%%%%%%%%%%%%%%%%%%%%%%%%%%%%
%%%%%%%%%%%%%%%%%%%%%%%%%%%%%%%%%%%%%%%%%%%%%
%%%%%%%%%%%%%%%%%%%%%%%%%%
%%%%%%%%%%%%%%%%%%%%%%%%%%%%%%%%%%%%%%%%%%%%%%
%%%%%%%%%%%%%%%%%%%%%%%%%%%%%%%%%%%%%%%%%%%%%%%%

\clearpage

\subsection{Confidence levels}
\label{sec:cl}

An important advantage of the Monte Carlo method used in the NNPDF approach
to determine PDF uncertainties is that, unlike in a Hessian approach,
one does not have to rely on linear error propagation. It is then
possible to test the implication of a non-gaussian
distribution of experimental data which were found in
Ref.~\cite{glazov} to be minor; and
and also to test for
non-gaussian distribution of the fitted PDFs even though our starting
data and data replicas are gaussianly distributed.

\begin{figure}[ht]
\begin{center}
\epsfig{width=0.49\textwidth,figure=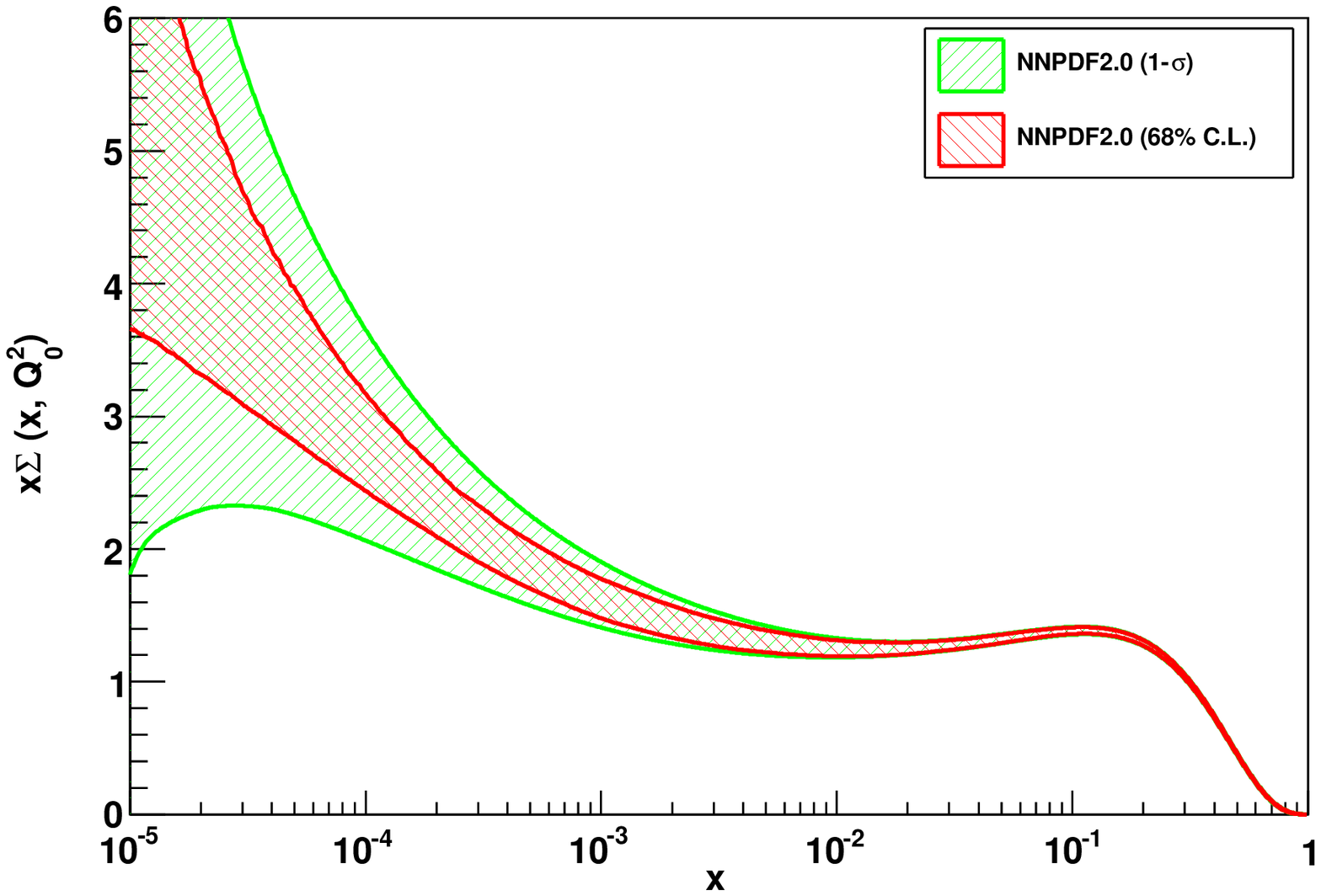}
\epsfig{width=0.49\textwidth,figure=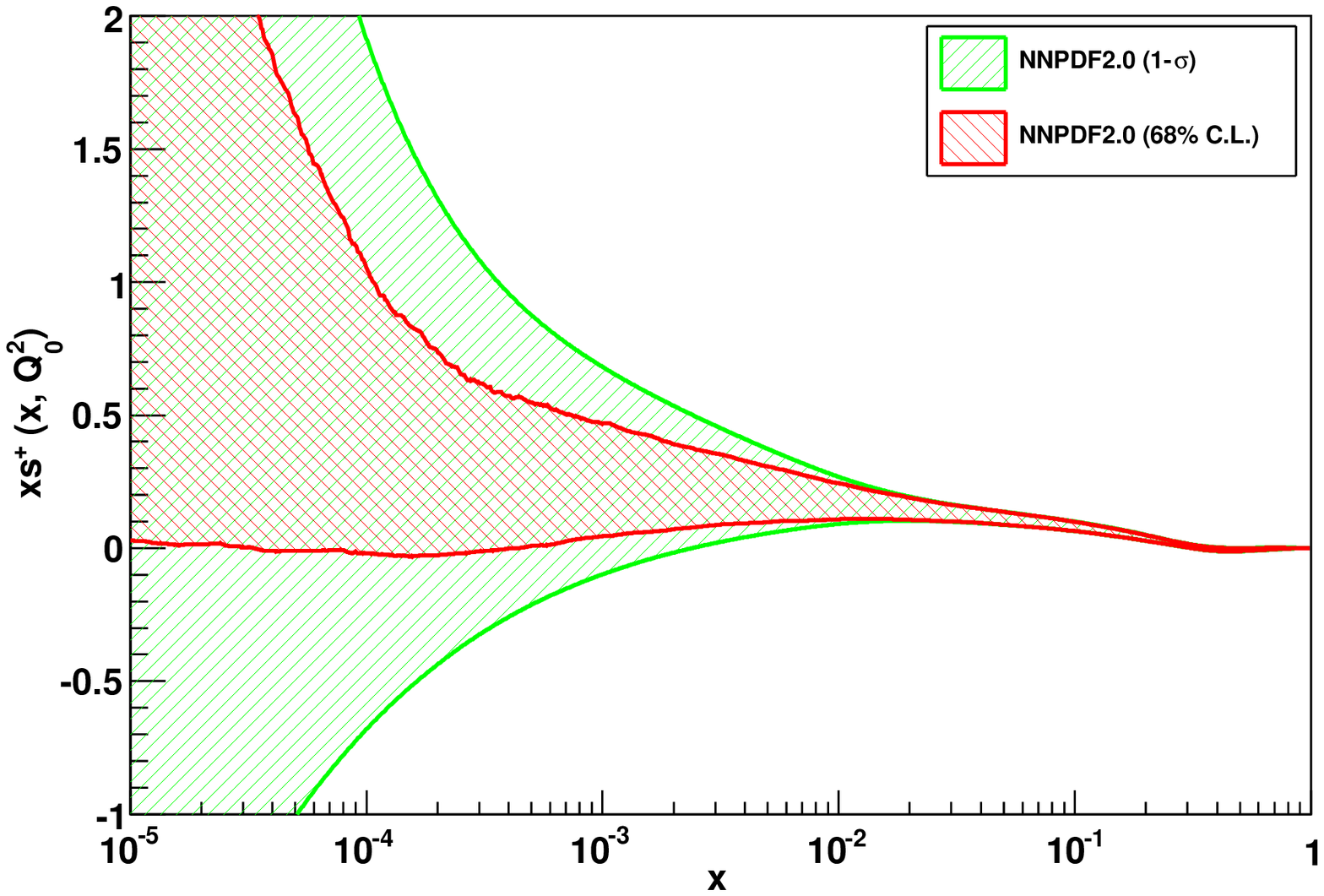}
\epsfig{width=0.49\textwidth,figure=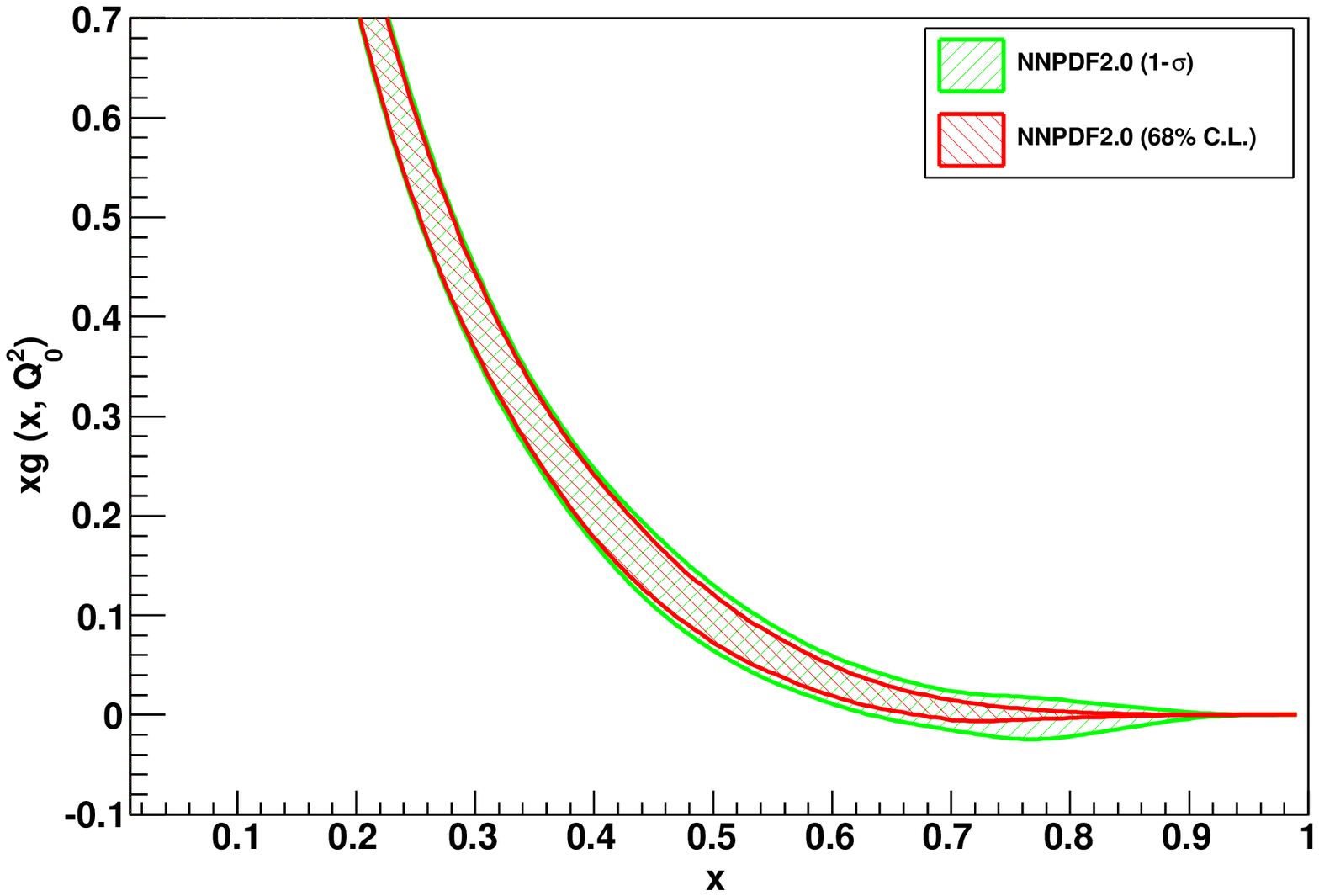}
\epsfig{width=0.49\textwidth,figure=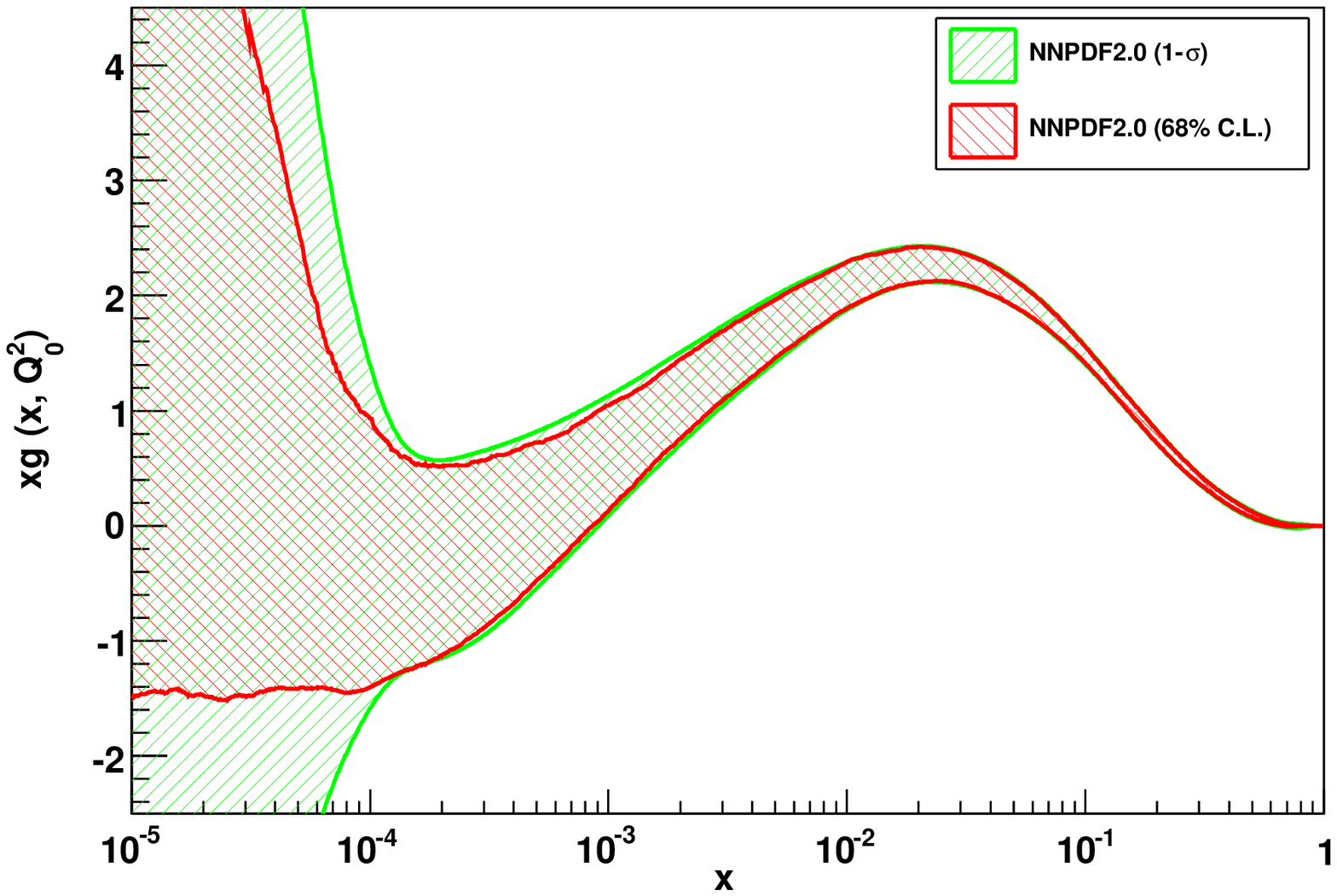}
\epsfig{width=0.49\textwidth,figure=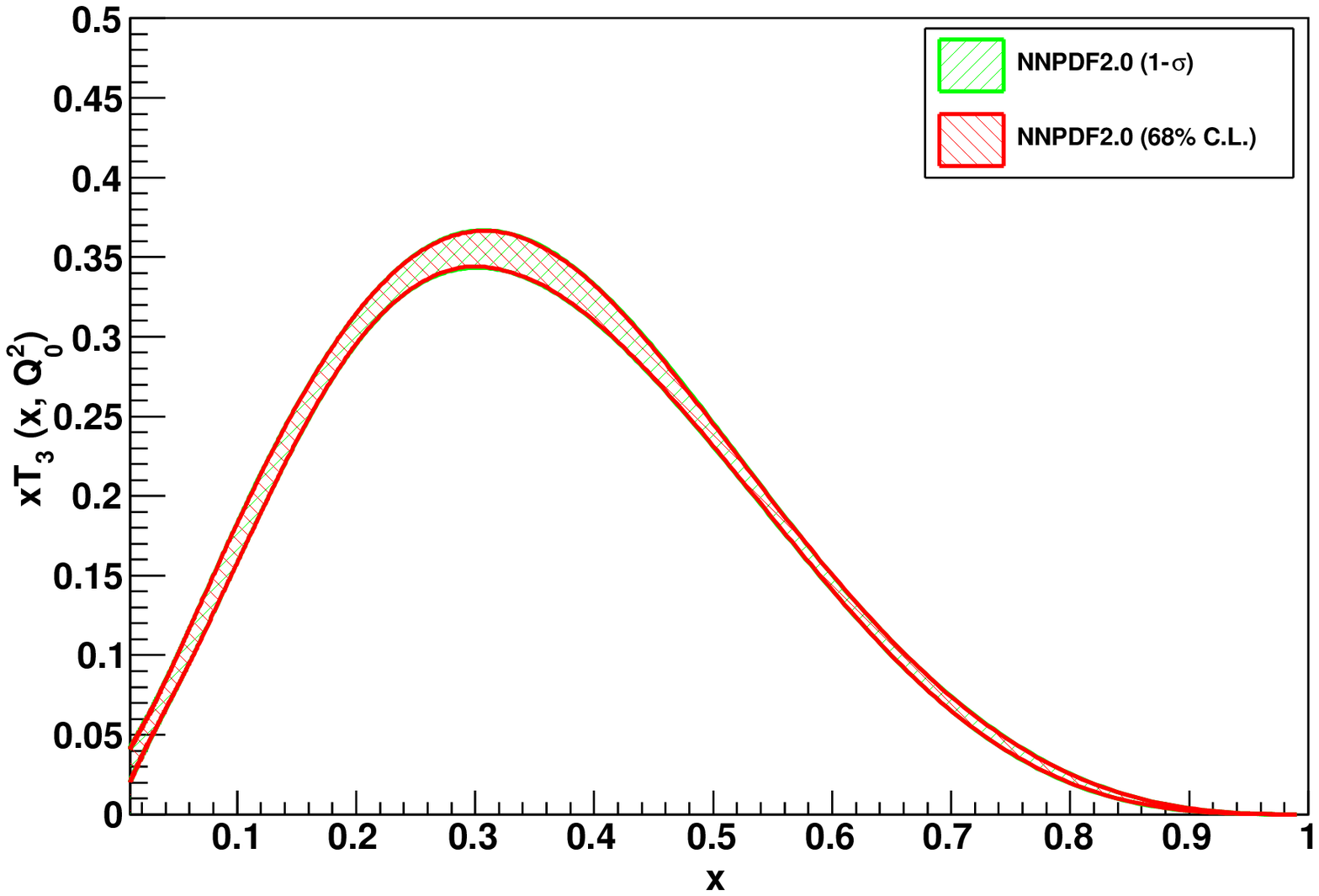}
\epsfig{width=0.49\textwidth,figure=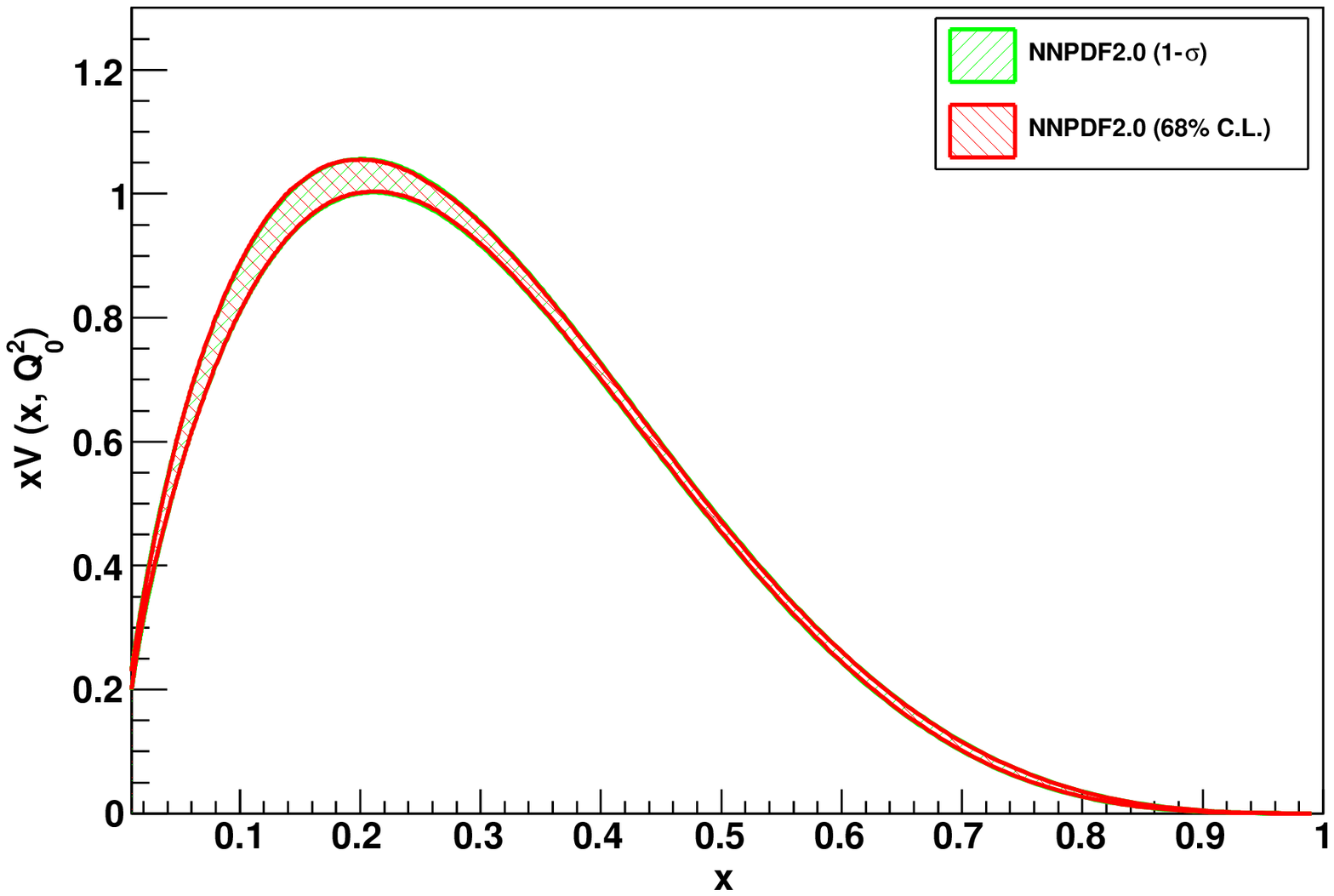}
\epsfig{width=0.49\textwidth,figure=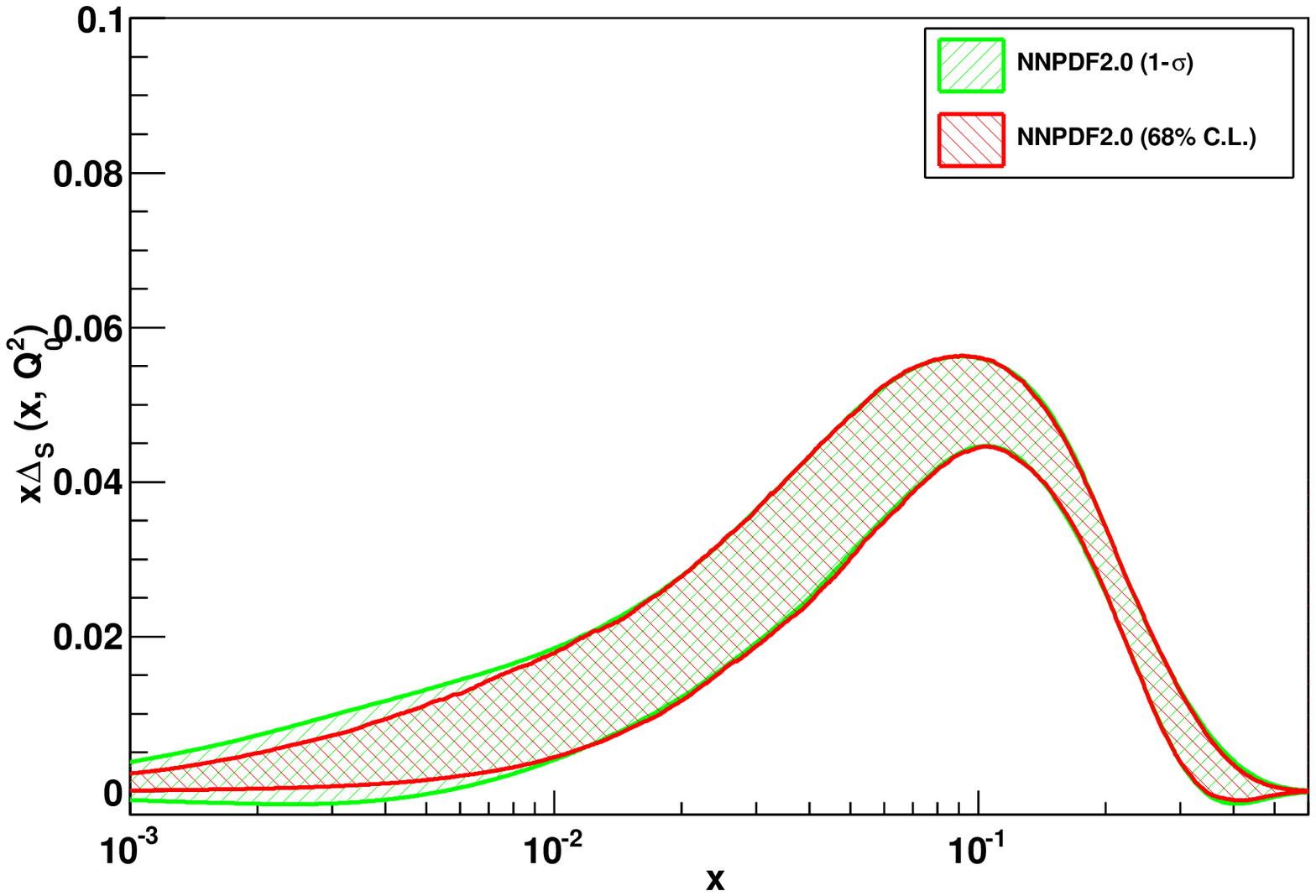}
\epsfig{width=0.49\textwidth,figure=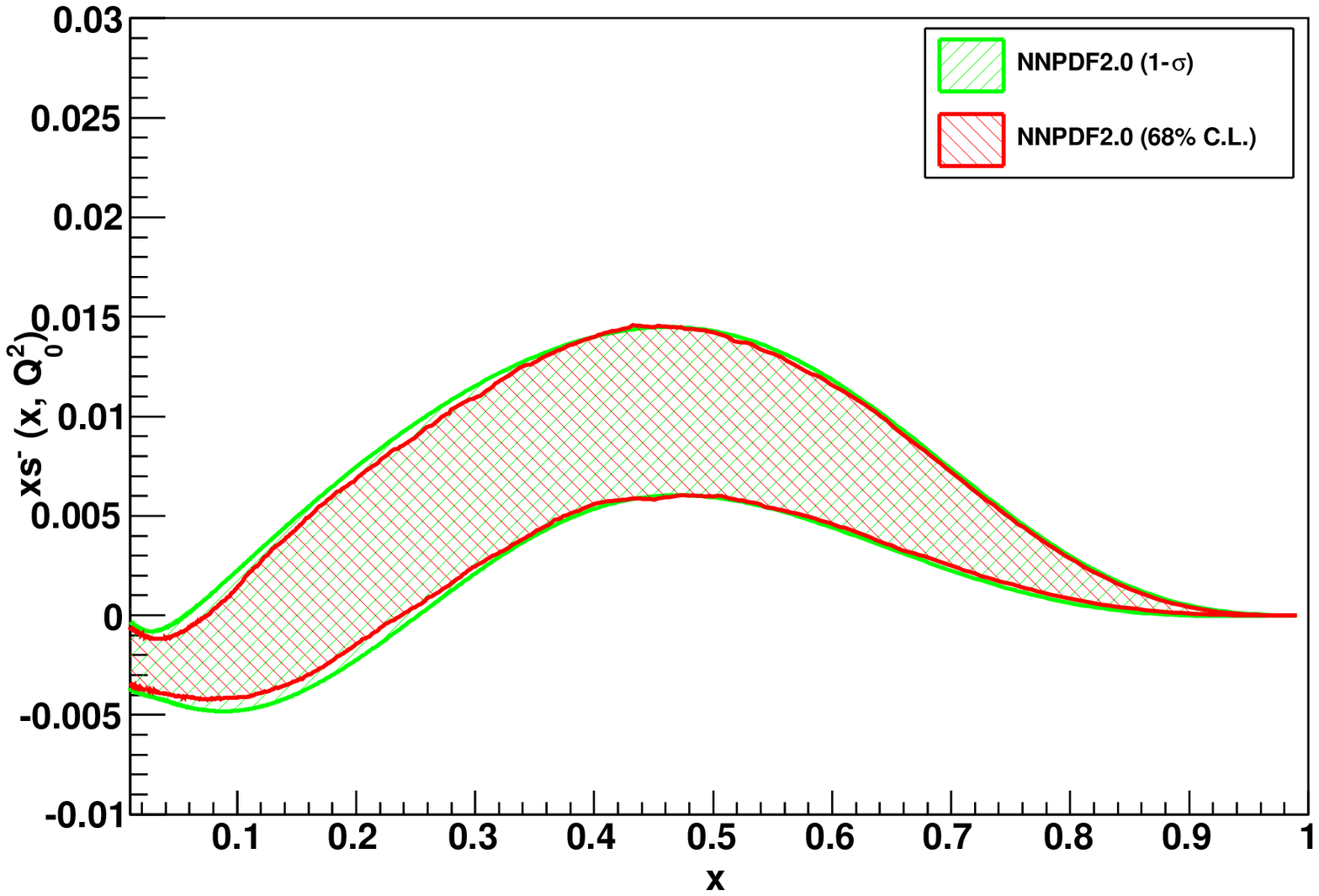}
\caption{\small Comparison of 68\% confidence level and one--sigma
  intervals for  NNPDF2.0 PDFs at the initial scale.
\label{fig:pdferrorscl}} 
\end{center}
\end{figure}

A simple way to test for non-gaussian behaviour for some quantity 
is to compute a 68\% confidence
level for it (which is straightforwardly done in a Monte Carlo
approach), and   compare the result to
the standard deviation.
This method was  used in Ref.~\cite{Ball:2009mk} to
identify large departures from gaussian behaviour in the
strange over non-strange momentum ratio.
In Fig.~\ref{fig:pdferrorscl} 
this comparison is shown for all NNPDF2.0 PDFs at the initial scale as
a function of $x$.

Figure~\ref{fig:pdferrorscl} shows that 
in the regions in which the PDFs are constrained by experimental data
the standard deviation and the  68\% confidence
levels coincide to good approximation, thus suggesting gaussian
behaviour. However, in the extrapolation region for most PDFs
deviations from gaussian behaviour are sizable.
This is
especially noticeable for the gluon at small $x$, and for the quark 
singlet and total strangeness  both
at small and large $x$. Deviations from gaussian behaviour are sometimes 
 related to positivity constraints Eq.~(\ref{eq:poscon}): for instance
positivity of $F_L$ and the dimuon
cross--section limits the possibility for
the small--$x$ gluon and strange sea PDFs respectively to go negative,
thereby leading to an asymmetric
uncertainty band. The impact
of positivity constraints on PDFs will be discussed in greater detail
in 
Sect.~\ref{sec:res:positivity}.
%%%%%%%%%%%%%%%%%%%%%%%%%%%%%%%%%%%%%%%%%%%%%%%%%%%%%%%%%%%%%%%%%%%
%%%%%%%%%%%%%%%%%%%%%%%%%%%%%%%%%%%%%%%%%%%%%%%%%%%%%%%%%%%%%

\clearpage
\begin{table}
\centering
\scriptsize
\begin{tabular}{|c|c|c|c|c|c|c|}
\hline  
Fit & NNPDF1.2 & NNPDF1.2+IGA & NNPDF1.2+IGA+$t_0$&2.0 DIS &  2.0 DIS+JET 
& NNPDF2.0 \\
\hline 
\hline
$\chi^{2}_{\tot}$ & 1.32  & 1.16 & 1.12 & 1.20 & 1.18  &  1.21  \\
$\la E \ra $   & 2.79 & 2.41 & 2.24 & 2.31 &  2.28        &  2.32\\
$\la E_{\rm tr} \ra $& 2.75 & 2.39 & 2.20 &  2.28  & 2.24   & 2.29   \\
$\la E_{\rm val} \ra$& 2.80  & 2.46  & 2.27 & 2.34 & 2.32   &  2.35 \\
\hline
$\la \chi^{2(k)}\ra  $  & 1.60   & 1.28 & 1.21 &1.29  & 1.27    & 1.29    \\
\hline
\hline
 NMC-pd    &  1.48  & 0.97 & 0.87& 0.85&  0.86 & 0.99\\
\hline
NMC             &  1.68 & 1.72& 1.65& 1.69 &  1.66  & 1.69\\
\hline
SLAC            &  1.20  & 1.42 & 1.33& 1.37 & 1.31& 1.34\\
\hline
BCDMS           & 1.59  & 1.33 & 1.25 & 1.26 &  1.27 & 1.27\\
\hline
HERAI        & 1.05   & 0.98 & 0.96 & 1.13 & 1.13  & 1.14\\
\hline
CHORUS          & 1.39  & 1.13 & 1.12& 1.13& 1.11& 1.18\\
\hline
FLH108          &  1.70  & 1.53 & 1.53& 1.51 &  1.49 & 1.49\\
\hline
NTVDMN          &  0.64 & 0.81 & 0.71& 0.71&  0.75 & 0.67\\
\hline
ZEUS-H2         & 1.52  & 1.51 & 1.49 & 1.50 &  1.49 & 1.51\\
\hline
DYE605          & {\it 11.19} & {\it 22.89}& {\it 8.21}& {\it 7.32} & {\it 10.35}  & 0.88 \\
\hline
DYE866          & {\it 53.20} & {\it 4.81} & {\it 2.46}& {\it 2.24}&  {\it 2.59}  & 1.28\\
\hline
CDFWASY         & {\it 26.76} & {\it 28.22}& {\it 20.32}& {\it 13.06}& {\it 14.13} &  1.85\\
\hline
CDFZRAP         & {\it 1.65} &{\it 4.61}& {\it 3.13}& {\it 3.12}&  {\it 3.31} & 2.02\\
\hline
D0ZRAP          &  {\it 0.56 }& {\it 0.80}& {\it 0.65}& {\it 0.65}&  {\it 0.68} & 0.57\\
\hline
CDFR2KT         &  {\it 1.10 }& {\it 0.95}& {\it 0.78}&{\it 0.91}&  0.79 & 0.80\\
\hline
D0R2CON         &  {\it 1.18} & {\it 1.07}& {\it 0.94}& {\it 1.00} & 0.93 & 0.93\\
\hline
\end{tabular}
\caption{\small \label{tab:estdataset1} Statistical estimators
  for the sequence of fits that take from  NNPDF1.2 to NNPDF2.0. The 
 estimators shown for NNPDF1.2 are as in Tab.~5-6 of
 Ref.~\cite{Ball:2009mk} and those for  NNPDF2.0 
are as in Tab.~\ref{tab:estfit1}--\ref{tab:estfit2}. Estimators are
shown for the total datasets in the upper part of the table, while
the lower part of the table shows the $\chi^2$ for each individual
experimental dataset. Values of the
$\chi^2$ for data not included in any given fit are shown in italic; the
total
  $\chi^2_{\rm tot}$ shown in the first line does not include the contribution
from these data.  The value of the $\chi^2$ in the  HERAI 
  line refers in the first three columns of the table 
to the weighted sum of the H1 and ZEUS data, and in
the latter three columns to the combined dataset, according to 
  which data has been included in the fit. }
\end{table}

\subsection{Detailed comparison to NNPDF1.2: methodology and dataset}
\label{sec:res:dataset}

As seen in Sect.~\ref{sec:res:pdfs} the quality of the
NNPDF2.0 fit is rather better than that of NNPDF1.2, despite the wider
dataset. We now perform a detailed comparison of these two fits, which
differ both in procedural aspects and in dataset.  In order to
elucidate the impact on the fit of each of these, we have produced a
sequence of PDF determinations that take us from NNPDF1.2 to NNPDF2.0 by
varying one by one each of the procedural aspects, then 
each of the datasets inclusions, as follows 

(i) we start from NNPDF1.2; 

(ii)  we switch to the
improved genetic algorithm and minimization of Sect.~\ref{sec-minim}
(IGA); 

(iii) we
introduce the improved treatment of normalization uncertainties of
Ref.~\cite{t0} ($t_0$ method); 

(iv) we replace the separate H1 and ZEUS
data with the new combined HERA-I dataset: this gives the NNPDF2.0 set,
but with DIS data only (2.0-DIS); 

(v) we add jet data (2.0-DIS+jet); 

(vi) we add the
DY data, thereby obtaining the NNPDF2.0 fit.

The statistical estimators for this sequence of fits are shown in
Table~\ref{tab:estdataset1} (including the NNPDF2.0 estimators already
shown in Tab.~\ref{tab:estfit1}--\ref{tab:estfit2}).
We will now discuss each of these subsequent fits in
turn by examining its general features, and determining and
understanding 
the distance 
(as defined in
Appendix~\ref{sec:distances}) between PDFs obtained in each pair of
subsequent fits.
 
\begin{enumerate}

\item {\it Effect of the improved genetic algorithm and stopping criterion
  (IGA).}\\\noindent\nobreak The improvement in neural network training leads to a
  significant improvement in fit quality: each replica fits better the
  corresponding data replica (lower $\langle E\rangle$), and also each
  replica neural network is more efficient in subtracting the
  statistical noise from data (lower $\langle \chi^{2\,(k)}\rangle$), thereby
  leading to a better global fit (lower $\chi^2_{\rm tot}$).
The improvement is due to the improvement in fit quality of 
fixed--target DIS experiments
(NMC, BCDMS and CHORUS) which probe the valence region which has more
structure, and which moreover are
known~\cite{Forte:2002fg,DelDebbio:2004qj,Pumplin:2009sc}  to have a
certain amount of data inconsistency, without change in fit quality
for other experiments: this means that the new algorithm is more
efficient in leading to a balanced fit quality between experiments,
without some data being underlearnt while others are overlearnt.

%%%%%%%%%%%%%%%%%%%%%%%%%%%%%%%%%%%%%%%%%%%%%%%%%%%%%%%
\begin{figure}[t!]
\begin{center}
\epsfig{width=0.99\textwidth,figure=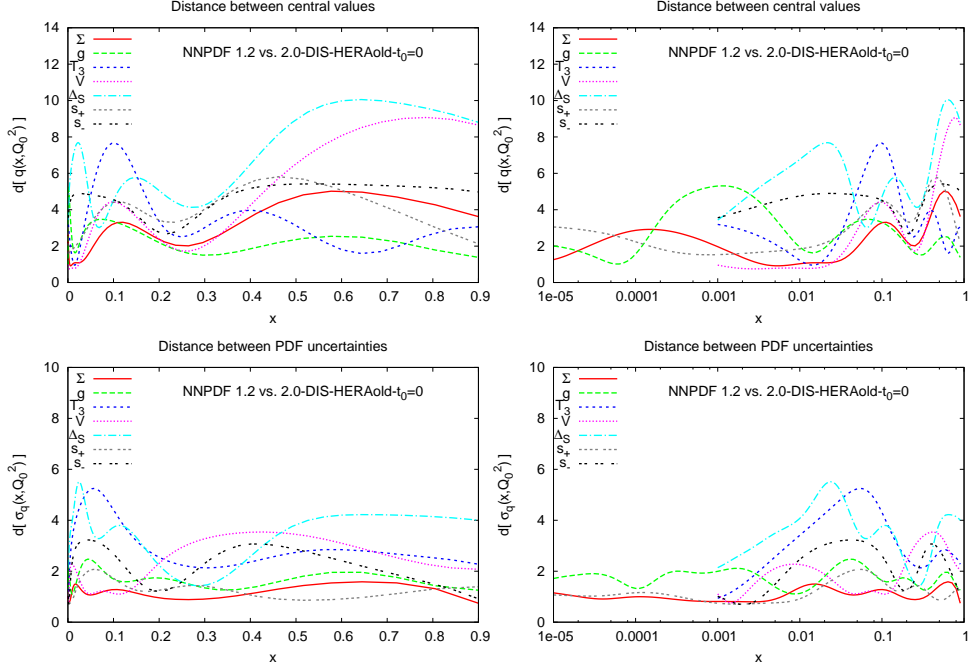}
\caption{\small Distance between
the NNPDF1.2 fit and  a fit to the same data but improved genetic
algorithm and stopping (IGA).
\label{fig:stabtab-12-20-dis-heraold-not0}}
\end{center}
\end{figure}
%%%%%%%%%%%%%%%%%%%%%%%%%%%%%%%%%%%%%%%%%%%%%%%%%%%%
%%%%%%%%%%%%%%%%%%%%%%%%%%%%%%%%%%%%%%%%%%%%%%%%%%%%%%%
\begin{figure}[t!]
\begin{center}
\epsfig{width=0.49\textwidth,figure=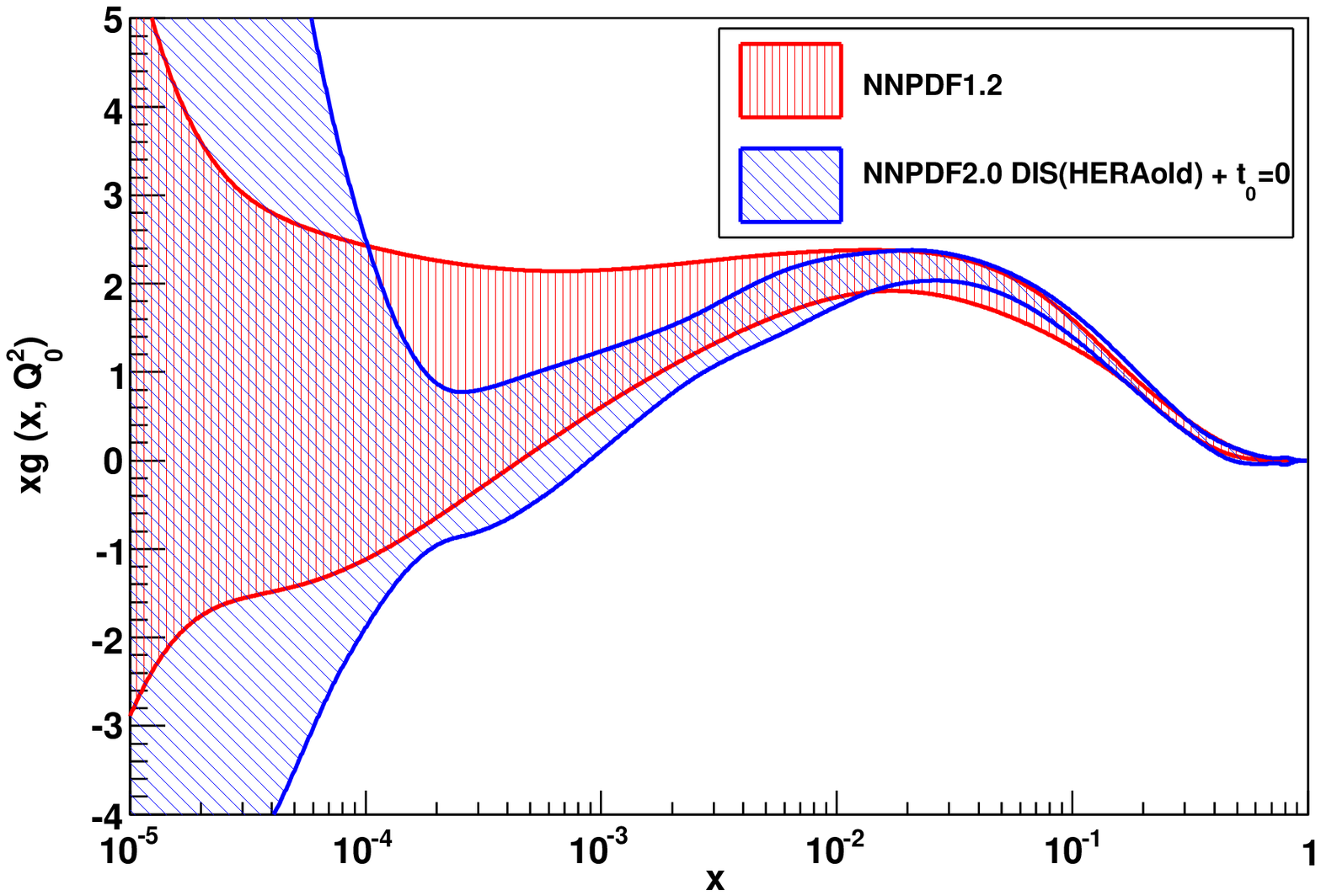}
\epsfig{width=0.49\textwidth,figure=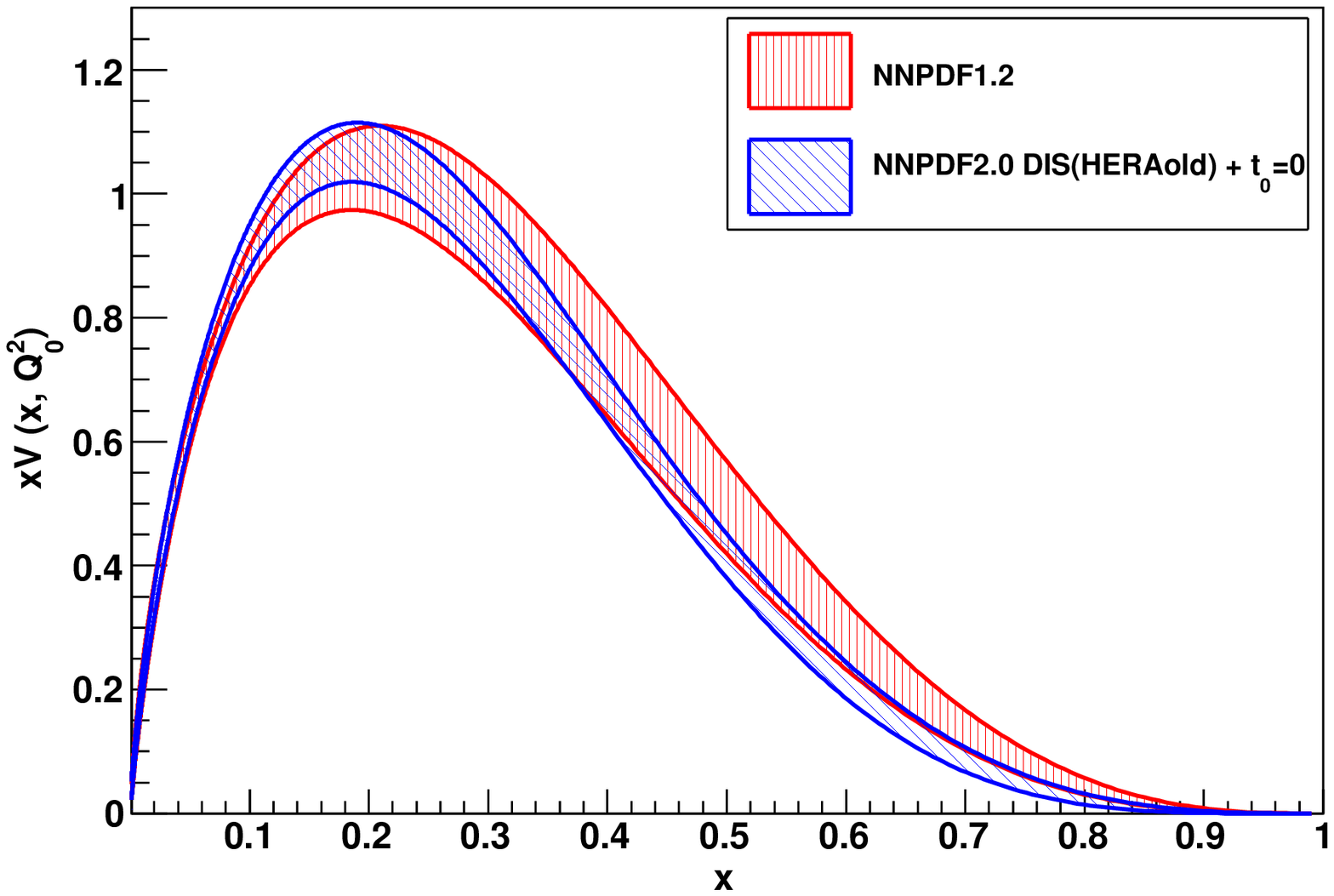}
\end{center}
\begin{center}
\epsfig{width=0.49\textwidth,figure=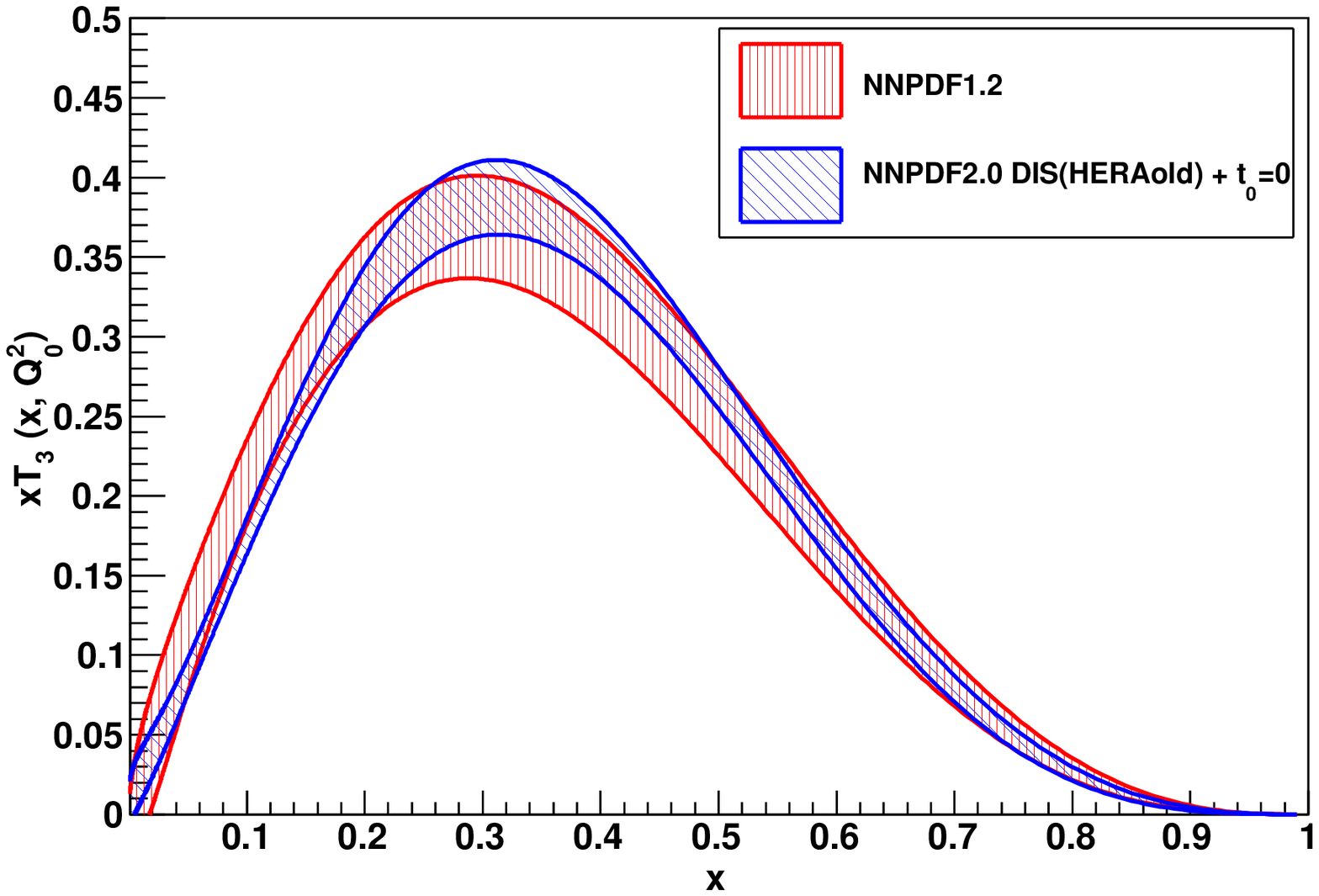}
\caption{\small Comparison between PDFs from the NNPDF1.2 fit and a 
fit to the same data but improved genetic
algorithm and stopping (IGA) (the distances
  are shown 
in Fig.~\ref{fig:stabtab-12-20-dis-heraold-not0}): small-$x$ gluon,
valence and triplet (from left
to right). 
\label{fig:pdfplots-12-20-dis-heraold-not0}}
\end{center}
\end{figure}
%%%%%%%%%%%%%%%%%%%%%%%%%%%%%%%%%%%%%%%%%%%%%%%%%%%%
The distance between NNPDF1.2 and this fit, which only differs from it
because of the IGA, is shown in
 Fig.~\ref{fig:stabtab-12-20-dis-heraold-not0}: the IGA affects
essentially all PDFs by reducing their uncertainties, the two fits are always consistent
at the 1-$\sigma$ level. The individual PDFs which
are more affected are the triplet, the valence and the gluon at
small-$x$, which are shown
Fig.~\ref{fig:pdfplots-12-20-dis-heraold-not0}.

\item {\it Impact of the treatment of normalization uncertainties.}\\\noindent\nobreak
The IGA fit is now repeated by also using the
improved $t_0$ method of Ref.~\cite{t0} for the treatment of
normalization uncertainties. This leads to a further small but not
negligible improvement in fit quality, mostly due to the fixed--target
DIS experiments which have largest normalization uncertainties. The
distances between the
two fits, which only differ in the treatment of normalizations, are shown
in 
Fig.~\ref{fig:stabtab-20-dis-heraold-20-dis-heraold-not0}. The PDFs
which are most affected are the small--$x$ singlet and gluon and the
triplet. A more detailed discussion of the impact of the treatment of
normalization uncertainties on fits to the NNPDF1.2 dataset was
presented in Ref.~\cite{t0} and will not be repeated here.

%%%%%%%%%%%%%%%%%%%%%%%%%%%%%%%%%%%%%%%%%%%%%%%%%%%%%%%
\begin{figure}[ht!]
\begin{center}
\epsfig{width=0.99\textwidth,figure=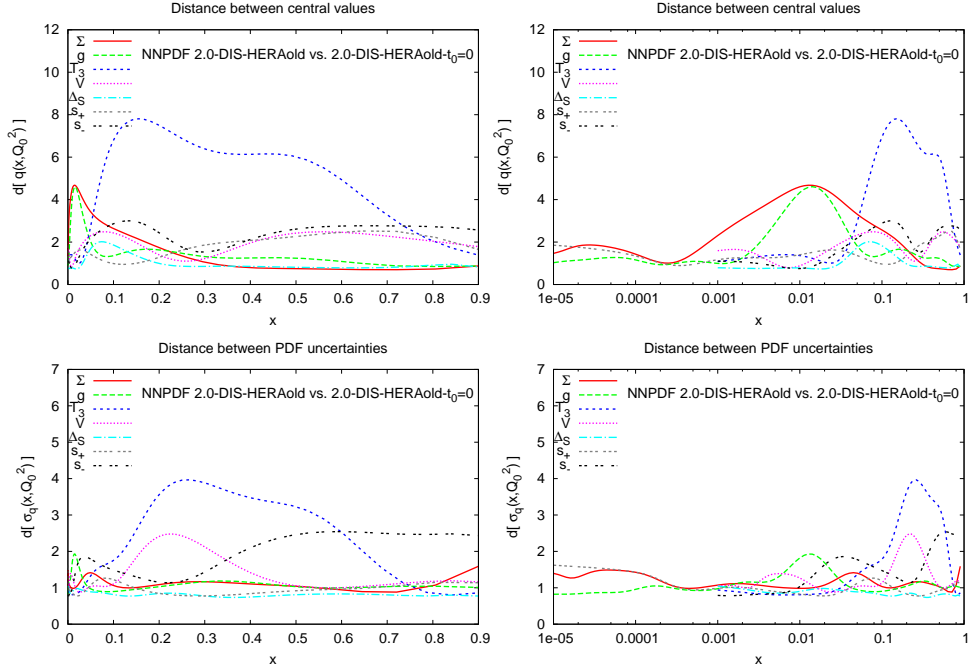}
\caption{\small Distance between the IGA fit of Fig.~\ref{fig:stabtab-12-20-dis-heraold-not0} and a fit
  with improved treatment of normalization uncertainties (IGA+$t_0$).
\label{fig:stabtab-20-dis-heraold-20-dis-heraold-not0}}
\end{center}
\end{figure}
%%%%%%%%%%%%%%%%%%%%%%%%%%%%%%%%%%%%%%%%%%%%%%%%%%%%

%%%%%%%%%%%%%%%%%%%%%%%%%%%%%%%%%%%%%%%%%%%%%%%%%%%%%%%
\begin{figure}[ht!]
\begin{center}
\epsfig{width=0.49\textwidth,figure=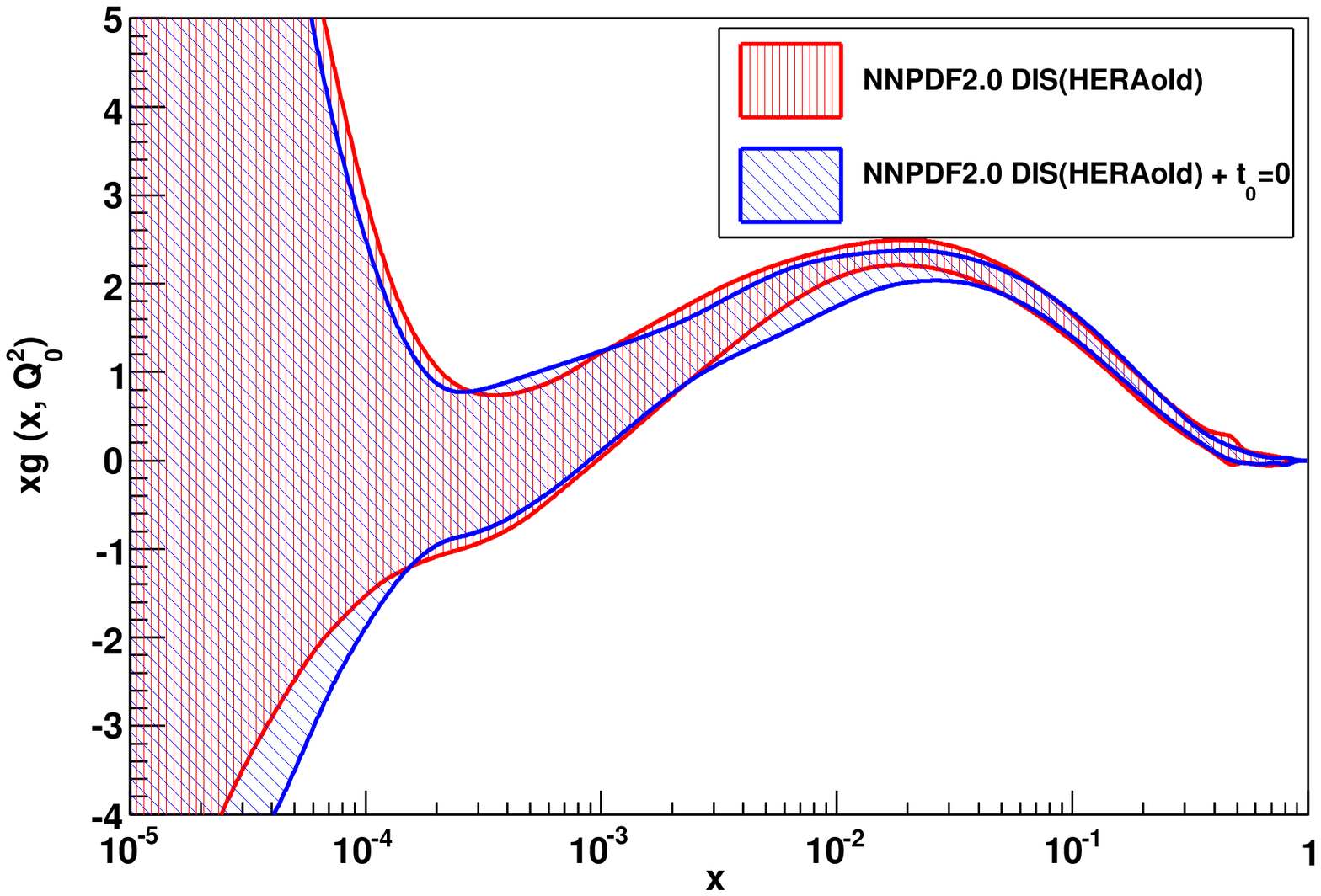}
\epsfig{width=0.49\textwidth,figure=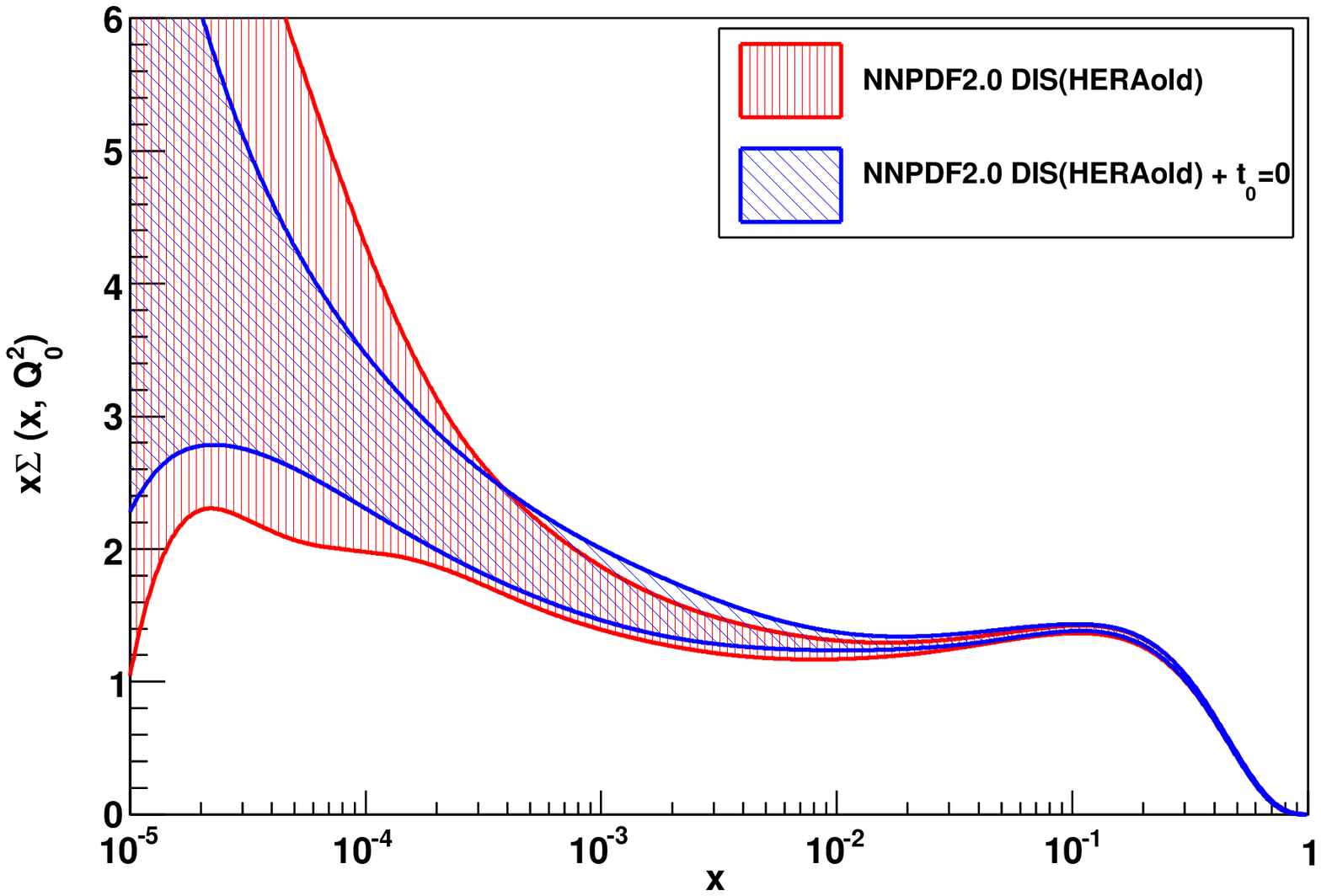}
\end{center}
\begin{center}
\epsfig{width=0.49\textwidth,figure=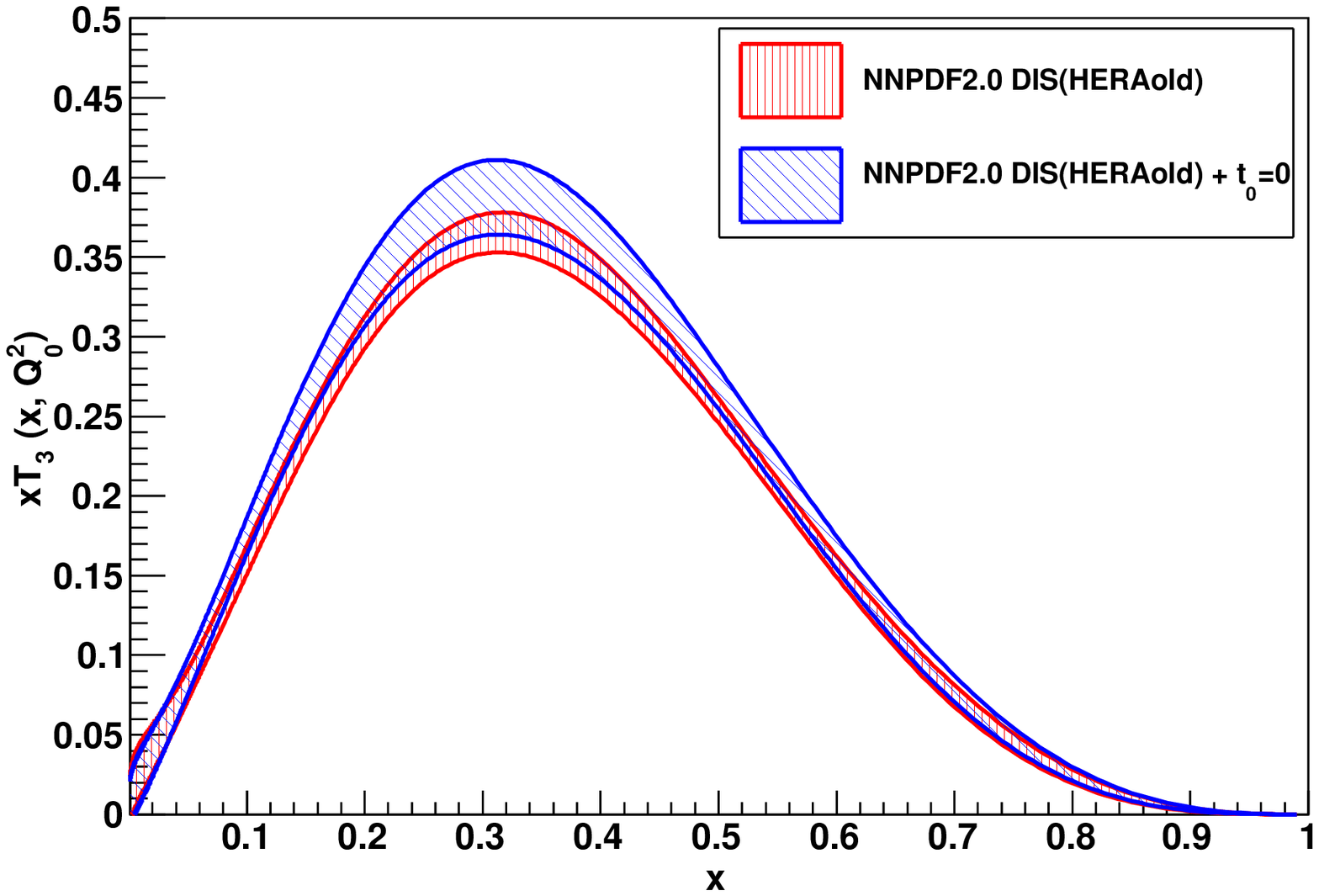}
\caption{\small 
Comparison between PDFs from the IGA fit of 
Fig.~\ref{fig:stabtab-12-20-dis-heraold-not0} and a fit
with improved treatment of normalization uncertainties 
(IGA+$t_0$) (the distances
  are shown 
in Fig.~\ref{fig:stabtab-20-dis-heraold-20-dis-heraold-not0}): small-$x$ gluon,
small $x$ singlet and triplet (from left
to right). 
\label{fig:pdfplots-20-dis-heraold-20-dis-heraold-not0}}
\end{center}
\end{figure}
%%%%%%%%%%%%%%%%%%%%%%%%%%%%%%%%%%%%%%%%%%%%%%%%%%%%

\item {\it Impact of the combined HERA-I data.}\\\noindent\nobreak
The previous IGA+$t_0$ fit is now repeated replacing the ZEUS and H1
data with the new combined HERA-I dataset of
Ref.~\cite{H1:2009wt}. This fit is now identical to the NNPDF2.0 fit,
but with only DIS data (i.e. no hadronic data) included (2.0-DIS).
The inclusion of the very precise HERA-I data leads to a slight
deterioration of fit quality, which remains however still better than
that of NNPDF1.2. This deterioration is concentrated in the HERA data
themselves, with the quality of the fit to all other data
unchanged. This suggests good consistency of the HERA and fixed target
data, but with the accuracy of the combined HERA-I data now exceeding
the accuracy of the theory used to describe them in NNPDF2.0: specifically, the
lack of inclusion of charm mass corrections, but also possibly
deviations from NLO DGLAP at small $x$~\cite{Caola:2009iy}, or 
possible evidence
for NNLO corrections at larger $x$. A particularly interesting aspect
of this fit is that the quality of the fit to Drell-Yan data (not
fitted),  which
was poor in all previous fits, improves
considerably, especially for the $W$ asymmetry. 
This suggests that the accuracy of the
charged--current data in the HERA-I combined set is now sufficient to provide
some handle on the flavour decomposition of the sea at large $x$ which
is only weakly constrained by neutral current DIS data, and strongly
constrained by DY data.

The distances between these fits is shown in 
Fig.~\ref{fig:stabtab-20-dis-20-disheraold}: the impact of the
combined  HERA data is a moderate but generalized improvement in
accuracy at small $x$. The effect on 
the singlet and the gluon at small-$x$ is shown in 
Fig.~\ref{fig:pdfplots-20-dis-20-disheraold}. The sizable
error reduction in the small $x$ singlet is specially interesting.

%%%%%%%%%%%%%%%%%%%%%%%%%%%%%%%%%%%%%%%%%%%%%%%%%%%%%%%
\begin{figure}[ht]
\begin{center}
\epsfig{width=0.99\textwidth,figure=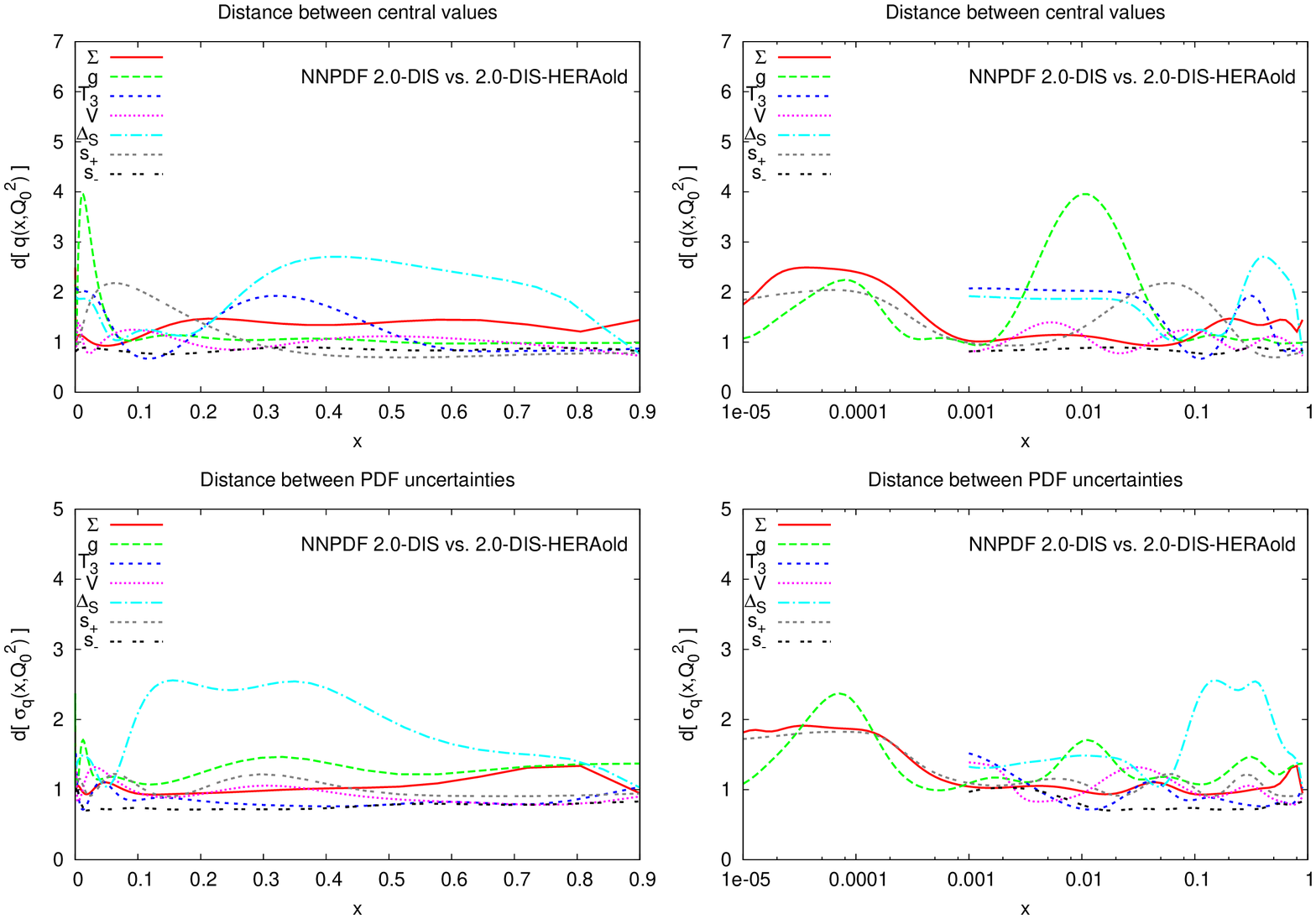}
\caption{\small Distance between the IGA+$t_0$ fit of
  Fig.~\ref{fig:stabtab-20-dis-heraold-20-dis-heraold-not0} and a fit
  in which the separate H1 and ZEUS data are replaced by the combined
  HERA-I DIS data (NNPDF2.0 DIS).
\label{fig:stabtab-20-dis-20-disheraold}}
\end{center}
\end{figure}
%%%%%%%%%%%%%%%%%%%%%%%%%%%%%%%%%%%%%%%%%%%%%%%%%%%%

%%%%%%%%%%%%%%%%%%%%%%%%%%%%%%%%%%%%%%%%%%%%%%%%%%%%%%%
\begin{figure}[ht]
\begin{center}
\epsfig{width=0.48\textwidth,figure=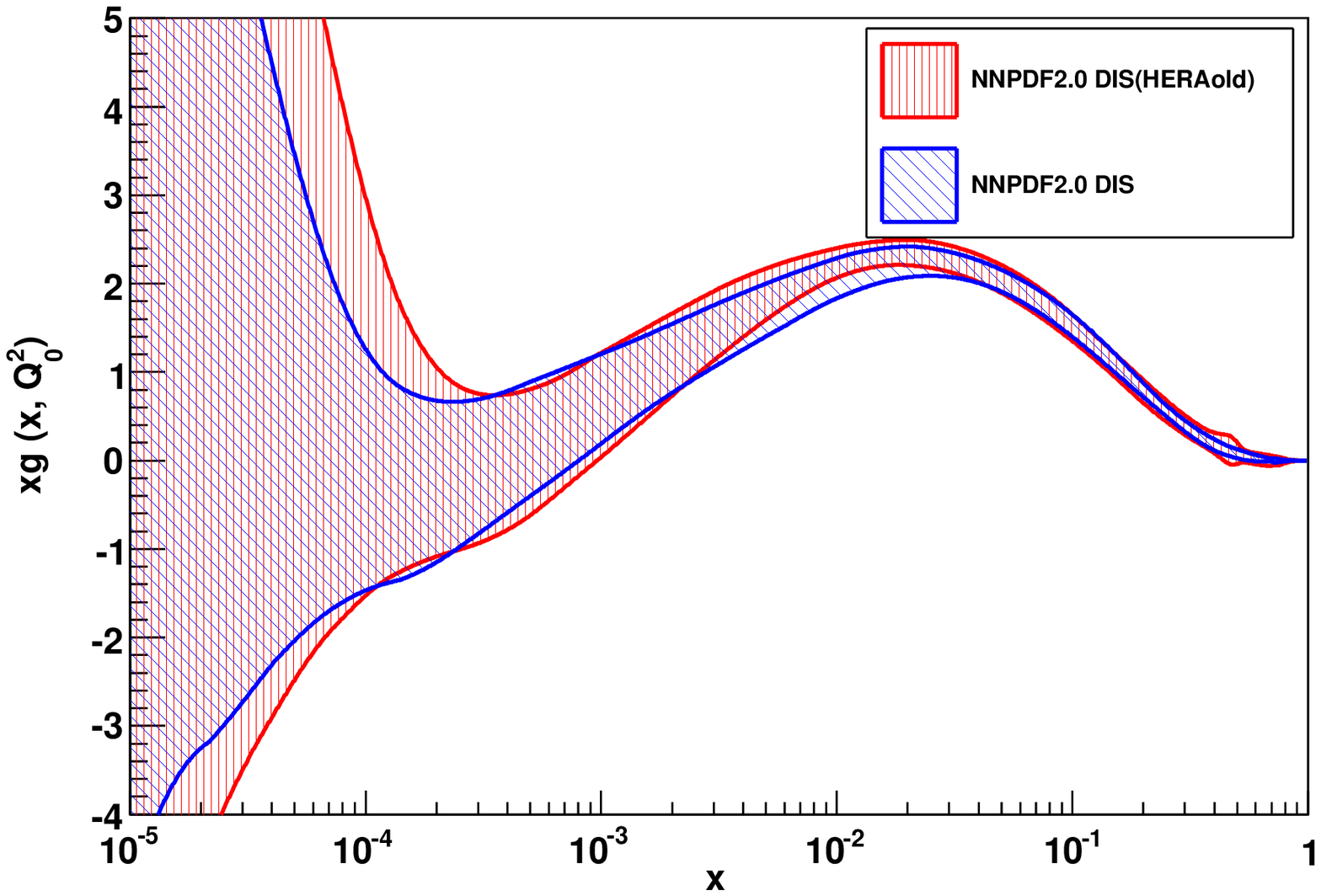}
\epsfig{width=0.48\textwidth,figure=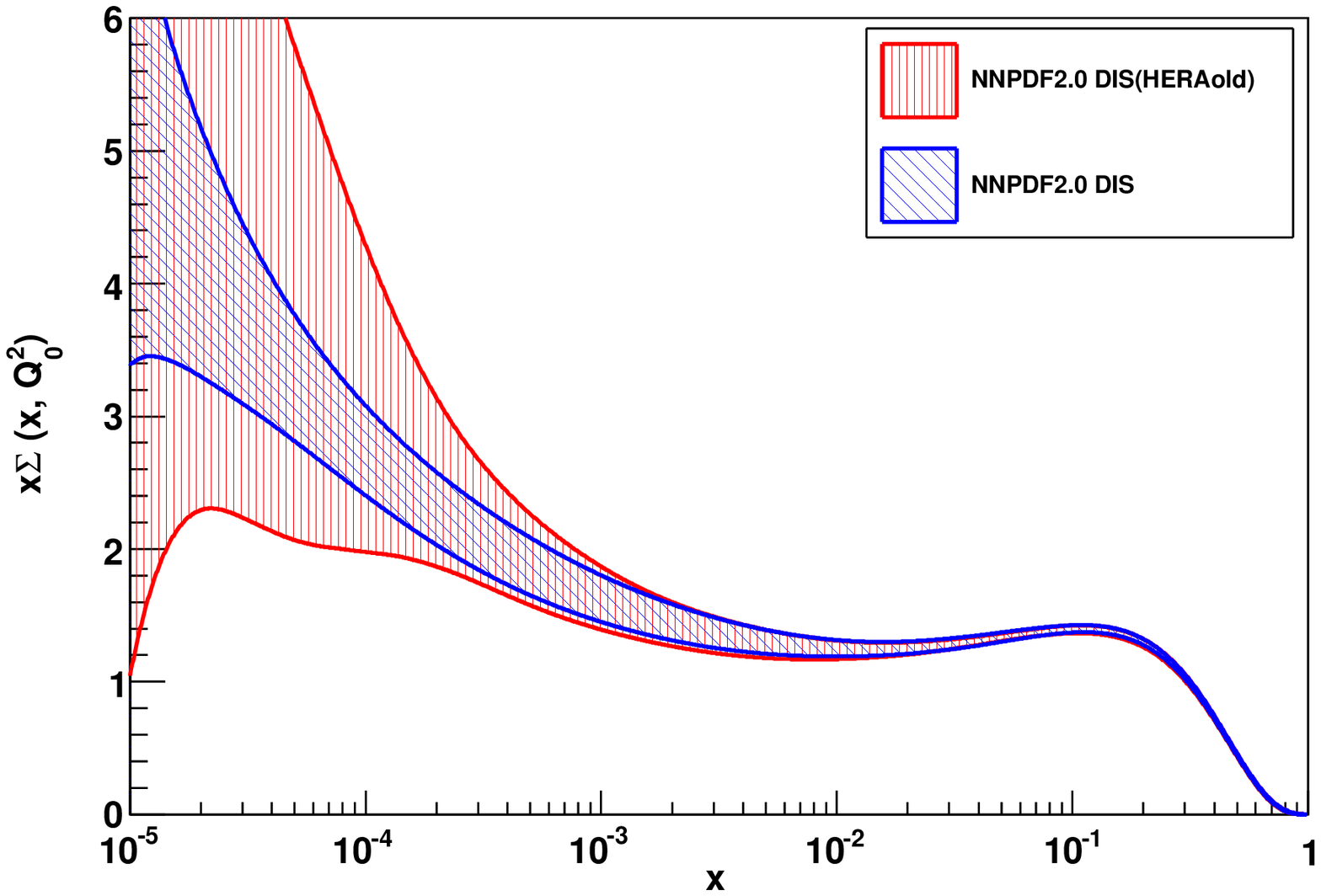}
\caption{\small 
Comparison between PDFs from IGA+$t_0$ fit of
  Fig.~\ref{fig:stabtab-20-dis-heraold-20-dis-heraold-not0} and a fit
  in which the separate H1 and ZEUS data are replaced by the combined
  HERA-I DIS data (NNPDF2.0 DIS) (the distances
  are shown 
in Fig.~\ref{fig:stabtab-20-dis-20-disheraold}): small-$x$ gluon and
small $x$ singlet (from left
to right). 
\label{fig:pdfplots-20-dis-20-disheraold}}
\end{center}
\end{figure}
%%%%%%%%%%%%%%%%%%%%%%%%%%%%%%%%%%%%%%%%%%%%%%%%%%%%

\item {\it Impact of jet data.}\\\noindent\nobreak
The addition of jet data to the 2.0-DIS fit leaves the quality of
the global fit unchanged. This demonstrates the perfect compatibility of
jet data with DIS data: in fact, the quality of the fit to jet data was
quite good even in all previous fits, in which they were not included
in the fitted dataset.
The distance between the 2.0-DIS and 2.0-DIS+JET fits, displayed in
Fig.~\ref{fig:stabtab-20-dis+jet-20-dis}, shows that these data affect
almost only the gluon, as one would
expect~\cite{Pumplin:2009nk}, leading to a better determination of it
at medium and large $x$. This is shown in
Fig.~\ref{fig:pdfplots-20-dis+jet-20-dis}, where the gluons of 2.0-DIS
and 2.0-DIS+JET are compared.

%%%%%%%%%%%%%%%%%%%%%%%%%%%%%%%%%%%%%%%%%%%%%%%%%%%%%%%
\begin{figure}[ht]
\begin{center}
\epsfig{width=0.99\textwidth,figure=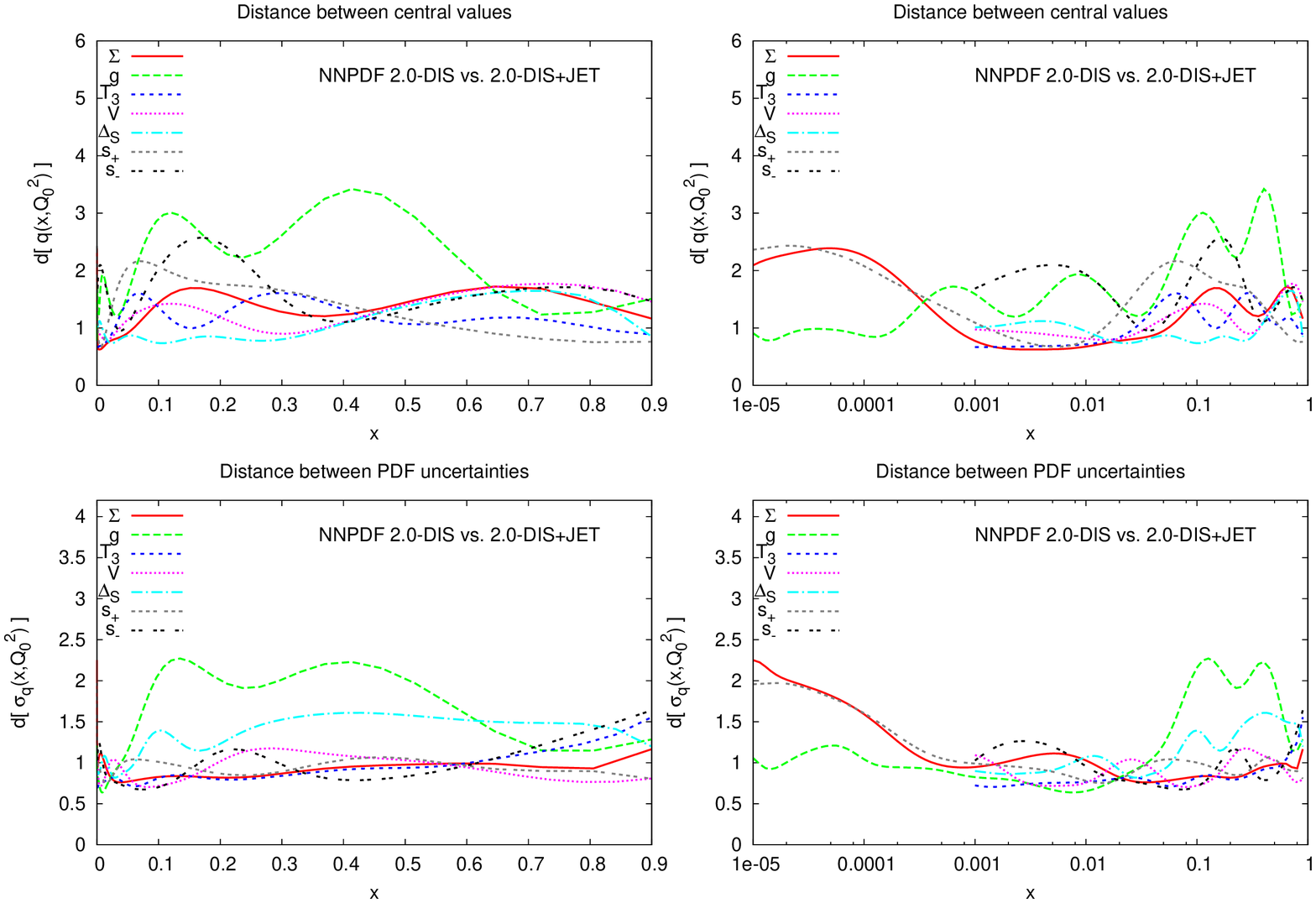}

\caption{\small Distance between the NNPDF2.0 DIS fit 
of
  Fig.~\ref{fig:stabtab-20-dis-20-disheraold} and a fit
  in which jet data are also included (NNPDF2.0 DIS+JET).
\label{fig:stabtab-20-dis+jet-20-dis}}
\end{center}
\end{figure}
%%%%%%%%%%%%%%%%%%%%%%%%%%%%%%%%%%%%%%%%%%%%%%%%%%%%

%%%%%%%%%%%%%%%%%%%%%%%%%%%%%%%%%%%%%%%%%%%%%%%%%%%%%%%
\begin{figure}[ht]
\begin{center}
\epsfig{width=0.48\textwidth,figure=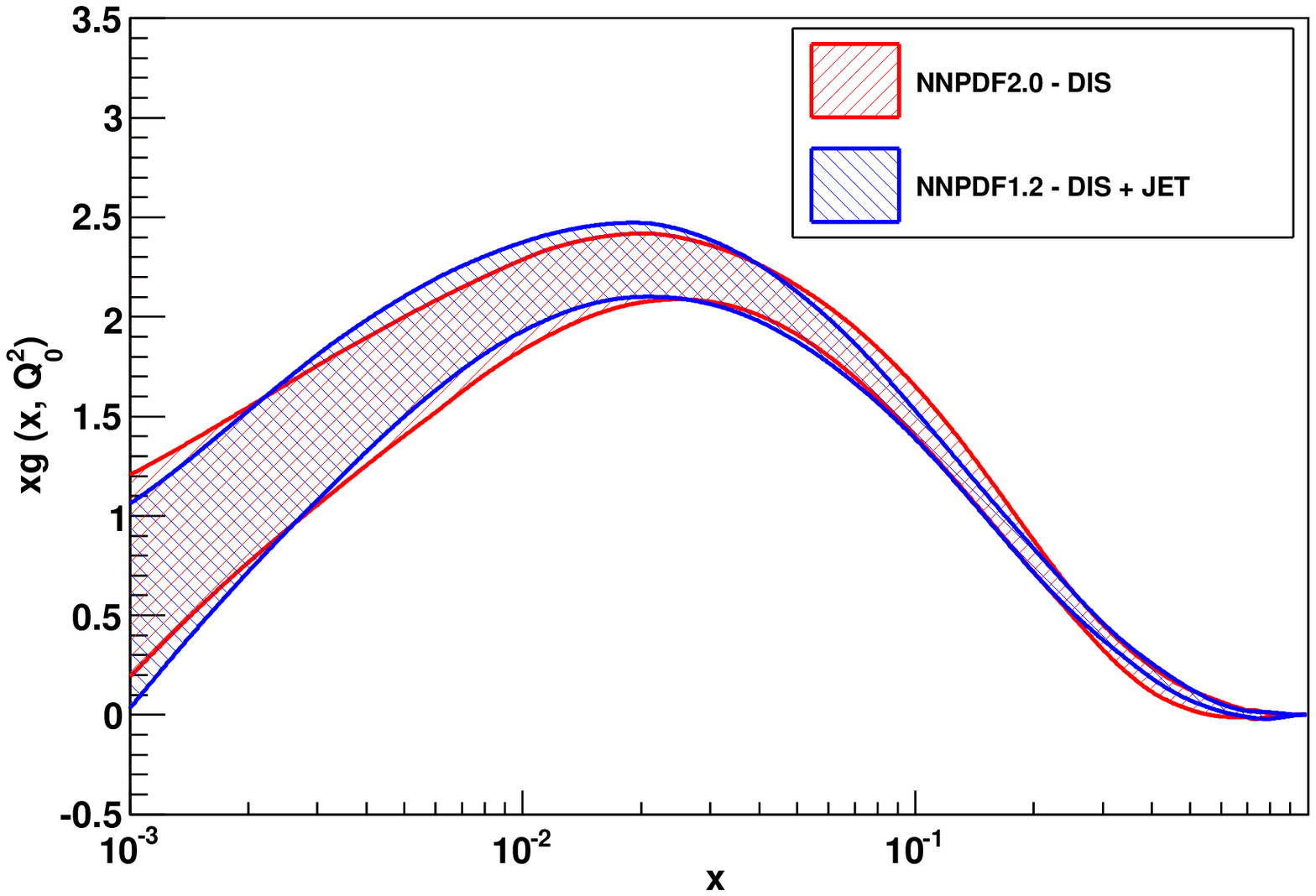}
\epsfig{width=0.48\textwidth,figure=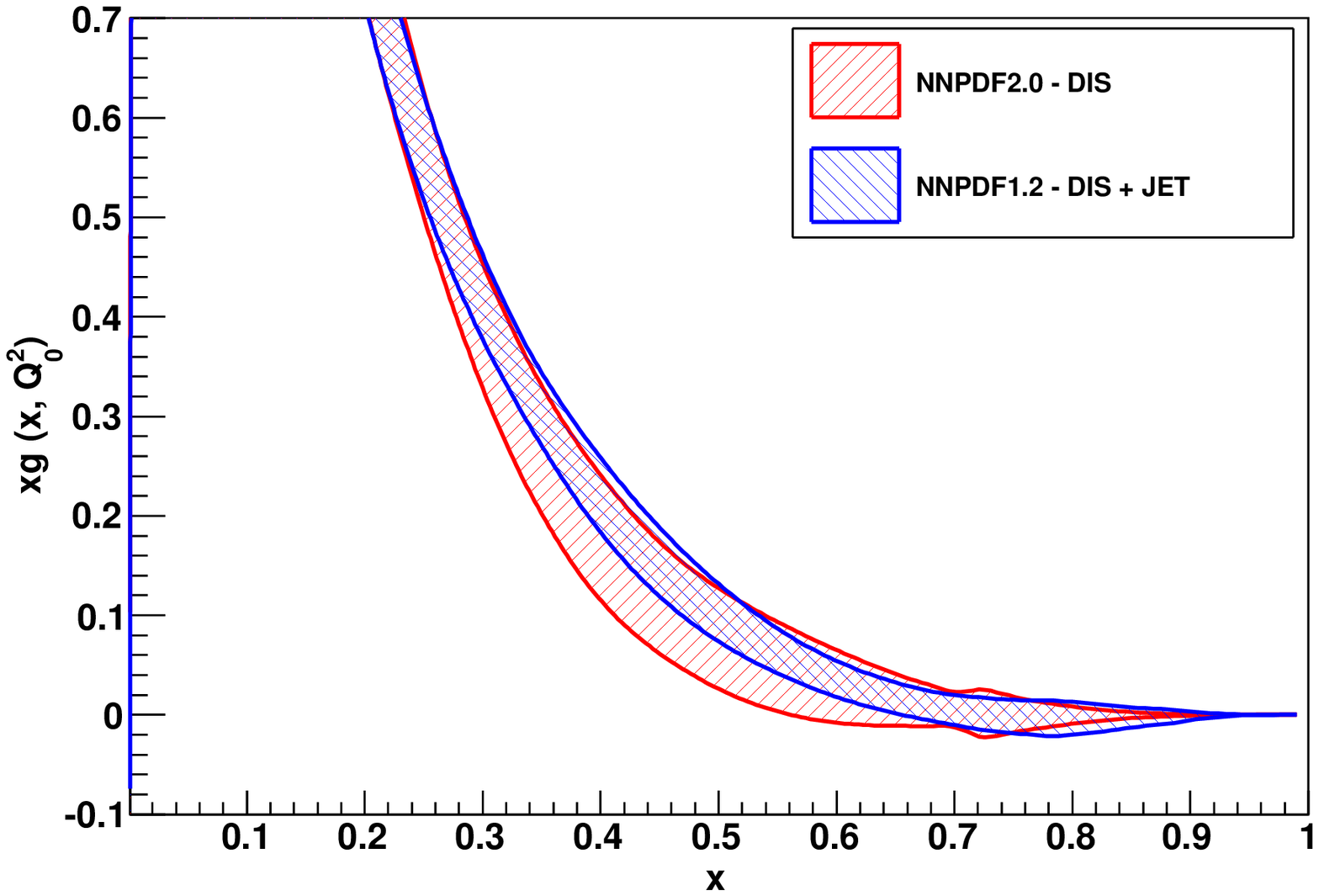}
\caption{\small Comparison between PDFs from NNPDF2.0 DIS fit 
of
  Fig.~\ref{fig:stabtab-20-dis-20-disheraold} and a fit
  in which jet data are also included (NNPDF2.0 DIS+JET) (the distances
  are shown 
in Fig.~\ref{fig:stabtab-20-dis+jet-20-dis}): the gluon at small and
large $x$ (from left
to right). 
\label{fig:pdfplots-20-dis+jet-20-dis}}
\end{center}
\end{figure}
%%%%%%%%%%%%%%%%%%%%%%%%%%%%%%%%%%%%%%%%%%%%%%%%%%%%

\item {\it Impact of Drell-Yan data}.\\\noindent\nobreak
The addition of Drell-Yan data to the 2.0-DIS+JET fit leaves the quality of
the global fit unchanged. Taken together with the previous comparison
of the 2.0-DIS and 2.0-DIS+JET data, this shows that DIS data and
hadronic data are fully compatible, and furthermore the two classes of
hadronic data included here, DY and inclusive jets, are compatible
with each other.  Minor incompatibilities only appear within each
dataset (typically due to some subset of data points or, in the case
of Drell-Yan to the CDF W asymmetry and Z rapidity distribution data).
However, the quality of the fit to Drell-Yan data was generally poor
when they were not included in the fit, due to the fact that they are
sensitive to the separation of individual flavours at
large $x$ which is only very weakly constrained by other data.

The distances between the 2.0-DIS+JET and the full NNPDF2.0 fits, displayed in
Fig.~\ref{fig:stabtab-20-20-dis+jet}, show the sizable impact of the
Drell-Yan data on
all valence--like PDF combinations
at medium and large-$x$: the triplet, the valence,
the sea asymmetry and the strangeness asymmetry. The significant
improvement in accuracy on all these PDFs is apparent in
Fig.~\ref{fig:pdfplots-20-dis+jet-20-dis}.  
The remarkable improvement in the accuracy of the determination 
of the strangeness
asymmetry $s^-(x)$ will turn out to have relevant phenomenological
implications for the so--called NuTeV
anomaly, as we discuss in Sect.~\ref{sec:pheno}.

%%%%%%%%%%%%%%%%%%%%%%%%%%%%%%%%%%%%%%%%%%%%%%%%%%%%%%%
\begin{figure}[ht]
\begin{center}
\epsfig{width=0.99\textwidth,figure=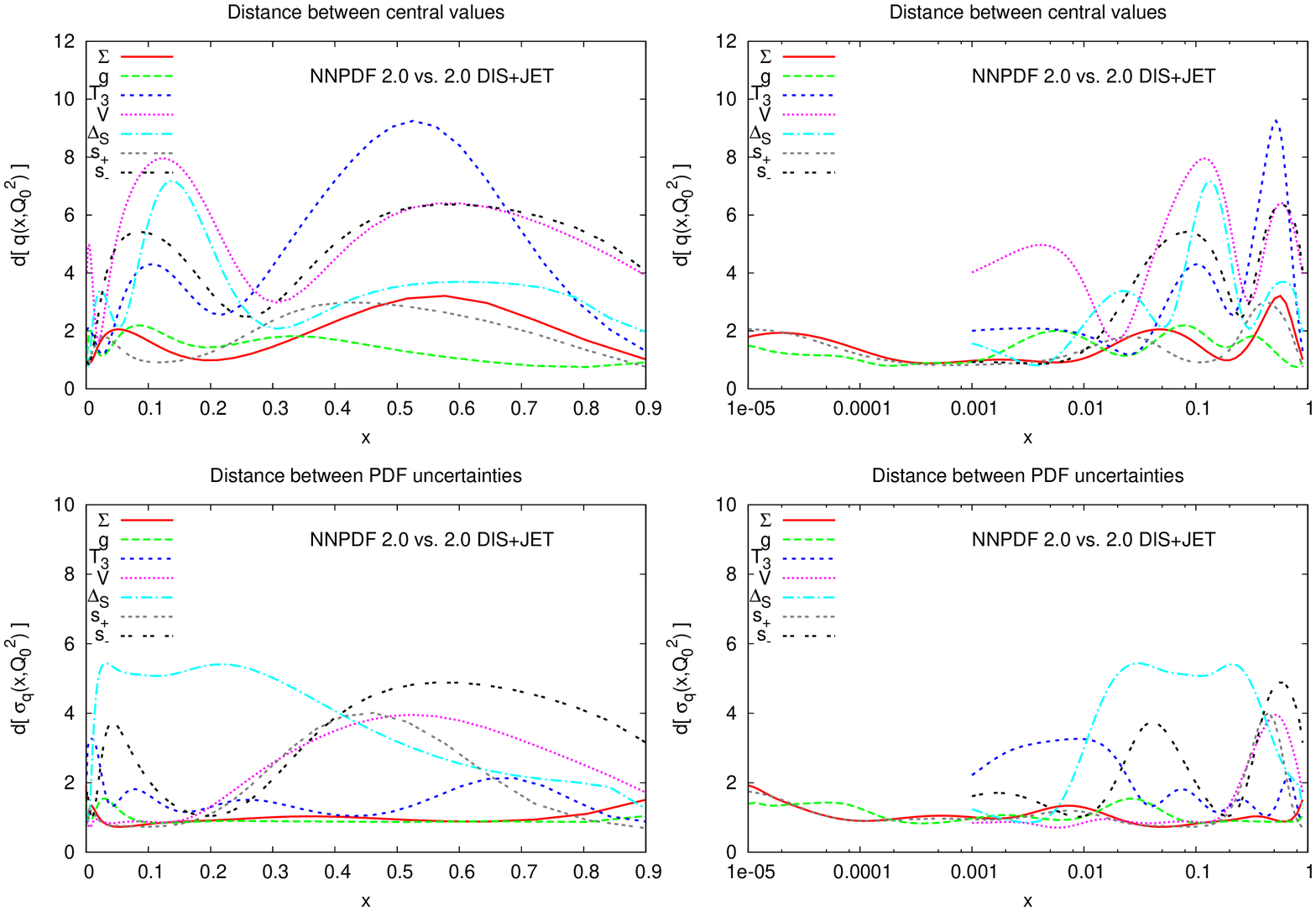}
\caption{\small Distance between the NNPDF2.0 DIS+JET fit
of
  Fig.~\ref{fig:stabtab-20-dis+jet-20-dis} 
and the reference NNPDF2.0 fit (Drell-Yan data also included).
\label{fig:stabtab-20-20-dis+jet}}
\end{center}
\end{figure}
%%%%%%%%%%%%%%%%%%%%%%%%%%%%%%%%%%%%%%%%%%%%%%%%%%%%

%%%%%%%%%%%%%%%%%%%%%%%%%%%%%%%%%%%%%%%%%%%%%%%%%%%%%%%
\begin{figure}[ht]
\begin{center}
\epsfig{width=0.48\textwidth,figure=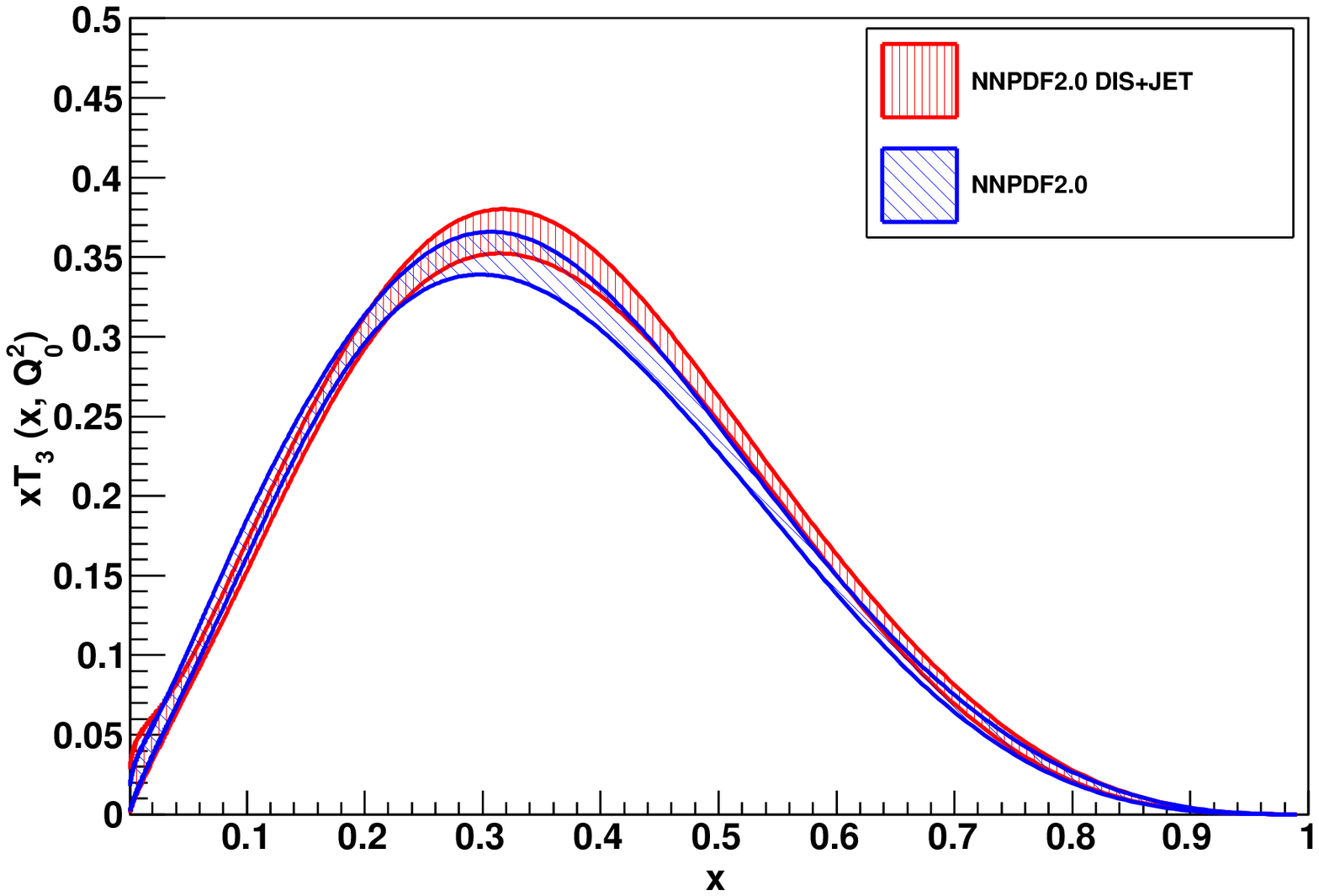}
\epsfig{width=0.48\textwidth,figure=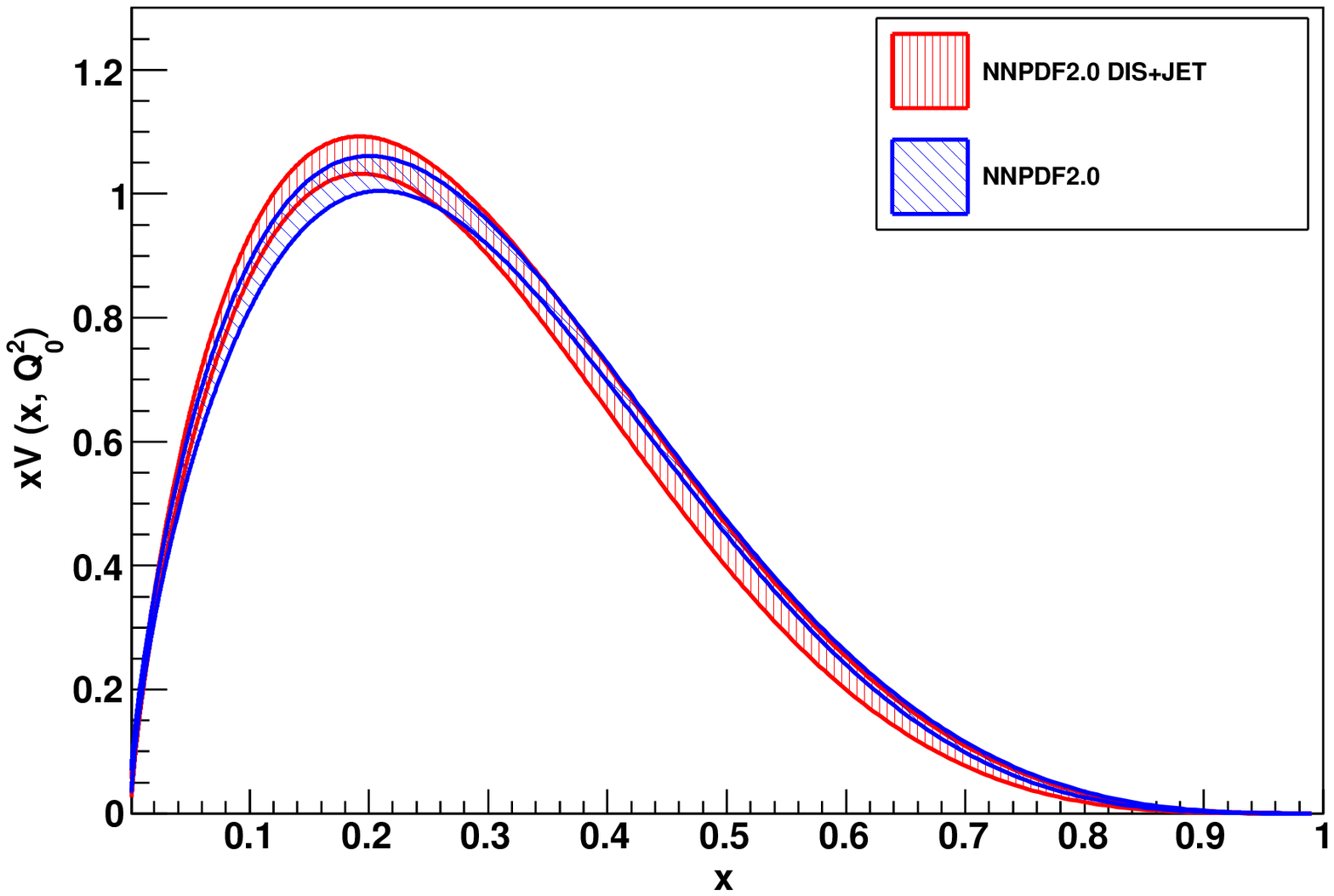}
\epsfig{width=0.48\textwidth,figure=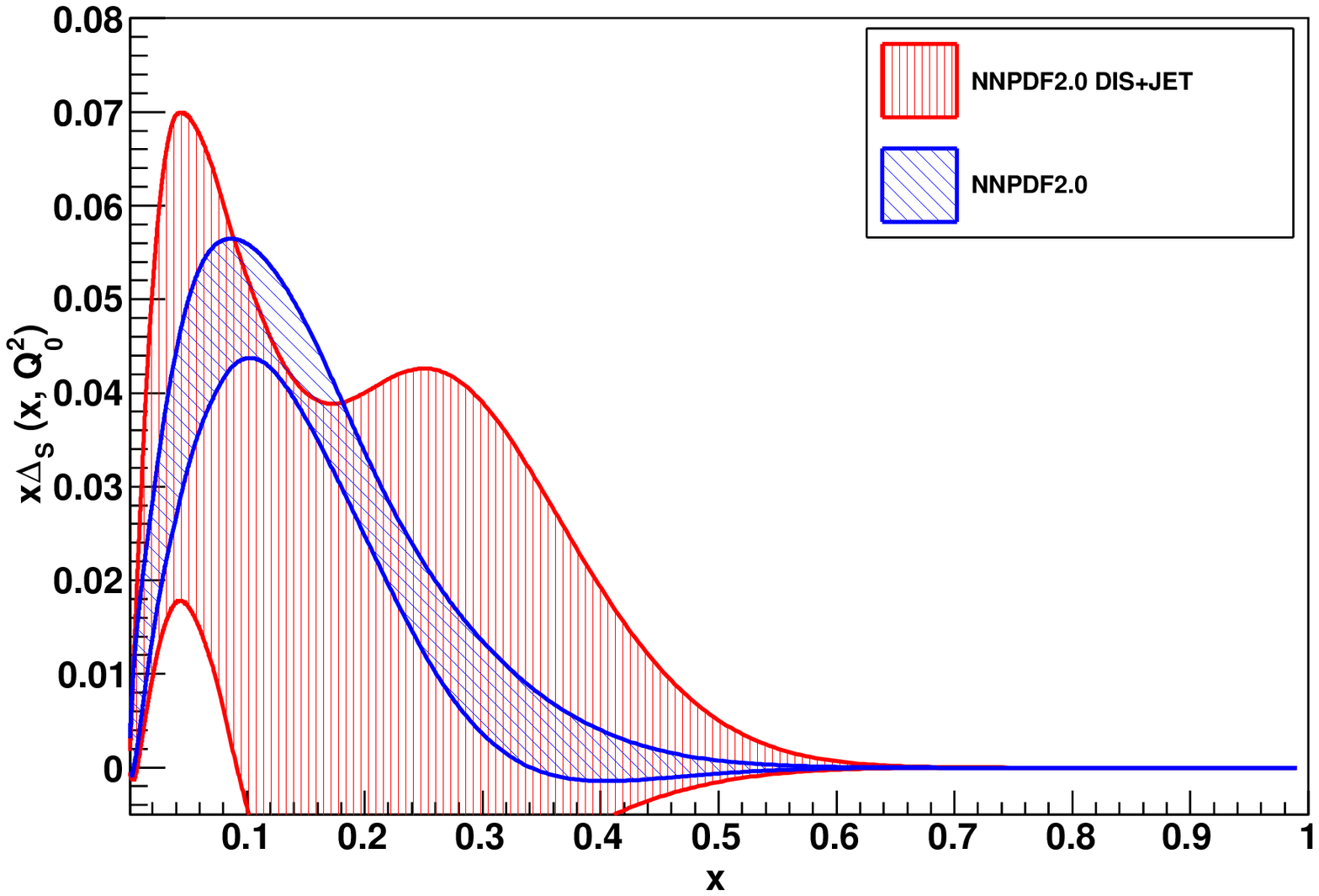}
\epsfig{width=0.48\textwidth,figure=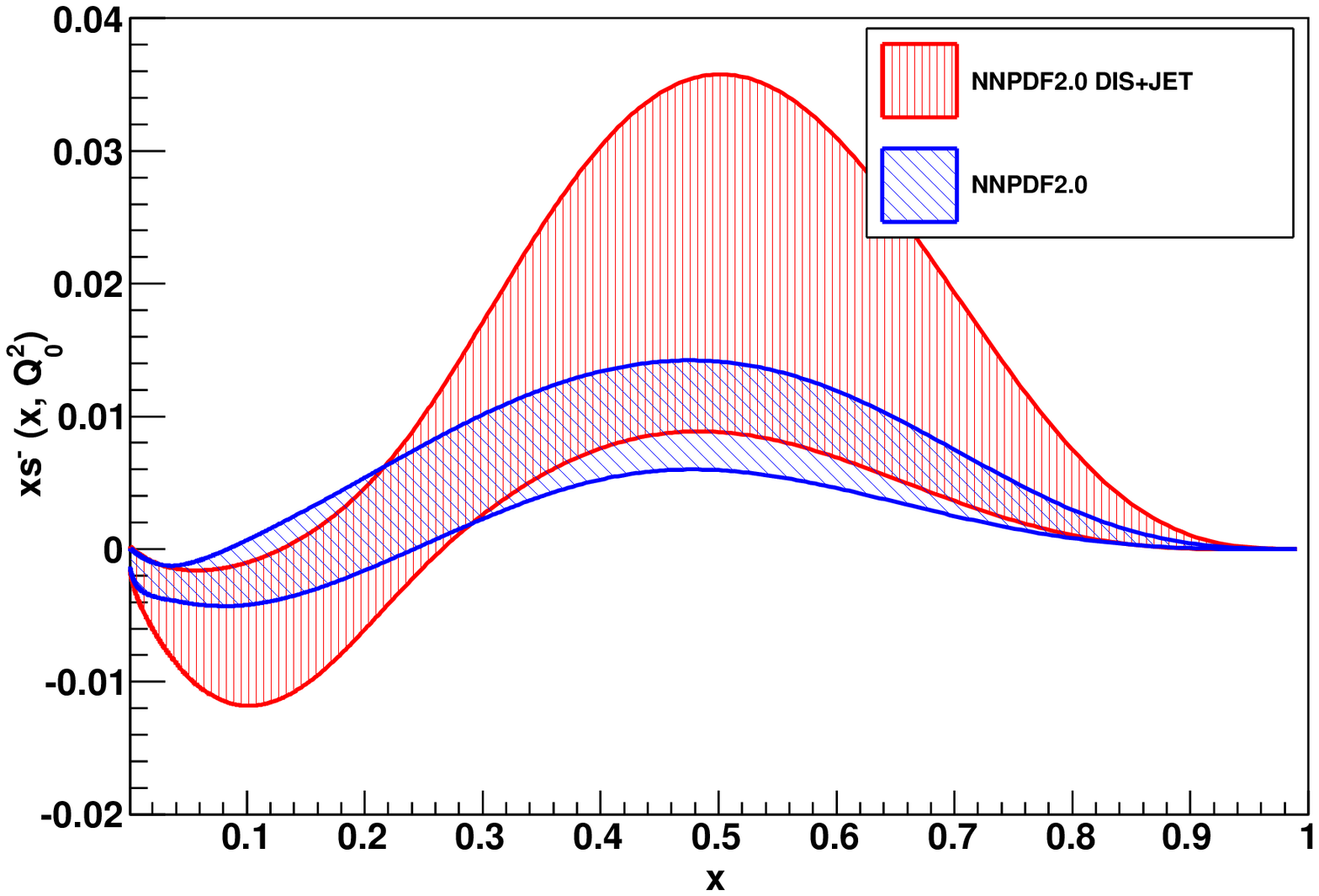}
\caption{\small  Comparison between PDFs the NNPDF2.0 DIS+JET fit
of  Fig.~\ref{fig:stabtab-20-dis+jet-20-dis} 
and the reference NNPDF2.0 fit (Drell-Yan data also included) (the distances
  are shown 
in Fig.~\ref{fig:stabtab-20-20-dis+jet}): triplet, valence, sea asymmetry and
strange valence (from left
to right and from top to bottom). 
\label{fig:pdfplots-20-20-dis+jet}}
\end{center}
\end{figure}
%%%%%%%%%%%%%%%%%%%%%%%%%%%%%%%%%%%%%%%%%%%%%%%%%%%%

\end{enumerate}

Finally, we have produced two further PDF sets: one with the full NPDF2.0
dataset, but with HERA-I combined DIS
data replaced by the previous separate H1 and ZEUS data; and the other
with DIS+DY data only. In both cases, we see that the impact of the new
data is independent of the dataset to which they are added: so for
instance the improvement in accuracy in the valence sector due to DY
data is independent of their being added to a dataset that does or
does not contain jet data. 

The main conclusion of this analysis is that we see no sign of tension
between datasets. To understand this, consider what would happen if,  
say, jet data were incompatible
with Drell-Yan data: then,  we should see a daterioration of the
quality of the fit to Drell-Yan when jets are included, and also we
should see that the impact of jet data is bigger when Drell-Yan data
are not included and more moderate when they are included. None of
these effects is observed, for any of the combinations that have been
tried here. Deterioration of the fit quality to each individual data
set upon global fitting has been discussed in detailed in
Ref.~\cite{Pumplin:2009sc}: whereas small data incompatibilities may
only be revealed by the more sensitive method used in this reference,
we see no evidence for the sizable incompatibilties found there.

%%%%%%%%%%
%%%%%%%%%%%

\clearpage
\subsection{Positivity constraints}
\label{sec:res:positivity}

As discussed in Sect.~\ref{sec-minim}, positivity of physical
observables has been imposed, in particular for the
longitudinal structure function $F_L(x,Q^2)$ and for
 the dimuon cross section through a Lagrange multiplier 
Eq.~(\ref{eq:poscon}).
In order to assess quantitatively the effect of the positivity
constraints, 
we have repeated the NNPDF2.0 parton determination without imposing
positivity, i.e.
setting $\lambda_{\rm pos}=0$ in Eq.~(\ref{eq:poscon}).

\begin{figure}[ht]
\begin{center}
\epsfig{width=0.49\textwidth,figure=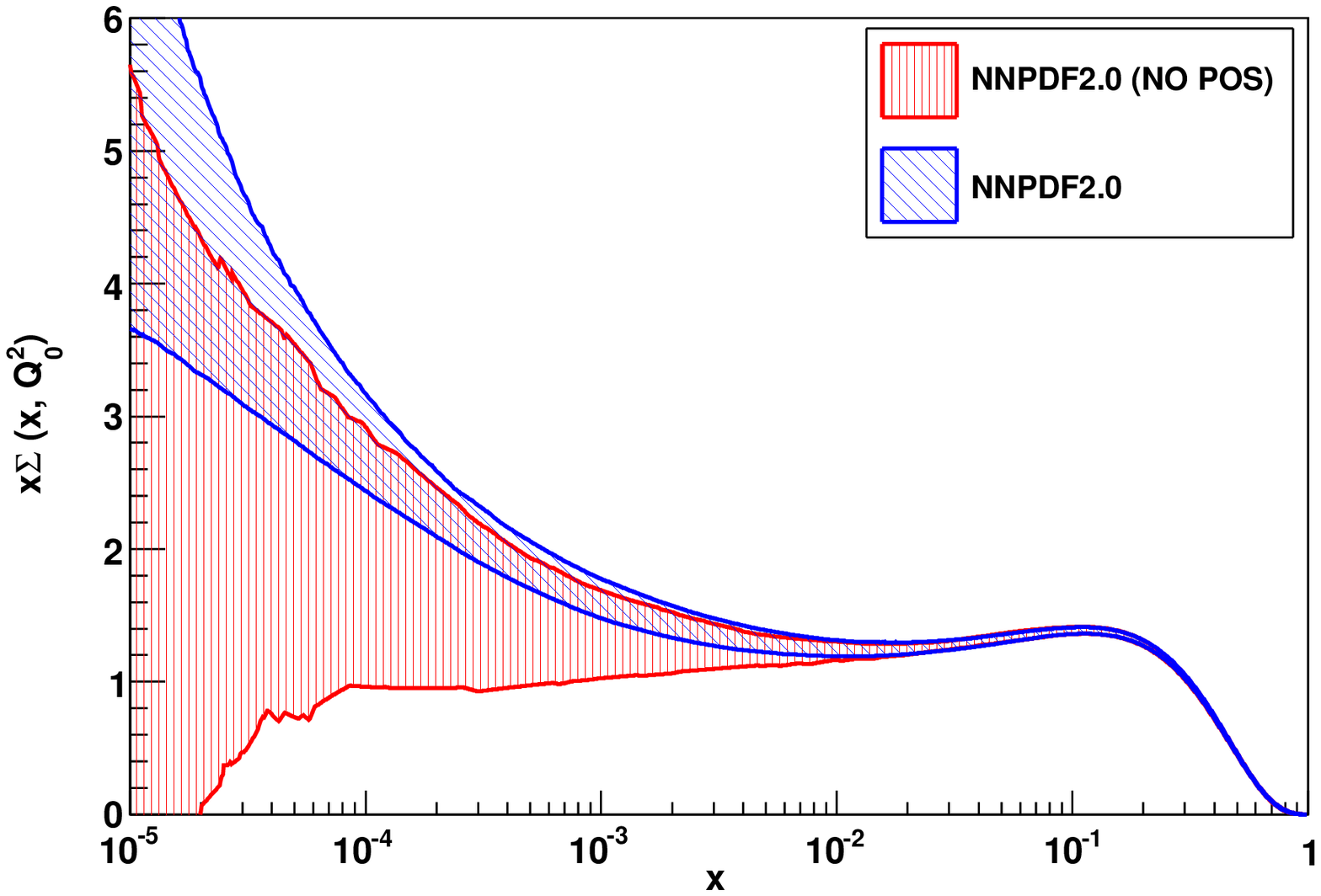}
\epsfig{width=0.49\textwidth,figure=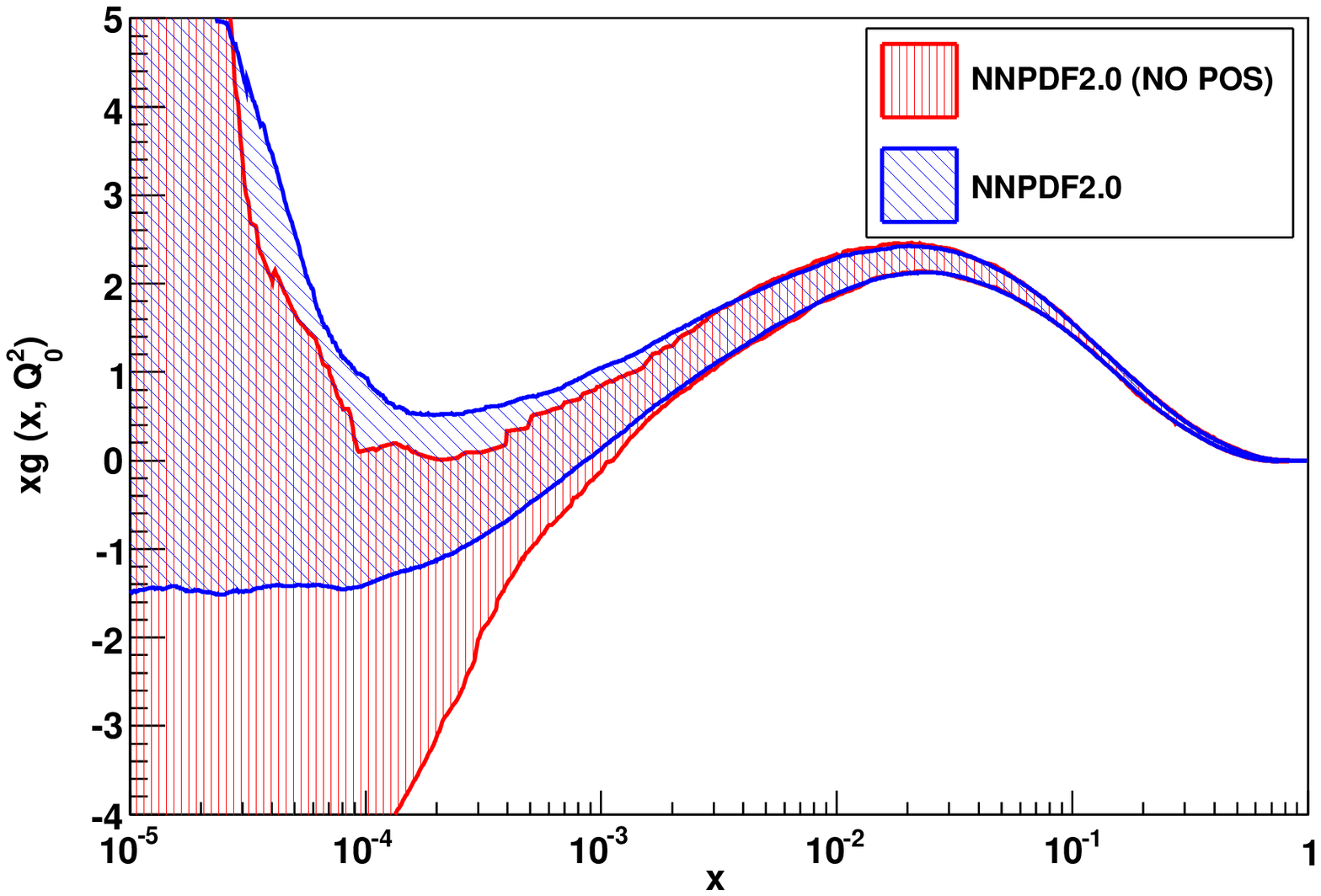}
\epsfig{width=0.49\textwidth,figure=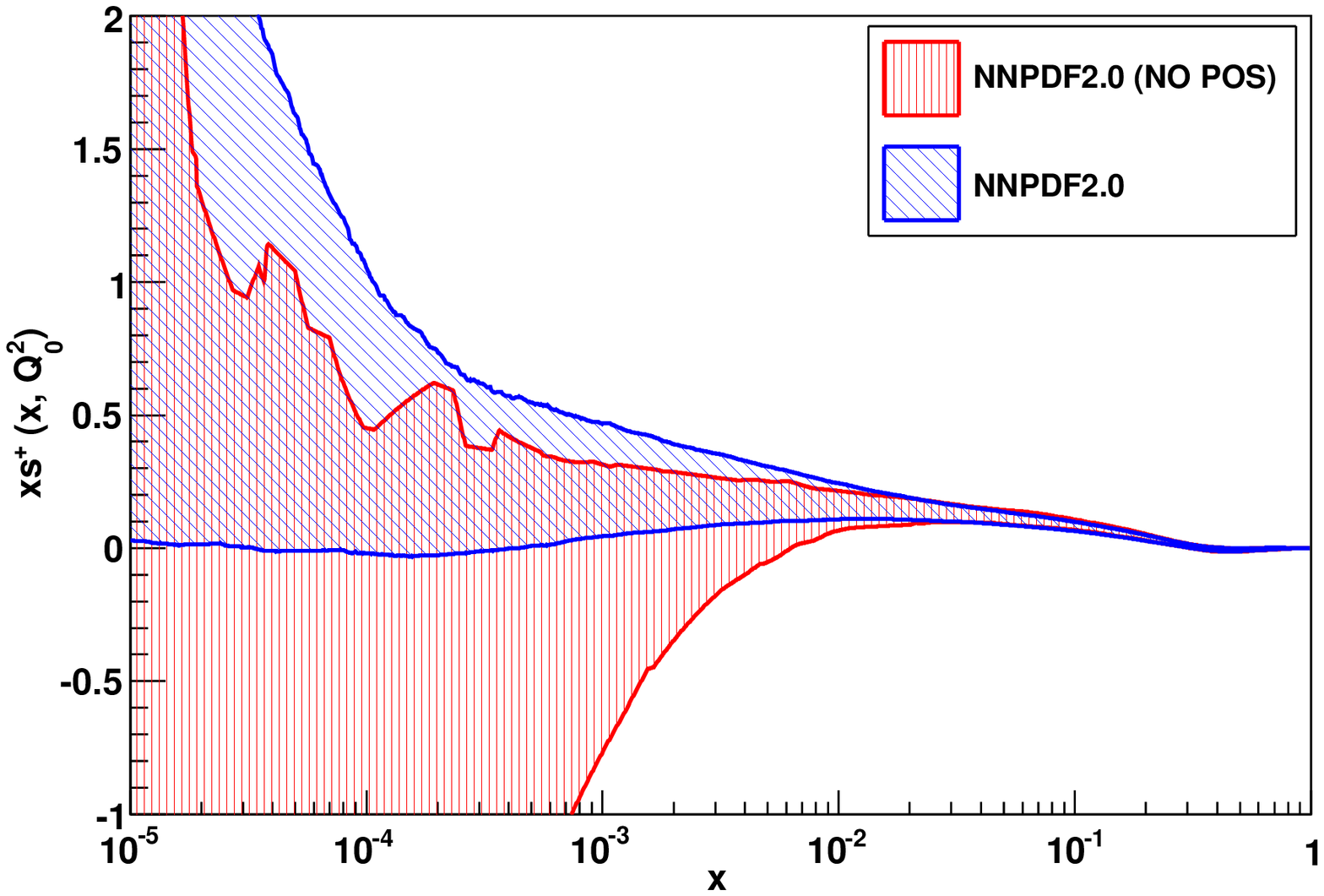}
\epsfig{width=0.49\textwidth,figure=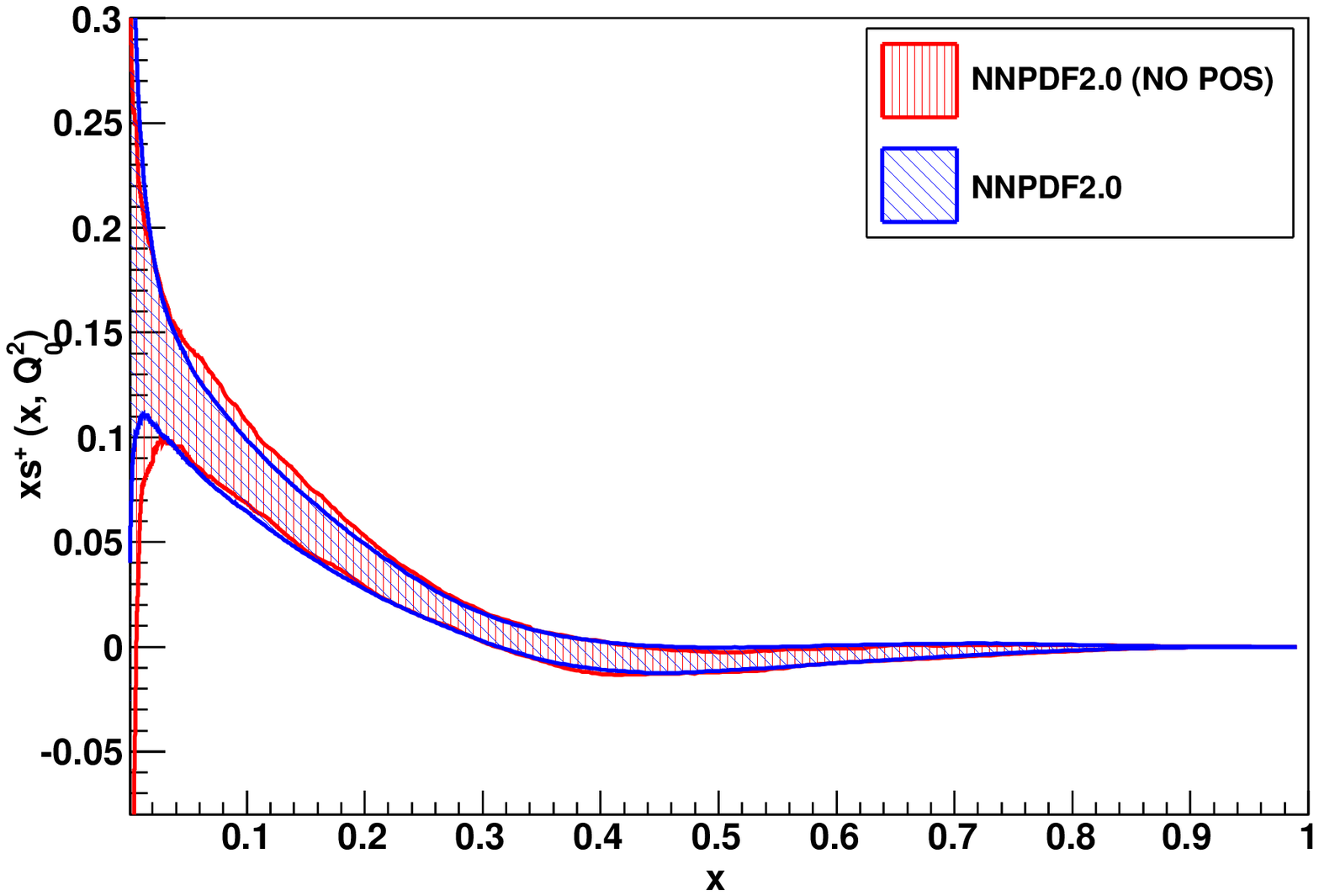}
\caption{\small NNPDF2.0 PDFs with and without positivity constraints:
  singlet, gluon and total strangeness at small $x$ and total
  strangeness at large $x$.
All uncertainty bands are 
determined as 68\% confidence levels. PDFs not shown here are
not affected by the positivity constraints.
\label{fig:pdfnopos}} 
\end{center}
\end{figure}

In Fig.~\ref{fig:pdfnopos} PDFs with uncertainties determined as 68\%
confidence levels with and without positivity constraints are
compared. As discussed in Sect.~\ref{sec:cl}, it is important to
perform the comparison with uncertainties determined as confidence
levels rather than standard deviations, because imposing positivity
can lead to deviations from gaussian behaviour. Clearly positivity
of $F_L(x,Q^2)$  leads to substantial  uncertainty reduction in the
small-$x$ gluon. Note that  there is nevertheless a kinematic region in which
the gluon goes negative by a small amount, though $F_L$ remains
positive.
Also, removing positivity of the dimuon cross section would lead
to a much softer strange sea at small-$x$ with rather larger
uncertainties. This in turn leads to a softer small-$x$ singlet, also
with  larger uncertainties. This is due to the fact that below $x\lsim
0.01$, where no neutrino data are available, 
positivity
is the only constraint on the total strangeness $s^+$.

 Finally, it is
interesting to observe that positivity also has the effect of
stabilizing the replica sample: indeed, the 68\% confidence levels
computed without positivity display some visible fluctuations which
would only be smoothened out by using a significantly wider replica
sample. These fluctuations are absent when positivity is imposed,
meaning that such wide fluctuations in individual replicas are
removed by the constraint.

%%%%%%%%%%%%%%%%%%%%%%%%%%%%%%%%%%%%%%%%%%%%%%%%%%%%%%%%%%%%%%%%%%%
%%%%%%%%%%%%%%%%%%%%%%%%%%%%%%%%%%%%%%%%%%%%%%%%%%%%%%%%%%%%%

\clearpage

\subsection{Dependence on $\alpha_s$}

The central NNPDF2.0 fit has been performed with $\alpha_s(M_Z)=0.119$ in
order to ease comparison with the previous NNPDF1.0 and NNPDF1.2
fits, even though the current~\cite{Amsler:2008zzb} 
PDG average is $\alpha_s(M_Z)=0.118\pm0.002$. In order to study the
dependence of our results on this choice, we have repeated the fit
with $\alpha_s$ varied by one and two standard deviations about this
value, i.e. we have produced PDF sets with
$\alpha_s(M_Z)=0.115,\;0.117,\;0.121$ and $0.123$. 

In the previous  NNPDF1.0 and NNPDF1.2 parton sets the dependence of
PDFs on 
$\alpha_s$ 
was found~\cite{Ball:2008by,mariani,Higgs,LH} to be noticeable but weak: when
$\alpha_s$ was varied  by $\Delta\alpha_s=\pm0.002$ most PDFs
 were found to be statistically 
indistinguishable from those obtained 
with $\alpha_s$ fixed to its central value (i.e. to be at a distance
$d\approx 1$ from them). The gluon (and to a lesser extent the
singlet PDF) was found to change in a statistically
significant way, but still within its
uncertainty band when $\alpha_s$ was varied in this range.

\begin{figure}[ht]
\begin{center}
\epsfig{width=0.45\textwidth,figure=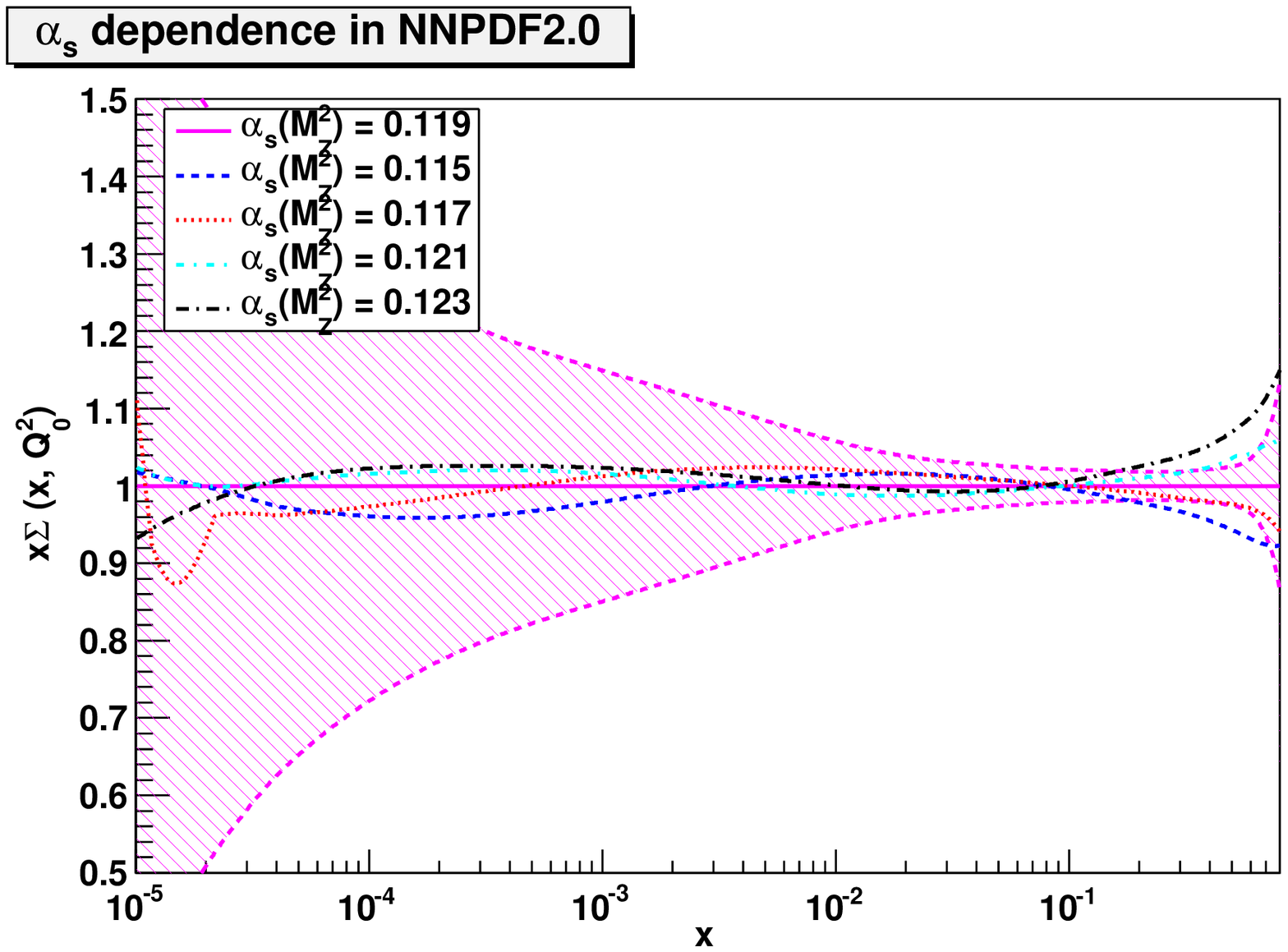}
\epsfig{width=0.45\textwidth,figure=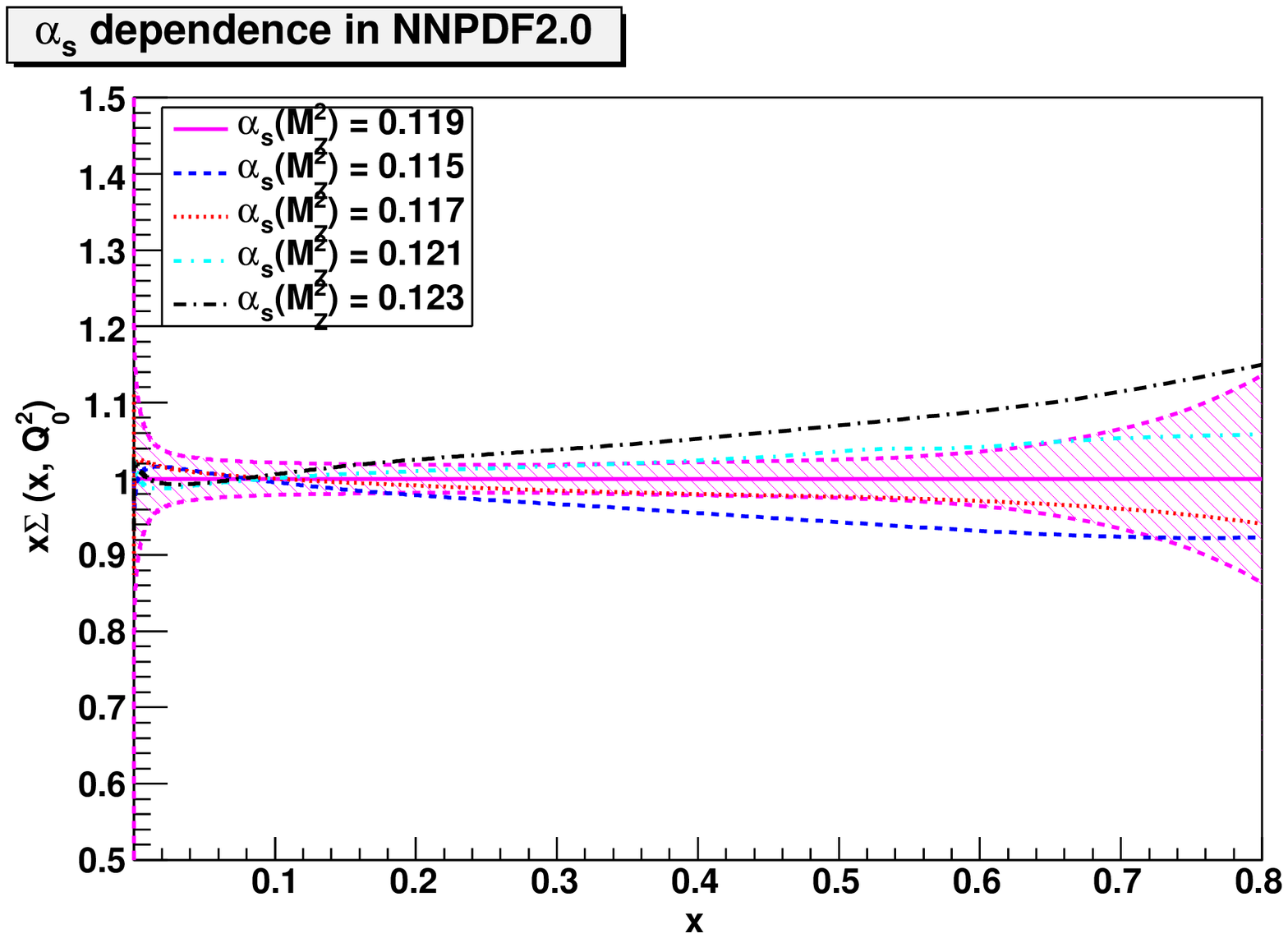}
\epsfig{width=0.45\textwidth,figure=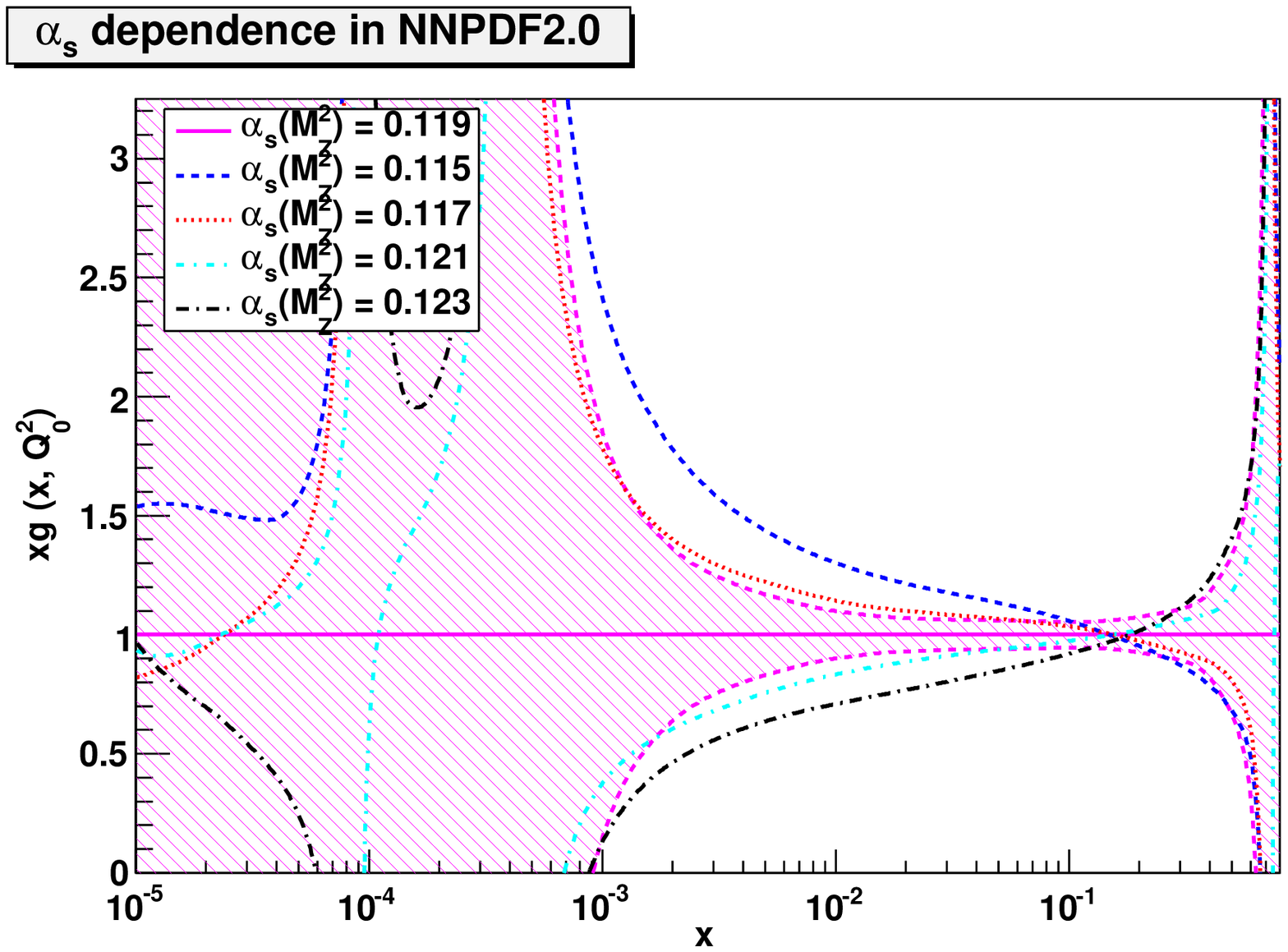}
\epsfig{width=0.45\textwidth,figure=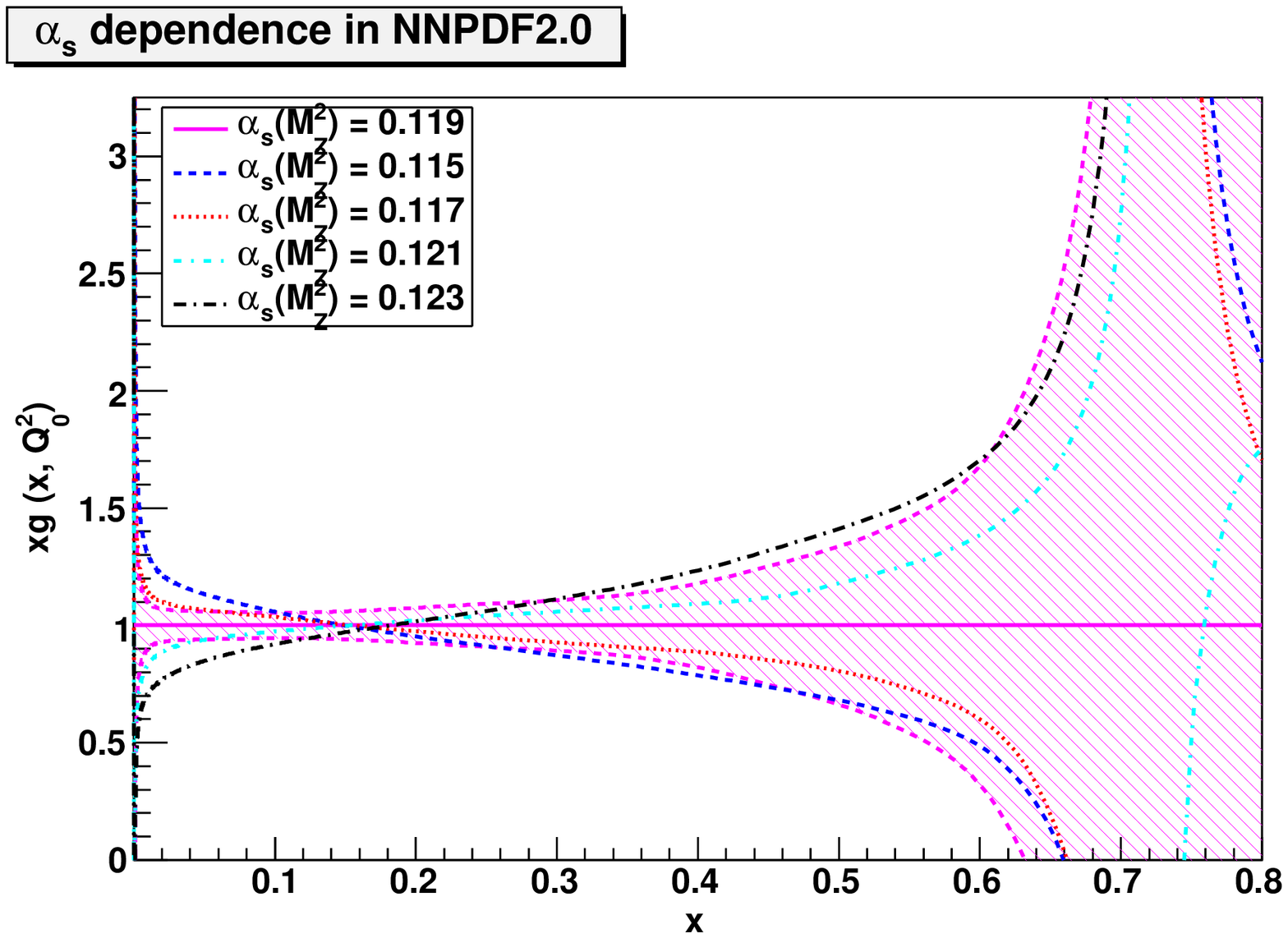}
\epsfig{width=0.45\textwidth,figure=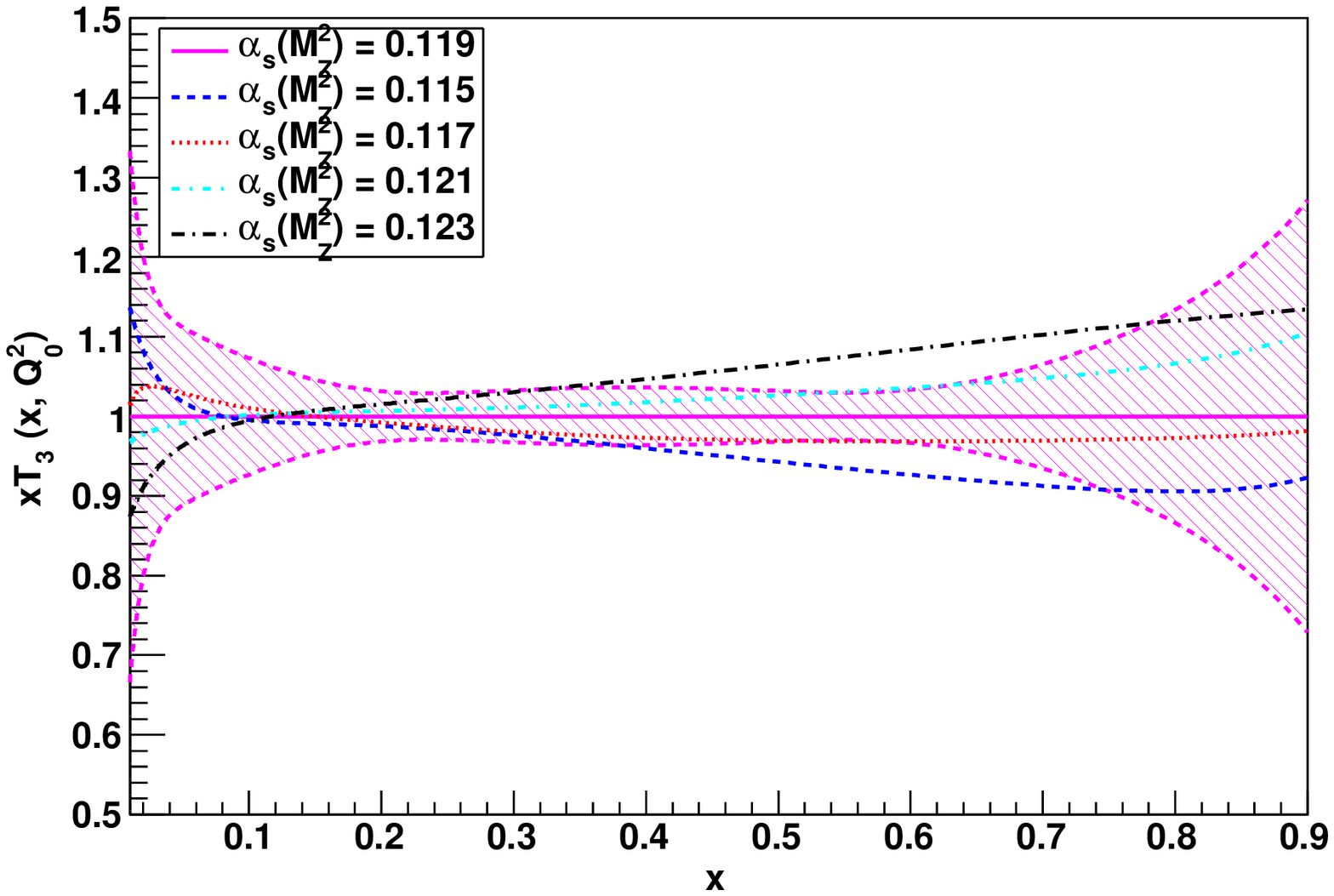}
\epsfig{width=0.45\textwidth,figure=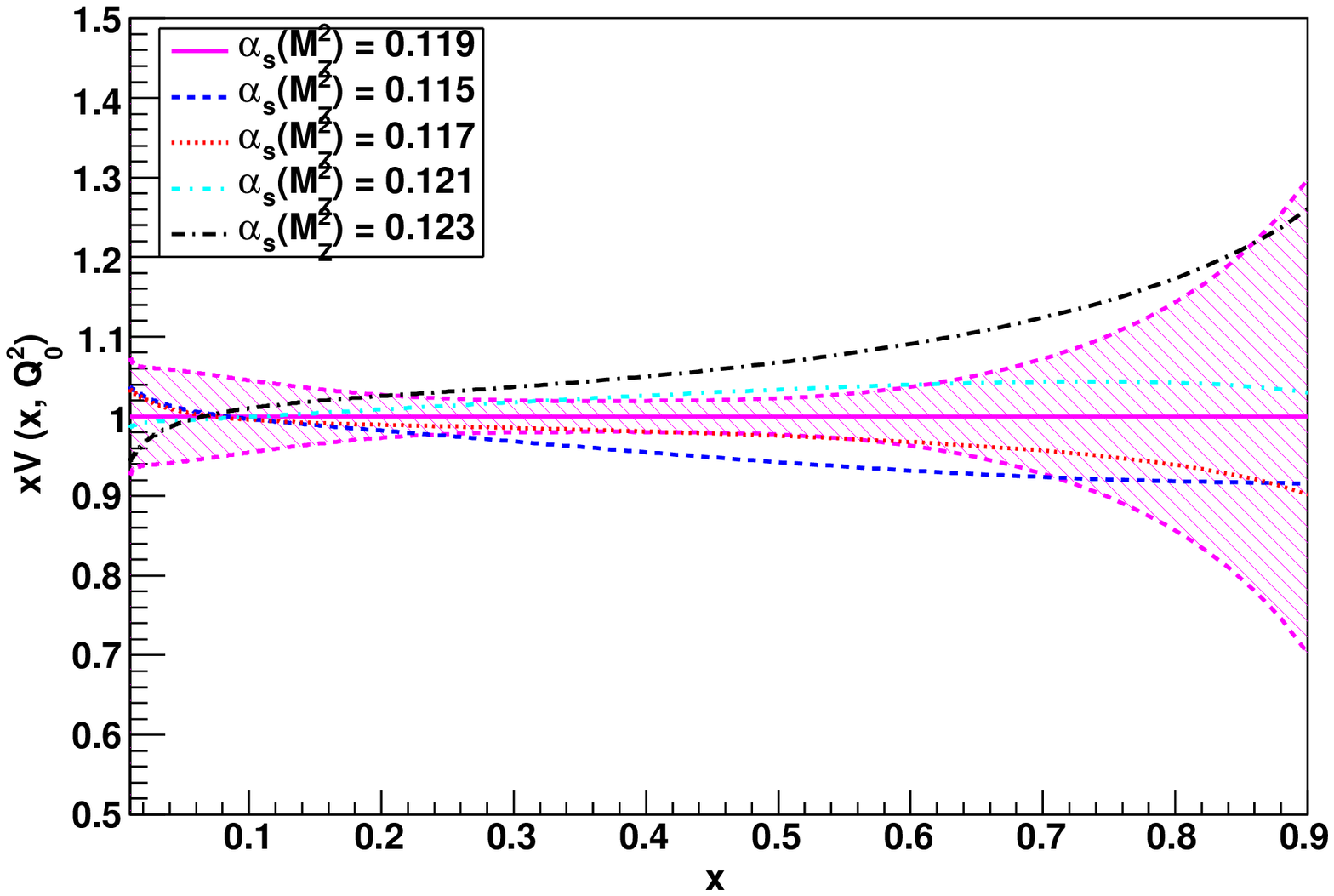}
\epsfig{width=0.45\textwidth,figure=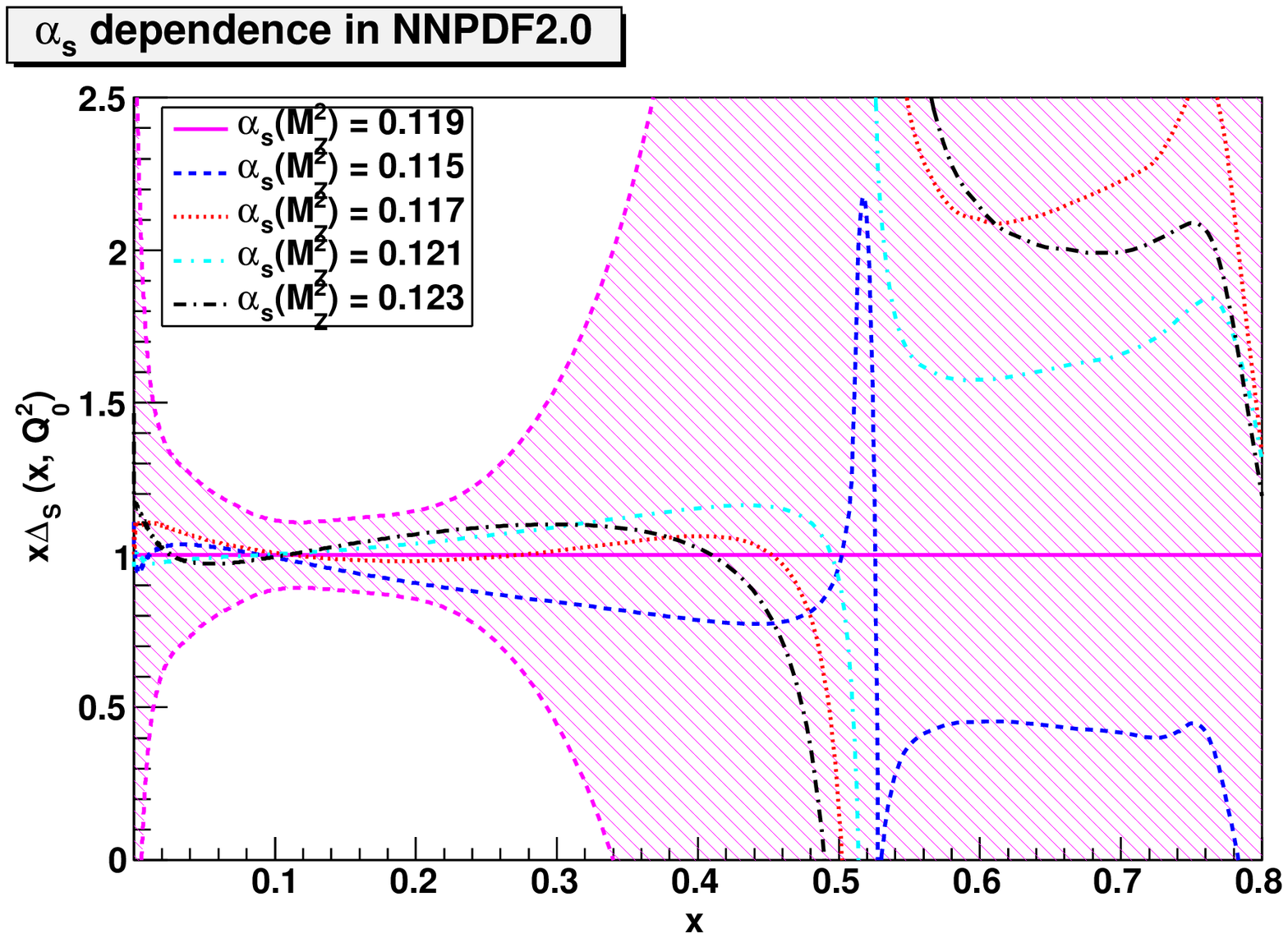}
\epsfig{width=0.45\textwidth,figure=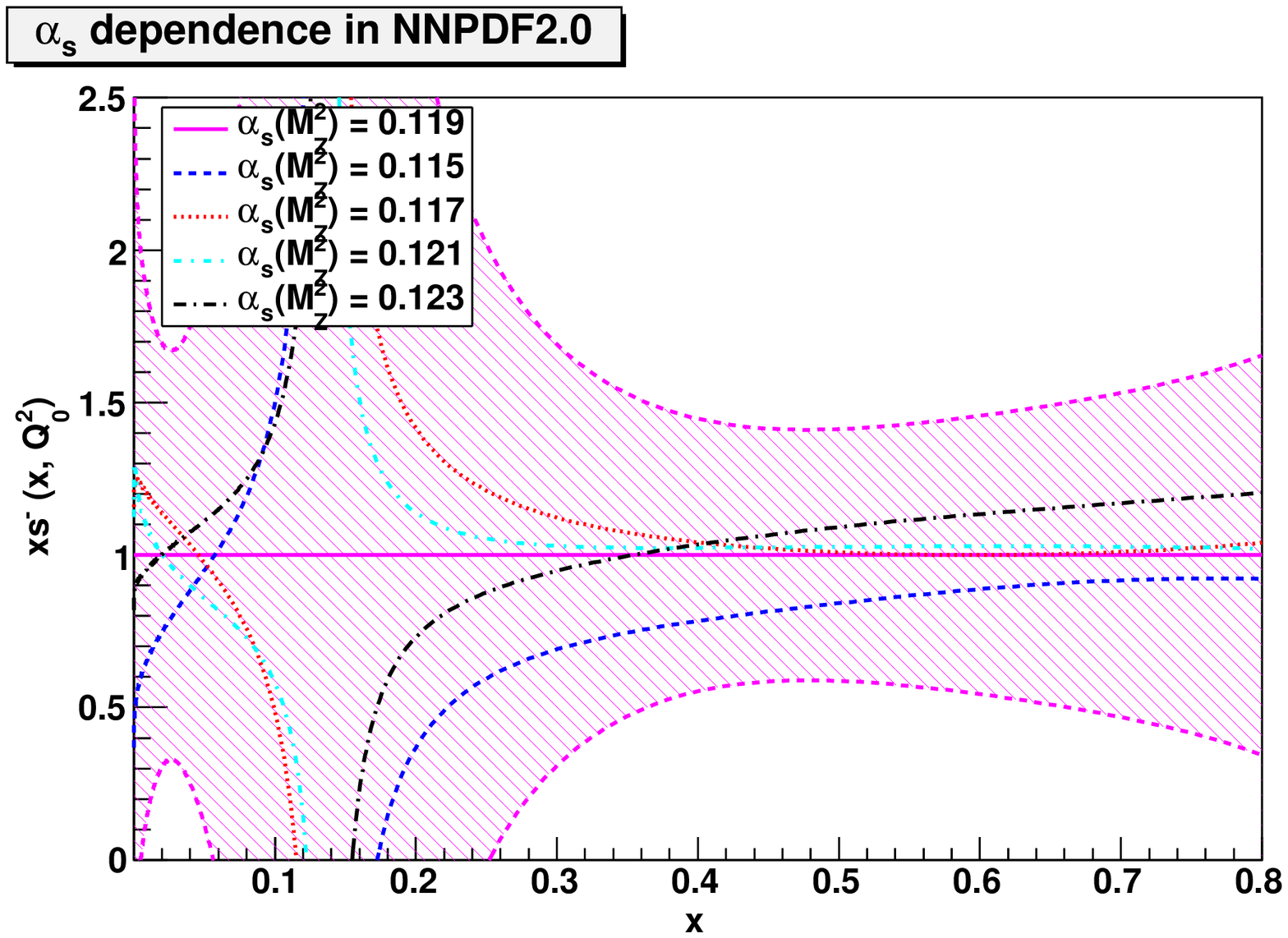}
\caption{\small Ratios of PDFs with $\alpha_s$ varied in the range
$0.115\le\alpha_s \le  0.123$ to the central NNPDF2.0
  determination, compared to the PDF uncertainty band:
 the singlet at small and large $x$, the gluon at small and large $x$
and the triplet, valence, sea asymmetry and strange valence (from top
to bottom).
\label{fig:asvariation}} 
\end{center}
\end{figure}

The dependence of NNPDF2.0 PDFs on $\alpha_s$ is shown
in Fig.~\ref{fig:asvariation}, where the ratio of the four $\alpha_s$ 
PDF sets to the central set are shown for all PDFs except the total
strangeness $s^+$ which is found not to vary significantly.
Clearly, all PDFs are still within the central uncertainty band when
$\Delta\alpha_s=\pm0.002$. However, there appears to be now somewhat
greater sensitivity to $\alpha_s$. Firstly, now not only the gluon but
also the triplet, singlet and valence, when $\alpha_s$ is varied in
the range $\Delta\alpha_s=\pm0.002$, move close to
the edge of the one--$\sigma$ range for the central PDF. This 
corresponds to a distance $d\approx 7$, well above the threshold of
statistical significance, and even for the gluon it is a somewhat
larger variation than  observed in NNPDF1.2. 
Furthermore, the triplet, which as discussed in
Sect.~\ref{sec:res:pdfs} is now determined very accurately, 
appears to be  as sensitive as the gluon to the value
of $\alpha_s$.
The increased sensitivity of quark distributions to the value of
$\alpha_s$  is likely a consequence of the inclusion of Drell-Yan data,
which undergo large NLO corrections and are thus sensitive to
$\alpha_s$.

This increased sensitivity with respect to $\alpha_s$ suggests that
the strong coupling could be determined from the global
PDF analysis with competitive accuracy, following
a procedure similar to that used to obtain the
accurate determination of the CKM matrix element $|V_{\rm cs}|$
of Ref.~\cite{Ball:2009mk}. 

%%%%%%%%%%%%%%%%%%%%%%%%%
%%%%%%%%%%%%%%%%%%%%%%%%%%%%%%%

%% -------------------------------------------------------

\section{Phenomenological implications}
\label{sec:pheno}

A full phenomenological study of the implications of 
NNPDF2.0 PDFs  is beyond the scope
of this paper. 
In this section we present some preliminary investigations:
we compare to the experimental data which has been included in the fit, 
then we discuss the implications for the proton strangeness and
in particular to the NuTeV anomaly, and finally we present
predictions for some LHC standard candles.

\subsection{Comparison to experimental data}

\begin{figure}[ht!]
\begin{center}
\epsfig{width=0.49\textwidth,figure=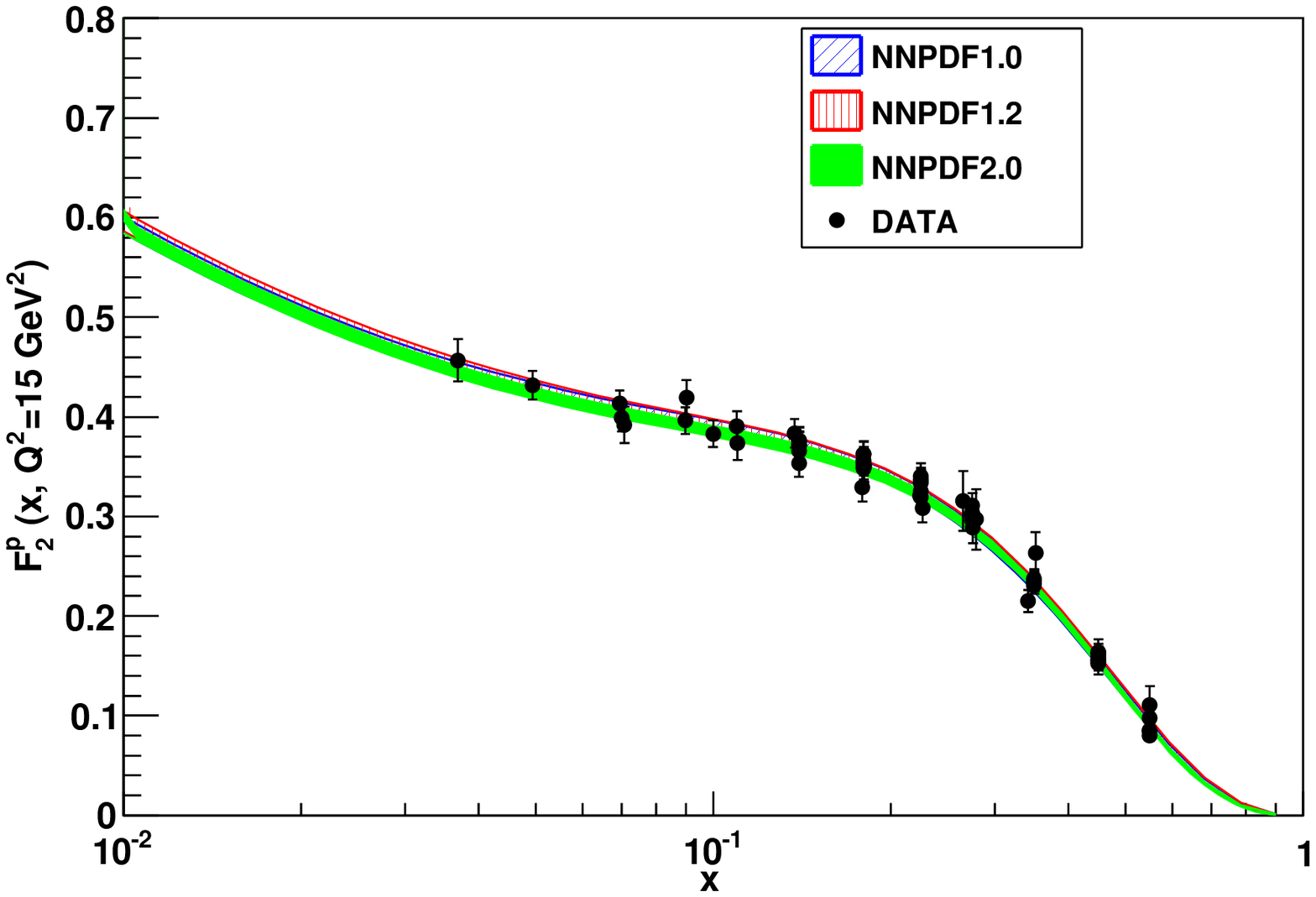}
\epsfig{width=0.49\textwidth,figure=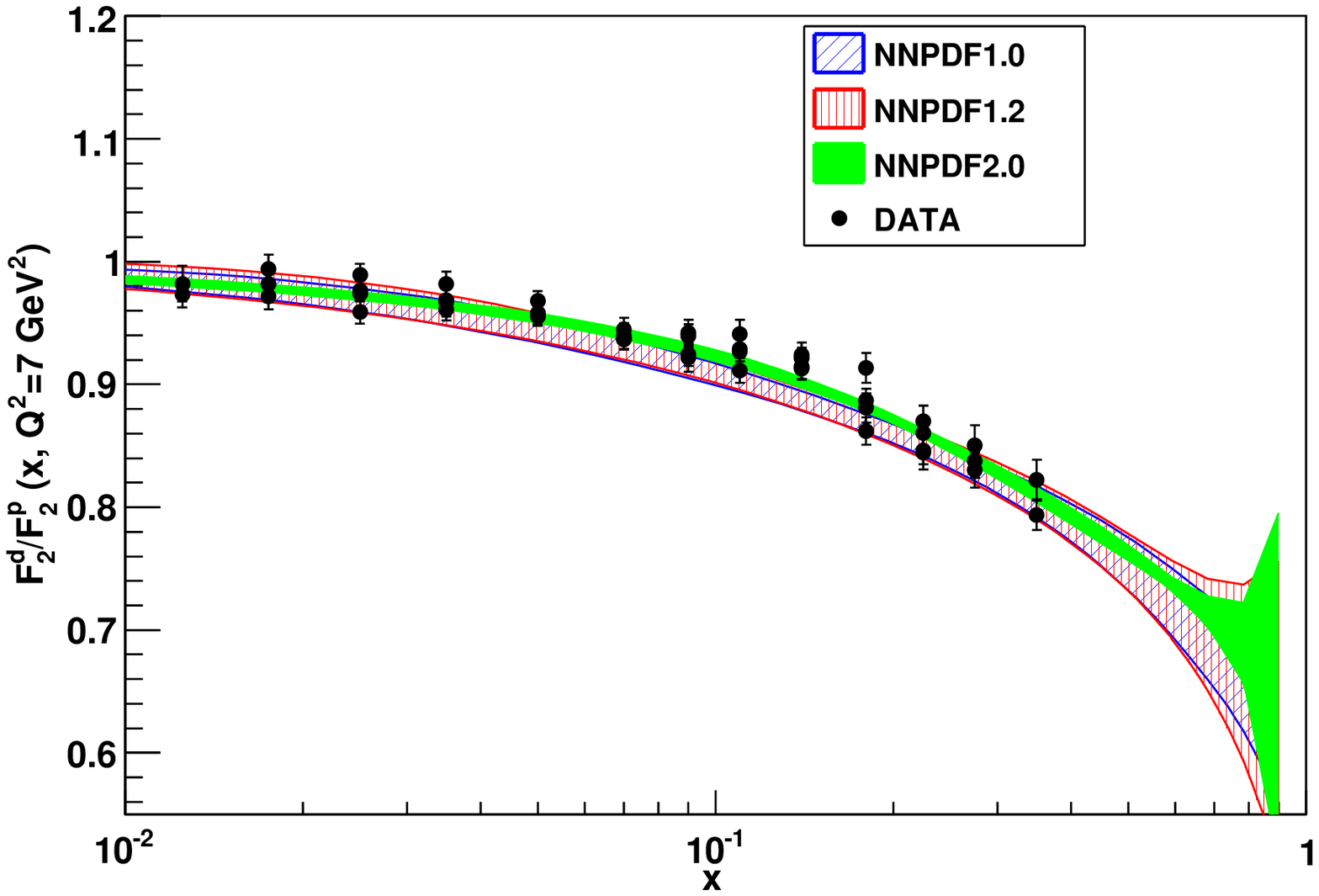}
\epsfig{width=0.49\textwidth,figure=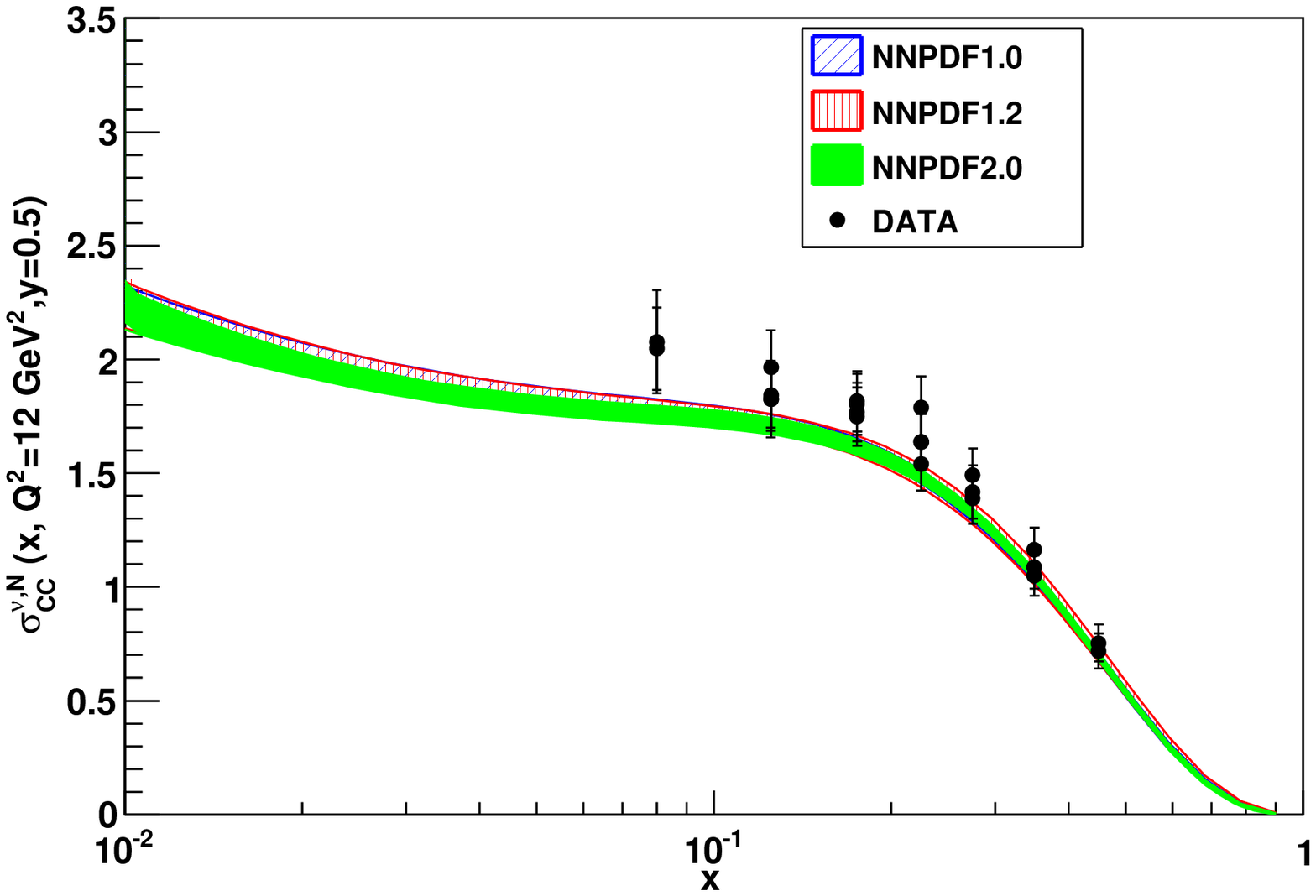}
\epsfig{width=0.49\textwidth,figure=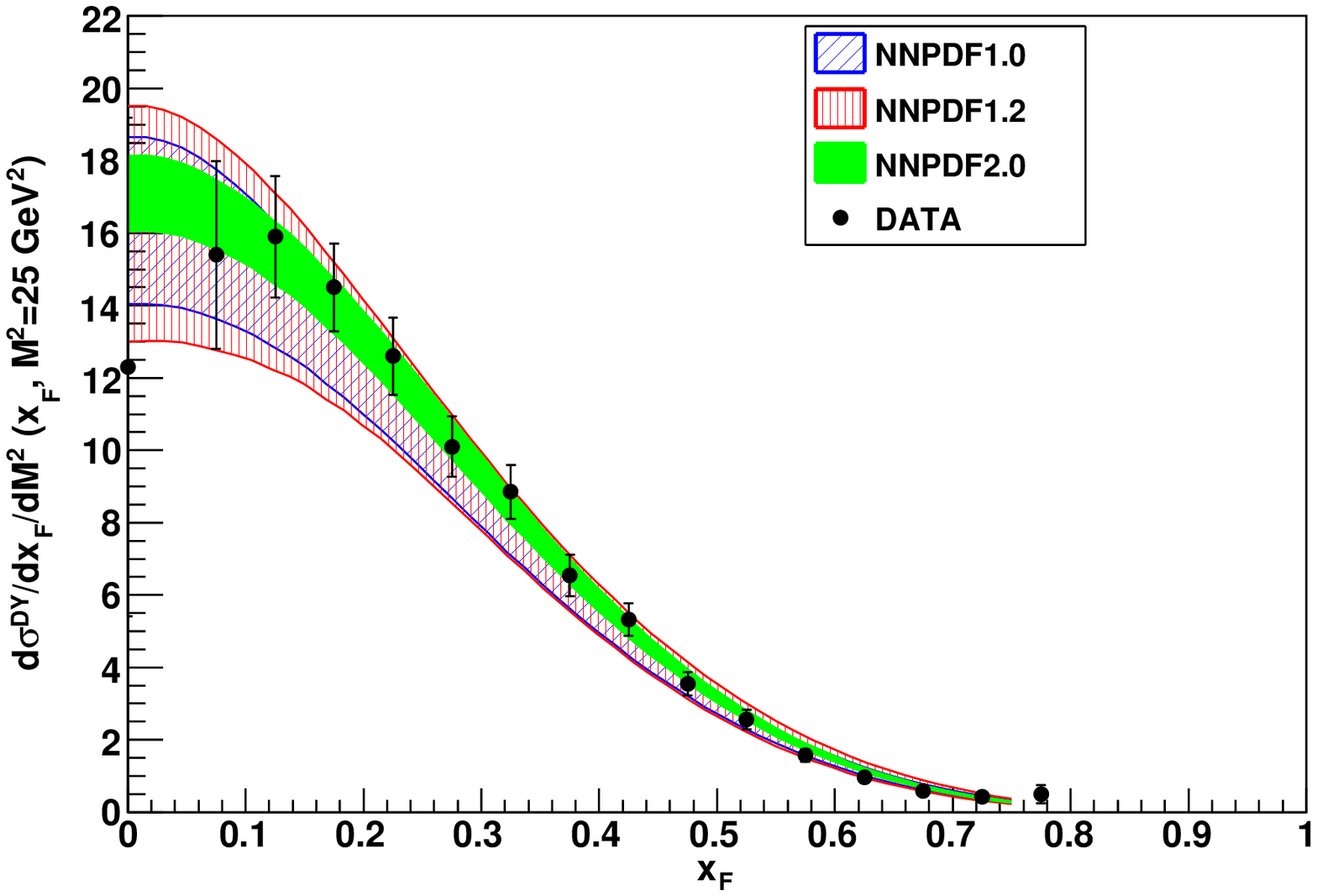}
\epsfig{width=0.49\textwidth,figure=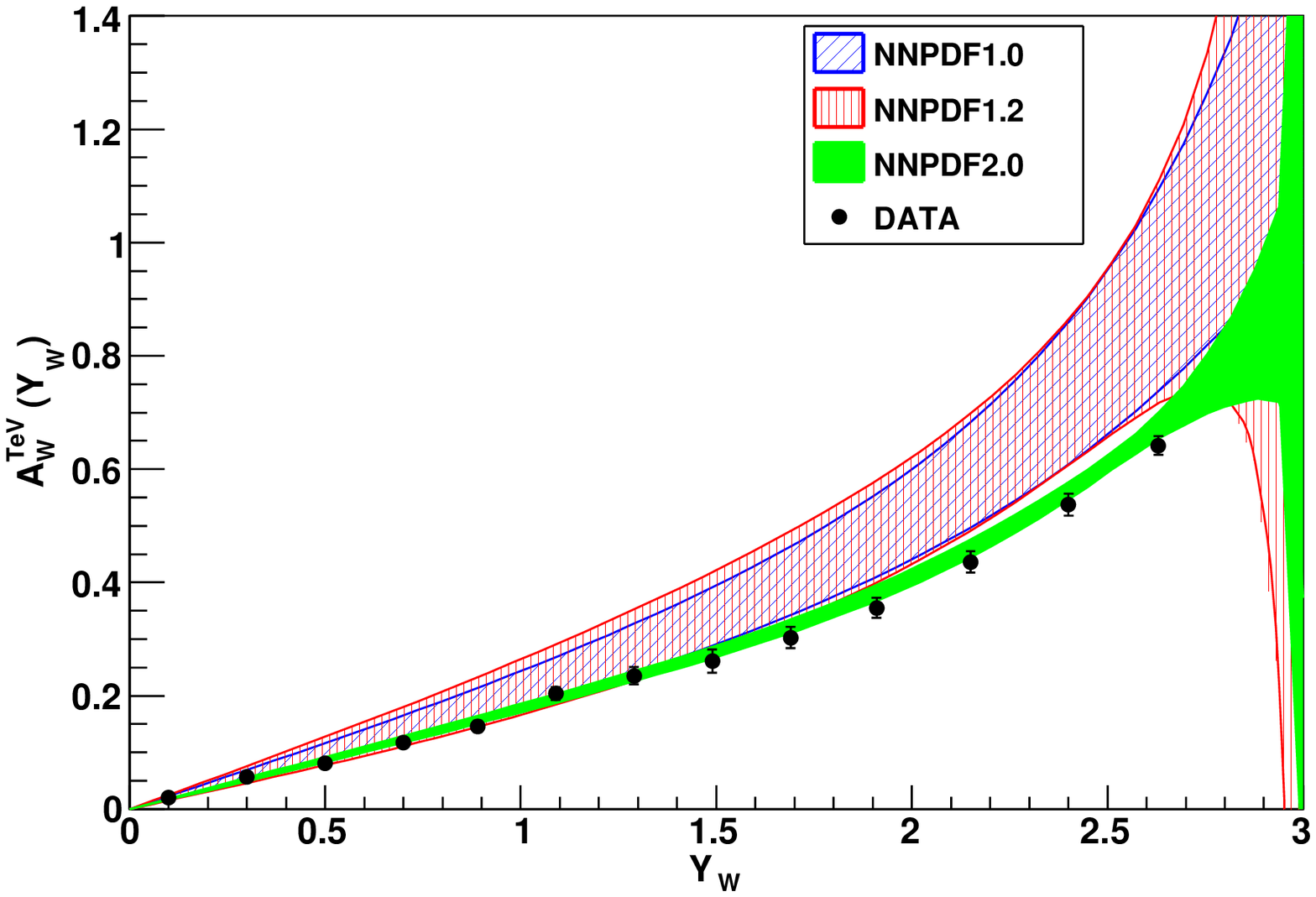}
\epsfig{width=0.49\textwidth,figure=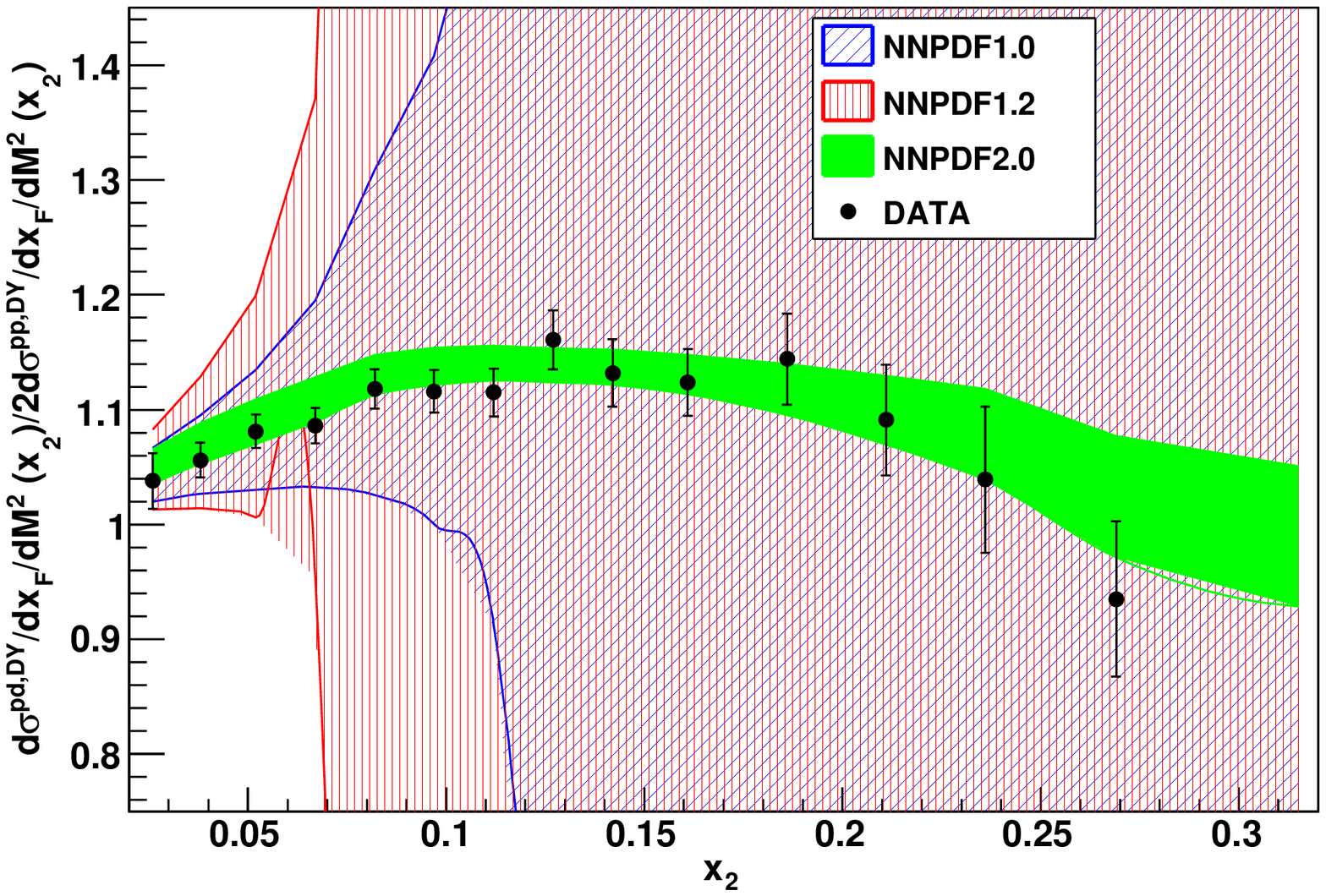}
\caption{ \small Comparison between data and NLO predictions obtained
  using NNPDF1.0, NNPDF1.2 and NNPDF2.0 PDFs,
for several DIS and Drell--Yan observables included in the NNPDF2.0 fit.
From top to bottom and from
left to right: the $F_2^p$ structure function and the
$F_2^d/F_2^p$ (NMC), the inclusive neutrino
cross-section (CHORUS), the Drell--Yan rapidity distribution (E866p),
the $W-$asymmetry (CDF) 
and the Drell--Yan p/d ratio (E866). For the purposes of this plot only,
experimental statistical and systematic 
uncertainties have been added in quadrature. \label{fig:obs}} 
\end{center}
\end{figure}

\begin{figure}[ht!]
\begin{center}
\epsfig{width=0.48\textwidth,figure=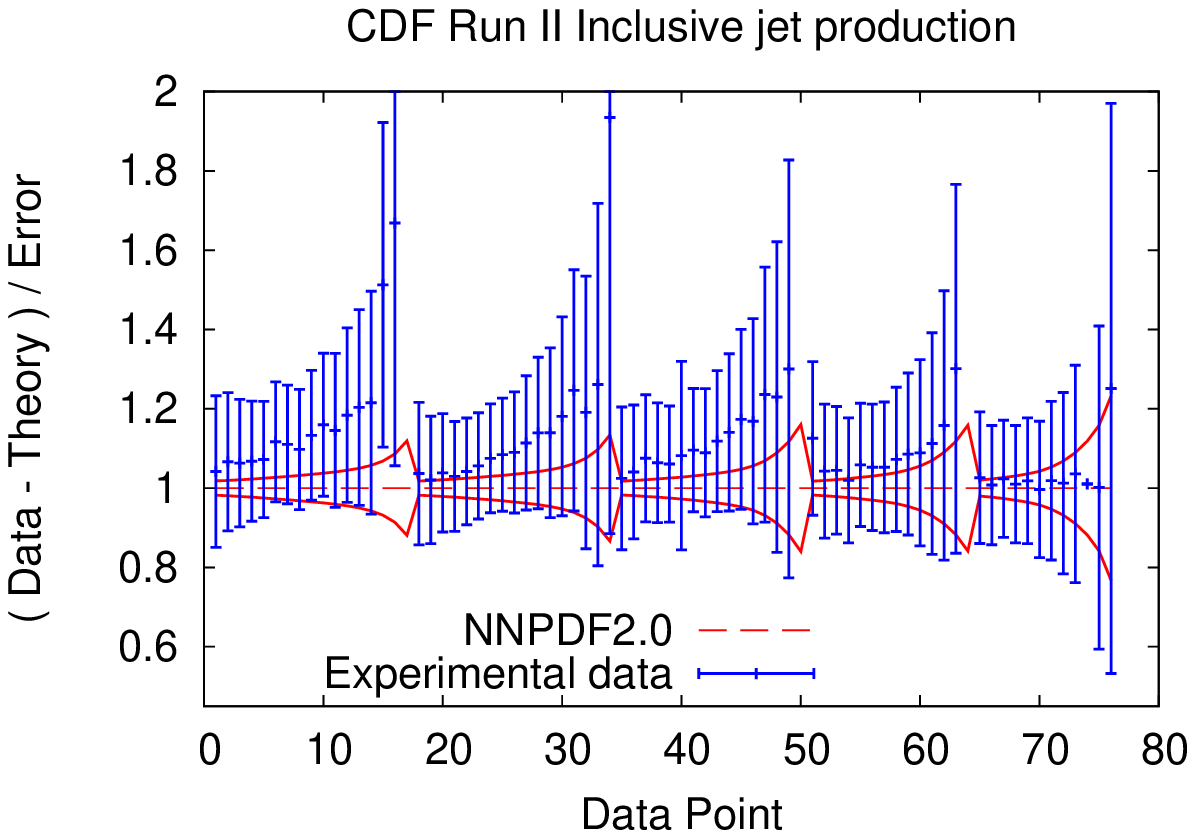}
\epsfig{width=0.48\textwidth,figure=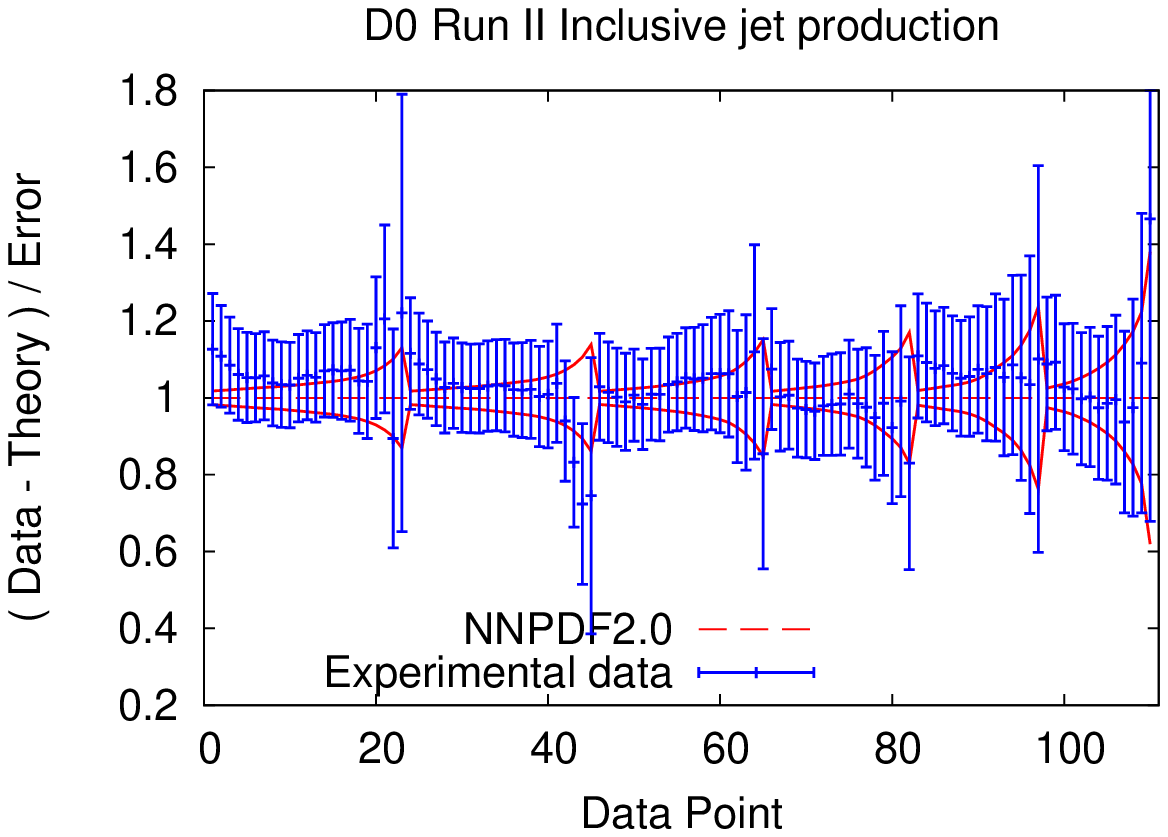}
\caption{ \small 
Comparison between data and NLO predictions obtained
  using NNPDF2.0 PDFs,
for inclusive jet production from D0 and CDF Run II.
Data points are  ordered in rapidity
and in transverse momentum from left to right.
 Experimental
statistical and systematic uncertainties have been
added in quadrature for this plot. The NLO theoretical
prediction has been obtained using the
FastNLO code.
 \label{fig:obsjet}} 
\end{center}
\end{figure}

The general quality of predictions obtained using NNPDF2.0 PDFs
for the observables  which have been included in the fits has already
been summarized in Table~\ref{tab:estfit2} and Fig.~\ref{fig:chi2histo}.
A direct comparison of the data with theoretical predictions for some
of these observables are shown in
Fig.~\ref{fig:obs} (DIS  and Drell-Yan) and Fig.~\ref{fig:obsjet} 
(inclusive jets).

In Drell-Yan observables, 
the improvement in accuracy of the prediction when going from NNPDF1.2
to NNPDF2.0 is apparent: in particular,
the sea asymmetry,
virtually unconstrained from DIS, is now very well constrained
by the E866 ratio data. Also the uncertainty reduction in the
$W$-asymmetry measurement shows the increase in the precision of the
determination of the quark decomposition  in NNPDF2.0.
In jet data, the excellent agreement between data and theory seen from
the $\chi^2$ of Tab.~\ref{tab:estfit2} is seen to hold
through the whole kinematical range
for all bins in transverse
momentum and rapidity.

\subsection{The proton strangeness revisited}

In Ref.~\cite{Ball:2009mk} a detailed study of the strangeness
content in the proton was performed, with particular 
emphasis on the precision determination
of electroweak parameters.
 The addition of fixed--target Drell--Yan data in the NNPDF2.0 PDF
 determination,  together with other
 improvements in the fit that have been discussed in
 Sect.~\ref{sec:res:dataset}, leads to significantly stricter 
constraints on the
 shape of the strange distributions $s^{\pm}(x)$ PDFs, as shown in
 Figs.~\ref{fig:singletPDFs-nnpdf}-\ref{fig:valencePDFs-nnpdf}: while remaining consistent with the NNPDF1.2
 result, the new determination of $s^+$  and especially $s^-$ 
at large $x$ have a much reduced
 uncertainty. 

Indeed, the strange momentum fraction $K_S=\frac{
   S^+}{U^++D^+}$ and strangeness asymmetry
$R_S=\frac{2
   S^-}{U^-+D^ -}$~\cite{Ball:2009mk}  at $Q^2=20$~GeV$^2$ are
\begin{eqnarray}
K_S&=&
\begin{cases} 
0.71 {{}^{+0.19}_{-0.31}}^{\rm stat}\pm 0.26^{\rm  syst}
& \qquad ({\rm NNPDF1.2})\label{eq:ksval} \\
0.503 \pm 0.075^{\rm stat}; & \qquad ({\rm NNPDF2.0})\label{eq:ksval-20}
\end{cases}\\ 
 R_S&=& 
\begin{cases}
0.006\pm0.045^{\rm stat} \pm 0.010^{\rm syst} &({\rm NNPDF1.2})
\label{eq:rsval} \\
0.019 \pm 0.008^{\rm stat} &({\rm NNPDF2.0}) 
 \label{eq:rsval-20},
\end{cases}
\end{eqnarray}
i.e. the PDF uncertainty on $K_S$ is reduced by more than a factor
two, while that  on $R_S$ is reduced by a factor 5, with all results
consistent within uncertainties. We have made no attempt to provide a
new determination of  systematic and theoretical
uncertainties on $R_S$, which are now comparable
to  the reduced statistical uncertainties, but they should be similar
to those determined in Ref.~\cite{Ball:2009mk} and quoted in
Eqs.~(\ref{eq:ksval}-\ref{eq:rsval}).
 
%%%%%%%%%%%%%%%%%%%%%%%%%%%%%%%%%
%------------------------------------------------------------
\begin{figure}[ht!]
\begin{center}
\epsfig{width=0.75\textwidth,figure=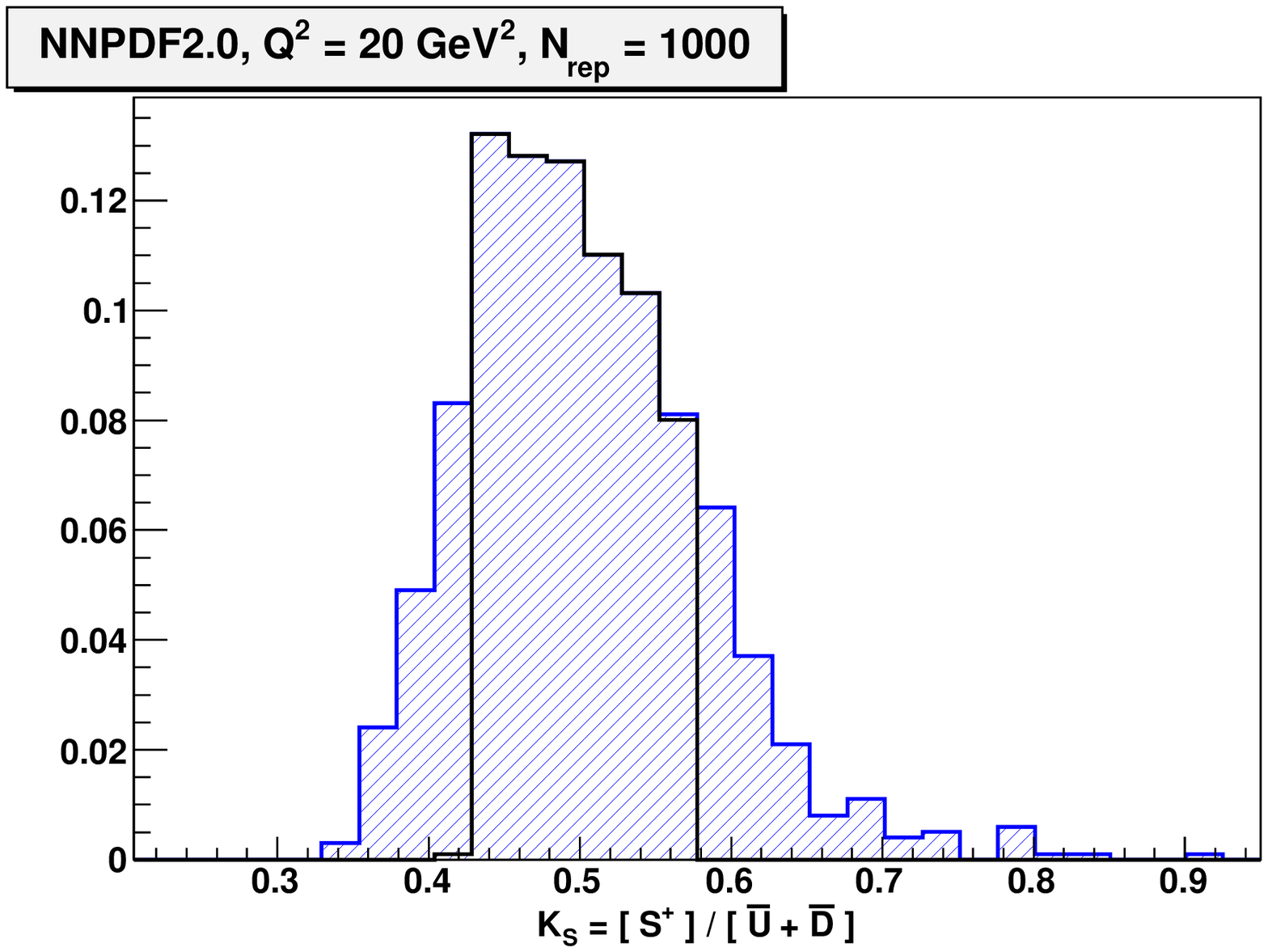} 
\end{center}
\caption{\small Distribution of
$K_S$  at $Q^2=20$~GeV$^2$ computed from the reference set of  $N_{\rm
    rep}=1000$ NNPDF2.0 PDF replicas. The central region
corresponds to the 68\% confidence 
interval, $K_S\lp Q^2=20 \,{\rm GeV}^2\rp=0.503 \pm 0.075~ ({\rm stat})$,
which coincides with the 1--$\sigma$ interval Eq.~\ref{eq:ksval-20}).
}
\label{fig:strangePDFdists}
\end{figure}
%------------------------------------------------------------------
%%%%%%%%%%%%%%%%%%%%%%%%%%%%%%%%%
The distribution of $K_S$ values for 1000 NNPDF2.0 replicas is shown 
in Fig.~\ref{fig:strangePDFdists}: in comparison to the analogous plot
in Ref.~\cite{Ball:2009mk} the narrower distribution which we now get 
is closer to gaussian and indeed, unlike in Ref.~\cite{Ball:2009mk}, 
we now find no difference between the
68\% confidence level and (symmetric) one--$\sigma$ intervals.

%%%%%%%%%%%%%%%%%%%%%%%%%%%%%%%%%%%%%%
\begin{figure}[ht!]
\begin{center}
\epsfig{width=0.75\textwidth,figure=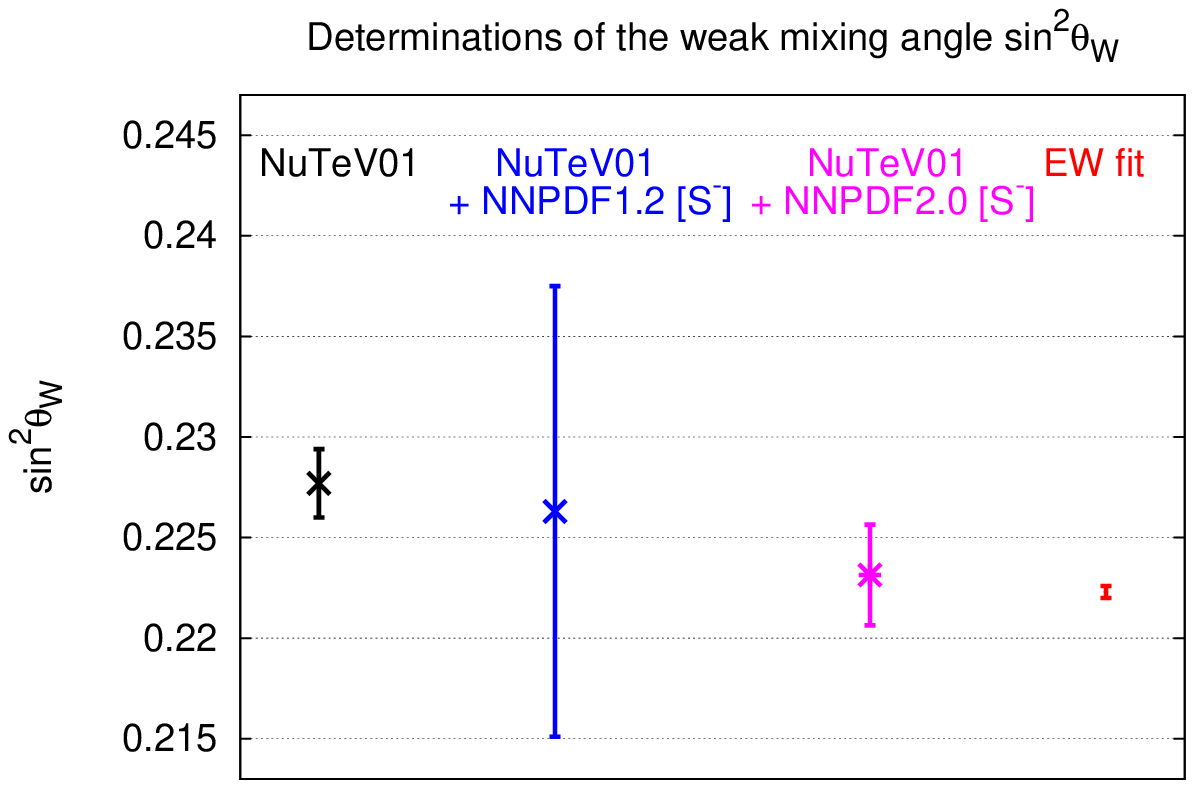}
\caption{\small Determination of the Weinberg angle from the
  uncorrected NuTeV data~\cite{Mason:2007zz}, with 
$\lc S^-\rc$ correction using NNPDF1.2 (Eq.~(\ref{eq:rsval})) and NNPDF2.0
  (Eq.~(\ref{eq:rsval-20})) results, and from
  a global electroweak fit~\cite{Flacher:2008zq}.
Note that that statistical
uncertainties only are included in the NNPDF2.0 correction.   
\label{fig:nutev}} 
\end{center}
\end{figure}
%%%%%%%%%%%%%%%%%%%%%%%%%%%%%%%%%%%%%%
The implication of the accurate determination Eq.~(\ref{eq:rsval-20}) of
the strangeness asymmetry $R_S$ for the so--called NuTeV
anomaly~\cite{Davidson:2001ji} are striking: 
in Fig.~\ref{fig:nutev} we compare the
NuTeV determination of the Weinberg angle~\cite{Mason:2007zz},
uncorrected or
corrected for strangeness
asymmetry as discussed in Ref.~\cite{Ball:2009mk}, using the values of
$R_S$  Eqs.~(\ref{eq:rsval}), and the result of a global electroweak
fit~\cite{Flacher:2008zq}.  The two corrected values, unlike the
uncorrected NuTeV value, are in perfect agreement with the electroweak
fit and with each other. However,  while the uncertainty on the
Weinberg angle with NNPDF1.2 correction
was considerably larger, the uncertainty
after NNPDF2.0 correction is comparable to that on the uncorrected
value. 
Indeed, Eq.~(\ref{eq:rsval}) 
provide a 2--$\sigma$ evidence for
a non-zero and positive strangeness asymmetry in the nucleon. While
such an asymmetry 
was previously advocated as a possible explanation of the NuTeV
anomaly~\cite{Davidson:2001ji}, evidence for 
it~\cite{Lai:2007dq,Alekhin:2008mb,Mason:2007zz,Martin:2009iq} was so
far inconclusive, and it is being established  here for the first time.

\subsection{Parton luminosities}
\label{subsec:lumi}

In order to highlight the impact of parton distributions
at LHC the parton--parton 
luminosities (also called partonic fluxes) are
relevant~\cite{Campbell:2006wx,Guffanti:2009xk}; of particular
interest are the
sizes of PDF uncertainties in parton luminosities
from different PDF sets.
 
%%%%%%%%%%%%%%%
\begin{figure}[t!]
  \centering
\epsfig{width=0.65\textwidth,figure=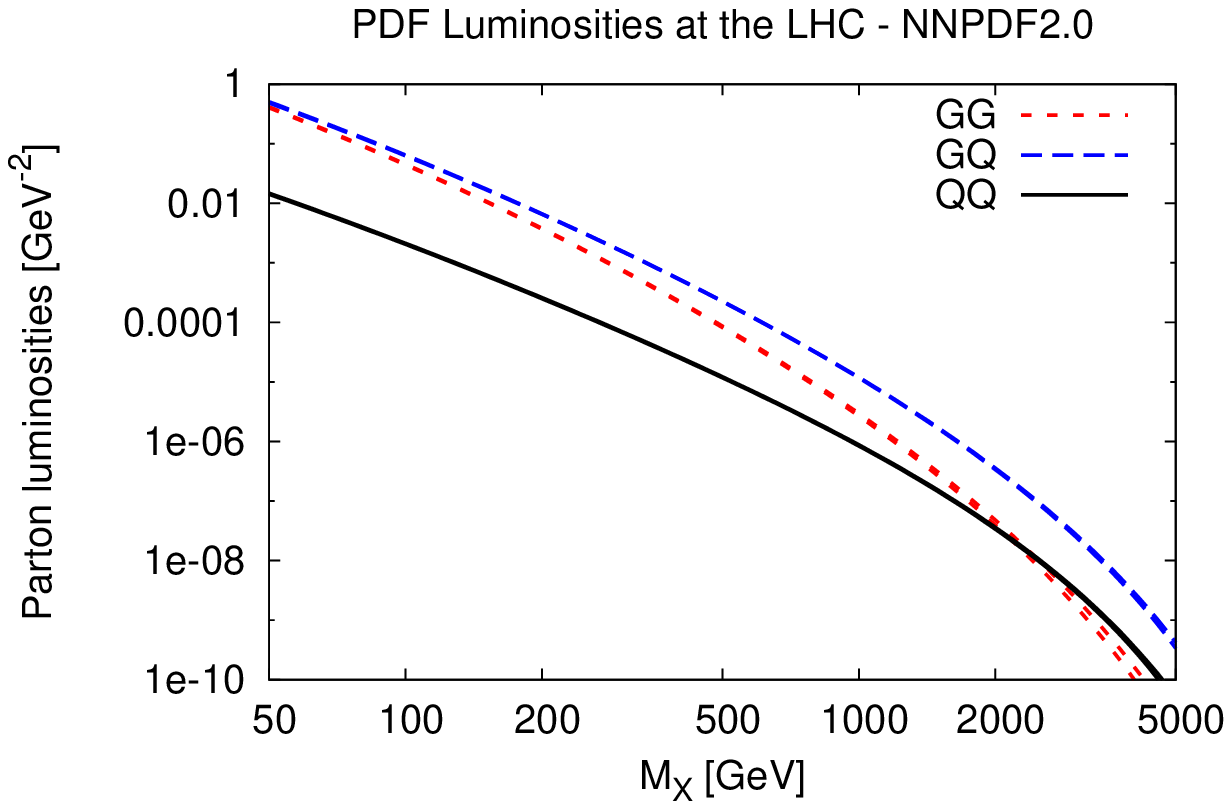}
\epsfig{width=0.65\textwidth,figure=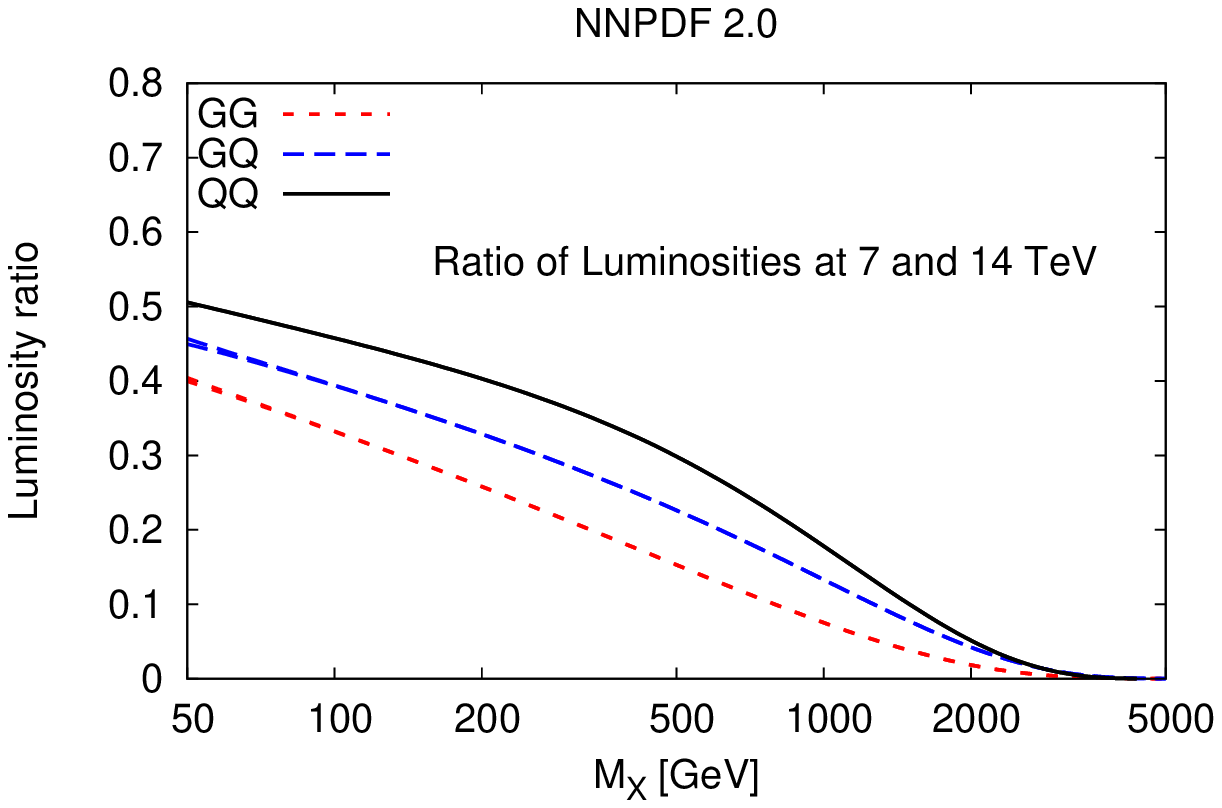}
\caption{\small Parton--parton luminosities
Eq.~(\ref{ref:fluxes}) 
in the various partonic channels, computed from
the NNPDF2.0 set at the LHC for $\sqrt{s}=$ 14 TeV (above) and ratio 
of results for 7 TeV and
14 TeV (below).
  \label{fig:fluxes2}}
\end{figure}
%%%%%%%%%%%%%%%%
We can define three relevant combinations of PDF luminosities
for the production of a massive object with mass $M_X$ in
hadronic collisions
as follows:
\bea
\Phi_{gg}\lp M_X^2\rp &=& \frac{1}{s}\int_{\tau}^1
\frac{dx_1}{x_1} g\lp x_1,M_X^2\rp g\lp \tau/x_1,M_X^2\rp \ ,\nonumber\\
\Phi_{gq}\lp M_X^2\rp &=& \frac{1}{s}\int_{\tau}^1
\frac{dx_1}{x_1} \lc g\lp x_1,M_X^2\rp \Sigma\lp \tau/x_1,M_X^2\rp
+ (1\to 2) \rc \ , \label{ref:fluxes}\\
\Phi_{qq}\lp M_X^2\rp &=& \frac{1}{s}\int_{\tau}^1
\frac{dx_1}{x_1} \sum_{i=1}^{N_f} \lc q_i\lp x_1,M_X^2\rp \bar{q}_i\lp \tau/x_1,M_X^2\rp
+  (1\to 2) \rc \nonumber\ ,
\eea
with $\tau \equiv M_X^2/s$ and $\sqrt{s}$ the center of
mass energy of the hadronic collision.

In Fig.~\ref{fig:fluxes2} we show the various partonic
luminosities Eq.~(\ref{ref:fluxes}) at the LHC as computed
with the NNPDF2.0 set. It is clear that at low masses the GG
and GQ channels are both important, while at large masses the GQ
channel dominates. Also in Fig.~\ref{fig:fluxes2} we show the
ratio of partonic luminosities between LHC 14 TeV and 7 TeV. 
While at small masses the loss in partonic luminosity is roughly
a factor two, it can be as large as a factor ten or more at large
masses. The gluon-gluon luminosity is the channel which suffers 
the greatest reduction.
%%%%%%%%%%%%%%%%
\begin{figure}[h!]
  \centering
\epsfig{width=0.48\textwidth,figure=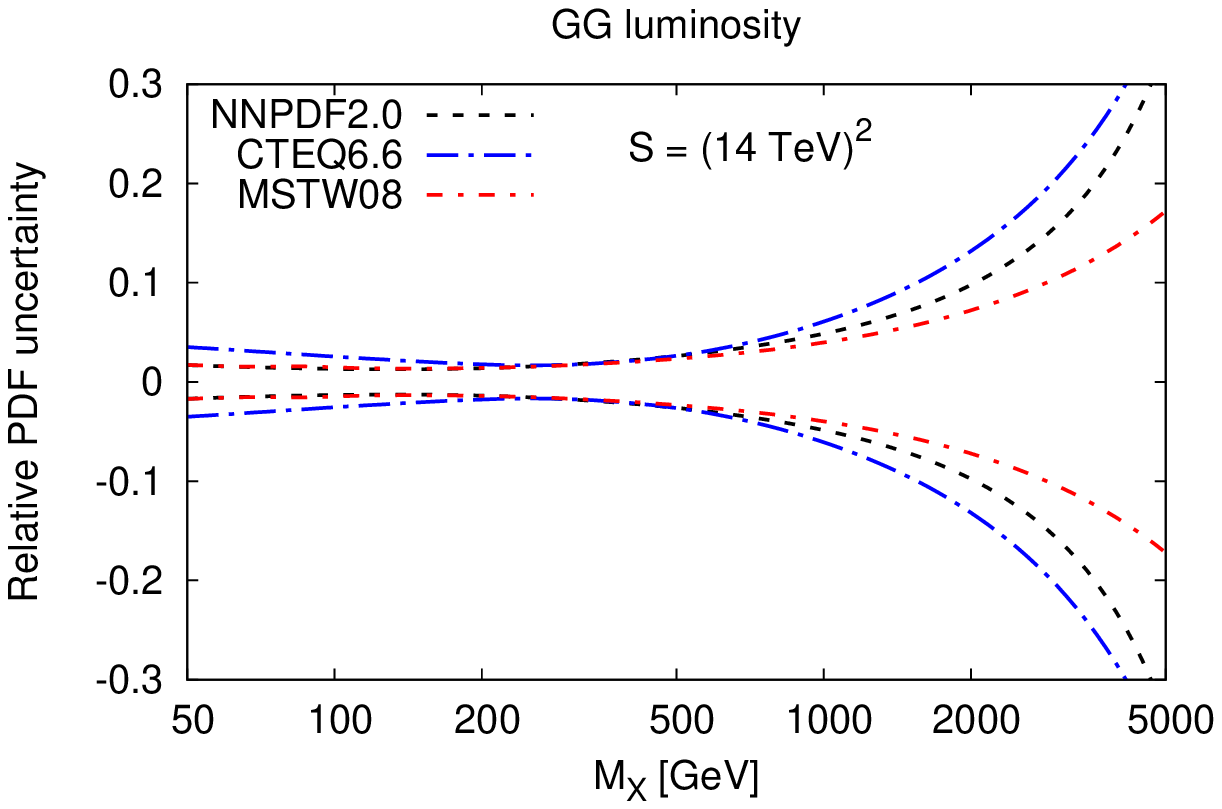}
\epsfig{width=0.48\textwidth,figure=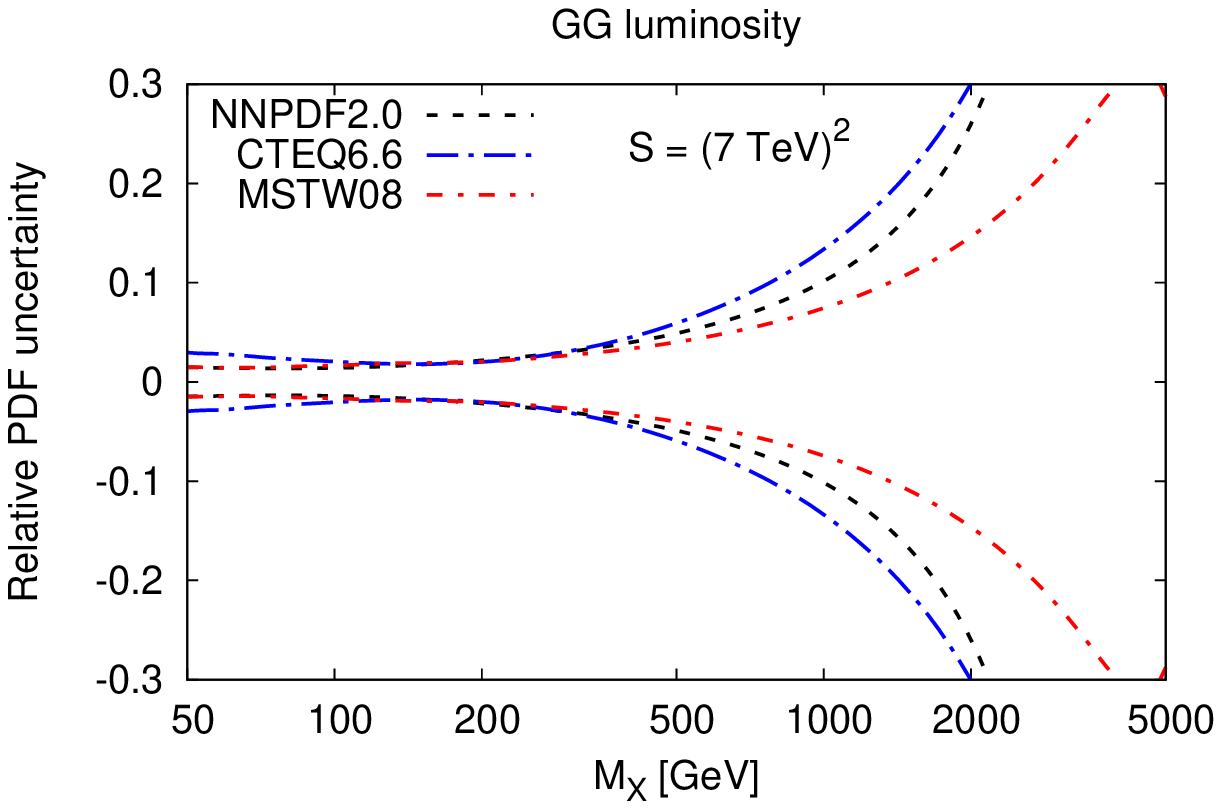}
\epsfig{width=0.48\textwidth,figure=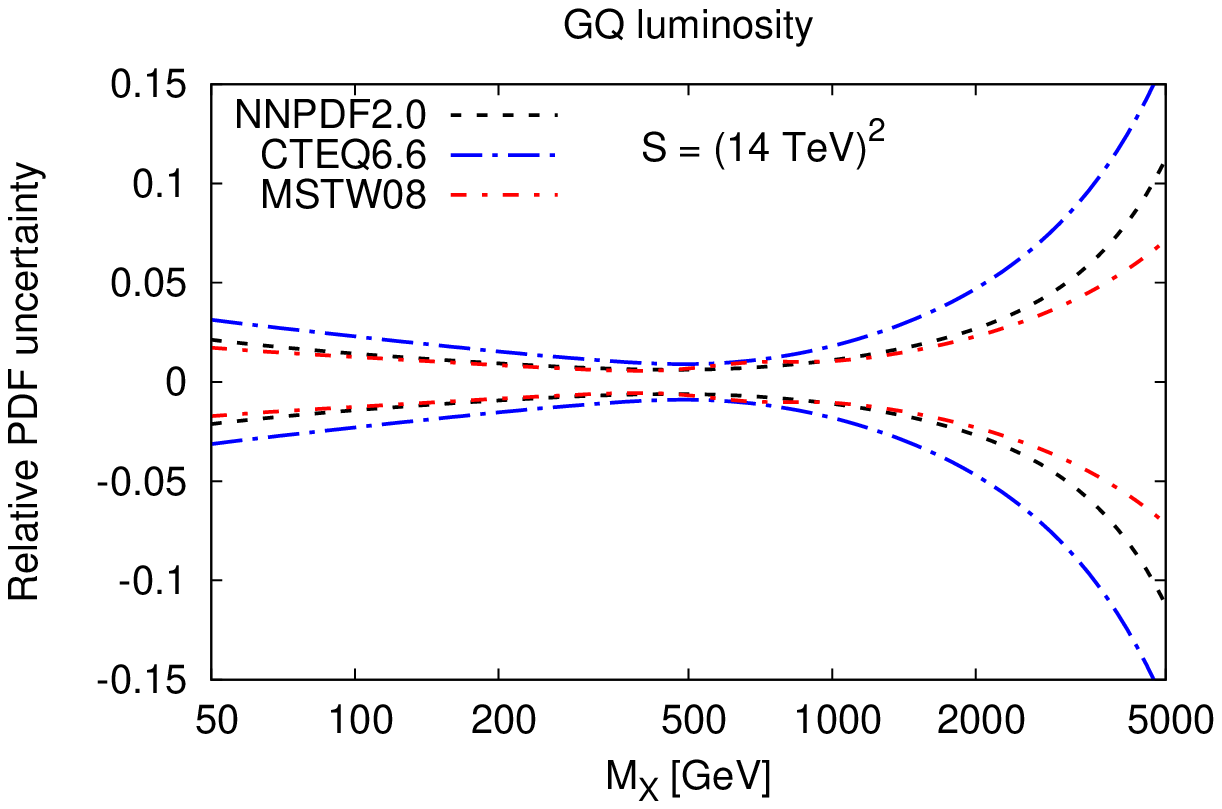}
\epsfig{width=0.48\textwidth,figure=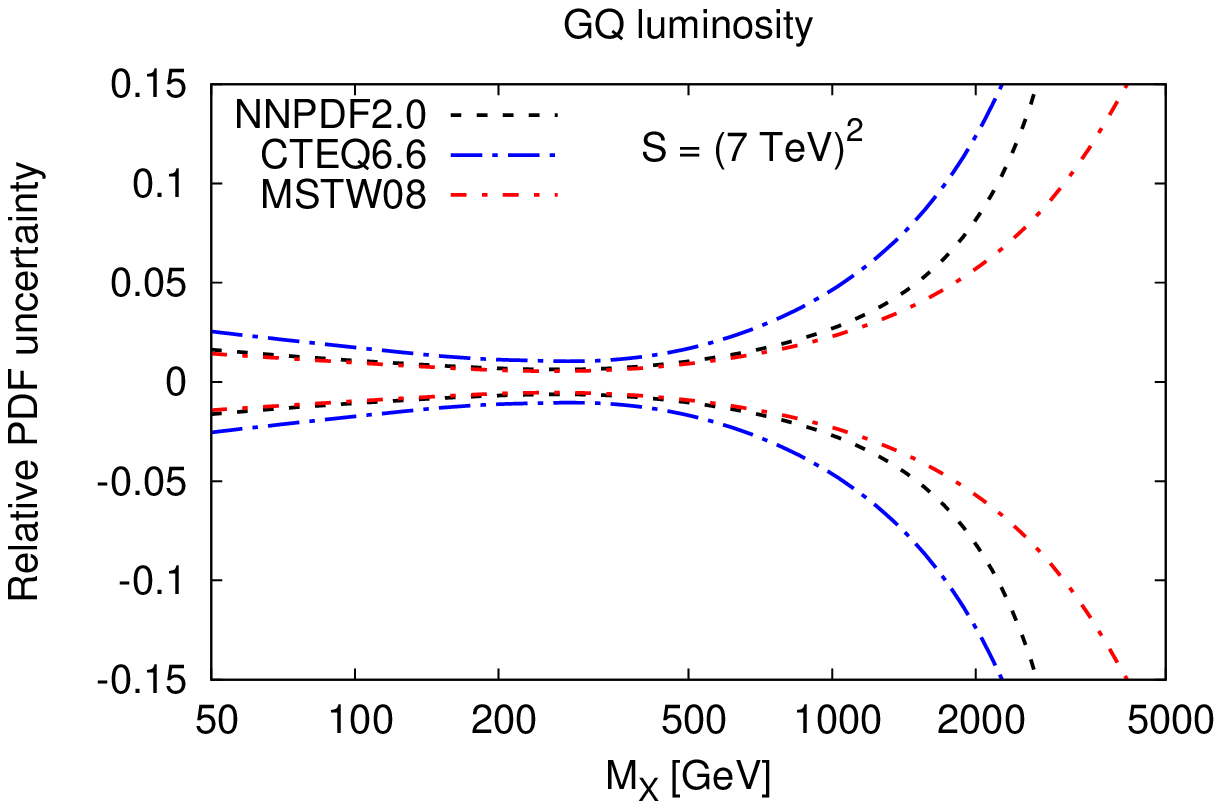}
\epsfig{width=0.48\textwidth,figure=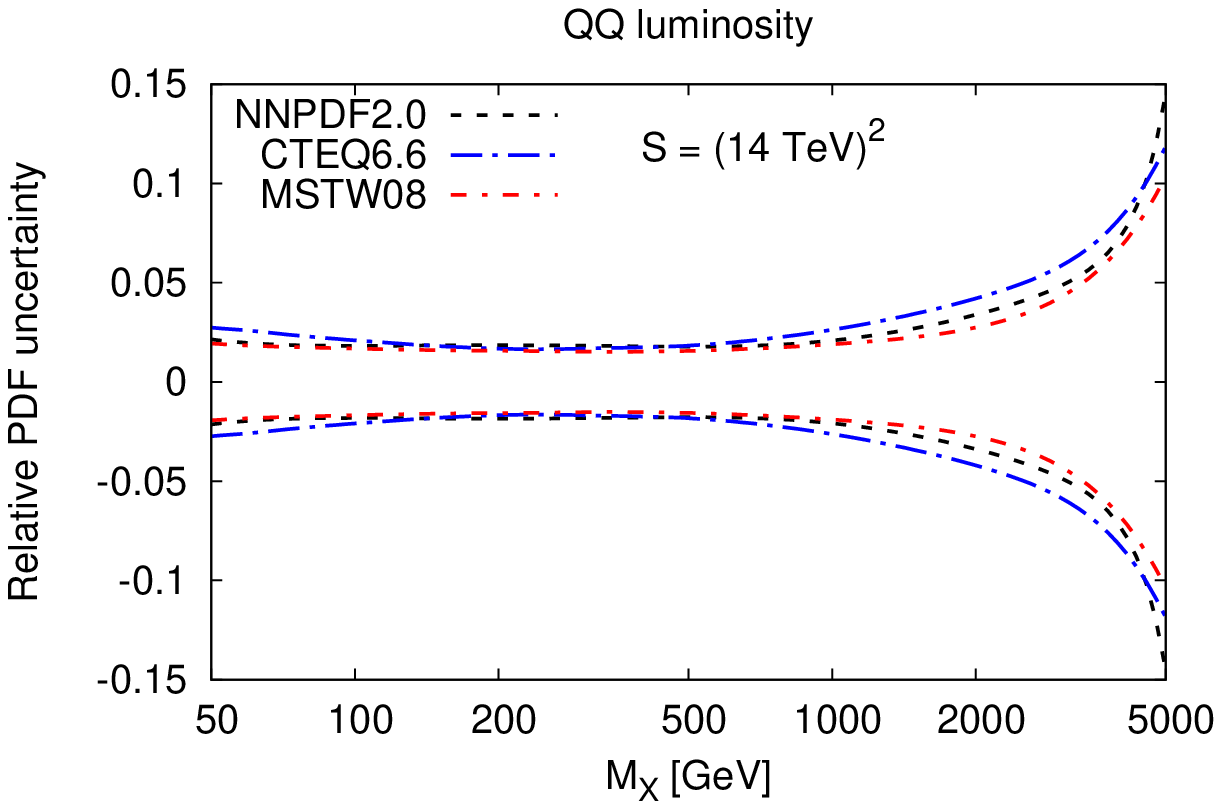}
\epsfig{width=0.48\textwidth,figure=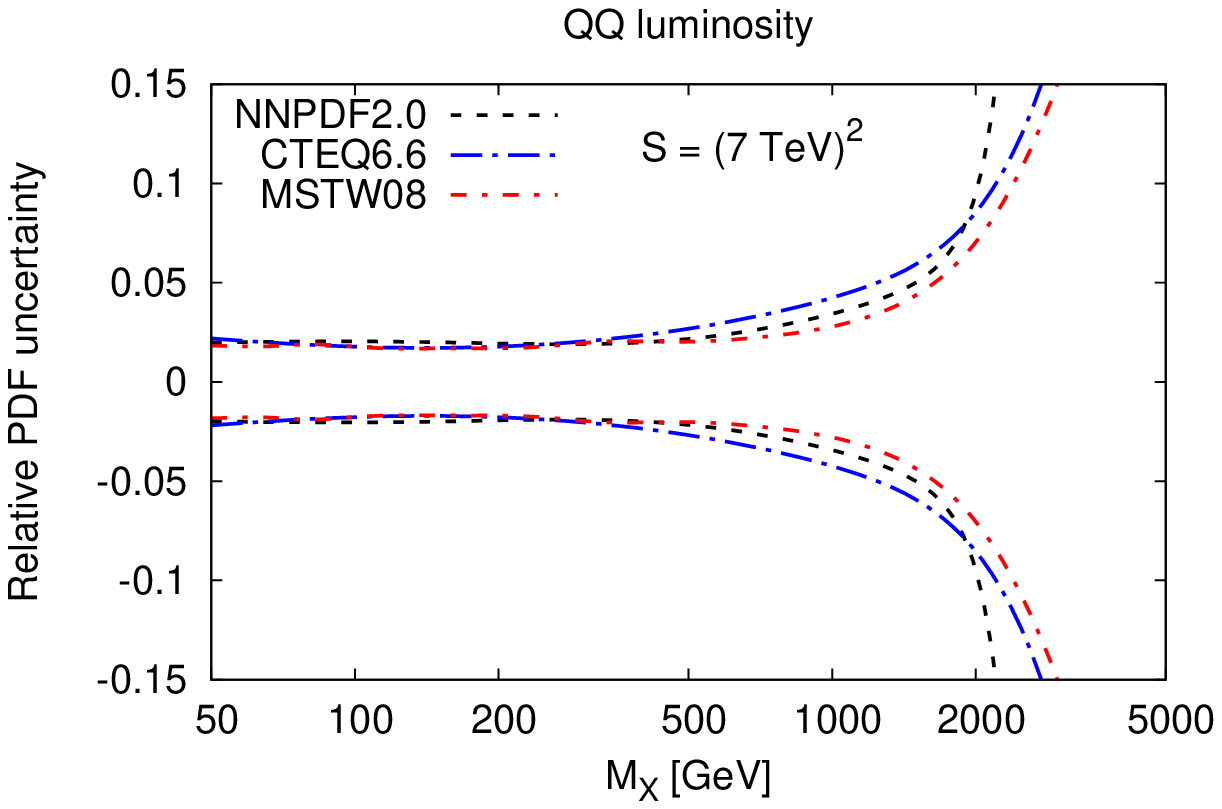}
\caption{\small Relative PDF uncertainties on
parton--parton luminosities  Eq.~(\ref{ref:fluxes}) for the
NNPDF2.0, CTEQ6.6 and MSTW2008 PDF sets, as function of
the mass of the produced heavy object $M_X$ at the
LHC for 14 TeV (left) and 7 TeV (right). From top
to bottom, the gluon-gluon luminosity, the gluon-quark
luminosity and the quark-quark luminosity are shown.}
  \label{fig:fluxes3}
\end{figure}
%%%%%%%%%%%%%%%%
Turning now to the   uncertainties on parton  luminosities due to PDFs,
in Fig.~\ref{fig:fluxes3} we compare the
relative PDF uncertainties (normalized to the respective central set)
in various channels of PDF luminosity for the NNPDF2.0,
CTEQ6.6 and MSTW08 sets. In the GG channel, all PDF sets agree
in the central mass region, and NNPDF2.0 is close to MSTW08
in general. In the QQ channel all PDF sets yield very similar
uncertainties at small and medium masses. It is also clear
from  Fig.~\ref{fig:fluxes3} that at 7 TeV the 
%reduced phase space
restricted $x$-range of the partons 
leads to sizably larger PDF uncertainties at large values
of $M_X$.

\subsection{LHC standard candles}

\begin{figure}[ht]
\begin{center}
\epsfig{width=0.48\textwidth,figure=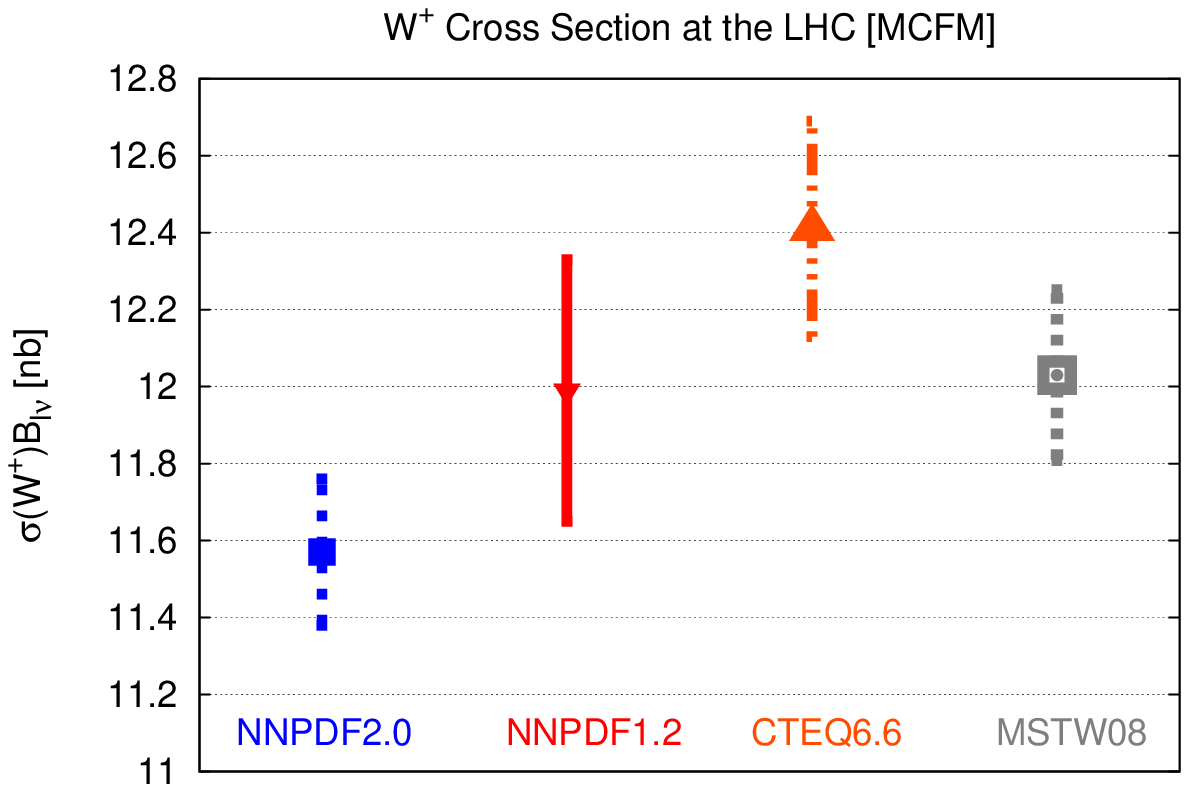}
\epsfig{width=0.48\textwidth,figure=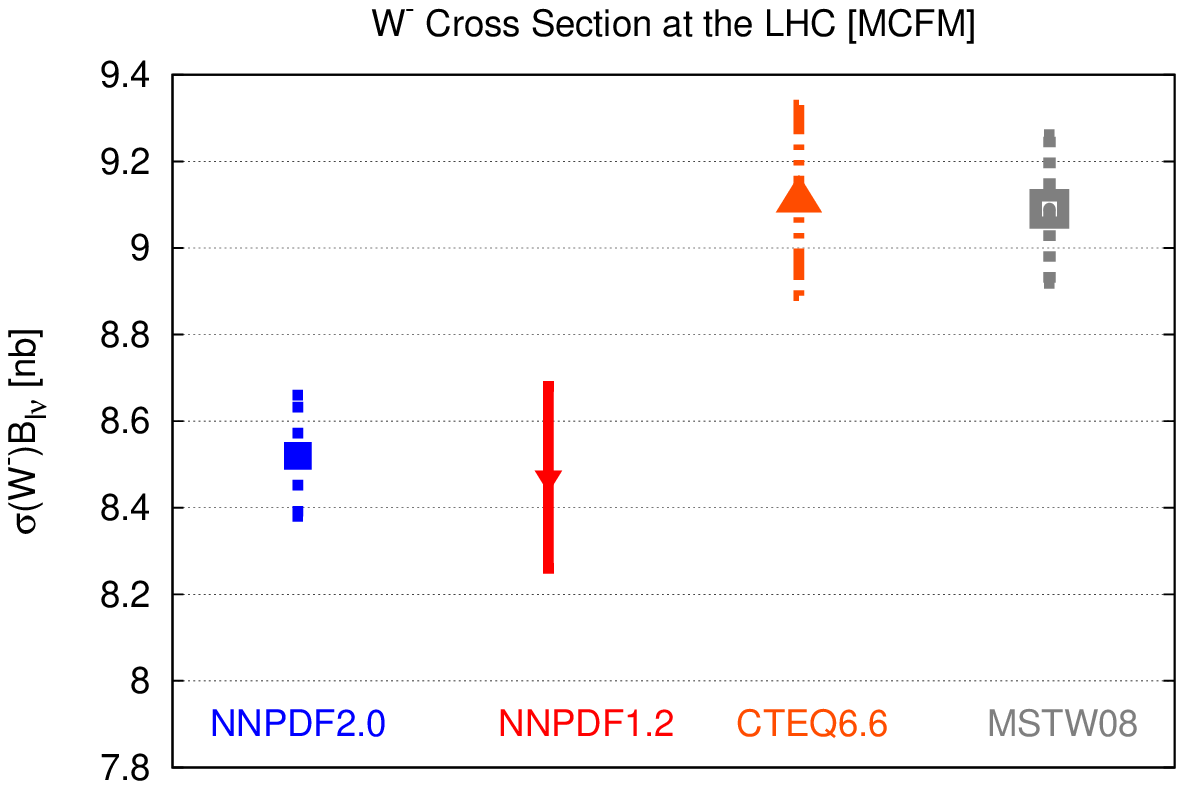}
\epsfig{width=0.48\textwidth,figure=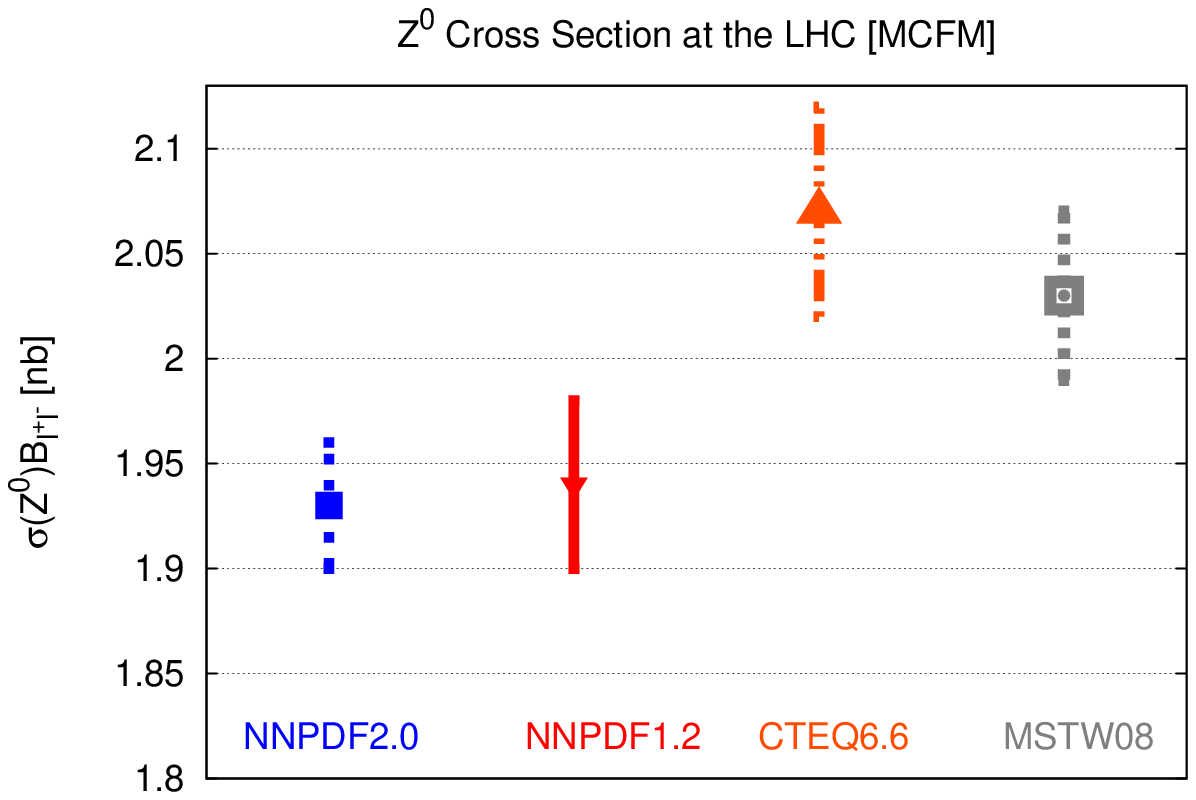}
\epsfig{width=0.48\textwidth,figure=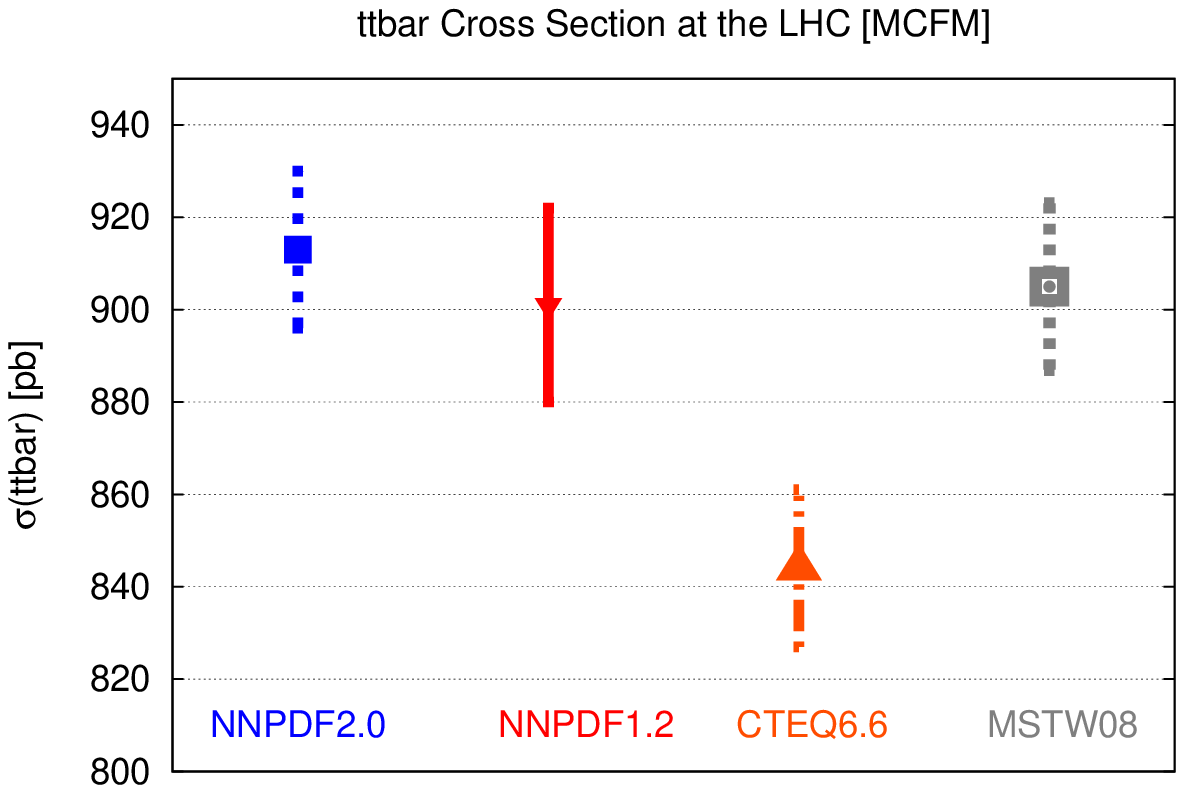}
\epsfig{width=0.48\textwidth,figure=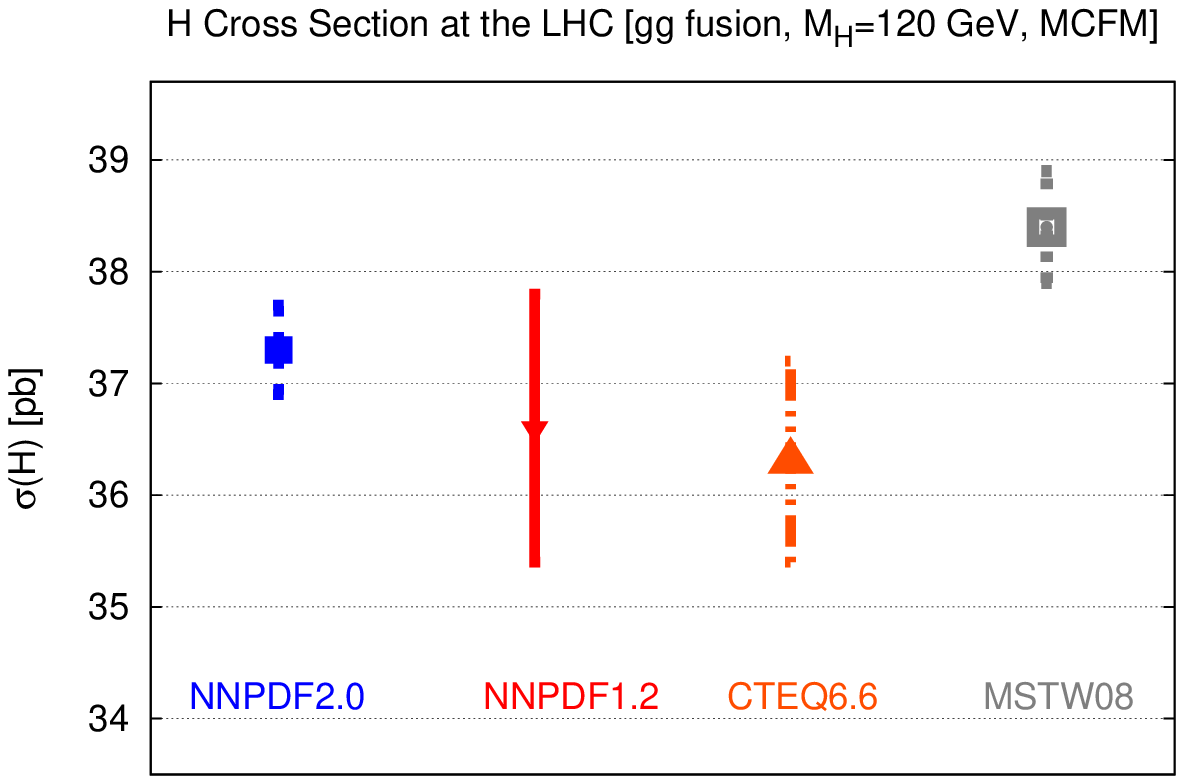}
\caption{\small Graphical comparison of the cross sections from
  Table~\ref{tab:LHCobs}. 
 \label{fig:LHCobs}} 
\end{center}
\end{figure}

The total cross sections at the LHC with $\sqrt s=14$~TeV 
for $W$, $Z$, $H$ and $t\bar{t}$ 
production computed at
NLO with
MCFM~\cite{Campbell:2000bg,Campbell:2002tg,Campbell:2004ch,MCFMurl}
and   NNPDF2.0, NNPDF1.2., CTEQ6.6, and MSTW08 PDFs are compared
in Table~\ref{tab:LHCobs} and Fig.~\ref{fig:LHCobs}.
Values obtained using NNPDF2.0 are in excellent agreement with those
from NNPDF1.2, with significantly smaller uncertainties.
The predictions from previous NNPDF sets were discussed 
in~\cite{Ball:2009mk}.

%%%%%%%%%%%%%%%%%%%%%%%%%%%%%%%%%%%%%%%%%%%%%%%%%
\begin{table}
  \centering
  {\small
  \begin{tabular}{|c|c|c|c|}
    \hline
        & $\sigma(W^+){\rm Br}\lp W^+ \to l^+\nu_l\rp\,$ 
         & $\sigma(W^-){\rm Br}\lp W^- \to l^+\nu_l\rp\,$ 
         & $\sigma(Z^0){\rm Br}\lp Z^0 \to l^+l^-\rp\,$ \\
    \hline
    NNPDF1.2  & $11.99 \pm 0.34$ nb &
$8.47 \pm 0.21$ nb  & $1.94 \pm 0.04$ nb \\
    \hline
    NNPDF2.0  & $11.57 \pm 0.19$ nb & 
$8.52 \pm 0.14$ nb & $1.93 \pm 0.03$ nb \\
   \hline
   CTEQ6.6  & $12.41 \pm 0.28$ nb &  $9.11 \pm 0.22 $ nb & 
$2.07 \pm 0.05 $ nb \\
  \hline
   MSTW08  & $12.03 \pm 0.22 $ nb & $9.09 \pm 0.17 $ nb &
$2.03 \pm 0.04$ nb \\
    \hline
  \end{tabular}}\\
\vspace{0.3cm}
 {\small
  \begin{tabular}{|c|c|c|}
    \hline
        & $\sigma(t\bar{t})$ 
          & $\sigma(H,m_H=120\,{\rm GeV})$
          \\
    \hline
   NNPDF1.2 &  $901\pm 21$ pb &  $36.6\pm 1.2$ pb \\
    \hline
   NNPDF2.0 &  $913\pm 17$ pb  &  $37.3  \pm 0.4 $ pb \\
    \hline
CTEQ6.6 & $844 \pm 17$ pb  & $36.3 \pm 0.9$  pb\\
 \hline
MSTW08 & $905 \pm 18$ pb  & $38.4 \pm 0.5 $ pb \\
 \hline
  \end{tabular}}
  \caption{\small Cross sections for W, Z, $t\bar{t}$ and Higgs production
at the LHC at $\sqrt{s}=14$ TeV. All quantities have been computed at NLO using
    MCFM~\cite{Campbell:2000bg,Campbell:2002tg,Campbell:2004ch,MCFMurl}
with default settings
for the NNPDF1.2, NNPDF2.0, CTEQ6.6 and MSTW08 PDF sets. All 
uncertainties shown are one--sigma. The Higgs cross section corresponds to the
gluon-gluon fusion production channel. \label{tab:LHCobs}}
\end{table}
It was already observed in Ref.~\cite{Ball:2008by} that NNPDF results
for W and Z production agree with those of CTEQ6.1, but undershoot the
CTEQ6.5 and CTEQ6.6 predictions by more than 5\%. The main
  difference between CTEQ6.5/CTEQ6.6 and CTEQ6.1 is that charm mass
  effects are included in the former pair of fits, but not in the
  latter, and are also not included in all available NNPDF fits. This
  suggests that charm mass effects  be responsible for the
discrepancy between the CTEQ6.6   and   NNPDF predictions for $W$
and $Z$ cross sections.  It should be noticed however that NNPDF1.0
results do agree~\cite{Ball:2008by} with MRST01~\cite{Martin:2002aw}, 
and do not agree with
MSTW08 (as it is clear from  Table~\ref{tab:LHCobs}) despite the fact that
charm mass effects are included both in MRST01 and MSTW08.
The pattern for Higgs and $t\bar{t}$ production is even less clear, with
NNPDF in good agreement with MSTW08 but not CTEQ6.6 for the former,
and in good agreement with CTEQ6.6 but not MSTW08 for the latter.

%%%%%%%%%%%%%%%%%%%%%%%%%
%%%%%%%%%%%%%%%%%%

Note however that most of these cross sections are quite sensitive to the
value of $\alpha_s$, and some of them extremely sensitive: for
example, the contribution to the 
Higgs cross section from gluon-gluon fusion varies by
about 5\% when $\alpha_s$ is varied by 2\%. The results shown
in Table~\ref{tab:LHCobs} and Fig.~\ref{fig:LHCobs} have been obtained
with the default settings of MCFM, and in particular with  the value of
$\alpha_s$ corresponding to each group's central parton
fit, namely $\alpha_s(M_Z)=0.118$ for CTEQ6.6 and
$\alpha_s(M_Z)=0.120$ for MSTW08 (and $\alpha_s(M_Z)=0.119$ for
NNPDF2.0). Hence, benchmarking of these cross sections
with the same value of all parameters including $\alpha_s$
should be performed before conclusions can be drawn. 

It should finally be noticed that some approximations used in the
MSTW08 and CTEQ6.6 PDF determinations but not by NNPDF
could have an impact on these
observables, such as the use of $K$--factors in fitting Drell-Yan data
by both MSTW and CTEQ,
the use of a restrictive small $x$ parametrization of the gluon by
CTEQ, and the use of very restrictive parametrizations of strangeness
by both MSTW and CTEQ.
In summary, while the lack of inclusion of heavy quark terms may be
responsible for some of the discrepancies observed in
Table~\ref{tab:LHCobs} and Fig.~\ref{fig:LHCobs} it cannot be the only
explanation (it cannot account for cases in  which NNPDF agrees with
CTEQ but not MSTW or conversely). The issue should be re-examined
after the inclusion of heavy quark mass effects in NNPDF, ideally
within a systematic benchmarking of parton distributions.

%
%-------------------------------------------------
%-----------------------------------------------

\section{Conclusions and outlook}
\label{sec:conclusions}
The NNPDF2.0 parton determination is the first global parton
determination based on NNPDF methodology, and it is also the first
global parton determination in which NLO QCD is used consistently throughout, 
without resorting to the $K$--factor approximation. We have seen that the NNPDF
methodology can accommodate a complex combination of DIS and hadronic 
datasets without any particular
difficulties: in fact, the only bottleneck in the implementation of
the NNPDF2.0 global fit has been computational, requiring the
development of the FastKernel method discussed in
Sect.~\ref{sec:evolution} in order for DGLAP evolution and the
computation of physical observables to be fast enough. 

In previous NNPDF work it was shown that NNPDF parton determinations
behave in a statistically consistent way upon the subsequent inclusion of
new data, without any adjustment being required as the new data are
included, and with uncertainties decreasing upon the addition of new
information, or at most remaining constant when inconsistent data are
added. Here we have seen that this remains true when hadronic and
deep--inelastic data are combined. In fact, we have found complete
consistency between DIS and hadronic data, with some hadronic data
(jets) being reasonably well predicted by the DIS fit and leading to
small improvements, and other hadronic data (Drell-Yan) introducing new
information which allows a quantitative determination of some PDF
combinations that were determined with moderate or poor accuracy by DIS
data, such as the light quark sea asymmetry.

Progress has been made recently towards the inclusion of 
heavy quark mass effects in the NNPDF framework~\cite{Forte:2010ta}
and in the benchmarking of different approaches for the inclusion of
heavy quark mass effects~\cite{LH}. Once these are included in a
global NNPDF fit accurate and reliable NLO phenomenology at the LHC will
be possible.
\bigskip
\bigskip
\begin{center}
\rule{5cm}{.1pt}
\end{center}
\bigskip
\bigskip
The NNPDF2.0 PDFs (sets of $N_{\rm rep}=100$ and 1000 replicas), as
well as several of the sets based on reduced or different datasets
discussed in Sect.~\ref{sec:res:dataset}
(old HERA-I data, DIS only, DIS+JET only,
DIS+DY only, sets of $N_{\rm rep}=100$ replicas), and also sets
determined using all values of $0.114\le\alpha_s(M_Z)0.124$ in steps
of $\Delta\alpha_s(M_Z)=0.001$
are available from the
NNPDF web site,
\begin{center}
{\bf \url{http://sophia.ecm.ub.es/nnpdf}~.}
\end{center}
They are also available
through the LHAPDF interface~\cite{Bourilkov:2006cj}.

\vfill\eject

{\bf\noindent  Acknowledgments \\}

We would like to thank J.~C.~Webb for help with the E866 data, 
E.~Halkiadakis for help with the CDF direct
$W$ asymmetry production data, H.~Schellman for assistance with 
D0 electroweak data, M.~Mart\'\i nez-P\'erez
for discussions about Run II inclusive jet production data 
and A.~Cooper-Sarkar
for discussions concerning the combined HERA-I data. We also thank 
A.~Accardi for information on the CTEQ6.X fits. We are especially
grateful to A.~Vicini and G.~Ridolfi for providing us with the
Drell-Yan code used in the FastKernel code benchmarking.
We acknowledge extensive discussions and correspondence 
with members of the PDF4LHC workshop, in particular 
A.~de~Roeck, 
A.~Glazov, J.~Huston, R.~McNulty, P.~Nadolsky, F.~Olness,
and especially J.~Pumplin and
R.~Thorne whom we thank for raising issues related to the proper
learning of neural networks.
This work was partly supported by a Spanish 
MEC FIS2007-60350 grant
and by the European network HEPTOOLS under contract
MRTN-CT-2006-035505. L.D.D. is funded by an STFC Advanced Fellowship and
M.U. by a SUPA graduate studentship.
We would like to acknowledge the use of the computing resources provided 
by the Black Forest Grid Initiative in Freiburg and by the Edinburgh Compute 
and Data Facility (ECDF) (http://www.ecdf.ed.ac.uk/). The ECDF is partially 
supported by the eDIKT initiative (http://www.edikt.org.uk).

%------------------------------------------------------

\appendix

% -----------------------------------------------

\section{Distances between PDFs: definition and meaning}
\label{sec:distances}

Given two sets of $N^{(1)}_{\mathrm{rep}}$ and $N^{(2)}_{\mathrm{rep}}$
replicas, one is often interested in knowing whether they correspond
to different instances of the same underlying probability distribution,
or whether instead they come from different underlying
distributions. Of course, for finite $N^{(i)}_{\mathrm{rep}}$ this
question can only be answered in a statistical sense.
To this purpose, we define the square distance between two estimators based
on the given samples as the square difference  between the estimators
divided by its expectation value, i.e. divided by the corresponding
standard deviation. By construction, the expectation value of the
distance is one.

The following cases are of particular interest:

\begin{itemize}
  
\item {\bf Expected value}\\\noindent 

Given a set of $N^{(k)}_{\mathrm{rep}}$ 
  replicas $q^{(k)}_i$ of some
  quantity $q$, the estimator for the expected (true) value of $q$ is
  the mean
  \begin{equation}
\label{eq:expval}
    \langle
    q^{(k)}\rangle_{(i)}=\frac{1}{N^{(i)}_{\mathrm{rep}}}\sum_{i=1}^{N^{(i)}_{\mathrm{rep}}}
    q^{(k)}_i.
  \end{equation}
  The square 
  distance between the two estimates of the expected value obtained from sets
  $q^{(1)}_i$, $q^{(2)}_i$  
  is then
  \begin{equation}
    \label{eq:d2}
    d^2\left(\langle q^{(1)}\rangle ,\langle q^{(2)}\rangle\right)=
    \frac{\left(\langle q^{(1)}\rangle_{(1)} - \langle q^{(2)}\rangle_{(2)}\right)^2}
    {\sigma^2_{(1)}[\langle q^{(1)}\rangle] +
      \sigma^2_{(2)}[\langle q^{(2)}\rangle]}
  \end{equation}
  where the variance of the mean is given by
  \begin{equation}
    \label{eq:stdmean}
    \sigma^2_{(i)}[\langle q^{(i)}\rangle]=
    \frac{1}{N^{(i)}_{\mathrm{rep}}} \sigma^2_{(i)}[q^{(i)}]
  \end{equation}
  in terms of the variance  $\sigma^2_{(i)}[q^{(i)} ]$ of the variables
  $q^{(i)}$ (which a priori could come from two distinct probability
  distributions). We
  estimate the variance of the mean from the variance of the replica sample as
  \begin{equation}
    \label{eq:std}
    \sigma^2_{(i)}[
    q^{(i)}]=\frac{1}{N^{(i)}_{\mathrm{rep}}-1}
    \sum_{k=1}^{N^{(i)}_{\mathrm{rep}}} \left(q^{(i)}_k-\langle q^{(i)}\rangle\right)^2,
  \end{equation}
  with $\langle q^{(i)}\rangle$ given by Eq.~(\ref{eq:expval}).

\item {\bf Uncertainty}\\\noindent 

Given a set of $N^{(k)}_{\mathrm{rep}}$ 
  replicas $q^{(k)}_i$ of some
  quantity $q$, the estimator for the square uncertainty of $q$ is
  the variance of the replica sample given by Eq.~(\ref{eq:std}). The
  distance between the two estimates of the square uncertainty 
  obtained from sets
  $q^{(1)}_i$, $q^{(2)}_i$  
  is then
  \begin{equation}
    \label{eq:d2_pdf}
    d^2(\sigma^2_{(1)},\sigma^2_{(2)})=
    \frac{\left(\bar\sigma^2_{(1)} -\bar\sigma^2_{(2)}\right)^2}
    {\sigma^2_{(1)}[\bar\sigma^2_{(1)}] + \sigma^2_{(2)}[\bar\sigma^2_{(2)}]}
  \end{equation}
  where for brevity we have defined
  \begin{equation}\label{eq:sigbardef}
    \bar\sigma^2_{(i)}\equiv\sigma^2_{(i)}[q^{(i)}]. 
  \end{equation}
  The 
  variances  $\sigma^2_{(i)}[\bar\sigma^2_{(i)}]$ of the square
  uncertainties could also be estimated from the replica sample, by
  computing the variance from various subsets of  the given replica
  sample, and then  the variance of these  resulting variances as the
  subset is varied; for finite number of replicas 
  this may lead to loss of statistical accuracy. For simplicity here we
  use instead the expression~\cite{Amsler:2008zzb}
  \begin{equation}
    \label{eq:ases}
    \sigma^2_{(i)}[\bar\sigma^2_{(i)}]= 
    \frac{1}{N_{\mathrm{rep}}^{(i)}}\left[m_4[q^{(i)}]
      -\frac{N_{\mathrm{rep}}^{(i)}-3}{N_{\mathrm{rep}}^{(i)}-1}
      \left(\bar\sigma^2_{(i)}\right)^2\right], 
  \end{equation}
  where as above $\bar\sigma^2_{(i)}$ is estimated using Eq.~(\ref{eq:std}),
  while the fourth moment $m_4$ of the probability distribution is
  estimated from the corresponding moment of the replica sample (which
  provides an estimate of it which is only asymptotically unbiased):
  \begin{equation}\label{eq:mfour}
    m_4[q^{(i)}]=\frac{1}{N^{(i)}_{\mathrm{rep}}}
    \sum_{k=1}^{N^{(i)}_{\mathrm{rep}}} \left(q^{(i)}_k-\langle q^{(i)}\rangle\right)^4.
  \end{equation}
\end{itemize}

In practice, for small--sized replica samples
the distances defined in Eq.~(\ref{eq:d2}) and Eq.~(\ref{eq:d2_pdf}) display
sizable statistical fluctuations. In order to stabilize the result, all
distances computed in this paper are determined as follows: we 
randomly pick $N^{(i)}_{\mathrm{rep}}/2$ out of the
$N^{(i)}_{\mathrm{rep}}$ replicas for each of the two subsets. The
computation of the square distance Eq.~(\ref{eq:d2}) or Eq.~(\ref{eq:d2_pdf})
is then repeated for 
$N_\mathrm{part}=100$ (randomly
generated)  choices of $N^{(i)}_{\mathrm{rep}}/2$ replicas, and the
result is averaged: this is sufficient to bring the statistical fluctuations of
the distance at the level of a few percent.
 The distances shown in Sect.~\ref{sec:results} are
the square root of this average, computed taking for $q^{(i)}$ the
value of some PDF at fixed $x$ and $Q^2$ obtained from a given pair of fits.
Through Sect.~\ref{sec:results} the choice $Q^2=Q^2_0=2$ GeV$^2$
is always adopted.

The distance defined in this way measures whether the given samples do
or do not come from the same underlying probability distribution, and
in particular Eq.~(\ref{eq:d2}) and Eq.~(\ref{eq:d2_pdf}) test
whether the two distributions from which the two samples are taken
have respectively the same mean and the same standard deviation. By
construction, the probability distribution for the distance coincides
with the $\chi^2$ distribution with one degree of freedom, and thus it
has mean $\langle d\rangle= 1$, and $d\lsim 2.3$ at 90\% confidence
level. 

Note that asking whether two PDF determinations come from
the same underlying distribution is much more restrictive than asking  
whether they are consistent within
uncertainties. Consider for instance the case of a pair of PDF
determinations, such that the dataset on which one of the two is based
is a subset of the dataset of the other, and such that all data are
consistent with each other. These two determinations will clearly not
come from the same underlying distribution, because the distribution
of PDFs obtained from the wider dataset will have smaller
uncertainty. However, if the data are consistent they will remain
nevertheless consistent within uncertainties. 

In particular, 
the determination of moments of the
underlying distribution becomes more precise as  as the
number of replicas is increased: e.g. the accuracy in determination of
the expectation value scales as $1/\sqrt{N_{\mathrm{rep}}}$,
compare Eq.~(\ref{eq:stdmean}),  so if the underlying probability
distributions are different the distance will grow as
$\sqrt{N_{\mathrm{rep}}}$ in the large $N_{\mathrm{rep}}$
limit. In this limit (in which the central values of the underlying
distribution  are accurately
estimated by mean over the replica sample) the distance between
central values is given by the distance rescaled by $\sqrt{N_{\mathrm{rep}}}$: 
otherwise stated, if $N^{(1)}_{\mathrm{rep}}=N^{(2)}_{\mathrm{rep}}=N_{\mathrm{rep}}$, 
then
\begin{equation}
  \label{eq:deltadef}
  \delta(\sigma^2_{(1)},\sigma^2_{(2)})\equiv 
  \frac{1}{\sqrt{N_{\mathrm{rep}}}}
  d(\sigma^2_{(1)},\sigma^2_{(2)})
\end{equation}
provides (in the large $N_{\mathrm{rep}}$) limit, the difference
between central values in units of the standard deviation. It follows
that because of 
the halving of  the size of the sample required for averaging as
discussed above, for all distances shown in Sect.~\ref{sec:results},
and computed with $N_{\mathrm{rep}}=100$ replicas, one sigma
corresponds to $d=\sqrt{50}\approx7$.

% ------------------------------------------------------

%----------------------------------------

\section{Drell--Yan observables}
\label{sec:dyobservables}

We provide here the full  expressions for Drell-Yan observables 
included into the NNPDF2.0 analysis (both virtual photon and vector
boson production).
We adopt the notations and conventions of
Refs.~\cite{Gehrmann:1997ez,Gehrmann:1997pi}.  
For explicit expression of the inclusive jet cross--sections, we can
refer  to the documentation of the FastNLO project from which we took
the precomputed tables~\cite{Kluge:2006xs}. 

\subsection{Rapidity and $x_F$ distributions}

The leading order parton kinematics was given in Eq.~(\ref{eq:ydef}).
The rapidity distribution for the DY process can be then expressed at NLO as
\begin{eqnarray}
\label{eq:dyrap}
\frac{\d   \sigma}{\d M^2 \d y}(M^2,y) & = & \frac{4\pi \alpha^2}{9 M^2 s} 
\sum_i e_i^2 \int_{x_1^0}^1 \d x_1 \int_{x_2^0}^1 \d x_2 \nonumber \\
& & \hspace{-3.2cm}
\times \Bigg\{\left[D_{q\bar{q}}^{(0)}(x_1,x_2) +\frac{\alpha_s}{4\pi}
D_{q\bar{q}}^{(1)}\left(x_1,x_2,\frac{M^2}{\mu_F^2}\right)\right]
\Big\{   q_i(x_1,\mu_F^2) \bar{q}_i(x_2,\mu_F^2) 
%\nonumber \\
%& & \hspace{0.1cm}
+ \bar{q}_i(x_1,\mu_F^2)  q_i(x_2,\mu_F^2) \Big\} \nonumber\\
& & \hspace{-2.2cm} + \Bigg[ \frac{\alpha_s}{4\pi}
 D_{g\bar{q}}^{(1)} \left(x_1,x_2,
\frac{M^2}{\mu_F^2}\right)   g(x_1,\mu_F^2) \left\{ 
  q_i(x_2,\mu_F^2) + 
  \bar{q}_i (x_2,\mu_F^2) \right\} 
%\nonumber \\
%& & \hspace{0.3cm}
+ (1 \leftrightarrow 2)\Bigg] \Bigg\} \ .
\label{eq:ymaster}
\end{eqnarray}

The LO coefficient functions for this distribution are given  by
\be
\label{eq:coeffdy1}
D_{q\bar{q}}^{(0)} (x_1,x_2) = \delta (x_1-x_1^0)\, 
\delta (x_2-x_2^0);
\ee
the NLO contribution is explicitly given in Ref.~\cite{Gehrmann:1997ez}.

For their practical implementation we exploited the following standard 
identities:
\bea
{\rm Li}_2(x) &=&-\int_0^x dt \frac{\ln (1-t)}{t}\\
\int_x^1 dt \frac{f(t)}{(t-x)_+} &=& \int_x^1 dt \frac{f(t)-f(x)}{t-x}\\
\int_x^1 dt f(t) \lc \frac{\ln (1-x/t)}{t-x}\rc_+ &=&
\int_x^1 dt \lp f(t) - f(x) \rp \lc \frac{\ln (1-x/t)}{t-x}\rc\\
\int_{x_1}^1 dt_1 \int_{x_2}^1 dt_2 \frac{f(t_1,t_2)}{
\lc (t_1-x_1)(t_2-x_2)\rc_+} &=&\nn\\
&&\hspace{-3.0cm}\int_{x_1}^1 dt_1 \int_{x_2}^1 dt_2 
\frac{f(t_1,t_2) - f(t_1,x_2) 
-f(x_1,t_2)+ f(x_1,x_2)}{
\lc (t_1-x_1)(t_2-x_2)\rc}
\eea

%\clearpage
%{\bf\noindent Drell-Yan $x_F$ distributions}

Distributions in terms of Feynman  $x_F$ are also frequently used:
the leading order parton kinematics was given in
Eq.~(\ref{eq:xfdef2}).
The Drell--Yan $x_F$ distribution of lepton pairs  at NLO is given by
\begin{eqnarray}
\label{eq:dyxf}
\frac{\d^2 \sigma}{\d M^2 \d x_F} & = & \frac{4\pi \alpha^2}{9 M^2 s} 
\sum_{i} e_i^2 \int_{x_1^0}^1 \d x_1 \int_{x_2^0}^1 \d x_2 \nonumber \\
& & \hspace{-2.7cm}
\times \Bigg\{\left[ \widetilde{D}^{(0)}_{q\bar{q}}(x_1,x_2)
 +\frac{\alpha_s}{4\pi}
\widetilde{D}^{(1)}_{q\bar{q}}\left(x_1,x_2,\frac{M^2}{\mu_F^2}\right)\right]
\Big\{   q_i(x_1,\mu_F^2) \bar{q}_i(x_2,\mu_F^2) 
%\nonumber\\
%& & \hspace{0.1cm}
+ \bar{q}_i(x_1,\mu_F^2)  q_i(x_2,\mu_F^2) \Big\} \nonumber \\
& & \hspace{-2.2cm} + \Bigg[ \frac{\alpha_s}{4\pi}
\widetilde{D}^{(1)}_{g\bar{q}} \left(x_1,x_2,
\frac{M^2}{\mu_F^2}\right)   g(x_1,\mu_F^2) \left\{ 
  q_i(x_2,\mu_F^2) + 
  \bar{q}_i (x_2,\mu_F^2) \right\} 
%\nonumber \\
%& & \hspace{0.3cm}
+ (1 \leftrightarrow 2)\Bigg] \Bigg\},
\label{eq:xfmaster}
\end{eqnarray}
where the sum over $i$ runs over all $N_f$ quark flavours.

The LO coefficient function is given by
\be
\widetilde{D}^{(0)}_{q\bar{q}}\left(x_1,x_2\right)
= \frac{\delta (x_1-x_1^0)\, 
\delta (x_2-x_2^0)}{x_1^0+x_2^0}.
\ee
The NLO contribution coming from $q\bar{q}$ annihilation is explicitly given in 
Ref.~\cite{Gehrmann:1997pi}. 

\subsection{Vector boson production}

For vector boson production at hadron colliders, the cross section is
differential in a single variable $y$, the rapidity of the vector
boson.
The unpolarized vector boson production  cross sections at NLO
is
\begin{eqnarray}
\label{eq:wprod}
\frac{\d \sigma}{\d y} & = &  \frac{\pi G_F M_V^2\sqrt{2}}{3 s} \sum_{i,j} c_{ij} 
\int_{x_1^0}^1 \d x_1 \int_{x_2^0}^1 \d x_2 \nonumber \\
& &  \hspace{-1.6cm}
\times \Bigg\{\left[ 
D_{q\bar q}^{(0)} (x_1,x_2,\xaa,\xbb)  +\frac{\alpha_s}{4\pi}
D_{q\bar{q}}^{(1)}
\left(x_1,x_2,\xaa,\xbb,\frac{M^2}{\mu_F^2}\right)\right]\nonumber \\
& & \hspace{-0.4cm} \times
\Big\{ q_i(x_1,\mu_F^2) \bar{q}_j(x_2,\mu_F^2) +  
\bar{q}_i(x_1,\mu_F^2) q_j(x_2,\mu_F^2) \Big\} \nonumber \\
& & \hspace{-1.2cm} +  \frac{\alpha_s}{4\pi}
D_{gq}^{(1)} \left(x_1,x_2,\xaa,\xbb, 
\frac{M^2}{\mu_F^2}\right) g(x_1,\mu_F^2) \left\{ 
q_j(x_2,\mu_F^2) + 
\bar{q}_j (x_2,\mu_F^2) \right\} \nonumber \\
& & \hspace{-1.2cm} + \frac{\alpha_s}{4\pi}
D_{qg}^{(1)} \left(x_1,x_2,\xaa,\xbb, 
\frac{M^2}{\mu_F^2}\right) \left\{ 
q_i(x_1,\mu_F^2) + 
\bar{q}_i (x_1,\mu_F^2) \right\} g(x_2,\mu_F^2) \Bigg\}\; ,
\end{eqnarray}
where $c_{ij}$ are the electroweak couplings defined in Eq.~(\ref{eq:ewcoup})
The coefficient functions in Eq.~(\ref{eq:wprod}) are identical to
those in the Drell-Yan rapidity distribution Eq.~(\ref{eq:ymaster}).

Note that for proton--antiproton collisions (such as at the Tevatron)
one of the two parton distributions refers to a proton and the other
to an antiproton, i.e. in practice one should replace
$ q_i(x_2)\to  \bar q_i(x_2)$ and conversely in the above expression.
Similarly for proton-nucleus collisions, where isospin symmetry
of the nucleus target should be taken into account.

\bibliography{nnpdf20}

\end{document}